%% file: main.tex
\pdfoutput=1 
\documentclass[11pt,phd,a4paper,oneside]{ucl_thesis}

\input{mainpackages}

\usepackage{booktabs}
\usepackage{xcolor}
\usepackage{subcaption}
\usepackage{algorithm}
\usepackage{algorithmicx}
\usepackage{algpseudocode}
\usepackage{afterpage}
\usepackage{cite}
\usepackage{tabularx}
\usepackage{longtable}
\usepackage{rotating}
\usepackage{multirow}
\usepackage[british]{babel}

\DeclareMathOperator*{\argmax}{arg\,max}

\MakeRobust{\Call}

\newcommand{\twodots}{\mathinner {\ldotp \ldotp}}

\input{linksandmetadata}


\input{variables}
\title{Reinforcement Learning and Tree Search Methods for the Unit Commitment Problem}
\date{August 2022}
\author{Patrick de Mars}

\begin{document}

\include{preamble}



\chapter{Introduction} \label{introduction}
\input{01-introduction/introduction}

\chapter{Literature Review} \label{literature}
\input{02-literature/literature}

\chapter{Methodology} \label{methodology}
\input{03-methodology/methodology}

\chapter{Guided Tree Search} \label{ch1}
\input{04-chapter1/chapter1}

\chapter{Informed and Anytime Search} \label{ch2}
\input{05-chapter2/chapter2}

\chapter{Case Studies: Curtailment and Outages} \label{ch3}
\input{06-chapter3/chapter3}

\chapter{Conclusion} \label{conclusion}
\input{07-conclusion/conclusion}

\bibliography{references.bib}

\end{document}

%% file: mainpackages.tex
\usepackage{blindtext}

\usepackage{emptypage}

\usepackage{graphicx}

\usepackage{float}

\usepackage{amsmath}

\usepackage{gensymb}
\usepackage{textcomp}

\usepackage{setspace}
\usepackage{multirow}

\usepackage{bibentry} 

\usepackage[format=hang,font=small,labelfont=bf]{caption}

\usepackage{etoolbox}

\usepackage{microtype}

%% file: linksandmetadata.tex
\usepackage{bibentry}
\makeatletter\let\saved@bibitem\@bibitem\makeatother
\usepackage[pdftex,hidelinks]{hyperref}
\makeatletter\let\@bibitem\saved@bibitem\makeatother
\makeatletter
\AtBeginDocument{
    \hypersetup{
        pdfsubject={Reinforcement Learning for Unit Commitment},
        pdfkeywords={reinforcement learning, tree search, unit commitment},
        pdfauthor={Patrick de Mars},
        pdftitle={Reinforcement Learning and Tree Search Methods for the Unit Commitment Problem},
    }
}
\makeatother

%% file: preamble.tex

\title{Reinforcement Learning and Tree Search Methods for the Unit Commitment Problem}
\author{Patrick de Mars}
\department{Bartlett School of Environment, Energy and Resources}

\maketitle
\makedeclaration

\begin{abstract} 

The unit commitment (UC) problem, which determines operating schedules of generation units to meet demand, is a fundamental task in power systems operation. Existing UC methods using mixed-integer linear programming are not well-suited to highly stochastic systems. Approaches which more rigorously account for uncertainty could yield large reductions in operating costs by reducing spinning reserve requirements; operating power stations at higher efficiencies; and integrating greater volumes of variable renewables. A promising approach to solving the UC problem is reinforcement learning (RL), a methodology for optimal decision-making which has been used to conquer long-standing grand challenges in artificial intelligence. This thesis explores the application of RL to the UC problem and addresses challenges including robustness under uncertainty; generalisability across multiple problem instances; and scaling to larger power systems than previously studied. To tackle these issues, we develop \emph{guided tree search}, a novel methodology combining model-free RL and model-based planning. The UC problem is formalised as a Markov decision process and we develop an open-source environment based on real data from Great Britain’s power system to train RL agents. In problems of up to 100 generators, guided tree search is shown to be competitive with deterministic UC methods, reducing operating costs by up to 1.4\%. An advantage of RL is that the framework can be easily extended to incorporate considerations important to power systems operators such as robustness to generator failure, wind curtailment or carbon prices. When generator outages are considered, guided tree search saves over 2\% in operating costs as compared with methods using conventional $N-x$ reserve criteria. The strategies adopted by guided tree search improve our understanding of the problems studied and demonstrate that RL is a rich methodology for solving the UC problem, offering practical value to system operators through more intelligent operation of complex and uncertain power systems.

\end{abstract}

\begin{impactstatement}

    There is active interest from electricity network operators in applications of artificial intelligence methods including reinforcement learning (RL) to improve the stability and efficiency of power systems. These interests are in part motivated by the increasing complexity in power systems deriving from increasing renewables penetration, electrification of end-use sectors and increasing decentralisation of generation, among other trends. This thesis shows that RL methods can be applied to solve the unit commitment (UC) problem, a fundamental task in power systems operation which has received comparatively little attention from RL researchers. The guided tree search methods presented in this thesis show substantial improvements in solution quality as compared with current mathematical optimisation methods, which could benefit system operators through significant operating cost reductions if applied at scale. There are further benefits of our method in terms of system security, shown through experiments studying high levels of uncertainty from wind generation, demand and generator outages. These studies reflect the challenges faced by system operators in current and future power systems, which demand novel solution methods to ensure reliable and cost-effective supply of electricity.
        
    Applied RL research requires problem-specific environments, the development of which is typically time-intensive and requires expert knowledge of the problem domain. The open-source UC environment developed for this research enables researchers from both artificial intelligence and power systems backgrounds to approach the UC problem, accelerating research in this area. This research has significance for RL research more broadly, as it bridges the gap between state-of-the-art methods and practical applications. It is widely recognised that significant challenges remain in developing RL methods that are applicable in the real world; this thesis addresses these problems in the context of power systems. In particular, UC is a challenging problem for existing RL methods due to its very large discrete action space. Guided tree search addresses this issue using a novel method to reduce the solution space, and could be applied in other fields sharing this characteristic where existing RL methods do not succeed such as vehicle routing, employee scheduling or portfolio optimisation. 
    
    With future advances in RL methods and due to the profound importance of UC for the effective operation of power systems, there will undoubtedly be further research conducted in this area. This thesis provides the foundation for future work, showing the value of combining model-free and model-based methods; developing techniques for incorporating domain knowledge; and developing the software to facilitate rapid development of new solution methods.

\end{impactstatement}

\begin{acknowledgements}

    Thank you to Aidan O'Sullivan, who has supported me since we first met during my bachelor's degree in 2016. This thesis would certainly not have been written if not for your backing. I would also like to thank my supervisors Ilkka Keppo and Paul Dodds, whose advice and perspective were valuable in determining the direction of this research. I would also like to thank Andreas Sch{\"a}fer, who has been a great inspiration and mentor.
    
    I am grateful to all those who have supported me since I started in 2017. In particular, I would like to thank Jonno Bourne, Connor Galbraith and Ayrton Bourn, who were brilliant colleagues at UCL and from whom I have learned so much. Thank you to my great friend Josh; you helped me see the bigger picture and patiently helped me through my more intricate problems over coffee at Fork. Thank you to all my family and friends who kept my feet on the ground. And finally, special thanks to Ellen, without whom this would have been so much harder.
    
    Above all, this thesis is dedicated to my dad, who was consistently encouraging of my PhD but sadly did not get to see me finish.

\end{acknowledgements}

\setcounter{tocdepth}{2} 

\tableofcontents
\listoffigures
\listoftables

%% file: 01-introduction/introduction.tex
\section{Background} \label{introduction:background}

Electricity is vital to modern societies, driving essential sectors of the global economy including communication, healthcare and finance. As countries transition towards net zero energy systems, electricity will play a key role as it can be decarbonised more easily than other energy carriers \cite{davis2018net}, leading to the electrification of further end-use sectors such as heat and transport \cite{luderer2022impact}. The provision of secure, cost-effective and low-carbon electricity is therefore increasingly important for the prosperity of societies and the mitigation of climate change \cite{williams2012technology}.

Power systems are the physical infrastructure connecting electricity generation with demand. A fundamental task for operators of the system, who must ensure the equal balance of supply and demand, is unit commitment (UC), determining the on/off schedules of a fleet of generators for a future period \cite{wood2013power}. Improving UC solutions can have a significant impact on system operating costs by operating thermal generators at higher efficiencies, reducing requirements of spinning reserve and integrating larger volumes of renewable energy generation. The introduction of the current state-of-the-art in mixed-integer linear programming (MILP) methods is estimated to have saved over \$1 billion annually in total operating costs in North America, through more efficient scheduling of generation \cite{o2017computational}. In its simplest form and assuming perfect information, a UC problem with $N$ generators and $T$ discrete commitment periods can be described as the following cost minimisation problem:

\begin{gather}
  \min \sum_{t=1}^{T} \sum_{i=1}^{N} C_i(t)
  \label{intro:eq:duc_objective}
\end{gather}

subject to

\begin{align}
  & \sum_{i=1}^{N} p_i (t) = D(t) \quad & \forall t \in \{1 \twodots T\} \label{intro:eq:duc_load_balance} \\
  & \boldsymbol{p}_i \in \Pi_i \quad & \forall i \in \{1 \twodots N\}\label{intro:eq:duc_operating constraint} 
\end{align} 

where $C_i(t)$ is the operating cost of generator $i$ at time $t$; $p_i(t)$ is the power output of unit $i$ at time $t$; $D(t)$ is the demand at time $t$; and $\Pi_i$ is the region of operating levels that obey generator operating constraints. The optimisation therefore requires that supply of electricity is balanced with demand in Equation \ref{intro:eq:duc_load_balance}, while respecting constraints on generator operation in Equation \ref{intro:eq:duc_operating constraint}. 

The complexity of the UC problem stems from non-linear and non-convex generator cost functions $C_i(t)$ and inter-temporal generator constraints such as minimum up/down times, which make the UC problem NP-hard \cite{bendotti2019complexity}. Furthermore, while the optimisation problem in Equations \ref{intro:eq:duc_objective}--\ref{intro:eq:duc_operating constraint} assumes perfect foresight, in practice UC decisions must be made based on forecasts of demand, renewables generation and other power system variables. Due to the long time horizons - typically several hours or days - over which the UC problem is typically solved \cite{knueven2020mixed}, forecasts used to inform UC decisions carry high levels of uncertainty. While the UC problem is primarily concerned with integer decision variables determining the on/off status of generators, most approaches to the UC problem simultaneously solve for the real-valued generator power outputs. This latter problem of determining the optimal dispatch of generators given a commitment decision is known as the economic dispatch (ED) problem, and is closely related to the UC problem. The ED problem is an essential task for short-term operational decision making such as real-time balancing or re-dispatch of generators in response to deviations in demand from forecasts. While the commitment of generators in Great Britain (GB) and other regions with similar market structures is not determined by a central operator, the cost-minimising UC problem given in Equations \ref{intro:eq:duc_objective}--\ref{intro:eq:duc_operating constraint} remains the most widely-studied, archetypal formulation, and is studied throughout this thesis.

Methods for formulating and solving the UC problem have been studied for decades \cite{padhy2004unit} and the current industry practice is to use deterministic MILP formulations \cite{bertsimas2012adaptive}. The UC problem must typically be solved within minutes in order to exploit the most up-to-date forecasts and allow sufficient time for the system operator to conduct security analyses \cite{knueven2020mixed, chen2016improving}. As power systems transition towards net zero carbon emissions, growing volumes of variable renewable generation introduce additional uncertainty to the UC problem \cite{kiviluoma2011impact, holttinen2011impacts}. This has motivated research into new solution methods which more rigorously account for uncertainty while remaining computationally tractable in short computing times \cite{papavasiliou2014applying, bertsimas2012adaptive, tuohy2009unit, barth2006stochastic, bouffard2008stochastic}. Deterministic formulations are not a natural framework for handling uncertainty, relying on reserve capacity that is determined by heuristic methods such as the widely-used $N-1$ criterion protecting against the loss of the single largest infeed on the network. Research has shown that deterministic approaches give variable levels of system security \cite{aminifar2009unit} and probabilistic treatments of the UC problem using stochastic optimisation techniques are better-suited to systems with high levels of uncertainty, achieving lower operating costs and more consistent system reliability \cite{tuohy2009unit, ruiz2009uncertainty, bertsimas2012adaptive}. However, much larger computational requirements have prevented practical applications of stochastic UC \cite{bertsimas2012adaptive, papavasiliou2014applying}. 

The results of studies using stochastic optimisation to solve the UC problem have shown that deterministic UC methods are sub-optimal, motivating research into alternative solution methods. This thesis is focused on the application of artificial intelligence (AI) methods to the UC problem. Driven by a growing abundance of computational resources, advanced optimisation algorithms and widespread digitalisation, AI algorithms are the state of the art in challenging domains such as image recognition \cite{dai2021coatnet}, machine translation \cite{vaswani2017attention} and protein folding \cite{jumper2021highly}. The sub-field of reinforcement learning (RL) is a framework for goal-directed decision-making under uncertainty which has surpassed the performance of existing methods for a number of complex control tasks including games-playing and robotics \cite{mnih2013playing, silver2018general, schrittwieser2020mastering}. In RL, a decision-making agent learns by trial-and-error to maximise a numerical performance signal called the reward. Through repeated interactions with an environment representing the problem domain, the agent learns to improve its operational strategy, known as a policy \cite{sutton1998introduction}. 

Although RL has been recognised as a promising framework for solving the UC problem due to its ability to learn optimal policies in complex, uncertain environments \cite{glavic2017reinforcement, rolnick2019tackling, perera2021applications}, RL methods make up only a small fraction of the existing UC literature. An RL solution to the UC problem offers the possibility of achieving the solution quality of stochastic optimisation methods with a fraction of the computational cost at decision time by off-loading computation to a training period. Furthermore, combining deep learning with RL (known as deep RL) offers a powerful and flexible framework to improve decision-making by automatically extracting relevant features from large amounts of data such as weather forecasts, market trends and smart meter readings. Much of the existing research in RL for UC was conducted prior to significant breakthroughs in RL including deep RL \cite{mnih2013playing}, state-of-the-art model-based algorithms \cite{silver2018general} and several policy optimisation techniques which have been shown to significantly improve training efficiency \cite{van2016deep, schulman2015trust, schulman2017proximal, haarnoja2018soft}. It is worthwhile revisiting the UC problem in light of recent advances in RL.

Currently, the dominant RL methods are model-free \cite{schrittwieser2020mastering}, where the optimal policy is estimated by interactions with the environment alone. Such methods have achieved state-of-the-art performance in several challenging sequential decision-making problems \cite{mnih2013playing, levine2016end, lillicrap2015continuous}, and are capable of rapidly producing high quality solutions. However, the combinatorial nature of UC decisions means that the action space contains $2^N$ unique actions for a system of $N$ generators, posing a formidable challenge for existing model-free methods \cite{dulac2015deep}. Other characteristics of the UC problem such as a high-dimensional state space, extreme penalties for lost load (blackouts) and long time dependencies make UC an extremely difficult problem for model-free RL. For these reasons, existing model-free RL research has been limited to UC problem instances in small power systems \cite{jasmin2009reinforcement, jasmin2016function, navin2019fuzzy, li2019distributed}. 

Model-based RL, which combines planning methods such as tree search with experience-driven learning, is a promising approach to solving the UC problem that has not received attention in the existing literature. Such methods combine the generalisability properties of model-free RL, training by trial-and-error in a simulated environment, with precise lookahead capabilities of tree search and are the state-of-the-art in challenging games-playing domains \cite{silver2016mastering, silver2018general, schrittwieser2020mastering}. Notably, the model-based RL algorithm AlphaGo was used to beat the number one ranked player in the board game Go, a long-standing grand challenge of AI \cite{silver2016mastering}. In the context of the UC problem, the ability to anticipate contingencies using model-based lookahead strategies is highly advantageous due to the crucial requirement of maintaining high levels of system security. Model-based methods can offer greater robustness in real-world contexts, and can also offer greater levels of explainability \cite{dulac2019challenges}. However, existing model-based RL methods are not well-suited to the UC problem, and new solution methods are required. 

In this thesis we develop an RL framework for solving the UC problem that combines model-free and model-based methods. This research addresses several practical challenges that have limited previous applications of RL to the UC problem and demonstrates the superior solution quality of RL methods over conventional deterministic optimisation approaches using MILP. In Chapter \ref{ch1}, we tackle the scaling issues that have prevented the application of RL to larger UC problem instances by incorporating an RL-trained policy into traditional planning methods in \emph{guided tree search}. This significantly improves the run time complexity with negligible impact on solution quality and enables the application of guided tree search to problem instances of 100 generators in Chapter \ref{ch2}. The variable run time of planning methods across problem instances is tackled by employing anytime methods, enabling solutions to be iteratively improved within a time budget. We address issues of trust in RL in Chapter \ref{ch3} by examining more complex UC problem instances involving power generation outages. We show that RL provides greater levels of system security than deterministic mathematical programming methods and can provide system operators with a tool to develop novel and robust strategies for dispatching generators. This research makes significant contributions in the field of RL for power systems and provides a foundation for further research in this area. In the next section, we will summarise the key contributions of this thesis.


\section{Contributions}

The contributions of this thesis are summarised under five themes: 

\begin{enumerate}
    \item Power System Environment for the UC Problem
    \item Guided Tree Search
    \item Informed and Anytime Tree Search Methods
    \item Large-Scale Applications
    \item Adaptability of Guided Tree Search
\end{enumerate}

\subsubsection{Contribution 1: Power System Environment for the UC Problem}

As we discuss in Section \ref{literature:duc:benchmarks} and Section \ref{literature:suc:benchmarks}, research into the UC problem has been limited by a lack of benchmark problem instances, which limits the extent to which solution methods can be robustly compared with one another. In addition, a suitable open-source simulation environment for RL research into the UC problem is not available in the existing literature. A significant contribution of this thesis is the development of an open-source power systems simulation environment and deterministic UC benchmark solutions. The simulator enables RL methods to be applied to the UC problem and enables the UC problem to be studied by the wider RL research community. The development of accessible environments has led to significant research outputs in RL applied in other energy domains, including building control \cite{pinto2021coordinated} using the CityLearn environment \cite{vazquez2019citylearn}; and electricity network operation \cite{yoon2020winning} using the Grid2Op environment \cite{marot2020learning}. These environments are valuable not only in enabling the application of RL in energy domains, but also for studying the effectiveness of RL methods more generally and their ability to tackle complex real-world problems, in addition to widely-studied AI problems such as games-playing. 

The RL environment developed in this thesis can be used to unify the evaluation of methods from across the optimisation literature reviewed in Chapter \ref{literature} as well as RL methods presented in this thesis. We show in Section \ref{ch1:env:benchmarks} how UC solutions can be evaluated in terms of expected costs by Monte Carlo simulations with the RL environment, enabling fair comparison of mathematical programming with RL methods. The environment is based on historical wind generation and demand data from the GB power system to create realistic and varied UC problems. This is essential to determining the generalisability of solution methods across problem instances with different characteristics, and our experiments find that some UC problems are more challenging to solve with a given solution method. In Chapter \ref{ch3}, we further develop the environment to add a carbon price, wind curtailment and generator outages. These cover prominent challenges faced by system operators in current and future power systems \cite{brouwer2015operational, burke2011factors, murphy2018resource}. 

\subsubsection{Contribution 2: Guided Tree Search}

To solve the UC problem, we develop guided tree search in Chapter \ref{ch1}, combining model-free RL with model-based planning. While most recent RL research has focused on model-free methods \cite{schrittwieser2020mastering}, the UC problem poses significant challenges for model-free methods due to the large, constrained, combinatorial action space; high-dimensional state space; long time dependencies; and the extreme nature of lost load events. The existing research into RL for the UC problem, reviewed in Section \ref{literature:rl}, has shown promising results, outperforming baseline solutions for small-scale problems \cite{jasmin2016function} but RL has not previously been successfully applied to larger scale problems. We find in Section \ref{ch1:guided_ucs:policy_training} that a model-free RL approach using a policy trained with PPO is not an effective approach to solving the UC problem; the agent is not able to maintain satisfactory levels of system security, resulting in high costs due to lost load. Despite vast improvements in model-free RL methods in recent years \cite{schulman2015trust, schulman2017proximal, mnih2016asynchronous}, applying these methods to the UC problem remains a significant challenge. Our research finds that employing model-based methods significantly improved quality of solution, and is essential to maintaining grid security with uncertain forecasts. 
In order to introduce model-based planning, tree search methods have been used successfully in previous research \cite{silver2016mastering, silver2018general, anthony2017thinking}. Traditional tree search methods without RL have been applied to the UC problem in \cite{dalal2015reinforcement} and shown to outperform benchmark solutions, but also face scaling issues due to the same problems of curses of dimensionality. Our experiments applying uniform-cost search (UCS) to UC problem instances in Section \ref{ch1:experiments:ucs_comparison} corroborate the scaling difficulties of traditional tree search methods, with exponential time complexity in the number of generators. 

While neither model-free RL nor model-based tree search methods alone are successful in solving the UC problem beyond small problem instances, combining both approaches in guided tree search is shown to be a powerful methodology that is competitive with state-of-the-art deterministic approaches. In Section \ref{ch1:experiments:milp_comparison} we show that Guided UCS is capable of outperforming MILP benchmarks for problems of up to 30 generators, significantly larger than studied elsewhere in the literature. Operating costs are found to be between 0.3--0.9\% lower using Guided UCS than deterministic UC formulated using MILP. The improved scaling of Guided UCS and other guided tree search algorithms relative to conventional planning methods applied in \cite{dalal2015reinforcement} is achieved by exploiting an RL-trained policy as a guide to reduce the branching factor of the search tree. We show that run time is held roughly constant with increasing numbers of generators, while operating costs are similar to those produced by exhaustive tree search methods without RL. Guided tree search contributes to the growing literature combining model-free RL with tree search, most notably with AlphaGo \cite{silver2016mastering, silver2018general}, and shows this approach is applicable to a real-world problem in power systems operation. 
 
\subsubsection{Contribution 3: Informed and Anytime Tree Search Methods}

In Chapter \ref{ch2}, we apply more advanced tree search methods, using informed and anytime algorithms. The two novel algorithms developed in this chapter, Guided A* and Guided Iterative-Deepening A* (IDA*), are significantly more effective in solving the UC problem than the general-purpose algorithm, Guided UCS, used in Chapter \ref{ch1}. Using Guided A* search and a novel heuristic function based on priority list solution methods \cite{senjyu2003fast}, run times are reduced by up to 94\% as compared with Guided UCS, with negligible ($<0.1$\%) impact on operating costs. These results demonstrate the value of domain expertise in designing solution methods for UC and other real-world problems. While a large proportion of RL literature has been applied in games-playing domains, applications in real-world contexts has progressed more slowly \cite{dulac2015deep}. Our results show that combining domain expertise with state-of-the-art RL can improve solution methods for specific applications, accelerating the adoption of RL methods for practical benefit. 

The UC problem is typically highly time-constrained, and must usually be solved within minutes \cite{knueven2020mixed}. The variable and unpredictable run times of fixed-depth tree search methods such as UCS and A* therefore pose practical problems for UC. As a result, we develop an anytime algorithm Guided IDA* to mitigate run time variability, constraining the run time to a fixed computational budget. We show that Guided IDA* achieves up to 1\% reduction in operating costs for similar computational budgets as compared with Guided UCS, comparable to the cost savings shown by stochastic optimisation over deterministic methods \cite{ruiz2009uncertainty, tuohy2009unit, bertsimas2012adaptive}. Anytime methods are shown to be particularly well-suited to the UC problem, enabling more reliable generation of high-quality solutions as compared with fixed-depth tree search.

In Chapter \ref{ch1} we show that the exponential time complexity of UCS can be overcome by using Guided UCS, employing RL to reduce the branching factor of the search tree. Chapter \ref{ch2} shows that Guided UCS can be further improved for UC by modifying the algorithm to exploit properties of the problem in Guided A* and Guided IDA*. While many tree search methods are impractical for UC applications, this thesis shows that problem-specific modifications through RL and advanced search methods can enable tree search methods to be successfully applied, producing high quality solutions.

\subsubsection{Contribution 4: Large-Scale Applications}

While many real-world UC problem instances involve small numbers of generators, such as the problems solved by generating companies in self-dispatching power markets, scaling characteristics of UC solution methods are nevertheless important to assess applications to larger problems such as those solved by system operators. Guided IDA* is applied to a problem of 100 generators in Section \ref{ch2:100gen} and found to produce operating costs that are competitive with deterministic MILP approaches (0.1\% lower using Guided IDA*). The only RL-based study of a similar scale studied a 99-generator problem, but made significant simplifications to the problem formulation by preventing intra-day commitment changes, making it an unrealistic point of comparison and limiting solution quality \cite{dalal2016hierarchical}. In addition, no comparison was made with state-of-the-art methods. The 100-generator experiment in Chapter \ref{ch2} is therefore the largest in the existing literature by number of generators, and the first to show that RL is a viable methodology for solving the UC problem at scale. 



\subsubsection{Contribution 5: Adaptability of Guided Tree Search}

A significant advantage of RL over mathematical optimisation methods is the ability to learn fundamentals of the given problem \emph{tabula rasa}. This has been shown by general-purpose games-playing algorithms such as MuZero, which achieved superhuman levels of performance in Go, Chess and Shogi, with no modification of the solution method. To demonstrate the generalisability of guided tree search to different problem instances and variants of the UC problem, in Chapter \ref{ch3} we develop two variations of the power system environment. The first includes curtailment decisions and carbon pricing; the second includes generator outages. We show that RL can be simply applied to solve variants of the UC problem by changing the underlying dynamics of the environment and retraining the agent. By contrast, mathematical programming methods may require significant expertise and development time to develop efficient representations of the problem variant. In some instances, this may include substantive adjustments to the problem itself, such as convexifying or linearising elements of the problem. 

In the first case study of Chapter \ref{ch3}, we introduce a carbon price to the reward function in addition to a wind curtailment action. Carbon pricing is already an important policy mechanism to promote low-carbon electricity \cite{klenert2018making}, and curtailment presents both a challenge and opportunity for system operators to effectively manage large penetrations of variable renewable energy \cite{burke2011factors}. We show that the RL framework enables different operating behaviours to be incentivised by modification of the carbon price. Operating patterns are responsive to changes in the carbon price with gas displacing coal as base-load generation as the carbon price is increased. This shows that using Guided IDA* system operators can dynamically adjust the objective function in response to current system demands, with more general support than is possible using MILP which requires a linear objective function. Guided IDA* also learns to curtail wind generation to manage large swings in demand and wind generation, improving system robustness. These strategies enable the identification of challenging or uncertain decision periods, providing insight into the nature of the UC problem instances studied. By comparison with MILP methods, introducing the additional curtailment action is straightforward, and does not require reformulation of the algorithm.

Outages of generation and other transmission assets can have catastrophic impacts on power systems and large economic consequences; the 2003 North American blackout led to the loss of power for 50 million people with an estimated total cost of \$6 billion \cite{minkel20082003}. The traditional $N-1$ criterion protects system security against the largest loss of infeed, but does not necessarily account for coincident outages, which was the cause of the recent blackout in England impacting over 1 million customers \cite{bialek2020does}. Achieving adequate levels of system security while maintaining lower system operating costs and avoiding high levels of carbon intensive spinning reserve requirements requires UC methods which more robustly account for the joint distribution of outages. The generator outages case study shows that heuristic reserve requirements based on $N-x$ reserve criteria are sub-optimal for the problem instances examined. Guided IDA* learns economic and robust reserve margins during training, taking into account a much larger number of uncertain parameters than in previous experiments without outages. The capacity of Guided IDA* to learn effective reserve margins in this case is indicative of its value in operating increasingly complex power systems with uncertainties stemming from multiple sources. In the context of increasing penetration of geographically-distributed variable renewable energy, large numbers of uncertain parameters are required in order to fully capture effects on the transmission network and the correlated forecast errors. We show in this thesis that RL can easily incorporate deep learning to learn such inter-dependencies and develop suitable strategies for managing this uncertainty. Our results find that irrespective of the problem variant studied, guided tree search is an effective solution method for UC.


    
    
    
    





\section{Thesis Structure} 

This thesis is structured in seven chapters. Chapters \ref{ch1}, \ref{ch2} and \ref{ch3} are results chapters containing the original contributions of this thesis. Chapters \ref{literature}, \ref{methodology} and \ref{conclusion} are Literature Review, Methodology and Conclusion chapters, respectively. 

\medskip

In \textbf{Chapter \ref{literature}}, we conduct a literature review of the state-of-the-art in UC research. We review in detail deterministic methods, which are the most widely-used approaches to solving the UC problem and are used to benchmark the guided tree search methods in later chapters. In addition, we review scenario-based stochastic optimisation and robust optimisation methods for the UC problem, which more rigorously account of uncertainties and have been shown to outperform deterministic methods, albeit at much higher computational cost. Finally, we review existing research using RL and tree search to solve the UC problem. 

\medskip

\textbf{Chapter \ref{methodology}} provides a background to the RL and tree search methods which form the basis of our methodology. We focus particularly on policy gradient RL methods, which are best suited to the UC problem for their ability to handle high-dimensional action spaces and to learn stochastic policies. We describe three tree search algorithms: uniform-cost search, A* search and iterative-deepening, which we apply in the guided tree search framework in Chapter \ref{ch1} and Chapter \ref{ch2}. In addition, we cover mathematical optimisation methods for power systems, which are used to develop the power system simulation environment in Chapter \ref{ch1} and provide benchmark solutions. In particular, we provide a background to MILP and methods for solving the related economic dispatch problem, which is an integral component of the simulation environment.

\medskip

\textbf{Chapter \ref{ch1}} is the first of three results chapters. In this chapter, we present the power systems environment developed for this research, which is used to conduct experiments solving UC problems with MILP and guided tree search methods. We use data from the GB power system to create several years' worth of training problems and a generative model for simulating forecast errors. We formulate the UC problem with stochastic demand and wind generation as a Markov decision process, suitable for RL methods. We then apply the traditional tree search algorithm uniform-cost search (UCS) \cite{dijkstra1959note} to solve small UC problem instances, showing competitive solution quality as compared with MILP approaches but practical limitations due to exponential run time complexity in the number of generators. To enable tree search methods to be applied at scale, we present \emph{guided expansion}; the key innovation of guided tree search. Using guided expansion, an RL-trained policy can be used to reduce the branching factor of a search tree so that it can be solved with traditional methods. We apply this approach in \textbf{Guided UCS}, the first of three guided tree search algorithms, and show that run time remains roughly constant with increasing numbers of generators, while achieving similar operating costs to traditional UCS. Compared with deterministic MILP benchmarks for systems of 10, 20 and 30 generators, Guided UCS consistently achieves lower operating costs. The results of this chapter show that guided tree search is an economic and scalable approach to solving the UC problem.

\medskip

The field of tree search encompasses multiple classes of algorithms with different characteristics which may be better suited to particular problem domains. \textbf{Chapter \ref{ch2}} extends the guided tree search framework to informed search and anytime search algorithms \cite{russellnorvig}, and compares these methods to Guided UCS. Three UC-specific heuristics are developed which are used in the informed search algorithm A* search \cite{hart1968formal} to achieve greater search efficiency. Using the heuristics, \textbf{Guided A*} search is shown to be up to an order of magnitude faster than Guided UCS with no significant impact on solution quality. We then present \textbf{Guided IDA*}, an anytime algorithm which can be terminated when a time budget is spent. The anytime property assures that computational resources are fully exploited, resulting in deeper search depths and improved solution quality in practice. These improvements, culminating in Guided IDA*, enable us to solve UC problems for a larger power system of 100 generators, achieving similar operating costs to deterministic MILP benchmarks. This is the largest simulation study applying RL and/or tree search to solve the UC problem in the existing literature. 

\medskip

In \textbf{Chapter \ref{ch3}}, the power system environment is modified in two case studies that demonstrate the flexibility of the guided tree search framework across heterogeneous power system contexts. In the first case study, we introduce a \textbf{curtailment action} and \textbf{carbon price}, studying the adaptability of Guided IDA* to more heterogeneous actions and the impact of reward shaping. Guided IDA* responds flexibly to changes in the problem environment, adjusting curtailment rates and utilisation of different fuel types to manage carbon costs. In the second case study, we study a system with \textbf{generator outages}, and compare Guided IDA* with MILP using typical $N-x$ security criteria, which are widely-used to handle such contingencies. We show that Guided IDA* adaptively allocates reserves and achieves lower operating costs and better security of supply overall. In both case studies, Guided IDA* exhibits novel operational strategies that uncover properties of the problem itself and demonstrate the value of guided tree search as a decision support tool for system operators.

\medskip

\textbf{Chapter \ref{conclusion}} discusses the results, contributions and limitations of the thesis as a whole and proposes further research topics.

%% file: 02-literature/literature.tex
\section{Introduction}

The UC problem is a large-scale, stochastic, mixed-integer optimisation problem which, even when reformulated as a deterministic problem, is NP-hard \cite{anjos2017unit, bendotti2019complexity}. Due to the large and growing size of power systems globally, there are significant economic and environmental incentives for improving UC solution methods, which has motivated a large body of research. The UC problem has been formulated as a deterministic \cite{carrion2006computationally}, stochastic \cite{takriti1996stochastic} and robust \cite{bertsimas2012adaptive} optimisation problem; formulations that use different approaches to managing uncertainty. Furthermore, a large number of solution methods have been applied, incorporating domain-specific heuristic approaches \cite{johnson1971large}, mathematical optimisation \cite{cohen1983branch, merlin1983new} and metaheuristics \cite{kazarlis1996genetic, snyder1987dynamic}. Facilitated by improvements in commercial solvers such as CPLEX and Gurobi, industry-standard approaches use the deterministic formulation, solved with mixed-integer linear programming (MILP) techniques \cite{carrion2006computationally}. This approach was first adopted by the PJM interconnection in 2005 \cite{ott2010evolution} and is estimated to have been responsible for operating cost savings of more than \$1 billion per year in North American markets compared with previous methods based on Lagrangian relaxation \cite{o2017computational}. However, the deterministic formulation has been shown to yield economically sub-optimal solutions as it is outperformed by stochastic optimisation methods which more rigorously account for uncertainties \cite{tuohy2009unit, ruiz2009uncertainty, bouffard2008stochastic}. Stochastic formulations are substantially more expensive to solve than deterministic ones \cite{papavasiliou2014applying, morales2018robust} and cannot be used in most practical contexts as UC problems must typically be solved within the order of minutes to satisfy market and operational constraints \cite{knueven2020mixed}. As a result, there are significant incentives for further research into novel UC solution methods and formulations that can outperform current deterministic approaches within practical computational budgets.

The purpose of this chapter is to present existing research into the UC problem, covering the principle formulations and solution methods and the current state of the art. The literature review will be organised around the most common UC formulations. Section \ref{literature:duc} describes deterministic formulations of the problem, which are the most widely used in practical applications \cite{bertsimas2012adaptive}. Section \ref{literature:suc} reviews the stochastic UC literature, which more rigorously accounts for uncertainty by minimising expected operating costs over scenarios. Section \ref{literature:ruc} reviews the robust UC literature, which aims to minimise the worst-case operating costs. Finally, Section \ref{literature:rl} reviews the small body of literature that has used reinforcement learning (RL) to solve the UC problem, which is the topic of this thesis.

\section{Deterministic Unit Commitment} \label{literature:duc}

The most widely-used approach to UC formulates the problem deterministically, minimising operating costs under point forecasts of demand and generation \cite{bertsimas2012adaptive}. Uncertainties are managed by enforcing a reserve constraint which is determined separately, either by statistical methods or by heuristics. Based on \cite{carrion2006computationally}, for a power system with $N$ generators and $T$ decision periods the deterministic UC problem can be formulated as:

\begin{gather}
  \min \sum_{i=1}^{N} \sum_{t=1}^{T} C_i(t) 
  \label{literature:eq:duc_objective}
\end{gather}

subject to

\begin{align}
  & \sum_{i=1}^{N} p_i (t) = D(t) \quad & \forall t \in \{1 \twodots T \} \label{literature:eq:duc_load_balance} \\
  & \sum_{i=1}^{N} \bar{p}_i(t) \geq D(t) + R(t) \quad & \forall t \in \{1 \twodots T \} \label{literature:eq:duc_reserve} \\
  & \boldsymbol{p}_i \in \Pi_i \quad & \forall i \in \{1 \twodots N \} \label{literature:eq:duc_operating constraint} 
\end{align} 

Equation \ref{literature:eq:duc_objective} is the objective function, which is to minimise the sum of generator operating costs $C_i(t)$ (including fuel and startup costs) over all periods $t$. Equation \ref{literature:eq:duc_load_balance} gives the load balance constraint, requiring that the sum of generator power outputs $p_i(t)$ equals the forecast demand $D(t)$. Equation \ref{literature:eq:duc_reserve} is the reserve constraint, ensuring that the sum of available capacities $\bar{p}_i$ exceeds the sum of forecast demand $D(t)$ and the reserve constraint $R(t)$. Equation \ref{literature:eq:duc_operating constraint} requires that generators operate within $\Pi_i$, the region of feasible operating levels for generator $i$. The operating constraints include minimum and maximum operating limits as well as inter-temporal constraints such as minimum up/down times, limiting the frequency of startups and shutdowns.

The deterministic UC problem is NP-hard \cite{bendotti2019complexity}, and even finding sub-optimal solutions is a challenging and expensive optimisation problem. The fuel costs included in $C_i(t)$ are usually modelled with quadratic cost curves \cite{wood2013power}, which require linear approximations for many solution methods such as MILP \cite{wu2011tighter}. In addition, startup costs are often modelled as a function of time using a step \cite{kazarlis1996genetic} or exponential function \cite{carrion2006computationally}. The presence of binary variables denoting commitment decisions makes the problem non-convex. Brute force solution methods are not scalable to large power systems \cite{padhy2004unit}, and numerous methods have been proposed including heuristic-based approaches, conventional mathematical optimisation methods and metaheuristics which are discussed in Section \ref{literature:duc:solution_methods}. Performance of different solution methods are evaluated on benchmark problems, defining generator cost functions and operating constraints, as well as demand and reserve profiles. The following subsection describes existing benchmark problems for the deterministic UC problem.

\subsection{Benchmark Problems} \label{literature:duc:benchmarks}

UC research uses test problems to evaluate the performance of solution methods. These problems vary substantially from small-scale systems of $<10$ generators \cite{jasmin2009reinforcement}, to large problems representing regional-scale transmission grids such as the North American mid-continent network \cite{chen2016improving}. Additional features include pumped hydropower \cite{borghetti2008milp, johnson1971large, amjady2013hydrothermal} and consideration of additional problem features such as transmission constraints \cite{pandvzic2013comparison} or ramping constraints \cite{patra2008differential, viana2013new}. Early UC research was generally conducted on proprietary data \cite{johnson1971large, lowery1966generating, merlin1983new}, making it difficult to directly compare solution methods. Benchmark problems have since been developed which have been used more widely across the UC literature \cite{kazarlis1996genetic, ostrowski2011tight, knueven2020mixed}. 

The benchmark problem proposed by Kazarlis et al. \cite{kazarlis1996genetic}, specifying cost functions and constraints for 10 generators and a 24-hour demand profile, has been extensively used in the literature \cite{carrion2006computationally, cheng2000unit, swarup2002unit, senjyu2003fast, ongsakul2004unit, zhao2006improved, ting2006novel, simopoulos2006unit, jeong2010new, singhal2011dynamic, quan2015improved}. The reserves are ƒixed to be 10\% of demand. By duplicating generators, the Kazarlis problem has been scaled to up to 1000 generators \cite{quan2015improved} with proportional scaling of the demand and reserve profiles. It has also been amended in \cite{ostrowski2011tight} which duplicates generators in varying combinations to generate a more diverse set of problem instances. Selected solutions to the 100-generator Kazarlis benchmark are shown in Table \ref{literature:tab:duc_kazarlis}, demonstrating the wide range of methods which have achieved better solution quality (i.e. lower operating costs) and lower run times as compared with the original genetic algorithm (GA) solution \cite{kazarlis1996genetic}. A limitation of the Kazarlis benchmark is that it includes a single 24-hour profile for demand and reserve, making it difficult to evaluate the generalisability of solution methods across problems. Algorithms can be tuned aggressively to achieve very low operating costs on a single problem instance, and it is difficult to compare solution methods by comparing solution quality on the Kazarlis benchmark alone. 

\afterpage{
\input{02-literature/kazarlis_table}
}

Knueven et al. \cite{knueven2020mixed} propose three benchmark systems specifically for comparing deterministic mixed-integer linear programming (MILP) formulations which aim to address this research limitation by creating a large number of diverse problem instances. The systems are based on data from the California ISO (CAISO), Federal Energy Regulatory Commission (FERC) \cite{krall2012rto} and the IEEE RTS-GMLC \cite{barrows2019ieee} test system, each with multiple demand and renewables generation profiles. This benchmark system has been used to systematically compare MILP-based approaches to the UC problem in \cite{nair2020solving, knueven2020mixed, kuttner2021ramping}. While the benchmark problems proposed in \cite{knueven2020mixed} are well-suited to comparing deterministic formulations of the UC problem, the literature lacks more general benchmarks for comparing non-deterministic solution methods on a variety of problem instances. In Section \ref{ch1:env} we address this research gap, introducing a simulation environment that can be used to evaluate solution methods on a diverse set of UC problems. 

\subsection{Solution Methods} \label{literature:duc:solution_methods}

The computational challenges and NP-hardness \cite{bendotti2019complexity} of the UC problem has motivated a large body of research proposing solution methods that achieve good solution quality in short computing times. In this section, we will cover the most widely-studied heuristic, mathematical optimisation and metaheuristic methods for solving the UC problem, which are summarised in Table \ref{literature:tab:duc_methods}. Direct comparison between solution methods is challenging, in part due to the limitations of existing benchmarks that were discussed in Section \ref{literature:duc:benchmarks}. This section will discuss the properties of each method and make side-by-side comparisons based on benchmark problems where appropriate.

\begin{table}[t]\centering
\renewcommand{\arraystretch}{1.5}
\scriptsize
\begin{tabular}{p{0.9in} >{\raggedright}p{0.9in} p{1.5in} p{1.5in}}\toprule
\textbf{Type} &\textbf{Method} &\textbf{Advantages } &\textbf{Disadvantages} \\\midrule
\textbf{Heuristic} &Priority list &Low computational cost and scalable to large power systems &Strong reliance on problem-specific heuristics; typically poor solution quality; no optimality gap \\
\midrule
\multirow{3}{*}{\parbox{0.9in}{\textbf{Mathematical optimisation}}} &Dynamic programming &Guaranteed to converge to an optimal and feasible solution &Exponential time complexity in the number of generators means heuristics are required to achieve tractability \\
&Lagrangian relaxation &Good solution quality; scalable to large power systems; measurable duality gap &Not guaranteed to produce a feasible solution without heuristics \\
&Branch and bound &Excellent solution quality; generally quick to execute using commercial solvers; measurable optimality gap &Can be slow to find a feasible solution under certain conditions or parameters \\
\midrule
\multirow{3}{*}{\parbox{0.9in}{\textbf{Metaheuristic}}} &Genetic algorithms &\multirow{3}{*}{\parbox{1.55in}{Potential for excellent solution quality; can be quick to execute}} &\multirow{3}{*}{\parbox{1.55in}{Generally requires extensive parameter tuning; no convergence guarantee or optimality gap}} \\
&Simulated annealing & & \\
&Particle swarm optimisation & & \\
\bottomrule
\end{tabular}
\caption{Summary of reviewed solution methods for the deterministic UC problem.}
\label{literature:tab:duc_methods}
\end{table}

\subsubsection{Priority List} 

\begin{figure}
    \centering
    \includegraphics[width=\textwidth]{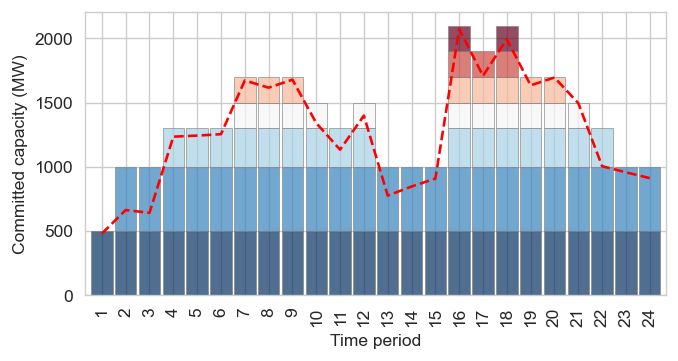}
    \caption{Example priority list unit commitment schedule. Generators are ordered in a priority list of decreasing preference (e.g. by fuel cost) and committed in this order until demand plus a reserve constraint (red line) is met. Inter-temporal constraints such as minimum up/down times are not guaranteed to be met with this algorithm and must typically be fixed using heuristic methods \cite{senjyu2003fast}.}
    \label{literature:fig:priority_list_diagram}
\end{figure}

Priority list (PL) methods are among the most conceptually simple and computationally inexpensive approaches to solving the UC problem and have been used extensively \cite{baldwin1959study, kerr1966unit, johnson1971large, shoults1980practical, sheble1990solution, senjyu2003fast, burns1975optimization, lee1988short, quan2015improved, senjyu2006emerging}. A summary of the main advantages and drawbacks of PL methods is given in Table \ref{literature:tab:duc_methods}. PL methods were among the first computational methods for UC, superseding manual approaches \cite{baldwin1959study, kerr1966unit, johnson1971large, shoults1980practical}. PL methods order generators in terms of capacity, startup costs, or fuel costs \cite{senjyu2003fast, elsayed2017new, quan2015improved}, and for each period commit generators in decreasing order of preference until the reserve constraint in Equation \ref{literature:eq:duc_reserve} is met. Figure \ref{literature:fig:priority_list_diagram} shows an example UC schedule produced by a priority list method. By committing units independently for each period, this simple algorithm neglects inter-temporal constraints such as minimum up/down time constraints, and hence solutions are not guaranteed to be feasible, i.e. satisfying Equation \ref{literature:eq:duc_operating constraint}. Therefore, the initial solutions are typically `fixed' by further heuristics in order to produce a feasible schedule \cite{johnson1971large}. As a result, a large number of PL variants have been proposed combining different heuristics and optimisation techniques to improve the run time and solution quality \cite{senjyu2003fast, quan2015improved, senjyu2006emerging, sheble1990solution}. The most recent PL-based approach in the surveyed literature combines MILP to generate an initial solution and neighbourhood search to fix the inter-temporal constraints \cite{quan2015computational}. This is the best performing solution on the Kazarlis benchmark problem, surveyed in Table \ref{literature:tab:duc_kazarlis}, achieving 0.49\% lower operating costs as compared with the original GA approach with a 2000-fold decrease in run time. The PL method proposed in \cite{senjyu2003fast} also achieves lower operating costs than the original GA solution with a 240-fold speed-up.  

Historically, PL methods replaced manual commitment methods and were reported to have achieved roughly 1\% annual operating cost savings in an industrial-scale application to the Connecticut power system \cite{johnson1971large}. Due to their low computational cost, PL methods are highly scalable to larger power systems, as evidenced by applications to systems of $\geq100$ generators from 1971 \cite{johnson1971large} and 1988 \cite{lee1988short} which are significantly larger than most contemporary research. PL methods rely heavily on problem-specific heuristics to improve solution quality. As a result, while the PL solutions reported in Table \ref{literature:tab:duc_kazarlis} are among the best performing, extensive algorithmic tuning can be used to inflate performance of PL methods for specific problem instances and the results do not indicate generalisability to other power systems or demand profiles. In general, PL methods are considered to have low solution quality relative to more general mathematical optimisation approaches \cite{abujarad2017recent, viana2013new}. In addition, they lack a measurable optimality gap for measuring solution quality (e.g. relative to an estimated lower bound of operating costs) that is offered by mathematical optimisation methods such as branch and bound and Lagrangian relaxation. These reasons have limited practical applications of PL methods since their use in early power systems \cite{baldwin1959study, kerr1966unit, johnson1971large, shoults1980practical}.

\subsubsection{Dynamic Programming} \label{literature:duc:dp}

Dynamic programming (DP) was among the first mathematical optimisation methods employed to solve the UC problem \cite{lowery1966generating, guy1971security, pang1981evaluation, snyder1987dynamic, ouyang1991intelligent, lee1981load, singhal2011dynamic, patra2009fuzzy}. Unlike heuristic-based methods such as PL, traditional DP benefits from guaranteeing an optimal solution. However, the time complexity of DP is generally exponential in the number of generators and DP implementations also typically use heuristics in practice to reduce the solution space and improve scalability. For instance, studying a system of 14 generators, \cite{lowery1966generating} ignores inter-temporal constraints and uses DP to calculate the lowest cost commitment for demand levels between minimum and maximum capacity of the generation mix. This is calculated offline and used to look-up commitment decisions for levels of forecast demand. Another heuristic approach adopted in \cite{snyder1987dynamic} is to group generators based on their properties, and commit groups of generators using DP rather than individual units. Using this method, a problem of 33 generators is reduced to a tractable problem considering 7 subsets. In addition, a common approach only considers states which commit generators in PL order - known as DP sequential combinations (DP-SC) \cite{pang1981evaluation}. In this method, a system of $N$ generators will only have $N+1$ states: all generators off, and the $N$ combinations which consider the first $\{1, 2, \dots , N\}$ generators committed in PL order. Further problem-specific decision rules are used in \cite{ouyang1991intelligent} to achieve tractability for a system of 26 generators. 

Studies which directly compare PL and DP show conflicting results. To the best of our knowledge, DP has not been used to solve the Kazarlis benchmark. On a proprietary test case of 33 generators, \cite{snyder1987dynamic} finds a 0.36\% improvement in operating costs of DP over PL. \cite{sheble1990solution} and \cite{lee1988short} find 2\% and 4\% improvements of PL over DP, respectively. Three DP algorithms are compared with a simple PL algorithm in \cite{pang1981evaluation} for six test problems of up to 17 generators. Savings of between 0.20--0.79\% are found using the DP-SC approach described above as compared with PL, but with 20--50 times higher run times. Due to the strong reliance of both PL and DP on heuristics, the precise implementation of both methods evidently has a significant impact on solution quality. Both PL \cite{johnson1971large} and DP \cite{snyder1987dynamic} were employed in early industrial contexts. 

\subsubsection{Lagrangian Relaxation} \label{literature:duc:lr}

Motivated by the lack of guarantees on solution quality of PL and the potentially large run times of DP, Lagrangain relaxation (LR) methods were developed for the UC problem \cite{muckstadt1977application, merlin1983new, cohen1987method, zhuang1988towards, virmani1989implementation, cheng2000unit, ongsakul2004unit, balci2004scheduling, fu2005security, li2005price} and were widely used by utilities during the 1990s and 2000s \cite{padhy2004unit}. Compared with DP and PL, LR has better solution quality, as measured by operating costs, than PL methods and is more scalable to larger power systems than DP \cite{sheble1990solution}. LR is a mathematical optimisation method for finding approximate solutions to constrained optimisation problems. LR `relaxes' constraints by including them as terms in the objective function, penalised with Lagrange multipliers. This relaxed (`dual') problem is easier to solve than the original (`primal') problem. Iterative methods are used to update the Lagrange multipliers and converge to a solution which is close to the optimum of the primal problem \cite{everett1963generalized}. Applying LR to the UC problem, the most common approaches decompose the dual problem into individual generator sub-problems which are solved with DP \cite{muckstadt1977application, merlin1983new, cohen1987method, zhuang1988towards, virmani1989implementation}. However, this method is not guaranteed to produce a feasible solution to the primal problem, and heuristics are generally used to satisfy the violated constraints \cite{zhuang1988towards, virmani1989implementation}. Several papers have used further metaheuristic approaches such as particle swarm optimisation (PSO) \cite{balci2004scheduling} or genetic algorithms (GA) \cite{cheng2000unit} to update the Lagrange multipliers. 

Improved solution quality of LR over PL methods are shown in early research (1988) \cite{bard1988short}, which finds large cost reductions of 10--33\% using LR. In applications to the 100-generator Kazarlis benchmark summarised in Table \ref{literature:tab:duc_kazarlis} LR is applied in \cite{cheng2000unit, ongsakul2004unit, balci2004scheduling} and achieves lower operating costs in run times that are between 2--24\% of the original GA run time. In addition to higher solution quality, LR exhibits two practical advantages over DP and PL methods. First, it is more scalable in the number of generators without strong reliance on heuristics. Due to the decomposition into generator sub-problems, LR methods have roughly linear time complexity in the number of generators \cite{virmani1989implementation}. A proprietary, 172-unit test system for Electricite de France (EDF) was studied in \cite{merlin1983new} and 100-unit systems in \cite{zhuang1988towards, bard1988short} which are among the largest studies of their era. The second advantage of LR is that unlike PL and DP methods, LR benefits from having a known `duality gap' giving the difference between the feasible solution cost and a lower bound for the optimal solution cost. The duality gap is measured as the difference between the dual solution cost and the primal solution cost \cite{merlin1983new}. This can be used as a stopping criterion, and provides assurances for system operators regarding the solution quality, despite the reliance of LR on heuristics. For these reasons, LR was the dominant method for practical UC applications in the 1990s and 2000s \cite{carrion2006computationally, padhy2004unit}. 

\subsubsection{Metaheuristics} \label{literature:duc:metaheuristics}

Several metaheuristic methods have been proposed to solve the UC problem, including genetic algorithms (GA) \cite{kazarlis1996genetic, senjyu2002unit, damousis2004solution, dang2007floating, amjady2009unit}; other evolutionary algorithms \cite{srinivasan2004priority, lau2009quantum, chung2010advanced, juste1999evolutionary, patra2008differential, trivedi2015hybridizing, dhaliwal2018modified}; particle swarm optimisation (PSO) \cite{chakraborty2012unit, shukla2016advanced, chen2011two, jeong2010new, yuan2011unit, logenthiran2010particle, zhao2006improved}; simulated annealing (SA) \cite{zhuang1990unit, mantawy1998simulated, senjyu2006absolutely, dudek2010adaptive, simopoulos2006reliability}. These methods use probabilistic techniques and rules to iteratively improve a solution or population of solutions. Metaheuristics do not generally guarantee an optimal solution, but may achieve good sub-optimal solutions in short run times.

The survey of the 100-generator Kazarlis benchmark problem in Table \ref{literature:tab:duc_kazarlis} shows several metaheuristic approaches are among the best performing techniques both in terms of operating cost and run time \cite{jo2018improved, lau2009quantum, senjyu2006absolutely}. Many of these metaheuristic approaches include problem-specific rules to improve the convergence time and solution quality \cite{jo2018improved, zhao2006improved, senjyu2006absolutely, srinivasan2004priority, juste1999evolutionary}. As a result, good performance on the Kazarlis benchmark does not necessarily imply generalisability across other problem instances, as metaheuristic approaches can be aggressively tuned to improve solution quality and reduce solution time. This reliance on expert rules and often extensive parameter tuning have prevented practical applications of metaheuristics \cite{fu2005security}. 

\subsubsection{Mixed-Integer Linear Programming} 

Currently, the most prevalent solution methods use a mixed-integer linear programming (MILP) formulation which is solved by branch-and-bound techniques (described in detail in Section \ref{methodology:mathop:milp}) \cite{dillon1978integer, cohen1983branch, chen1993branch, carrion2006computationally, borghetti2008milp, viana2013new, knueven2020mixed, chen2016improving, morales2013tight}. Advanced optimisation techniques such as pre-solve and cutting planes are implemented in commercial solvers such as CPLEX and Gurobi, enabling large-scale deterministic UC problems to be solved even with limited hardware \cite{carrion2006computationally}. Like LR, branch-and-bound produces an optimality gap measuring the proximity of the current solution costs to a lower bound. 

Only two of the surveyed solutions to the 100-generator Kazarlis benchmark in Table \ref{literature:tab:duc_kazarlis} used MILP \cite{carrion2006computationally, viana2013new}. Although both solutions achieve similar operating costs, \cite{carrion2006computationally} has roughly 130-fold lower run time than \cite{viana2013new}. The variation in run time of MILP approaches to the UC problem is influenced by the tightness (size of the feasible solution region) and compactness (the problem size, such as number of constraints and decision variables) of the formulation \cite{morales2013tight}, as well as the efficiency of the solver. Designing tight and compact MILP formulations for the UC problem has been studied extensively \cite{morales2013tight, knueven2020mixed, ostrowski2011tight, carrion2006computationally, pan2017convex}. 41 MILP formulations are compared in \cite{knueven2020mixed}, using 68 problem instances from the 3 large-scale benchmarks described in Section \ref{literature:duc:benchmarks} to systematically evaluate the efficiency of MILP formulations. This study finds large variability in solution times, with relative solution times of different formulations also varying across problem instances. Machine learning has recently been used to improve the efficiency of branch-and-bound for MILP \cite{nair2020solving, balcan2018learning}. This approach is applied to solve the DUC problem in \cite{nair2020solving}, and is found to reduce the optimality gap by a factor of 2 for similar computing times as compared with a conventional solver.

As the dominant solution method, MILP has been widely used to solve variations on the deterministic UC problem and in planning studies. Problem variants have considered combined heat and power (CHP) \cite{mitra2013optimal}, storage assets \cite{borghetti2008milp} and UC in a microgrid setting \cite{hawkes2009modelling}. These studies have focused on deriving efficient MILP formulations to include additional decision variables and constraints. Formulating UC problem variants for MILP is often challenging due to the non-convexities and non-linearities arising from integer decision variables and inter-temporal constraints such as minimum up/down times and ramping constraints \cite{mitra2013optimal}. Due to its accuracy and scalability to large power systems, MILP is also appropriate for simulation studies on large transmission networks and has been used to estimate startup costs \cite{schill2017start} and curtailment rates \cite{mc2013much} in future power systems, as well as for power system planning models \cite{koltsaklis2016mid}. 

The efficiency improvements of MILP solvers and solution techniques as well as tighter and more compact deterministic MILP formulations have led to this becoming the most widely used approach for solving the UC problem in practical applications \cite{bertsimas2012adaptive}. The experiments in \cite{knueven2020mixed} which consider up to 939 generators in the FERC benchmark \cite{krall2012rto} and studies on the MISO grid of approximately 1400 generators \cite{chen2016improving} show that deterministic MILP is scalable to large power systems. In addition, the optimality gap assures high quality solutions are achieved. However, the deterministic UC problem formulation is limited by its consideration of the point forecasts alone and use of heuristic reserve constraints to manage uncertainties. In the following section, we review stochastic formulations of the UC problem. These formulations more rigorously account for uncertainties and achieve lower expected operating costs in practice, at the expense of higher computational requirements.

\section{Scenario-Based Stochastic Unit Commitment} \label{literature:suc}

In contrast with the deterministic UC formulation which manages uncertainties using reserve constraints, scenario-based stochastic formulations have been adopted where the objective function minimises expected cost over a finite number of scenarios \cite{takriti1996stochastic}. Using the same notation as for the deterministic UC problem in Equations \ref{literature:eq:duc_objective}--\ref{literature:eq:duc_operating constraint}, the stochastic UC problem is formulated as: 

\begin{gather}
  \min \sum_{s \in \mathcal{S}} P(s) \sum_{i=1}^{N} \sum_{t=1}^{T} C_i(t,s) 
  \label{literature:eq:suc_objective}
\end{gather}

subject to

\begin{align}
  & \sum_{i=1}^{N} p_i (t,s) = D(t, s) \quad & \forall t \in \{1 \twodots T \}, \forall s \in \mathcal{S} \label{literature:eq:suc_load_balance} \\
  & \boldsymbol{p}_i(s) \in \Pi_i \quad & \forall i \in \{1 \twodots N \}, \forall s \in \mathcal{S} \label{literature:eq:suc_operating constraint} \\
\end{align} 

Compared with the deterministic UC formulation defined in Equations \ref{literature:eq:duc_operating constraint}--\ref{literature:eq:duc_operating constraint}, the stochastic formulation introduces $\mathcal{S}$, a finite set of scenarios representing realisations of the uncertain processes such as demand and renewables generation. Scenarios are typically represented in a tree, where each node represents the realisation of uncertainties at a given decision period, branching into one or more child nodes at the next decision period. The objective function aims to minimise the weighted sum of operating costs under each scenario, with scenarios weighted by probability $P(s)$. Commitment decisions must remain fixed across scenarios, while the generator setpoints $p_i$ can differ between scenarios to satisfy the load balance constraint in Equation \ref{literature:eq:suc_load_balance}. This formulation is therefore a two-stage stochastic program, where the first-stage commitment decisions are scenario independent and the second-stage (also known as real-time) dispatch decisions are made on the basis of uncertain outcomes \cite{kim2016large}.

The stochastic formulation of the UC problem omits the reserve constraint in Equation \ref{literature:eq:duc_reserve}. Instead, reserve is implicitly allocated as solutions must satisfy the load balance constraints under all scenarios. If the set of scenarios $\mathcal{S}$ contains sufficiently extreme scenarios, solving the stochastic UC problem yields robust solutions. Like the deterministic UC problem, several extended formulations have been proposed, including those discussed in Section \ref{literature:suc:reserve} which add explicit reserve constraints to improve robustness where including large numbers of scenarios in the problem formulation is not tractable. Further formulations have incorporated hydropower scheduling \cite{caroe1997unit, nowak2005stochastic, dentcheva1998optimal} and the consideration of AC network constraints \cite{nasri2015network}; however, in this section we will focus on research studying the commitment of thermal power stations without network constraints.

\subsection{Benchmark Problems} \label{literature:suc:benchmarks}

Unlike the deterministic UC benchmarks described in Section \ref{literature:duc:benchmarks}, the stochastic UC problem lacks established problems that allow for solution methods to be systematically compared. Among the popular test systems is the IEEE RTS benchmark systems generators which has been studied with varying numbers of generators \cite{ruiz2009uncertainty, blanco2017efficient, nasri2015network, dvorkin2014comparison}. In addition, variations on the IEEE 118 network have been used in \cite{che2018two, wang2012stochastic, wang2015fully, bouffard2008stochastic}, the CAISO system in \cite{papavasiliou2014applying} and Irish system in \cite{tuohy2009unit, meibom2010stochastic, lowery2014reserves}. However, due to the random nature of the stochastic UC problem, even the results of simulation studies conducted on the same system cannot be easily compared. It is noted in \cite{ruiz2009uncertainty} that stochastic UC simulation studies usually report costs under the `in-sample' scenarios $\mathcal{S}$ which were included in the formulation. Differences in random seeds as well as methods for scenario generation and reduction \cite{dvorkin2014comparison} mean that the set $\mathcal{S}$ is typically unique for each stochastic formulation. Since the costs reported consider the `in-sample' scenarios $\mathcal{S}$, results are generally not comparable.

Instead of reporting costs under the in-sample scenarios $\mathcal{S}$, Monte Carlo methods for evaluating solution quality have been used. The second-stage costs are calculated under a large number of out-of-sample scenarios, and expected costs are estimated with the empirical mean \cite{ruiz2009uncertainty, dvorkin2014comparison, papavasiliou2014applying, constantinescu2009unit}. With a sufficiently large number of scenarios, results from different studies can be compared when using this approach provided the distributions of random processes are the same. Benchmarks have not so far been developed for systematically comparing solution methods using Monte Carlo approaches.

\subsection{Rolling Horizon Optimisation} \label{literature:suc:rolling_horizon}

In this literature review and throughout this thesis, we focus on solutions to the day-ahead UC problem. However, due to large computational costs associated with stochastic optimisation, a significant proportion of the stochastic UC literature has used a rolling horizon approach, where UC decisions are made for a limited time horizon, such as a 3, 6 or 12 hours, then rescheduled periodically \cite{tuohy2009unit, schulze2016value, sturt2011efficient, meibom2010stochastic, lowery2014reserves}. This greatly improves the computational tractability of stochastic UC, as the scenario tree of each stochastic program generally grows exponentially with the planning period. 

Several studies have used the WILMAR planning tool \cite{barth2006stochastic} to run rolling horizon stochastic UC models for the Irish power system \cite{tuohy2009unit, meibom2010stochastic, lowery2014reserves, denny2010impact}. These studies have mainly focused on simulating future power systems, for purposes such as assessing the sizing of reserve requirements \cite{lowery2014reserves} and the impact of increased penetration of renewables \cite{meibom2010stochastic, denny2010impact}. In \cite{tuohy2009unit}, the authors use a rolling horizon model to assess the benefits of stochastic optimisation with a rolling horizon in the Irish power system, in terms of operating costs and security supply, with comparison made against deterministic UC as well as optimisation with perfect foresight. However, rolling horizon methods are not applicable in day-ahead electricity markets, where decisions must be fixed for a 24-hour period. As most exchange-traded power demand is settled in day-ahead markets \cite{maciejowska2019day}, rolling horizon approaches are currently more limited than day-ahead methods in terms of practical applications. Next, we will describe solution methods that have been employed to solve the stochastic, day-ahead problem. 

\subsection{Solution Methods}

Unlike deterministic UC reviewed in Section \ref{literature:duc}, there is no dominant solution method for the stochastic UC problem. The scenario-based stochastic UC formulation is NP-hard, and considerably more expensive to solve than the deterministic version \cite{papavasiliou2014applying}. MILP approaches such as branch-and-bound, which are the state of the art in solving deterministic UC problems, have been used to solve the stochastic UC problem \cite{bouffard2008stochastic} but face challenges in large power systems due to run time and memory constraints \cite{papavasiliou2014applying}. Lagrangian relaxation approaches has been widely used to decompose the stochastic UC problem into generator sub-problems or scenario sub-problems \cite{takriti1996stochastic, nowak2000stochastic, wu2007stochastic, papavasiliou2014applying}. Further decomposition methods such as Benders' decomposition \cite{nasri2015network, zheng2013decomposition} and Dantzig-Wolfe decomposition \cite{shiina2004stochastic} have also been used to solve  large-scale stochastic UC problems. The decomposed problems can be solved in parallel across multiple machines, which offers promising opportunities for achieving tractability using high performance computing resources \cite{papavasiliou2014applying}. Nevertheless, the large computational cost of stochastic optimisation remains a barrier to the adoption of stochastic UC for practical applications.

Due to the large computational burden of scenario-based stochastic UC, determining the size and content of the scenario set $\mathcal{S}$ is an important decision in stochastic UC solution methods \cite{dvorkin2014comparison, wu2007stochastic}. The set should capture important statistical information about the underlying stochastic processes within a small number of scenarios to achieve computational tractability. Common approaches generate a set of scenarios using a Monte Carlo method and select a subset based on the Kantorovich distance, minimising the difference between the original and reduced scenario sets \cite{morales2009scenario, barth2006stochastic}. K-means clustering and importance sampling approaches are among those presented in a comparison of approaches \cite{dvorkin2014comparison}, which finds that the reduction method impacts both operating cost and run time. Nevertheless, due to the high computational requirements of solving stochastic formulations, the number of scenarios is generally small and may not capture extreme scenarios in cases with large numbers of uncertain parameters \cite{ruiz2009uncertainty}. For this reason, reserve requirements have been introduced to some stochastic UC formulations to improve solution security, as described in Section \ref{literature:suc:reserve}.

\subsection{Formulations with Reserve Constraints} \label{literature:suc:reserve}

A common extension to the stochastic UC formulation in Equations \ref{literature:eq:suc_objective}--\ref{literature:eq:suc_operating constraint} is to add an explicit reserve constraint to improve solution robustness \cite{ruiz2009uncertainty, asensio2015stochastic}. Since the number of scenarios $|\mathcal{S}|$ must usually be relatively small in order for the problem to be computationally tractable, it often cannot include extreme situations \cite{ruiz2009uncertainty}. As a result, including a reserve constraint as in the deterministic formulation (Equation \ref{literature:eq:duc_reserve}) decreases the probability of lost load under large deviations of demand or generation from their forecasts. Furthermore, the performance of stochastic UC models without reserve constraints is highly dependent on the statistical approach to modelling uncertainties and generating the scenario set $\mathcal{S}$. By including reserve constraints, solutions are made more robust against random generation of scenarios \cite{lowery2014reserves}. 

Using the Monte Carlo evaluation method described in Section \ref{literature:suc:benchmarks}, \cite{ruiz2009uncertainty} compares stochastic UC with and without reserve constraints in simulations on the IEEE RTS system with 32 generators. Solutions with reserve constraints are found to be more robust to changes in the underlying probability distributions, reflecting inaccurate models of the real world. In addition, solutions have lower expected costs when reserves are included, and the reserve requirements are found to be smaller than for deterministic UC approaches. The results of \cite{ruiz2009uncertainty} show that including reserve requirements can be used to reduce the computational burden of stochastic UC by requiring fewer scenarios to achieve robust solutions. However, this requires heuristic methods to determine reserve constraints, and comes at the expense of lower solution quality as the number of scenarios increases. In the following section, we compare the solution quality of stochastic UC approaches, with and without reserve requirements, to deterministic methods.

\subsection{Comparison with Deterministic UC} \label{literature:suc:duc_comparison}

In comparison with deterministic methods reviewed in Section \ref{literature:duc}, solutions to the stochastic UC problem have been shown to achieve lower operating costs due to its more rigorous consideration of uncertainties \cite{takriti1996stochastic, carpentier1996stochastic, bouffard2008stochastic, tuohy2009unit, ruiz2009uncertainty, papavasiliou2014applying, wang2011unit, lowery2014reserves, quan2014incorporating, schulze2016value}. Simulating one year's rolling operation of the Irish 2020 power system, \cite{tuohy2009unit} shows that total operating costs of a stochastic model are 0.25--0.9\% lower than a deterministic approach. In \cite{ruiz2009uncertainty}, in which reserve requirements are used, both demand uncertainty and generator outages are modelled, and costs of the stochastic approach are found to be 1.3\% lower than deterministic UC in experiments on the IEEE RTS system with 32 generators. Studying a representation of the GB power system, \cite{schulze2016value} uses a rolling planning framework to evaluate the relative costs of stochastic and deterministic UC over two years, considering transmission network constraints and pumped storage. Stochastic UC outperforms deterministic UC by 0.3\% in terms of operating costs.

Operating cost reductions compared to deterministic UC are shown to increase with uncertainty, for example resulting from increasing wind penetration \cite{bouffard2008stochastic, papavasiliou2014applying, wang2011unit, ruiz2009uncertainty, schulze2016value}. Deterministic UC solutions to the problem in \cite{bouffard2008stochastic} are unable to accommodate wind penetrations above 8\% without suffering lost load, while stochastic UC solutions are secure for penetrations up to 20\%. Cost reductions in \cite{tuohy2009unit} are attributed to less cycling of flexible generation, with more consistent use of base-load with fewer startups. In \cite{ruiz2009uncertainty}, which includes reserve constraints in the stochastic formulation, the required reserve levels are found to be lower as compared with deterministic UC, causing generators to operate at greater efficiencies. 

Stochastic UC has also been compared with UC methods with perfect foresight of demand and renewables generation in a rolling horizon optimisation context \cite{tuohy2009unit, meibom2010stochastic}. The cost of uncertain forecasts is quantified in \cite{meibom2010stochastic} by comparing the costs of commitment schedules with perfect forecast and stochastic UC: perfect foresight is found to result in between 0.05\% and 1.2\% lower operating costs in case studies for the Irish power system. A similar comparison is made in \cite{tuohy2009unit}, finding that scenario-based stochastic UC achieves lower 1.5\% higher operating than the optimisation with perfect foresight, while a deterministic UC with reserve constraints results in 1.75\% higher costs. Stochastic UC is therefore shown to mitigate the impact of uncertainty as compared with deterministic methods, but there are still significant cost impacts of uncertain forecasts. 

The primary drawback of scenario-based stochastic optimisation approaches are the computational requirements, which are much larger than those of deterministic methods \cite{papavasiliou2014applying}. Run times for stochastic UC with 12 scenarios and reserve requirements are found to be between 1 and 3 orders of magnitude larger than a deterministic approach in \cite{ruiz2009uncertainty}. Distributed implementations for solution methods which use decomposition techniques can be used to improve computational tractability, but still require significant high performance computing resources \cite{papavasiliou2014applying}. An additional drawback is that the underlying probability distributions of stochastic demand and renewables generation may be hard to obtain, making scenario generation difficult and solutions potentially sensitive to deviations from the modelled distributions \cite{bertsimas2012adaptive}. However, the stochastic UC literature has demonstrated that the dominant deterministic UC approaches described in Section \ref{literature:duc} are sub-optimal, and significant operating cost savings can be achieved by methods which aim to optimise the expected cost over scenarios, rather than considering only the point forecast. These improvements are of a similar magnitude to those achieved by significant advances in deterministic UC methods, such as the transition to MILP and branch-and-bound reviewed in Section \ref{literature:duc:solution_methods} which reportedly achieved annual operating cost reductions of \$1 billion in North America \cite{o2017computational}. The potential for such large improvements in solution quality that has been demonstrated by the stochastic UC literature is one of the principle motivations for further investigating UC solution methods. Robust optimisation, reviewed in Section \ref{literature:ruc}, mitigates some of the issues of stochastic UC, having lower computational costs while accounting for uncertainties without heuristic reserve constraints.

\section{Robust Unit Commitment} \label{literature:ruc}

Stochastic UC approaches have been shown to be effective in reducing operating costs, but practically difficult to implement due to high computational costs \cite{papavasiliou2014applying} and requiring knowledge of the distributions underlying stochastic processes \cite{bertsimas1997introduction, ruiz2009uncertainty}. Robust optimisation has been applied to the UC problem, which is typically less expensive to compute and makes fewer assumptions regarding the uncertain distributions \cite{bertsimas2012adaptive}. In general, robust optimisation tackles problems where solutions must minimise worst-case costs over a deterministic uncertainty set. The uncertainty set is the region of possible realisations of uncertain parameters, which is usually determined based on a desired level of robustness. Feasible solutions to robust optimisation problems must satisfy constraints for all realisations of the data in the uncertainty set while optimising for the worst-case outcome \cite{ben1998robust}. 

Using the same notation as in previous formulations, robust UC problem formulation is:

\begin{gather}
  \min \max_{d \in \mathcal{D}} \sum_{i=1}^{N} \sum_{t=1}^{T} C_i(t,d) 
  \label{literature:eq:ruc_objective}
\end{gather}

subject to:

\begin{align}
  & \sum_{i=1}^{N} p_i (t,d) = D(t, d) \quad & \forall t \in \{1 \twodots T \}, \forall d \in \mathcal{D} \label{literature:eq:ruc_load_balance} \\
  & \boldsymbol{p}_i(d) \in \Pi_i \quad & \forall i \in \{1 \twodots N \}, \forall d \in \mathcal{D} \label{literature:eq:ruc_operating constraint} \\
\end{align} 

The set $\mathcal{D}$ is a probabilistic uncertainty set, defining the region of realisations of uncertainty; $C_i(t,d)$ is the cost of generator $i$ at time $t$ under the realisation of uncertainties $d \in \mathcal{D}$. The objective of the robust UC formulation is to find a commitment schedule which minimises operating costs in the worst (most expensive) case realisation of uncertainties. Note that unlike the stochastic objective in Equation \ref{literature:eq:suc_objective}, the scenario probability $P(s)$ is not included in the objective function Equation \ref{literature:eq:ruc_objective}. Instead, the set is a deterministic uncertainty set. As a result, a model for the underlying distributions of uncertain parameters is not required for robust UC. Like the stochastic formulation, the robust formulation does not include a reserve constraint, with reserves being implicitly allocated to satisfy extreme demand deviations or other contingencies in the uncertainty set.

Robust UC problem formulations have been used widely in the UC literature to manage uncertainties \cite{bertsimas2012adaptive,blanco2017efficient,jiang2010two,jiang2011robust,jiang2013two,jiang2014two,lee2013modeling,li1997robust,liu2015robust,lorca2014adaptive,lorca2016multistage,velloso2019two,wang2013two,ye2016uncertainty,zhao2012robust,street2010contingency, morales2018robust}. Compared with stochastic UC, robust UC has the following benefits: (1) it is often more computationally tractable; (2) it does not require knowledge of underlying probability distribution governing stochastic processes like renewables generation; (3) it is not sensitive to scenario generation and reduction methods \cite{jiang2010two, bertsimas2012adaptive, jiang2011robust}. However, solutions tend to be more conservative as they optimise for the worst-case outcome in the uncertainty set, regardless of its probability, and may produce higher expected operating costs as a result \cite{zheng2014stochastic}. In Section \ref{literature:ruc:comparison} we review research which has compared robust UC with deterministic and stochastic UC. In the following subsection, we discuss alternative formulations which unify stochastic UC and robust UC, mitigating the over-conservatism of robust optimisation.

\subsection{Hybrid Stochastic and Robust Formulations}

UC formulations have been proposed which aim to combine the advantages of stochastic UC (minimising expected costs over scenarios) and robust UC (minimising costs under worst-case scenarios) \cite{blanco2017efficient, zhao2013unified, zhao2015data, morales2018robust}. These hybrid approaches produce solutions which are less conservative than traditional robust optimisation approaches. In \cite{blanco2017efficient}, it is assumed that the probability distribution over scenarios is known, and scenarios are sampled and then aggregated into a scenario tree, similar to stochastic UC approaches. A robust optimisation approach is then formed, aiming to minimise the worst-case costs under the sampled scenarios. Minimisation of a weighted sum of expected costs and worst-case costs are used in \cite{zhao2013unified}, creating a hybrid stochastic/robust objective function with a parameter controlling the level of robustness. This hybrid approach is shown to produce solutions which are less conservative and have lower expected operating costs than robust UC. However, a drawback of this approach is that, like stochastic UC, it assumes that the underlying probability distributions of uncertain variables are known. 

\subsection{Benchmark Problems}

Robust UC faces similar challenges regarding benchmark problems as those faced in stochastic UC research, described in Section \ref{literature:suc:benchmarks}. Namely, given the stochastic nature of the problem formulation, UC solutions must be evaluated against an equal set of possible realisations of uncertainties. A broad range of test systems has been studied including New England ISO network with 312 generators \cite{bertsimas2012adaptive}, variations of the IEEE 118 system \cite{jiang2013two, morales2018robust, lorca2016multistage, lee2013modeling}, IEEE RTS with 96 generators \cite{blanco2017efficient, liu2015robust} and the Polish transmission network \cite{lorca2016multistage}. Furthermore, studies have variously considered transmission constraints \cite{lee2013modeling, lorca2016multistage, bertsimas2012adaptive}, dispatchable wind \cite{morales2018robust} and pumped hydro resources \cite{jiang2011robust}. There has been no established benchmark in the literature that has allowed for robust UC methods to be rigorously evaluated against one another. 

Several research papers have implemented multiple solution methods and compared solution quality using the Monte Carlo approach described in Section \ref{literature:suc:benchmarks} \cite{bertsimas2012adaptive, jiang2013two, morales2018robust, liu2015robust}. Using this method the robustness of schedules is evaluated against out-of-sample scenarios \cite{morales2018robust}. In Section \ref{literature:ruc:comparison}, we present the results of experiments comparing robust UC with deterministic and stochastic UC approaches.

\subsection{Solution Methods} \label{literature:ruc:solution_methods}

Like the stochastic UC problem, branch-and-bound approaches using commercial MILP solvers cannot be effectively applied in most cases to solve the robust UC problem \cite{street2010contingency} and most research has focused on decomposition techniques, notably Benders' decomposition \cite{bertsimas2012adaptive, street2010contingency, wang2013two, jiang2011robust}. 

The computational tractability of robust UC depends to a large extent on the form of the uncertainty set $\mathcal{D}$. Demand and wind variation may be represented using continuous uncertainty regions (e.g. with polyhedral constraints) \cite{bertsimas2012adaptive} or with discrete sets representing scenarios \cite{zhao2012robust, jiang2011robust}. To account for generator outages, knapsack constraints restricting coincident outages to a maximum of $x$ are used to implement security criteria protecting against the loss of the $x$ largest infeeds \cite{street2010contingency, wang2013two}. While robust UC formulations are generally more computationally tractable than stochastic ones, robust UC has longer execution time than deterministic UC \cite{bertsimas2012adaptive}. In addition, large-scale problems with many uncertain parameters and complex uncertainty sets can also be intractably expensive to solve for practical use cases \cite{velloso2019two}. 

The potential for over-conservative solutions and consequently high operating costs has been managed in several cases by introducing an uncertainty budget parameter defining the extremity of scenarios included in the uncertainty set $\mathcal{S}$ \cite{bertsimas2012adaptive, jiang2013two}. Tuning the uncertainty budget is shown in \cite{bertsimas2012adaptive} to have a significant impact on performance and prevent over-conservative schedules. In the following section, we will compare the performance of robust UC approaches to the deterministic and stochastic UC formulations.

\subsection{Comparison with Deterministic and Stochastic UC} \label{literature:ruc:comparison}

Operating costs for solutions to the robust UC problem have been compared with deterministic UC in \cite{bertsimas2012adaptive, lee2013modeling,lorca2014adaptive,lorca2016multistage}. Comparisons of average operating costs for robust UC and deterministic UC approaches have found that while robust UC operating costs are often higher for nominal demand \cite{lee2013modeling} or demand profiles with low uncertainties \cite{bertsimas2012adaptive}, operating costs have lower variance under different scenarios. Comparing robust optimisation with deterministic UC and evaluating solutions with a Monte Carlo approach with 1000 scenarios, \cite{bertsimas2012adaptive} reports expected cost reductions of between 0.34--5.48\% on a 312-unit test system reflecting the New England ISO. The standard deviation of operating costs is also found to be between 8--14 times larger when using the deterministic approach. Operating costs for the robust UC approach under the nominal demand scenario are found to be 0.8\%  higher than for a deterministic solution in \cite{lee2013modeling}, demonstrating the relative conservatism of robust optimisation. However, for the worst-case scenario, the deterministic solution is found to be infeasible (i.e. resulting in lost load), whereas the robust approach is able to maintain system security. The impact of the robust UC uncertainty budget described in Section \ref{literature:ruc:solution_methods} which provides control over the conservatism of the solution is analysed in \cite{bertsimas2012adaptive}. A statistical method is proposed to tune the budget parameter to improve economic efficiency of solutions. 

Comparing robust UC with scenario-based stochastic methods, \cite{jiang2013two} finds that stochastic UC achieves lower expected operating costs but larger regret (deviation from optimal solution cost) on average under high levels of uncertainty. A stochastic approach is compared with robust optimisation in \cite{morales2018robust}, and again found to achieve lower average operating costs but with more frequent lost load events and higher worst-case operating costs. Increasing the number of scenarios for the stochastic UC approach reduces expected operating costs, but at 30 scenarios, stochastic UC is found to be 71 times slower to solve than robust UC. Studying a small 4-bus system with 14 generators, \cite{velloso2019two} finds that stochastic UC does not significantly outperform robust UC in terms of expected operating costs, even with low uncertainty budgets. In addition, the stochastic UC is not applied to a larger IEEE 118-bus system, due to intractable compute times.

In summary, robust UC is generally a more computationally tractable alternative to stochastic UC, with potential advantages over deterministic UC methods due to more rigorous consideration of uncertainties \cite{bertsimas2012adaptive}. Robust optimisation has not been widely applied on such large-scale power systems as deterministic methods. Most research has been conducted on small to medium size systems; larger simulation studies have considered the New England ISO \cite{bertsimas2012adaptive} and Polish transmission network \cite{lorca2016multistage} but these remain smaller than large-scale deterministic UC simulation studies, such as MISO \cite{chen2016improving} and FERC \cite{knueven2020mixed}. The principle disadvantage of robust UC compared with stochastic UC is that schedules may be overly conservative and hence economically sub-optimal \cite{jiang2013two, blanco2017efficient}. In addition, the computational cost may be relatively high as compared with deterministic methods when large numbers of uncertain parameters are considered \cite{velloso2019two}. 


\section{Reinforcement Learning} \label{literature:rl}

A small body of research has been dedicated to solving the UC problem with reinforcement learning (RL) \cite{jasmin2009reinforcement, jasmin2016function, li2019distributed, navin2019fuzzy, dalal2015reinforcement, dalal2016hierarchical}, which is the focus of this thesis. These papers are summarised in Table \ref{literature:tab:rl_uc}. In contrast with the other methods discussed in this chapter, RL methods rely on the formulation of the UC problem as a Markov Decision Process (MDP), which is used to solve sequential decision-making problems. An `agent' learns by trial-and-error, typically using a simulation of the real power system, to make decisions which minimise operating costs. We discuss MDPs and provide a background to RL more generally in Section \ref{methodology:background}.

\begin{sidewaystable}
\renewcommand{\arraystretch}{1.5}
\footnotesize
    \centering
\begin{tabular}{p{0.7in} p{1.2in} p{0.8in} p{1in} p{0.8in} p{0.8in} p{0.7in} p{1.3in}}
\toprule
\textbf{Authors} &\textbf{Method} &\textbf{Function Approximation} &\textbf{Max. Gens} &\textbf{Stochastic} &\textbf{Multiple Training Days} &\textbf{Unseen Test Days} &\textbf{Notes} \\\midrule
Jasmin et al., 2009 \cite{jasmin2009reinforcement} &Q-learning &No &4 &No &No &No & \\
Jasmin et al., 2016 \cite{jasmin2016function} &Q-learning &Yes &10 &Yes &No &No & \\
Li et al., 2019 \cite{li2019distributed} &Q-learning &Yes &10 &Yes &No &No & \\
Navin \& Sharma \cite{navin2019fuzzy} &Multi-Agent Fuzzy Q-Learning &Yes &10 &No &No &No & \\
Dalal et al., 2016 \cite{dalal2016hierarchical} &Cross-Entropy &Yes &99 &Yes &Yes &No &Day-ahead UC simplified to a single decision per day, choosing between 20 actions \\
Dalal \& Mannor, 2015 \cite{dalal2015reinforcement} &SARSA &Yes &8 &No &No &No & \\
Dalal \& Mannor, 2015 \cite{dalal2015reinforcement} &Tree Search &No &12 &No &NA &NA & \\
\bottomrule
\end{tabular}
\caption{Summary of research applying RL to the UC problem. We show the method used; the maximum problem size by number of generators; whether function approximation was used; whether stochastic demand or renewables generation are included in the problem setup; whether multiple days were used in training and whether testing was conducted on unseen test days.}
\label{literature:tab:rl_uc}
\end{sidewaystable}

Q-learning, a popular class of RL methods, has been applied to the UC problem in \cite{jasmin2009reinforcement, jasmin2016function, navin2019fuzzy, li2019distributed}. These papers have applied Q-learning in a tabular format \cite{jasmin2009reinforcement} and with function approximation \cite{jasmin2016function, li2019distributed, navin2019fuzzy} with applications to problems of up to 10 generators. As shown in Table \ref{literature:tab:rl_uc}, all studies consider optimisation for a single profile; training and testing on a single day. As a result, they do not demonstrate the ability of the trained policy to generalise to unseen problems. Two of the Q-learning studies include uncertainty in the problem setup through stochastic renewables generation \cite{li2019distributed, jasmin2016function}. Fuzzy Q-learning is used in \cite{navin2019fuzzy} to solve the widely-studied Kazarlis et al. benchmark problem with 10 generators, and is shown to outperform several existing deterministic UC solution methods. The Q-learning methods studied suffer from curses of dimensionality in the state and action spaces for the UC problem, which has limited their application to systems of up to 10 generators.

A larger, 99-generator system is studied in \cite{dalal2016hierarchical}, which solves a combined problem of day-ahead UC and real-time ED, represented using two interleaved MDPs. The UC component is solved with the cross-entropy RL method. This is by far the largest study applying RL to solve the UC problem. In addition, stochastic demand and wind generation are considered, and the agent is trained over multiple profiles. However, the UC component of this problem is simplified significantly to selecting a single commitment decision for each 24-hour period, with no intra-day commitment changes. In addition, the set of commitment decisions $\mathcal{X}$ is reduced to just 20 decisions (of a possible $2^{99}$) using a heuristic approach. Compared with simple heuristic solutions to the UC problem (e.g. committing a random subset of generators or the cheapest subset), the cross-entropy method employed is shown to result in lower total operating costs. However, by reducing the UC solution space to just 20 actions and not allowing intra-day commitment changes, this approach is not guaranteed to converge towards cost-optimal solutions and solution quality is not validated by comparison with mathematical optimisation approaches such as MILP. 

Three algorithms are proposed in \cite{dalal2015reinforcement}: one using the SARSA method \cite{rummery1994line} and two using tree search approaches. SARSA, which is a model-free RL algorithm, was found to be the least effective algorithm and could not be scaled beyond a problem of 8 generators due to slow convergence. To the best of our knowledge, \cite{dalal2015reinforcement} is the only research which applies tree search methods to solve the UC problem and the only research to formulate the UC problem as a search tree. Of the two tree search methods, a simple lookahead strategy with a limited time horizon was found to be the most effective approach, outperforming a metaheuristic solution by 27\% in terms of operating costs for a deterministic problem of 12 generators. Unlike the other methods reviewed, the tree search method proposed does not store a policy which is iteratively improved upon and hence it is not strictly an RL algorithm. However, by comparison with the RL algorithm SARSA, exploiting lookahead capability is shown to be a more effective strategy for the UC problem.

In summary, the existing research on RL for UC has shown some promising results, but curses of dimensionality in both state and action spaces has limited these methods to small-scale problems. The only large-scale application \cite{dalal2016hierarchical} simplifies the UC problem substantially by making a single commitment decision per day without intra-day commitment changes, significantly limiting the solution quality potential. Furthermore, no comparison has been made in the existing literature with the state-of-the-art in deterministic UC methods using MILP. RL approaches to the UC problem differ fundamentally from the other formulations discussed in this chapter. The goal is to learn a function mapping from problems to solutions by trial-and-error. The learned function can in principle generalise to unseen problems, without retraining. This property of RL, which offers the opportunity for most of the computational burden to be shifted to offline training, has not yet been explored by the existing literature which consider training and testing on the same profile. In addition, a number of recent breakthroughs in RL, including hybrid approaches using tree search \cite{silver2018general} and efficient policy optimisation algorithms such as proximal policy optimisation \cite{schulman2017proximal} have not been applied to the UC problem at the time of writing. Compared with the other approaches discussed in this chapter, there is significant scope for further research investigating novel solution methods applying RL to the UC problem. 

\section{Conclusion}

In this section we have reviewed the state of the art in UC methods. Deterministic UC approaches described in Section \ref{literature:duc} remain the most widely-used in industry due to their low computational cost in comparison with stochastic optimisation approaches. However, they have no capacity to integrate uncertainty in day-ahead forecasts into the problem formulation, instead relying on heuristic reserve constraints. Scenario-based stochastic UC formulations, which were developed to handle uncertainty in forecasts using a set of scenarios, have been shown to outperform deterministic UC in terms of operating costs \cite{ruiz2009uncertainty, tuohy2009unit, bouffard2008stochastic}, but are generally too computationally expensive for practical applications \cite{bertsimas2012adaptive}. Robust formulations discussed in Section \ref{literature:ruc} have been shown to exhibit advantages over deterministic methods, such as lower variance of operating costs \cite{bertsimas2012adaptive} and lower worst-case operating costs \cite{lorca2016multistage, lee2013modeling}. Robust optimisation techniques generally produce conservative solutions with high expected operating costs on average \cite{zheng2014stochastic}. In addition, large-scale problems with many uncertain parameters can be intractably expensive to solve \cite{velloso2019two}. There is motivation to develop new UC solution methods that achieve better solution quality than deterministic approaches while remaining computationally tractable. 

This literature review has shown that comparison between solution methods is difficult due to a lack of established benchmark problems with multiple problem instances. The problems proposed by Knueven et al. \cite{knueven2020mixed} have recently been proposed for the deterministic problem to address this issue, but have not yet seen widespread use. In addition, these benchmark problems do not provide stochastic models for Monte Carlo evaluation of stochastic or robust UC methods. Table \ref{literature:tab:duc_kazarlis} surveyed solutions to the 100-generator Kazarlis benchmark \cite{kazarlis1996genetic}, but this is limited to a single problem instance. For the stochastic, robust and reinforcement learning approaches, there are no widely-used benchmarks. In machine learning domains, the establishment of robust test problems such as ImageNet \cite{deng2009imagenet} for computer vision and SQUAD \cite{rajpurkar2016squad} has accelerated research and enabled algorithms to be systematically compared. The establishment of a universal benchmark for the UC problem which can be used to evaluate methods from across deterministic, stochastic, robust and RL-based UC literature would enable further progress in this domain.

Only a small body of research has investigated applying RL to solve the UC problem. These papers have generally investigated applications of model-free RL to small problem instances. In addition, no comparison has been made with the state-of-the-art in deterministic UC methods. Only one paper reviewed has applied tree search \cite{dalal2015reinforcement}, which was successfully applied in a deterministic UC environment for problem instances of up to 12 generators. The limitations of these studies has consistently been scalability to large power systems, which so far has not been shown. Nevertheless, significant methodological breakthroughs in the field of RL have yet to be applied in the UC problem, and offer potentially large improvements relative to existing literature as has been seen in other domains \cite{silver2018general, schrittwieser2020mastering}. In the next chapter, we will provide a background to RL and tree search, which will form the basis of the novel RL-aided tree search method presented in Chapter \ref{ch1}.

%% file: 02-literature/kazarlis_table.tex
\begin{table}
\centering
\scriptsize
\renewcommand*{\arraystretch}{1.5}
\begin{tabular}{lrrrr}
\toprule
\textbf{Method} &\textbf{Year} &\textbf{Cost ratio (\%)} &\textbf{Time ratio (\%)} &\textbf{Reference} \\\midrule
GA &1996 &100.00 &100.00 &\cite{kazarlis1996genetic} \\
EA &1999 &99.94 &38.90 &\cite{juste1999evolutionary} \\
LR &2000 &99.75 &25.71 &\cite{cheng2000unit} \\
GA &2002 &99.98 &22.54 &\cite{senjyu2002unit} \\
PL &2003 &99.66 &0.41 &\cite{senjyu2003fast} \\
LR &2004 &99.93 &4.64 &\cite{balci2004scheduling} \\
EA &2004 &99.65 &7.63 &\cite{srinivasan2004priority} \\
LR &2004 &99.61 &2.19 &\cite{ongsakul2004unit} \\
SA &2006 &99.83 &4.42 &\cite{simopoulos2006reliability} \\
PL &2006 &99.79 &2.38 &\cite{senjyu2006emerging} \\
SA &2006 &99.71 &1.31 &\cite{senjyu2006absolutely} \\
MILP &2006 &99.60 &0.78 &\cite{carrion2006computationally} \\
GA &2007 &99.77 &Not reported &\cite{dang2007floating} \\
EA &2008 &99.78 &12.84 &\cite{patra2008differential} \\
DP &2009 &99.69 &Not reported &\cite{patra2009fuzzy} \\
EA &2009 &99.68 &0.51 &\cite{lau2009quantum} \\
PSO &2011 &99.66 &1.88 &\cite{yuan2011unit} \\
EA &2011 &99.57 &1.86 &\cite{chung2010advanced} \\
PSO &2012 &99.53 &1.33 &\cite{chakraborty2012unit} \\
MILP &2013 &99.74 &99.31 &\cite{viana2013new} \\
EA &2015 &99.60 &2.52 &\cite{trivedi2015hybridizing} \\
PL &2015 &99.51 &0.05 &\cite{quan2015improved} \\
PSO &2016 &99.54 &4.20 &\cite{shukla2016advanced} \\
GA &2018 &99.54 &0.14 &\cite{jo2018improved} \\
EA &2018 &99.53 &Not reported &\cite{dhaliwal2018modified} \\
\bottomrule
\end{tabular}
\caption{Selected solutions to 100-generator Kazarlis benchmark problem \cite{kazarlis1996genetic} for deterministic UC. Included methods are genetic algorithms (GA), evolutionary algorithms (EA), Lagrangian relaxation (LR), priority list (PL), particle swarm optimisation (PSO), dynamic programming (DP), simulated annealing (SA), mixed-integer linear programming (MILP). Ratios compare operating costs and run time with the original GA solution in \cite{kazarlis1996genetic}.}
\label{literature:tab:duc_kazarlis}
\end{table}

%% file: 03-methodology/methodology.tex
\section{Introduction}

In this chapter, we will describe in detail the methods used in this thesis, providing the theoretical foundation for the novel UC solution methods developed and applied in Chapters \ref{ch1}--\ref{ch3}. In Section \ref{literature:rl} we reviewed the small body of research that has used RL to solve the UC problem. The existing literature has shown promising results for small power systems, but has been unsuccessful in overcoming the curses of dimensionality in state and action spaces required to scale to larger problems. Furthermore, recent RL methods which have been used to achieve state-of-the-art in challenging domains in AI research \cite{schulman2017proximal, silver2018general, schrittwieser2020mastering} have not yet been exploited to tackle the UC problem. Among these developments are RL approaches which use traditional planning methods such as tree search to enable lookahead strategies \cite{anthony2017thinking, silver2018general, schrittwieser2020mastering}. Combining RL with tree search to tackle the UC problem is the subject of this thesis, and forms the basis of the methodology developed in Chapter \ref{ch1} and Chapter \ref{ch2}.

Sections \ref{methodology:background}--\ref{methodology:pg} provides a review of RL material relevant to this thesis, with a particular focus on policy gradient methods which form the basis of the RL methods used in this thesis. In Chapter \ref{ch1}, we will introduce guided tree search, which combines policy gradient RL with tree search algorithms. Sections \ref{methodology:tree_search}--\ref{methodology:tree_search_algos} provide a background to tree search and describe in detail the tree search algorithms uniform-cost search, A* search, and iterative deepening methods. These algorithms are used in the guided tree search framework in Chapters \ref{ch1}--\ref{ch3}. Finally, Section \ref{methodology:mathop} provides a review of traditional mathematical optimisation techniques for power systems. We describe mixed-integer linear programming for the UC problem and the lambda-iteration method for economic dispatch, which are used to develop the power system environment and benchmark solutions in Section \ref{ch1:env:benchmarks}. 

\section{Background to Reinforcement Learning} \label{methodology:background}

This section provides a background to RL, reviewing the theory necessary for the original guided tree search methods developed in this thesis. For a comprehensive introduction to RL, we refer the reader to \cite{sutton1998introduction}. 

Along with supervised and unsupervised learning, RL is one of the three main paradigms of machine learning. RL is the task of learning, by trial-and-error, how to act in order to maximise a numerical reward signal \cite{sutton1998introduction}. This type of goal-oriented learning differs from supervised learning, which relies on labelled data for regression and classification tasks, and unsupervised learning, which aims to discover patterns in unlabelled data. In this section we will describe the general principles of RL, beginning with the concept of an \emph{agent} interacting with an \emph{environment}. 

\subsection{Agent-Environment Interaction} \label{methodology:background:agent_environment}

The problem of RL is often informally described in terms of a decision-making \emph{agent} interacting with its surroundings, the \emph{environment}. The agent interacts with the environment through \emph{actions}, which it chooses based on contextual information about the environment's \emph{state}. The agent's actions can have an impact on the state, akin to decisions in the real world. Each time the agent acts, it receives feedback known as a \emph{reward}, and a observation of the new state. These steps form a routine that is repeated in a discrete-time process: the agent observes a state, chooses an action, receives a reward, observes a new state, and so on. Each iteration of the (state, action, reward) routine is known as a \emph{timestep}. The agent's goal is to learn a behavioural strategy called a \emph{policy} that will maximise the sum of future rewards. The process is more formally described as a Markov Decision Process (MDP) which we cover in Section \ref{methodology:background:mdp}. 

Almost all real-world decision-making processes can be formulated in terms of an agent-environment interaction. Some examples of agent-environment interactions which have been studied in the RL literature include:

\begin{itemize}

\item Chess \cite{silver2018general}: the agent moves pieces (action) based on the board position (state). At the end of the agent registers whether the game was won or lost (reward). 
\item News recommender system \cite{zheng2018drn}: the agent recommends an article (action) based on information about the user (state). The agent is notified whether the user read the article (reward).
\item Microgrid management \cite{kuznetsova2013reinforcement}: the agent controls a battery to meet household electricity demand. A solar panel can be used to charge the battery at no cost, or the battery can draw from the grid. The agent receives a weather forecast (state) and charges and discharges the battery (action), aiming to maximally exploit the solar panel. The agents receives feedback in the form of an electricity cost (reward). 

\end{itemize}

The examples illustrate several features of RL problems. In some problems, such as chess, the agent receives feedback infrequently (i.e. at the end of the game). In these contexts, the agent must learn which moves contributed to the win (or loss), which may include strategic moves early in the game. This is known as the \emph{credit assignment problem}. In the recommender system context, the agent must trade-off recommending articles from topics the agent already knows the user is interested in, with new topics that the agent has rarely recommended before. This is an example of the \emph{exploration-exploitation} trade-off. While there are benefits to exploiting actions that are known to lead to high rewards, the agent will only improve its performance and achieve even higher rewards by exploring new actions. A characteristic of the microgrid management problem is uncertain feedback, making this a case of \emph{decision making under uncertainty}. Since the solar output depends on the weather, which is unpredictable, the agent may be fortunate when the day is sunnier than the forecast predicted, receiving a higher reward than on less sunny days. The agent must accumulate a wide range of experience in order to learn a strategy that maximises its \emph{expected} reward. 

\subsection{Markov Decision Processes} \label{methodology:background:mdp}

As described in Section \ref{methodology:background:agent_environment}, the agent-environment interaction involves the agent taking actions and observing a new state of the environment at each timestep. This can be formalised in a Markov decision process (MDP), which describes sequential decision-making problems of this kind. An MDP is defined by a set of states $\mathcal{S}$, set of actions $\mathcal{A}$, reward function $R(s,a)$, and a transition function $F(s', s, a)$. The transition function describes the dynamics of the environment, giving the probability of transitioning to state $s'$ after having taken action $a$ in state $s$:  $F(s',s,a) = \Pr(S_{t+1} = s' | S_t = s, A_t = a)$. In some cases, a distinction is made between what the agent perceives, the \emph{observation}, and the underlying definition of the environment, the state. MDPs where the agent does not observe the full state are known as partially-observable MDPs (POMDPs). 

An agent operating in the MDP decides which action $a$ to take given a state (or observation) $s$. The function that maps from states to actions is called a policy $\pi(a|s)$:

\begin{equation}
    \pi(a|s) = \Pr(A_t = a | S_t = s)
\end{equation}

As the agent acts following the policy $\pi(a|s)$, the following sequence develops \cite{sutton1998introduction}:

\begin{equation}
    S_0, A_0, R_1, S_1, A_1, R_2, S_2, A_2, \dots
\end{equation}

This sequence continues, until a special state called a \emph{terminal state} is reached, if one exists. Problems which have terminal states (such as checkmate in a game of chess) are known as \emph{episodic}, and each sequence is called an \emph{episode}. In this thesis, we will only consider episodic problems. We refer the reader to \cite{sutton1998introduction} for a discussion of continuing problems.

\subsection{The Objective of RL} \label{methodology:background:objective}

The task of RL is to learn a policy which maximises the sum of the agent's rewards in the long run. More precisely, the objective is to find a policy $\pi(a|s)$, giving the probability of taking action $a$ in state $s$, which will maximise the expected \emph{return} $G_t$, which is the discounted sum of rewards:

\begin{equation}
    G_t = \sum_{k=1}^{\infty} \gamma^{k-1} R_{t+k}
    \label{methodology:eq:return}
\end{equation}

where $0 \leq \gamma \leq 1$ is a discount factor. The discount factor determines the present value of future rewards \cite{sutton1998introduction}. If $\gamma=1$, the return is undiscounted and the agent assigns as much credit to distant rewards as to immediate ones; if $\gamma=0$, the agent only aims to maximise the immediate reward. Discounting is used to tackle the problem of credit assignment described in the context of chess in Section \ref{methodology:background:agent_environment}. The field of RL is concerned with algorithms that gradually improve $\pi(a|s)$ from experience in the environment; that is, sampling states, actions and rewards.

\subsection{Simulation Environments} \label{methodology:background:simulation}



In many contexts, it is not practical to generate enough experience to learn effective policies by taking actions in the real world. As a result, most RL training is conducted in simulation, with the trained policy then deployed in the real environment. Simulation environments are essential tools for training RL agents that can be deployed in the real world. Some environments with simple, deterministic dynamics can be easily simulated (such as board games), while others are much more challenging and may only roughly approximate the real world. In recent years, a wide range of simulation environments have been developed for the purpose of RL research in domains such as biomechanics \cite{todorov2012mujoco}, electricity distribution networks \cite{henry2021gym} and building management \cite{vazquez2019citylearn}. In Section \ref{ch1:env:benchmarks} we develop a simulation environment for the UC problem, enabling the application of RL methods.

\section{Taxonomy of RL Algorithms} \label{methdoology:rl_taxonomy}

Having introduced the problem framework of RL and some key concepts, this section will provide a general taxonomy of RL algorithms. A large number of RL algorithms exist which are suited to different problem domains, and several methods have already been applied to solve small-scale UC problem instances as described in Section \ref{literature:rl}. The choice of RL method is a key decision for our RL-based approach to the UC problem described in Chapter \ref{ch1}. In this section, we will first discuss value-based and policy gradient methods, and the relevant properties of each. Second, we will discuss methods which exploit or learn a model of the environment (model-based), and those which do not (model-free). 

\subsection{Value-Based and Policy Gradient Methods} \label{methodology:rl_taxonomy:value_policy}

Most RL methods can be divided into either value-based or policy gradient methods. Here we will first describe each type of method and then compare their properties, benefits and drawbacks.

\subsubsection{Value-Based Methods}

Many methods for solving the RL problem described in Section \ref{methodology:background:objective} involve estimating a $\emph{value function}$, describing the expected return from a given state or following a state-action pair. Learning value functions is a fundamental component of almost all RL algorithms, including both value-based and policy gradient methods.  

Two types of value functions are used in RL: state-value $V^{\pi}(s)$ and action value $Q^{\pi}(s,a)$. Given a policy $\pi(a|s)$, the state-value function for state $s$ is: 

\begin{equation}
    V^{\pi}(s) = \mathbb{E} [ G_t | S_t = s ]
    \label{methodology:eq:state_value}
\end{equation}

The state-value function gives the expected return $G_t$ (defined in Equation \ref{methodology:eq:return}) when starting in state $s$ and acting according to policy $\pi(a|s)$ thereafter. Similarly, the action-value function for a state-action pair $(s,a)$ is: 

\begin{equation}
    Q^{\pi}(s,a) = \mathbb{E} [ G_t | S_t = s, A_t = a ]
\end{equation}

This is the expected return of taking action $a$ in state $s$ and acting according to $\pi(a|s)$ thereafter. The optimal state-value and action-value functions are often denoted $V^*(s)$ and $Q^*(s,a)$, which are the value functions when $\pi$ is optimal: 

\begin{gather}
    V^*(s) = \max_{\pi} V^{\pi}(s) \\
    Q^*(s,a) = \max_{\pi} Q^{\pi}(s,a)  
\end{gather}

The basis of value-based RL methods is to iteratively improve the value function estimating $Q^{\pi}(s,a)$ or $V^{\pi}(s)$ for a given policy $\pi$, and then adjust the policy to be greedy with respect to the new value function (selecting the action with highest estimated action-value). This is known as $\emph{generalised policy iteration}$ (GPI), and is the basis of Q-learning methods \cite{watkins1992q}. Using Q-learning, the action-value function approaches the optimal $Q^*(s,a)$. Once the optimal action-value function $Q^*(s,a)$ is known, then finding an optimal policy $\pi^*(a|s)$ becomes trivial:

\begin{equation}
    \pi^*(a|s) = \argmax_{a} Q^*(s,a)
    \label{methodology:eq:q_policy}
\end{equation}

Experience (observed states, actions and rewards) is accumulated by the agent through interaction with the environment (as described in Section \ref{methodology:background:simulation}), which is used to update the estimated action-value function $Q^{\pi}(s,a)$ to be consistent with $\pi$, in a step known as policy evaluation. In simple problems, $Q^{\pi}(s,a)$ can be represented in tabular form with an entry for each $(s,a)$ combination. However, this approach is infeasible for most real-world tasks including those with continuous state or action spaces. Deep Q-learning, a widely used RL algorithm, represents the action-value function as a deep neural network (deep Q-network, or DQN) \cite{mnih2013playing}. Variants of this approach such as double Q-learning \cite{van2016deep}, dueling DQN \cite{wang2016dueling} and Rainbow DQN \cite{hessel2018rainbow} have become widely-used RL algorithms.

\subsubsection{Policy Gradient Methods}

Rather than deriving a policy from learned state-values or action-values, policy gradient methods aim to directly map from a numerical representation of the state onto a distribution over actions. The policy is parameterised as a function $\pi_{\theta}(a|s)$ with parameters $\theta$. Commonly, this function is represented by a neural network with weights $\theta$. Based on experience gained through interaction with the environment, policy gradient methods repeatedly estimate the gradient of the following performance measure with respect to the policy parameter $\theta$:

\begin{gather}
J(\theta) = \sum_{s \in \mathcal{S}} \mu^{\pi_{\theta}}(s) V^{\pi_{\theta}}(s)
\label{methodology:eq:performance}
\end{gather}

where $\mu^{\pi_{\theta}}(s)$ is the state probability distribution of the MDP when acting under policy $\pi_{\theta}$, $\mathcal{S}$ is the set of states and $V^{\pi_{\theta}}(s)$ is the state-value function acting under policy $\pi_{\theta}$ from state $s$. Gradient ascent is then used to update the policy to a stronger one. From the state-value function in Equation \ref{methodology:eq:state_value}, the performance measure gives the state probability-weighted expected return of acting under policy $\pi_{\theta}$. As policy gradient methods form the basis of the RL methods applied to the UC problem in Chapters \ref{ch1}--\ref{ch3}, we discuss policy gradient methods in more detail in Section \ref{methodology:pg}. 

\subsubsection{Comparison of Value-Based and Policy Gradient Methods}

The relative merits of value-based and policy gradient methods are problem-specific, and both have been used to achieve state-of-the-art performance in different domains. Policy gradient methods benefit from better convergence properties than many value-based methods, where the policy can change dramatically following an update due to the $\argmax$ evaluation in Equation \ref{methodology:eq:q_policy}. In addition, policy gradients can be used to learn stochastic policies, which value-based methods do not naturally accommodate \cite{sutton1998introduction}. Lastly, policy gradients are generally better suited to domains with large or continuous action spaces, since they do not require the evaluation of Q-values over all actions \cite{degris2012off}. However, policy gradient methods tend to require more samples, whereas value-based approaches can reuse experience more efficiently \cite{schulman2015high, schulman2015trust}. In contexts where experience is expensive to gather either in simulation or in the real-world, value-based methods may be more appropriate. Furthermore, there are practical challenges in reliably estimating the gradient of the performance measure $J(\theta)$, and policy gradient methods can converge prematurely to local optima. 

In practice, there is not always a clear distinction between value-based and policy gradient methods. The vast majority of policy gradient algorithms involve simultaneously learning a value function to stabilise training, the so-called actor-critic approach which we describe in Section \ref{methodology:pg:ac}. Other methods such as deep deterministic policy gradients (DDPG) \cite{silver2014deterministic} and twin-delayed DDPG (TD3) \cite{fujimoto2018addressing} combine Q-learning with policy gradients, allowing value-based methods to be applied in continuous action spaces.

\subsection{Model-Based and Model-Free Methods} \label{methdoology:rl_taxonomy:model_based_model_free}

An important dichotomy exists between model-based RL algorithms which exploit a learned or pre-determined model of the environment's dynamics versus model-free algorithms which do not. In model-based methods, transitions can be sampled using the model in order to estimate the outcome of actions before they are taken. This brings two advantages. First, it improves \emph{sample efficiency} - the number of interactions required in the environment in order to reach a given performance level - which can be particularly beneficial when access to the real environment is limited. Second, planning methods such as tree search can in some contexts be applied when a model is available, enabling greater action precision and benefits of foresight in decision-making \cite{schrittwieser2020mastering}. By employing planning, model-based methods can offer greater levels of interpretability and robustness \cite{dulac2019challenges}. However, since model-free methods do not require an environment model, they can be applied more generally than model-based ones and have been the subject of the majority of recent RL research. Nevertheless, model-based methods are the state of the art in several domains where they have been applied, outperforming model-free methods \cite{silver2018general, schrittwieser2020mastering, oh2017value, silver2017predictron}. 

In model-based methods, the environment dynamics may be learned or pre-determined offline based on knowledge of the problem. Due to the complexity of many problems, learning an environment model is often prohibitively difficult or computationally expensive, such as for MDPs with high dimensional state spaces. However, in simple instances, model-based algorithms such as Dyna \cite{sutton1991dyna} can be used which simultaneously learn reward and transition functions and exploit these to simulate additional experience without interacting with the environment. In addition, recent research has found that key features of complex environments such as the widely-studied Atari benchmarks problems can be learned using deep learning methods in order to apply model-based methods \cite{schrittwieser2020mastering, kaiser2019model}. Model-based approaches such as MuZero \cite{schrittwieser2020mastering} have been shown to significantly outperform state-of-the-art model-free methods and are substantially more sample efficient \cite{kaiser2019model}.

Aside from learning the environment dynamics, it is also possible in some instances to pre-determine the environment dynamics offline. This approach has been applied in simple games-playing domains such as Go, where the transition function is cheap to evaluate and is known a priori. The AlphaGo \cite{silver2016mastering} and AlphaZero \cite{silver2018general} algorithms used this approach to achieve state-of-the-art performance in the game of Go. A similar approach was also applied in \cite{anthony2017thinking}, which was used in the game Hex. 

A disadvantage of model-based compared with model-free methods is the additional computation required to evaluate the environment model \cite{moerland2020model}. In cases where the models are learned, this may be relatively inexpensive. However, in some cases where complex physical models are used (such as electricity network management) evaluating the environment model may add significant computational expense that partially or completely offsets the improved sample efficiency of model-based over model-free methods.

While RL research has focused largely on model-free methods due to their more general application, significant milestones in AI have been reached by exploiting knowledge of the model dynamics through tree search, most notably in games-playing domains \cite{schrittwieser2020mastering, silver2016mastering, silver2018general}. In Section \ref{ch1:guided_ucs} we show that exploiting model dynamics in guided tree search also improves on a purely model-free approach for the UC problem. Our methodology utilises model-free policy gradient methods for training, and model-based tree search methods in evaluation. In the follow section, we provide a background to policy gradient RL. 

\section{Policy Gradient Methods} \label{methodology:pg}

Section \ref{methodology:rl_taxonomy:value_policy} introduced value-based and policy gradient methods as two classes of RL algorithms. In Chapter \ref{ch1} we will develop a novel approach to solving UC problems based on the widely-used policy gradient method, proximal policy optimisation (PPO). This section will provide a background to policy gradient methods in general and a detailed description of PPO. 

\subsection{Estimating the Policy Gradient} 

Equation \ref{methodology:eq:performance} defined the performance measure $J(\theta)$ which policy gradient methods aim to improve. $J(\theta)$ is the sum of state-values over all states, weighted by their state-probability when acting under a policy with parameters $\theta$. The improvements are made by maximising the performance measure $J(\theta)$ using stochastic gradient ascent:

\begin{equation}
    \theta_{t+1} = \theta_t + \alpha \nabla J(\theta_t)
    \label{methodology:eq:pg_gradient_ascent}
\end{equation}

where $\nabla J(\theta_t)$ is the gradient of the performance measure with respect to the parameters $\theta$; i.e. the policy gradient. 

We will begin this section by describing the policy gradient theorem, which defines how $\nabla J(\theta)$ can be approximated using samples of states, actions and rewards observed through experience in the environment. The policy gradient theorem analytically expresses the gradient of the performance measure $J(\theta)$ with respect to $\theta$, and is fundamental to policy gradient RL methods. It states:

\begin{align}
    \nabla J(\theta) & \propto \sum_s \mu^{\pi_{\theta}}(s_t) \sum_a Q^{\pi_{\theta}}(s_t,a_t) \nabla \pi_{\theta}(a|s) 
    \label{methodology:eq:pg_propto}
    \\ 
    & = \mathbb{E}_{\pi_{\theta}} \big[ G_t \nabla \log \pi_{\theta}(a_t|s_t) \big]
    \label{methodology:eq:pg_expected}
\end{align}

as proved in \cite{sutton2000policy}. $\mu^{\pi_{\theta}}(s)$ is the state probability distribution when acting under policy $\pi_{\theta}$ and $G_t$ is the return, as defined in Equation \ref{methodology:eq:return}. Equation \ref{methodology:eq:pg_expected} provides a means of estimating the policy gradient based on samples of states, actions and rewards observed when following policy $\pi_{\theta}$. The expectation is thus taken over states and actions when following $\pi_{\theta}$.

Combining Equation \ref{methodology:eq:pg_gradient_ascent} and Equation \ref{methodology:eq:pg_expected} produces the REINFORCE \cite{williams1992simple} update, which is the simplest policy gradient method: 

\begin{equation}
    \theta_{t+1} = \theta_t + \alpha G_t \nabla \log \pi_{\theta}(a_t|s_t)
    \label{methodology:eq:pg_update}
\end{equation}

To implement REINFORCE, the agent simulates an episode following policy $\pi_\theta(a|s)$, generating a sequence of states actions and rewards: 

\begin{equation}
    S_0, A_0, R_1, \dots , S_{T-1}, A_{T-1}, R_T
\end{equation}

Then for each timestep $t$, the return $G_t$ is calculated using the discounted sum of future rewards defined in Equation \ref{methodology:eq:return}. When $\pi_{\theta}$ is a neural network, the gradient of $\log \pi_{\theta}(a_t|s_t)$ with respect to $\theta$ can be computed efficiently using backpropagation. The update increases the probability of actions which resulted in large returns $G_t$ in a way that is inversely proportional to its original probability. Low probability actions with high returns are therefore promoted most strongly. 

While REINFORCE is the simplest policy gradient method, by using Monte Carlo estimates of the return $G_t$ (i.e. without bootstrapping) it generally has very high variance and poor sample efficiency \cite{sutton1998introduction}. Implementations of REINFORCE and other policy gradient methods often sample a `batch' of multiple episodes and calculate the mean policy gradient over the batch to reduce the variance of updates. In addition, a simple approach to reducing the variance of updates is to learn an approximation of $V^{\pi}(s_t)$ (known as a baseline), and replace $G_t$ in the REINFORCE update (Equation \ref{methodology:eq:pg_expected}) with $G_t - V^{\pi}(s_t)$. Intuitively, $G_t - V^{\pi}(s_t)$ is an estimate of the quality of the action trajectory relative to what was expected. REINFORCE with baseline is still far from the state-of-the-art and has relatively high variance compared to more advanced actor-critic methods that use bootstrapping \cite{sutton1998introduction}.

\subsection{Actor-Critic Methods} \label{methodology:pg:ac}

Several policy gradient expressions that are analogous to Equation \ref{methodology:eq:pg_expected} have been proposed which achieve better training performance than REINFORCE or REINFORCE with baseline. These policy gradient expressions are of the form \cite{schulman2015high}:

\begin{equation}
    \nabla J(\theta) = \mathbb{E}_{\pi} \big[ \Psi_t \nabla \log \pi_{\theta}(a_t|s_t) \big]
    \label{methodology:eq:pg_general}
\end{equation} 

where $\Psi_t = G_t$ for REINFORCE and $\Psi_t = G_t - V^{\pi}(s_t)$ for REINFORCE with baseline. Those which involve learning a value function and use bootstrapping (where value estimates are based on the value estimates of future states) are known as \emph{actor-critic} methods, where the learned value function (critic) is used to calculate $\Psi_t$. The actor-critic framework forms the basis of a wide range of state-of-the-art policy gradient methods including soft-actor critic (SAC) \cite{haarnoja2018soft}, A3C \cite{mnih2016asynchronous}, trust region policy optimisation (TRPO) \cite{schulman2015trust}, and proximal policy optimisation (PPO) \cite{schulman2017proximal} which is used in our research and discussed in detail in Section \ref{methodology:pg:ppo}. 

One common policy gradient expression in the form of Equation 
\ref{methodology:eq:pg_general} uses an \emph{advantage function} $\Psi_t = A^{\pi}(s_t,a_t)$, which (similar to REINFORCE with baseline) quantifies the value of taking action $a_t$ relative to other actions from state $s_t$. The advantage function can be formulated in terms of action-value and state-value functions: 

\begin{equation}
        A^{\pi}(s_t,a_t) = Q^{\pi}(s_t,a_t) - V^{\pi}(s_t)
\end{equation}

Using the advantage function, the policy gradient encourages better-than-average actions (those having $A^{\pi}(s_t,a_t) > 0$). However, action-value and state-value functions are not easily known and must be approximated, often by a neural network. Since $Q(s_t,a_t) = R_{t+1} + \gamma V(s_{t+1})$, the advantage is often expressed in terms of the one-step return (i.e. the immediate reward) and the state-value function: 

\begin{equation}
    A^{\pi}(s_t,a_t) = R_{t+1} + \gamma V^{\pi}(s_{t+1}) - V^{\pi}(s_t)
    \label{methodology:eq:onestep_advantage}
\end{equation}

The learned state-value function is known as the \emph{critic}, which is used to estimate the advantage function. Note that $A^{\pi}(s_t,a_t)$ involves bootstrapping: the advantage is calculated by summing the one-step return and the value function for the following state, rather than the full return $G_t$ used in REINFORCE with baseline. By learning a value function which is used to bootstrap, using the advantage function in Equation \ref{methodology:eq:onestep_advantage} gives an actor-critic method, whereas REINFORCE with baseline (which does not use a value function (critic) for bootstrapping) is not. 

\subsubsection{Multi-step Advantage Estimation}

Equation \ref{methodology:eq:onestep_advantage} gives the one-step advantage of choosing $a_t$ over other actions from state $s_t$. That is, we use the immediate reward $R_{t+1}$ and the state-value function of the following state $V^{\pi} (s_{t+1})$. A more general advantage uses the $n$-step return, $G_{t:t+n}$:

\begin{gather}
    A^{\pi}(s_t,a_t) = G_{t:t+n} - V^{\pi}(s_t) \label{methodology:eq:nstep_advantage} \\ 
    G_{t:t+n} = \sum_{l=1}^n \gamma^{l-1} R_{t+l} + \gamma^n V^{\pi}(s_{t+n}) 
\end{gather}

With $n$-step return we take the discounted sum of the next $n$ observed rewards, and bootstrap the remaining steps of the episode using the state-value function $V^{\pi}(s_{t+n})$. Using the $n$-step return to calculate the advantage function as in Equation \ref{methodology:eq:nstep_advantage} and varying $n$ can be used as a way to trade-off bias and variance. Setting $n = \infty$ yields the REINFORCE with baseline update, which has low bias and high variance. With $n=1$, updates have high bias and low variance. 

\subsubsection{Generalised Advantage Estimation}

Instead of using the $n$-step return to calculate the advantage as in Equation \ref{methodology:eq:nstep_advantage}, it is common to use generalised advantage estimation (GAE) \cite{schulman2015high}. GAE uses the $\lambda$-return $G_t^{\lambda}$ \cite{sutton1998introduction}, which is an exponentially weighted average of $n$-step returns: 

\begin{equation}
    G_t^{\lambda} = (1-\lambda) \sum_{n=1}^{\infty} \lambda^{n-1}G_{t:t+n}
\end{equation}

The GAE is calculated by comparing the $\lambda$-return to the learned value function: 

\begin{gather} \label{methodology:eq:gae}
  A(s_t,a_t) = G_t^{\lambda} - V(s_t)
\end{gather}

GAE has been shown to result in more stable training than calculating the advantage using the $n$-step return \cite{schulman2015high} and is currently the standard technique for advantage estimation \cite{berner2019dota}.

\subsection{Proximal Policy Optimisation} \label{methodology:pg:ppo}

We will now introduce proximal policy optimisation (PPO) \cite{schulman2017proximal}, one of the most widely-used actor-critic algorithms. We use PPO in order to train the guided tree search algorithms developed in Chapters \ref{ch1} and \ref{ch2}. PPO prevents large policy updates from drastically impacting policy performance, a problem which has been observed in other policy gradient methods such as A3C \cite{mnih2016asynchronous}. For power system applications, high variance, unpredictable convergence during training can have substantial real-world impacts on system reliability.

As described in Section \ref{methodology:rl_taxonomy:value_policy}, policy gradient methods are generally less sample efficient than value-based methods. Many policy gradient methods improve sample efficiency by performing multiple gradient updates using Equation \ref{methodology:eq:pg_update} on a single batch of data. However, the performance $J(\theta)$ can be sensitive to small changes in the policy parameters $\theta$, and multiple updates can therefore destabilise training and lead to catastrophic performance losses in practice \cite{schulman2017proximal}. Trust region policy optimisation (TRPO) and proximal policy optimisation (PPO) are two state-of-the-art actor-critic algorithms which tackle this problem by enforcing a constraint on updates to ensure that the updated policy remains similar to the original. As a result, the performance is more resilient to multiple policy gradient steps.

TRPO enforces a constraint based on the Kullback-Leibler (KL) divergence between the original policy $\pi_{\theta_{\text{old}}}$ and the updated policy $\pi_{\theta}$. TRPO aims to maximise the surrogate objective:

\begin{equation}
    J^{\text{TRPO}}(\theta)  = \mathbb{E_{\pi_{\theta}}}[r_t(\theta) A_t]
    \label{methodology:eq:trpo_objective}
\end{equation}

subject to (\emph{trust region} constraint): 

\begin{equation}
    \mathbb{E} [\text{KL}(\pi_{\theta_{\text{old}}}, \pi_{\theta}) ] \leq \delta
\end{equation}

where $\delta$ is a constant tolerance parameter defining the size of the trust region. For brevity, the advantage function is written as $A_t = A(s_t, a_t)$. $r_t$ is the probability ratio between the two policies:

\begin{equation}
    r_t(\theta) = \frac{\pi_{\theta}(a_t | s_t)}{\pi_{\theta_{\text{old}}}(a_t | s_t)} 
\end{equation}

As a result, the TRPO objective in Equation \ref{methodology:eq:trpo_objective} measures how the new policy $\pi_{\theta}$ performs relative to the old policy $\theta_{\text{old}}$. As in Equation \ref{methodology:eq:pg_expected}, the expectation is taken over states and actions when following $\pi_{\theta}$.

Based on a similar principle, proximal policy optimisation (PPO) \cite{schulman2017proximal} was proposed with a similar goal of preventing catastrophic performance decreases while allowing for multiple updates to be performed on a single batch of transitions. The objective function for PPO is: 

\begin{equation}
    J^\text{PPO}(\theta) = \mathbb{E}[\min (r_t(\theta) A_t, \text{clip} (r_t(\theta), 1 - \epsilon, 1 +\epsilon) A_t) ]
\label{methodology:eq:ppo_objective}
\end{equation}

The PPO objective takes the minimum of unclipped and clipped objective functions. The unclipped objective  is identical to the TRPO objective (Equation \ref{methodology:eq:trpo_objective}). The clipped objective truncates $r_t(\theta)$ to be between $[1-\epsilon, 1+\epsilon]$. By taking the minimum of these two terms, PPO ensures that the update is `pessimistic'. That is, when $A_t > 0$, the probability ratio is clipped at $1+\epsilon$, while when $A_t < 0$, $r_t$ is clipped at $1-\epsilon$, thus preventing large, greedy updates. 

The PPO implementation is simpler and more general than TRPO, allowing for parameters to be shared between actor and critic networks. Empirically, it has been shown to perform as least as well as TRPO on a wide range of tasks \cite{schulman2017proximal} and has become a very widely used, state-of-the-art RL algorithm \cite{akkaya2019solving, heess2017emergence, berner2019dota, ecoffet2021first, mirhoseini2021graph}.

\subsection{Entropy Regularisation}

Policy optimisation often suffers from premature convergence to local optima with additional techniques to promote exploration. As a result, it is common to include an additional term to actor-critic objective functions to promote higher entropy policies. For a policy $\pi$, the entropy is defined as:

\begin{equation}
    H(\pi) = \mathbb{E}_{a \sim \pi (\cdot | s)}[-\log \pi(a|s)]
\end{equation}

\begin{figure}
    \centering
    \includegraphics[width=0.7\textwidth]{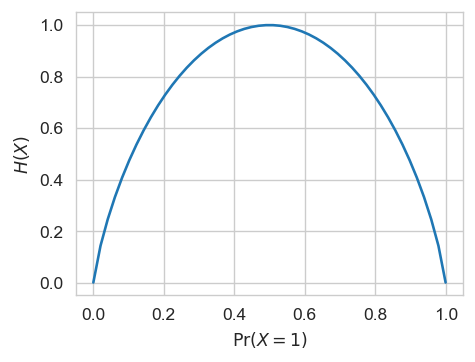}
    \caption{Entropy $H(X)$ of a Bernoulli random variable (i.e. a coin flip). Entropy is largest when $X$ is fair $\Pr(X=1)=0.5$, and decreases as $X$ becomes more biased.}
    \label{methodology:fig:binary_entropy}
\end{figure}

An additive term that is proportional to $H(\pi)$ may be included in any actor-critic objective function of the form described in Equation \ref{methodology:eq:pg_general}. The entropy regularised objective function for PPO is: 

\begin{equation}
    J^{\text{PPO}+H} = \mathbb{E}[J^{\text{PPO}}(\theta) + \beta H(\pi_{\theta})]
    \label{methodology:eq:ppo_entropy_objective}
\end{equation}

where $\beta$ is a constant parameter controlling the amount of entropy regularisation. The entropy of a random variable $X$ is defined as: 

\begin{equation}
    H(X) = - \sum_i^{n}(P(x_i)) \log P(x_i)
    \label{methodology:eq:entropy}
\end{equation}

where $P(x_i) = \Pr{X = x_i}$. The entropy function of a Bernoulli random variable (i.e. a coin flip) $X$ is shown in Figure \ref{methodology:fig:binary_entropy}. An unbiased coin (that is, a Bernoulli random variable with $\Pr(X=0)=\Pr(X=1)=0.5$) has maximum entropy, while a biased coin has lower entropy. Entropy regularisation in Equation \ref{methodology:eq:ppo_entropy_objective} promotes policies which more evenly distributed policy mass across actions and is generally used to encourage more stochastic behaviour in training, preventing the agent from converging prematurely to a local optimum \cite{mnih2016asynchronous}.

In Section \ref{ch1:guided_ucs:policy_training} we use PPO with entropy regularisation to train an RL agent to solve the UC problem. We then combine the trained agent with tree search methods, introduced in the next section, to produce `guided tree search'. 

\section{Background to Tree Search} \label{methodology:tree_search} 

Sections \ref{methodology:background}--\ref{methodology:pg} focused on RL, which is used to solve sequential decision-making problems formulated as MDPs. In particular, we focused on model-free policy gradient methods in Section \ref{methodology:pg} which are used to train the guided tree search algorithms developed in later chapters. In this section, we will focus on tree search, a class of model-based planning algorithms for decision-making and a key component of guided tree search. Rather than using trial-and-error as in RL, tree search uses one-shot decision-making and does not learn from prior experience. However, the fields of tree search and RL are closely related and have been combined in powerful algorithms such as AlphaGo \cite{silver2016mastering, silver2018general} and MuZero \cite{schrittwieser2020mastering}. In the context of the UC problem, exploiting a model via tree search methods enable the agent to evaluate the outcome of actions or action sequences with respect to a model during decision-making, allowing for more robust planning. We will discuss the theory relevant to this thesis and focus on the tree search algorithms employed in Chapters \ref{ch1}--\ref{ch3}. For a more comprehensive introduction to the broader field of tree search, we refer the reader to Chapters 3 and 4 of \cite{russellnorvig}.


\subsection{Definitions} \label{methodology:tree_search:definitions} 

Tree search problems are concerned with finding the lowest cost path from an initial node to a goal node \cite{luger2005artificial}. The path is a sequence of edges taken to traverse the tree, equivalent to the actions defined in the MDP setting. Tree search algorithms rely on formulating the problem as a \emph{search tree}, where nodes represent states and edges represent actions. The initial state is called the root node, and is the root of the search tree. A transition function (defined identically to MDPs), determines the state that follows a state-action pair. Crucially, a model of the transition function is always available to the decision-maker in a tree search problem, unlike in model-free RL where the transition function can only be sampled by interacting with the environment. In addition to states, actions and transitions, a tree search problem additionally requires the definition of a goal test function, which can be consulted to determine whether a node is a goal node. Examples of problems which can be formulated as tree search problems include: 

\begin{itemize}
    \item Route-finding \cite{goldberg2005computing}: the objective is to find the fastest route (least cost path) from an initial location (root node) to a destination (goal node). 
    \item Cargo loading (knapsack problem) \cite{frieze1976shortest}: beginning with an inventory of items with set value and weight (root node), load items until the total weight of the loaded cargo is as large as possible (goal node). The objective is to maximise the value of the cargo loaded (least cost path). 
    \item Rubik's Cube \cite{korf2001time}: beginning with a scrambled initial configuration of the Rubik's Cube (root node), twist the faces until all of the faces have a uniform colour (goal node). The optimal solution is the one which uses the least number of twists (least cost path). 
\end{itemize}

\begin{figure}
    \centering
    \includegraphics[width=0.7\textwidth]{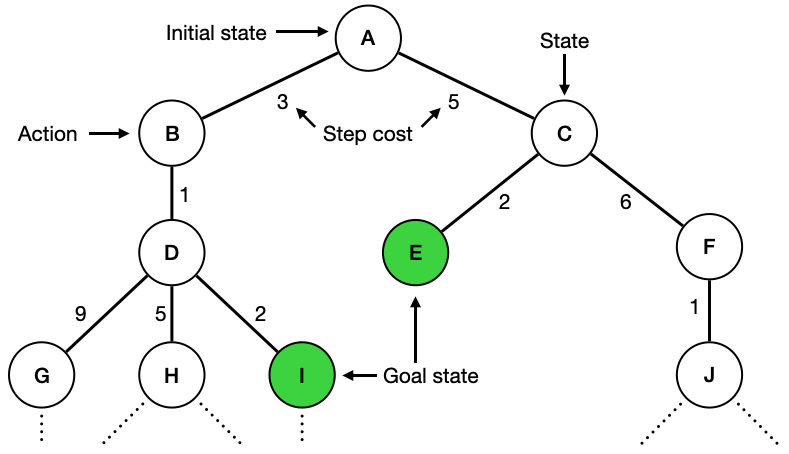}
    \caption{Example search tree. Nodes represent states, edges represent actions, numeric values represent costs. Dotted lines represent further branches that have not yet been added to the tree. The lowest cost path from the root node to a goal node takes the following branches, with corresponding step costs: [left (3), middle (1), right (2)].}
    \label{methodology:fig:example_search_tree}
\end{figure}

An example of a generic search tree is shown in Figure \ref{methodology:fig:example_search_tree}. The green nodes represent goal nodes, and the numerical values indicate step costs. Leaf nodes are those which have no children, and the node at the top of tree (that which has no parents) is the root node. In Figure \ref{methodology:fig:example_search_tree}, dotted lines indicate that the search tree is incomplete and further actions are available. The lowest cost path in the example search tree takes the following branches: [left (3), middle (1), right (2)], which has a total path cost of 6. Note that by convention, tree search generally considers costs of transitioning between states via actions, while MDPs uses rewards which are equivalent to negative costs. 

A tree is a special type of directed acyclic graph (DAG), one where each node has each exactly one parent. As a result, the path from the root to any other node is unique and can be found by following the path from child to parent, beginning at the destination node. A possible brute force solution to a tree search problem is therefore to enumerate all nodes, consult the goal test function to determine the goal nodes, and compute the lowest cost path that leads to a goal nodes. However, many problems contain an extremely large number of states, making this approach intractable. Tree search algorithms, like those described in Section \ref{methodology:tree_search_algos}, build a search tree beginning at the root in a principled way to efficiently find the optimal solution path, visiting as few nodes as possible.

Although the field of tree search is independent from RL, some MDPs used to represent problems for RL can be reformulated as a search tree and solved with tree search methods. Next, we discuss how MDPs can be formulated as search trees, allowing RL, tree search and hybrid methods to be applied.



\subsection{MDPs as Search Trees}

This thesis focuses on hybrid RL and tree search methods to solve the UC problem. As introduced in Section \ref{methodology:background:mdp}, RL requires an MDP formulation of the problem, while tree search methods require a search tree. The translation of MDPs to search trees has been employed in the games-playing literature, notably in \cite{anthony2017thinking, silver2016mastering, silver2018general, schrittwieser2020mastering}, which use Monte Carlo tree search to solve games such as Go. Similarly, we formulate the UC problem as an MDP in Section \ref{ch1:uc_mdp} and as a search tree in Section \ref{ch1:methodology:search_tree}. 

As in previous work \cite{anthony2017thinking, silver2016mastering, silver2018general, schrittwieser2020mastering}, we focus on MDPs with deterministic transition functions, and discrete states which can be represented by search trees of the form visualised in Figure \ref{ch1:fig:search_tree}. We discuss the deterministic representation of a search tree for the UC problem in Section \ref{ch1:methodology:search_tree}. States and actions can be mapped directly on to nodes and states in the search tree context under these circumstances, with terminal states represented by goal nodes. MDPs with stochastic transition functions can be represented by And/Or trees, which are solved with a different class of methods. We refer the reader to \cite{luger2005artificial} for a description of And/Or trees and solution methods. 

A further requirement of the tree search methods used in this thesis, which we describe in Section \ref{methodology:tree_search_algos} is for non-negative step costs, or non-positive rewards. This constraints is naturally satisfied in the UC problem context, which is typically represented as a cost minimisation problem such as in the deterministic, stochastic and robust formulations given in Chapter \ref{literature}. Methods for finding least cost paths in search trees with negative step costs include the Bellman-Ford algorithm \cite{bang2008digraphs}, which is typically slower than methods which assume non-negative step costs. 

The search tree representation of the UC problem is relied upon in later chapters to solve the UC problem with tree search methods. In the following section, we describe the taxonomy of methods used to solve tree search problems. 

\subsection{Taxonomy of Tree Search Methods} \label{methodology:tree_search:taxonomy}

Numerous tree search methods exist which trade-off generalisability across problem domains, optimality guarantees, search efficiency and other factors \cite{hart1968formal, dijkstra1959note, korf1985depth, korf1990real, kocsis2006bandit}. As a result, algorithms have benefits and drawbacks that depend on characteristics of the task. Tree search methods can be classified along several dimensions, such as their exploitation of domain-specific knowledge; whether they can be used for online decision making; and if they can be interrupted. Here we introduce three classifications which are used to inform the development of guided tree search methods in Chapter \ref{ch1} and Chapter \ref{ch2}. 

\subsubsection{Informed Search}

Informed search methods use domain-specific knowledge about the problem to more efficiently reach a solution. A heuristic function $h(n)$ is used, which approximates the cost of the optimal path from node $n$ to a goal node (sometimes called the \emph{cost-to-go}). Examples of informed tree search methods are greedy best-first search \cite{russellnorvig} and A* search \cite{hart1968formal}, which we describe in detail in Section \ref{methodology:tree_search:a_star}. Some informed search methods have the same optimality guarantees as uninformed methods but may be much more efficient in practice. However, by relying on domain-specific heuristics, informed search algorithms cannot be applied out-of-the-box across different problem domains.

\subsubsection{Real-Time Search}

In some cases, it is not possible for an agent to solve the entire tree search problem offline. This is a problem if the search tree is very large, or if the agent is presented with new information after each action, such that the search tree must be rebuilt from scratch. Real-time search algorithms repeatedly solve limited sub-problems (e.g. depth-limited or time-limited) at each timestep. After an action is taken, the search is run again to a greater depth \cite{russellnorvig}. By limiting the search horizon, real-time search does not carry optimality guarantees of other methods but may be much more practical to implement and can enable the application of tree search to problems with large state spaces.

\subsubsection{Anytime Search}

Anytime (or interruptible) algorithms can be terminated at any point and return a solution, with solution quality improving with run time \cite{dean1988analysis}. Non-anytime algorithms such as uniform-cost search and A* search terminate once a goal node is found. As a result, a solution cannot be retrieved until the algorithm has run to completion, which can be problematic for time-constrained problems. Many tree search methods for games-playing such as Monte Carlo tree search \cite{kocsis2006bandit} and iterative deepening (discussed in Section \ref{methodology:tree_search_algos:iterative_deepening}) are anytime. While anytime approaches may be necessary for real-world time-constrained applications, they do not guarantee optimal solutions.





\section{Tree Search Algorithms} \label{methodology:tree_search_algos}

Having introduced the key concepts and taxonomy of tree search algorithms in Section \ref{methodology:tree_search}, in this section we will provide a detailed description of the tree search algorithms employed in the novel guided tree search algorithms developed in Chapters \ref{ch1} and \ref{ch2}. We will cover uniform-cost search \cite{dijkstra1959note}, an uninformed algorithm; A* search \cite{hart1968formal}, an informed algorithm; and iterative deepening search algorithms \cite{korf1985depth}, a class of algorithms that can be considered anytime search algorithms.

\subsection{Uniform-Cost Search} \label{methodology:tree_search:ucs}

Uniform-cost search (UCS) \cite{dijkstra1959note} is a general-purpose, uninformed search algorithm that can be used to find the shortest path from an initial node to a goal node. Other simple uninformed methods such as breadth-first search (BFS) \cite{moore1959shortest} and depth-first search (DFS) \cite{even2011graph} are applicable only to search trees where all step costs are equal. Therefore, BFS and DFS are appropriate for tree search problems where the objective is to reach a goal node in the least number of actions. By contrast, UCS is also applicable to search trees with non-uniform step costs, which is an essential property of the search tree formulation of the UC problem, which we describe in Section \ref{ch1:methodology:search_tree}. 

UCS is derived from Dijkstra's algorithm \cite{dijkstra1959note}. Dijkstra's algorithm is used to find the shortest path from a root node to \emph{all} other nodes, whereas UCS finds the shortest path to a goal node only, and terminates once this is found. The UCS algorithm is described in Algorithm \ref{methodology:algo:ucs}. UCS relies on a \emph{priority queue} of nodes (sometimes known as the frontier) that orders nodes by their \emph{path cost} $g(n)$, the sum of costs of reaching $n$ from the root. At each iteration of the main loop, the node with lowest path cost is removed from the queue, and the goal test is applied. If the node $n$ is a goal node, the algorithm terminates, returning the path to $n$. Otherwise the node is \emph{expanded} by adding all subsequent child nodes from $n$ to the priority queue. The first goal node that is reached using UCS is guaranteed to return the optimal (i.e. lowest cost) path. Hence, UCS is an \emph{optimal} search algorithm \cite{russellnorvig}.

\begin{algorithm}[t]
\caption{Uniform-cost search algorithm for finding shortest path from root node $r$ to a goal node.}\label{methodology:algo:ucs}
\begin{algorithmic}
\Function{UniformCostSearch}{$r$}
  \State $q \leftarrow$ priority queue containing root node $r$
  \Loop
    \State remove first node from queue and assign to $n$
    \If{$n$ is a goal node}
      \State \Return path to $n$
    \EndIf
    \For{action $a$ available from $n$}
      \State $c \leftarrow$ child node following $a$ from $n$
      \State $g(c) \leftarrow$ cost of path to $c$
      \State add $c$ to priority queue $q$ with cost $g(c)$
    \EndFor
  \EndLoop
\EndFunction
\end{algorithmic}
\end{algorithm}

As an example, we will apply UCS to the search tree in Figure \ref{methodology:fig:example_search_tree}. Table \ref{methodology:tab:ucs_example} shows the selected node $n$ at the beginning of each iteration in Algorithm \ref{methodology:algo:ucs}, and the priority queue $q$ at the end of the iteration. Elements in $q$ are ordered by their path cost (in brackets). The algorithm begins with a priority queue consisting of only the root node $q=\text{[A(0)]}$. At each iteration, the first element of the queue is removed and assigned to $n$. The children of $n$ are added to the tree and to $q$, ordered by their path cost. Once node I is reached, the algorithm terminates as this is a goal node. The path to I is returned: [A $\rightarrow$ B $\rightarrow$ D $\rightarrow$ I], which is the lowest cost path.

\begin{table}[]
    \centering
    \begin{tabular}{lcl}
        \toprule
        \textbf{Iteration} & \textbf{Selected Node $n$} & \multicolumn{1}{c}{\textbf{Queue $q$}} \\
        \midrule
        0 (start) & - & [A(0)] \\ 
        1 & A & [B(3), C(5)] \\ 
        2 & B & [D(4), C(5)] \\ 
        3 & D & [C(5), I(6), H(9), G(13)] \\ 
        4 & C & [I(6), E(7), H(9), F(11), G(13)] \\ 
        5 & \color{green!45!black} \textbf{I} & [E(7), H(9), F(11), G(13)] \\
        \bottomrule
    \end{tabular}
    \caption{Uniform-cost search solution to the search tree in Figure \ref{methodology:fig:example_search_tree}. Each row is an iteration of the main loop in Algorithm \ref{methodology:algo:ucs}. The selected node $n$ corresponds to the node removed from the priority queue $q$ at the beginning of the looped routine. $q$ represents the priority queue at the end of the routine.}
    \label{methodology:tab:ucs_example}
\end{table}

In Section \ref{ch1:guided_ucs} we present Guided UCS, the first guided tree search algorithm presented in this thesis. Guided UCS is based on the UCS algorithm described in this section, using a RL-trained policy to reduce the branching factor of the search tree.

\subsection{A* Search} \label{methodology:tree_search:a_star}

A* search \cite{hart1968formal} is an \emph{informed} search method (see Section \ref{methodology:tree_search:taxonomy}) that uses a problem-specific heuristic to improve the search efficiency. The heuristic $h(n)$ estimates the optimal \emph{cost-to-go} from node $n$ to a goal node $h^*(n)$. A* search is very similar to UCS, except that the priority queue is ordered by $f(n)$, the sum of the path cost $g(n)$ and the heuristic estimate of the cost-to-go $h(n)$: 

\begin{equation}
    f(n) = g(n) + h(n)
\end{equation}

Nodes are expanded in an order that considers both the observed path cost $g(n)$ and the anticipated future costs $h(n)$. By contrast, UCS always chooses the node with the lowest observed path cost $g(n)$ to expand first and is more short-sighted as compared with A*. Note that UCS is a special case of A* search where $h(n) = 0$ for all $n$. Pseudocode for the A* search algorithm is shown in Algorithm \ref{methodology:algo:a_star}.

\begin{algorithm}[t]
\caption{A* search for finding shortest path from root node $r$ to a goal node.}\label{methodology:algo:a_star}
\begin{algorithmic}
\Function{A*Search}{$r$}
  \State $q \leftarrow$ priority queue containing root node $r$
  \Loop
    \State remove first node from queue and assign to $n$
    \If{$n$ is a goal node}
      \State \Return path to $n$
    \EndIf
    \For{action $a$ available from $n$}
      \State $c \leftarrow$ child node following $a$ from $n$
      \State $h(c) \leftarrow$ heuristic estimate of optimal cost-to-go
      \State $g(c) \leftarrow$ cost of path to $c$
      \State add $c$ to priority queue $q$ with cost $g(c) + h(c)$
    \EndFor
  \EndLoop
\EndFunction
\end{algorithmic}
\end{algorithm}

A* search is optimal if the heuristic $h(n)$ is \emph{admissible}, meaning it strictly underestimates the optimal cost-to-go $h^*(n)$ \cite{dechter1985generalized}:

\begin{equation}
    h(n) \leq h^*(n)
    \label{methodology:eq:admissibility}
\end{equation}

With an admissible heuristic, A* search will return the same solution as UCS, but can do so more efficiently if an accurate heuristic is used. To demonstrate this, we will apply A* search to the example search tree in Figure \ref{methodology:fig:example_search_tree}. Table \ref{methodology:tab:a_star} shows the iterations of the main loop in A* search, as we showed for UCS in Table \ref{methodology:tab:ucs_example}. Illustrative values for the estimated cost-to-go $h(n)$ of each node in the search tree are shown as a lookup table. Note that due to the large state space of many problems, a lookup table is not practical and the function $h(n)$ must be evaluated as-and-when it is needed. Compared with UCS, A* search requires one fewer iteration of the main loop, as node C is never expanded, but reaches the same solution. In terms of number of nodes expanded, A* is more efficient. However, an important consideration is the run time of the heuristic itself: if the heuristic function is slow to evaluate, then the run time reduction from fewer node expansions can be cancelled out by running the heuristic function. We discuss the heuristic properties impacting efficiency improvements of A* search relative to UCS in Section \ref{methodology:tree_search:heuristic_properties}. 

\begin{table}[]
    \centering
    \begin{tabular}{lcl | ll}
        \toprule
        \textbf{Iteration} & \textbf{Selected Node $n$} & \multicolumn{1}{c}{\textbf{Queue $q$}} & \multicolumn{2}{c}{\textbf{h($n$)}} \\
        \midrule
        0 (start) & - & A(5) & A: 5 & F: 2 \\ 
        1 & A & B(6), C(7) & B: 3 & G: 3 \\ 
        2 & B & D(5), C(7) & C: 2 & H: 4 \\ 
        3 & D & I(6), C(7), H(13), G(16) & D:1 & I: 0 \\ 
        4 & \color{green!45!black} \textbf{I} & C(7), H(13) G(16) & E: 0 & J: 1 \\ 
        \bottomrule
    \end{tabular}
    \caption{A* search solution to the search tree in Figure \ref{methodology:fig:example_search_tree}, as well as a lookup table for heuristic values $h(n)$. Note that A* search requires one fewer iteration to reach the same optimal solution as UCS (Table \ref{methodology:tab:ucs_example}).}
    \label{methodology:tab:a_star}
\end{table}

In Section \ref{ch2:methodology:a_star}, we present Guided A* search, a novel RL-aided algorithm based on A* search.

\subsection{Iterative Deepening Algorithms} \label{methodology:tree_search_algos:iterative_deepening}

Iterative deepening \cite{korf1985depth} is a general search strategy, whereby a tree search is conducted repeatedly up to a limited horizon, with the horizon increasing at each iteration. At the first iteration, a search is conducted up to a depth of $H=1$. At each subsequent iteration, the depth $H$ is incremented and the search tree is discarded and rebuilt from scratch. The solution proceeds until some stopping criterion (such as a run time limit or based on solution quality) is met. Psuedocode for applying iterative deepening to a generic tree search function $f$ such as UCS or A* search is shown in Algorithm \ref{methodology:algo:iterative_deepening}. By stopping the algorithm according to a time budget, iterative deepening algorithms can be made anytime (Section \ref{methodology:tree_search:taxonomy}). This approach has been widely used to create anytime games playing algorithms \cite{slate1983chess}. Section \ref{ch2:methodology:ida_star} describes Guided IDA* search.

\begin{algorithm}[t]
\caption{General purpose iterative deepening for a tree search function $f$, for a problem with initial state $r$. At each iteration, $f$ is used to solve the search tree up to a depth $H$ and the search horizon is incremented.}\label{methodology:algo:iterative_deepening}
\begin{algorithmic}
\Function{IterativeDeepeningSearch}{$r, f$}
  \State $H \leftarrow $ 1
  \Repeat
    \State solution $\leftarrow f(r, H)$
    \State $H \leftarrow H+1$
  \Until{stopping criterion is met}
  \State \Return solution
\EndFunction
\end{algorithmic}
\end{algorithm}



\subsection{Properties of Heuristics for A* Search} \label{methodology:tree_search:heuristic_properties}

The main characteristic of informed search methods such as A* search described in Section \ref{methodology:tree_search:a_star} is their use of heuristics. The choice of heuristic has a significant impact on performance \cite{sormaz2010comparison} as well as the optimality of informed algorithms due to the admissibility criterion in Equation \ref{methodology:eq:admissibility}. In this section we discuss the properties of effective heuristics for informed search.

Many problems have well-established heuristics. In route planning problems, a common heuristic is the straight-line distance from $n$ to the destination \cite{huang2007shortest}. Expert pattern databases have been used in some problems, such as the Rubik's cube puzzle \cite{korf2001time}. Supervised learning has also been used to learn $h(n)$ for route planning problems \cite{wang2019empowering, ernandes2004likely}. In unstudied domains without established heuristics, heuristic design is an important research topic that can have significantly impact performance in practice. In Section \ref{ch2:heuristics}, we develop problem-specific heuristics for the UC problem in order to apply A* search. 

The effectiveness of heuristics in improving search efficiency of A* search relative to UCS is dependent on the following heuristic properties: 

\paragraph{Run time}

Heuristic run time is the average time taken to evaluate $h(n)$. Complex methods may be used to accurately estimate $h(n)$, but may be impractically slow to calculate. Since informed search methods are used to improve search efficiency by reducing the number of node evaluations required to reach a solution, a large heuristic run time may offset efficiency improvements achieved from fewer node evaluations. Run time is therefore an important heuristic property impacting the efficiency improvements achieved in practice by informed search methods.

\paragraph{Admissibility}

As described in Section \ref{methodology:tree_search:a_star}, heuristic admissibility is a necessary condition for A* search to be optimal \cite{dechter1985generalized}. The admissibility criterion states that the heuristic must not overestimate the optimal cost-to-go $h^*(n)$:

\begin{gather}
    h(n) \leq h^*(n) \quad \forall n
\end{gather}

While admissibility is necessary to guarantee the optimality of A* search, in some contexts an inadmissible heuristic may still be effective in practice if optimal solutions are not required \cite{ernandes2004likely}. This is due to efficiency improvements resulting in lower execution time of A* search, which may be more valuable than higher solution quality in practice. 

\paragraph{Accuracy}

Heuristic accuracy measures how well $h(n)$ is able to approximate $h^*(n)$ \cite{russellnorvig} and is an important factor influencing the efficiency improvement achieved by A* search relative to UCS, its uninformed counterpart. Accuracy is measured by an error metric, such as mean absolute error or root mean squared error. Perfect heuristics where $h(n) = h^*(n)$ are oracles that can be used in A* to immediately find the optimal solution without exploring any sub-optimal sub-trees, yielding maximal efficiency improvement. By contrast, in the case where $h(n) = 0$, A* is equivalent to UCS and no efficiency improvement is achieved. 

\medskip

In conclusion, efficiency improvements achieved in practice by informed search relative to uninformed search depends collectively on all three properties: run time, admissibility and accuracy. Accurate and admissible heuristics can be impractical if the heuristic is slow to compute. Inadmissible heuristics can be effective if accuracy is high and run time is low due to the practical value of lower run times. In designing an effective heuristic, the three properties must usually be traded off and heuristics evaluated experimentally to determine effectiveness in a given use case.

We use the methods described in this section as the basis of guided tree search algorithms developed in this thesis. In the following section, we describe mathematical optimisation methods for the UC problem that are used to benchmark the performance of guided tree search methods.

\section{Mathematical Optimisation for Unit Commitment} \label{methodology:mathop}

RL and tree search covered in Sections \ref{methodology:background}--\ref{methodology:tree_search_algos} form the basis of the original solution methods developed in this thesis. In this section we will cover conventional mathematical optimisation methods to solve the UC problem. We begin by covering priority list methods, heuristic algorithms for UC. In Section \ref{methodology:mathop:milp}, we cover mixed-integer linear programming (MILP), which is the predominant method for solving the UC problem, as discussed in Section \ref{literature:duc:solution_methods}. We use MILP in Section \ref{ch1:env} to produce benchmark solutions to UC problem instances. Finally, in Section \ref{methodology:mathop:ed} we will describe the lambda-iteration for economic dispatch, the task of determining the lowest cost setpoints of generators to satisfy demand. The ED problem is an important component of the power systems simulation environment described in Section \ref{ch1:env}.

\subsection{Priority List} \label{methodology:mathop:pl}

Priority list (PL) methods for the UC problem are based on heuristics and use a simple ordering of generators (known as the PL) to commit generators, typically in order of start cost, fuel cost or capacity, initially ignoring inter-temporal constraints \cite{senjyu2003fast}. We reviewed literature applying PL methods to the UC problem in Section \ref{literature:duc:solution_methods}. The generator with highest priority (i.e. cheapest or largest capacity) is committed first at each time period, and further generators are committed until demand (sometimes with a reserve constraint) is met. Algorithm \ref{methodology:algo:PL} shows this procedure in pseudocode. The drawback of this algorithm is that it does not consider inter-temporal constraints such as minimum up/down time constraints, usually resulting in an infeasible schedule. An illustrative schedule produced by a PL algorithm was shown in Figure \ref{literature:fig:priority_list_diagram}, with generators committed in decreasing priority (blue to red). PL-produced schedules are typically `fixed' to obey constraints using expert rules or heuristics \cite{senjyu2003fast}. 

\begin{algorithm}[t]
\caption{Priority list algorithm for the UC problem with demand forecast $\boldsymbol{D}$ and reserve volumes $\boldsymbol{R}$ ($T$ decision periods), and generators $\mathcal{G}$.}\label{methodology:algo:PL}
\begin{algorithmic}
\Function{PriorityList}{$\boldsymbol{D}, \boldsymbol{R}, \mathcal{G}$}
  \State $\boldsymbol{x} \leftarrow [0]_{N \times T} $ (empty solution matrix)
  \For{$t$ in $\{1..T\}$}
    \State $p \leftarrow$ priority list ordering of generators $\mathcal{G}$
    \While{committed capacity $\leq D_t + R_t$}
        \State remove generator with index $g$ from $p$
        \State $x_{g,t} \leftarrow 1$
    \EndWhile
  \EndFor
  \State \Return $\boldsymbol{x}$
\EndFunction
\end{algorithmic}
\end{algorithm}

As discussed in Section \ref{literature:duc:solution_methods}, PL solutions to the UC problem are quick to calculate, but generally have higher operating costs than more advanced methods such as MILP. In Section \ref{ch2:heuristics}, we develop three PL-based heuristics for the UC problem to deploy in informed guided tree search methods based on A* search. Thanks to their simplicity and short run times, these heuristics are able to provide a rapid estimation of the optimal operating costs for a given power system state, which significantly improves search efficiency when applied in A* search. In the following section, we provide a background to MILP for the UC problem, which we use to benchmark the guided tree search methods applied in this thesis.

\subsection{Mixed-Integer Linear Programming for Unit Commitment} \label{methodology:mathop:milp}

In order to provide high quality benchmark solutions for the UC problem instances used in this thesis, we use mixed-integer linear programming (MILP) to solve the deterministic UC problem. This is the dominant solution method for practical UC applications \cite{knueven2020mixed}. MILP benefits from very efficient solution methods such as branch-and-bound, as well as highly optimised software implementations in commercial solvers such as CPLEX. We will compare MILP solutions to the UC problem with guided tree search in Chapters \ref{ch1}--\ref{ch3}. In Section \ref{ch1:env:benchmarks}, we use MILP to produce benchmark solutions to 20 UC problem instances for power systems of different sizes. This section provides a truncated description of the MILP formulation of the deterministic UC problem that we will use in this thesis. For a comprehensive introduction to MILP, we refer the reader to seminal texts \cite{wolsey1999integer, bertsimas1997introduction}. 

\subsubsection{Unit Commitment Formulation}

In this thesis we formulate the deterministic UC problem as an MILP, using the model given in \cite{morales2013tight}. The extensive review of MILP formulations of the UC problem in \cite{knueven2020mixed} identified this formulation as computationally efficient, capable of solving large-scale problems of 1000s of generators with modest hardware in practical run times. The formulation in Equations \ref{methodology:eq:milp_obj}--\ref{methodology:eq:milp_reserves} uses piecewise linear approximations of the fuel cost curves, which are typically quadratic. A full description of the model is given in \cite{morales2013tight}; here we will provide a truncated version for brevity:

\begin{align}
		& \text{min } \sum_{g \in \mathcal{G}} \sum_{t \in \mathcal{T}} ( c_g(t) + CP_g^1 \, u_g(t) + CS_g \delta(t) ) \label{methodology:eq:milp_obj} 
\end{align}

subject to (see \cite{morales2013tight} for full constraints):
		
\begin{align}
		& \sum_{g \in \mathcal{G}} \left( p_g(t) + \underline{P}_g u_g(t) \right) = D(t) & \quad \forall t \in \mathcal{T} \label{methodology:eq:milp_demand} \\
		& \sum_{g \in \mathcal{G}} r_g(t) \geq R(t) & \quad \forall t \in \mathcal{T} \label{methodology:eq:milp_reserves}
\end{align}

\paragraph{Parameters:}

\begin{itemize}
    \item $g \in \mathcal{G}$: the set of thermal generators 
    \item $t \in \mathcal{T}$: the set of time periods
    \item $\underline{P}_g$: minimum power output of generator $g$
    \item $CP^l_g$: cost of operating at piecewise generation point $l$ for generator $g$
    \item $CS_g$: startup cost for generator $g$ at time $t$
    \item $D(t)$: demand net of wind generation at time $t$
    \item $R(t)$: reserve requirement at time $t$
\end{itemize}

\paragraph{Variables:}

\begin{itemize}
    \item $c_g(t)$: cost of power produced over minimum for generator $g$ at time $t$
    \item $p_g(t)$: power output above minimum for generator $g$ at time $t$
    \item $u_g(t)$: commitment of generator $g$ at time $t$
    \item $\delta(t)$: startup status of generator $g$ at time $t$
    \item $r_g(t)$: reserve provided by generator $g$ at time $t$
\end{itemize}

The dominant solution technique for solving UC models such as that in Equations \ref{methodology:eq:milp_obj}--\ref{methodology:eq:milp_reserves} is branch-and-bound \cite{land1960}, which is implemented in several open-source and commercial solvers. Improvements in MILP solvers such as the use of cutting planes, parallelism and branching heuristics, have contributed to branch-and-bound becoming the dominant solution method for solving the UC problem \cite{knueven2020mixed}. In Section \ref{ch1:env:benchmarks} we use branch-and-bound to produce benchmark solutions to UC problem instances, using the open-source COIN-OR software \cite{johnjforrest_2020_3700700}. A useful property of branch-and-bound is that at any point, a lower bound on the objective function (in the minimisation case) is given. The difference between the objective function and this lower bound is the gap, which is commonly used to define stopping criteria. Once the gap falls below a threshold (e.g. 1\%), the current best solution is returned. However, in many contexts branch-and-bound may take many iterations to find a feasible solution and is therefore not a fully anytime algorithm. 

The MILP formulation of the UC problem simultaneously solves both the UC problem setting the integer decision variables, as well as the economic dispatch (ED) problem setting the real-valued generator setpoints. However, in some contexts, the ED problem is considered as a separate problem, such as in the context of real-time balancing or re-dispatch of generators in response to deviations in demand from forecasts. In the following section, we describe the lambda-iteration method for solving the ED problem.

\subsection{Economic Dispatch with Lambda-Iteration} \label{methodology:mathop:ed}

While the UC problem is concerned with integer-valued commitment decisions, optimising the real-valued power outputs of online generators is known as the economic dispatch (ED) problem \cite{wood2013power}. As can be seen in the MILP formulation in Equations \ref{methodology:eq:milp_obj}--\ref{methodology:eq:milp_reserves}, solutions to the UC problem typically solve both the UC problem and the ED problem simultaneously, since the objective function depends on the power outputs. However, in practice, the ED problem is often solved independently of the UC problem in the real-time operation of power systems. Generator dispatch may be required to change throughout the day in order to meet deviations in demand or renewables generation from forecasts, or other contingencies such as generator outages. The ED problem is an important component of the power system simulation environment described in Section \ref{ch1:env} as it is used to determine the total fuel costs of generators for a given commitment decision under realisations of demand and wind generation.  

The ED problem is concerned with finding the lowest-cost dispatch of online generators to meet a given demand, subject to generator operating limits. Mathematically, the ED problem ignoring ramping constraints and transmission losses can be described as follows: 

\begin{align}
    \text{minimise } \qquad & F = \sum_{i \in \mathcal{G}} F_i(p_i) \\ 
    \text{subject to}\qquad & \phi = D - \sum_{i \in \mathcal{G}} p_i = 0  
    \label{methodology:eq:lambda_iteration_demand_constraint} \\
    & p_i^{\text{min}} \leq p_i \leq p_i^{\text{max}}
    \label{methodology:eq:lambda_iteration_power_constraint}
\end{align}

where $\mathcal{G}$ is the set of online (committed) generators; $F_i$ is the fuel cost function for generator $i$; $p_i$ is the power output; $p_i^{\text{min}}$ and $p_i^{\text{max}}$ are minimum and maximum operating limits and $D$ is the demand. 

Whereas the UC problem is NP-hard, the ED problem is typically formulated as a constrained convex optimisation problem and can be solved quickly with a wide range of methods \cite{chowdhury1990review}. A common numerical method for solving the ED problem is the lambda-iteration method. In this section, we will describe the lambda-iteration method for solving the economic dispatch as described in \cite{wood2013power}. We consider quadratic fuel cost curves of the form:

\begin{equation}
    F_i(p_i) = a_i p_i ^2 + b_i p_i + c_i
    \label{methodology:eq:quadratic_fuel_cost}
\end{equation}

where $a_i, b_i, c_i$ are constant coefficients. Fuel costs are often modelled using quadratics \cite{kazarlis1996genetic}, and we use this model in our UC problem setting in Section \ref{ch1:env}. 

The lambda-iteration method begins by constructing a Lagrange function:

\begin{equation}
    \mathcal{L} = F + \lambda \phi
    \label{methodology:eq:lagrange_equation}
\end{equation}

where $\lambda$ is a Lagrange multiplier and $\phi$ is the load balance equality in Equation \ref{methodology:eq:lambda_iteration_demand_constraint}. Finding the stationary points of $\mathcal{L}$ with respect to the variables $p_i$ and the Lagrange multiplier is equivalent to finding the extreme of the objective function $F$ while observing the constraint $\phi$ \cite{beavis1990optimisation}. Using this fact, taking the partial derivative of $\mathcal{L}$ with respect to $p_i$ yields the following set of equations: 

\begin{gather}
    \frac{\partial \mathcal{L}}{\partial p_i} = \frac{dF_i(p_i)}{dp_i} - \lambda = 0  \\
    \lambda = \frac{dF_i}{dp_i} 
    \label{methodology:eq:lambda_equality}
\end{gather}



To account for the inequality constraints \ref{methodology:eq:lambda_iteration_power_constraint}, the following conditions are added: 

\begin{gather}
    \lambda \geq \frac{dF_i}{dp_i} \quad \text{when } p_i = p_i^{\text{max}}
    \label{methodology:eq:lambda_geq} \\
    \lambda \leq \frac{dF_i}{dp_i} \quad \text{when } p_i = p_i^{\text{min}}
    \label{methodology:eq:lambda_leq}
\end{gather}

Differentiating the quadratic fuel cost curve in Equation \ref{methodology:eq:quadratic_fuel_cost} and substituting into Equation \ref{methodology:eq:lambda_equality}: 

\begin{align}
    \lambda = a_i p_i + b_i \\
    p_i = \frac{\lambda - b_i}{a_i} \label{methodology:eq:p_lambda}
\end{align}

For a given value of $\lambda$, Equation \ref{methodology:eq:p_lambda} can be used to calculate the power output $p_i$ for each generator. Then, using the constraints in Equations \ref{methodology:eq:lambda_geq}, and \ref{methodology:eq:lambda_leq}, $p_i$ violating operating limit constraints in Equation \ref{methodology:eq:lambda_iteration_power_constraint} are updated without loss of optimality. That is, where $p_i \geq p_i^{\text{max}}$, we set $p_i = p_i^{\text{max}}$ and correspondingly for $p_i^{\text{min}}$. For fixed $\lambda$, these powers given by Equation \ref{methodology:eq:p_lambda} will provide the lowest fuel cost, being an extremum of the Lagrange equation (Equation \ref{methodology:eq:lagrange_equation}). However, not all values of $\lambda$ will satisfy the load balance constraint in Equation \ref{methodology:eq:lambda_iteration_demand_constraint}. Since the ED problem is convex, a local minimum is also a global minimum. It follows that the optimal solution to the ED problem can be found by searching for a value of $\lambda$ which yields generator outputs $p_i$ that satisfy the load balance constraint in Equation \ref{methodology:eq:lambda_iteration_demand_constraint}. 

\begin{figure}
    \centering
    \includegraphics[width=0.5\textwidth]{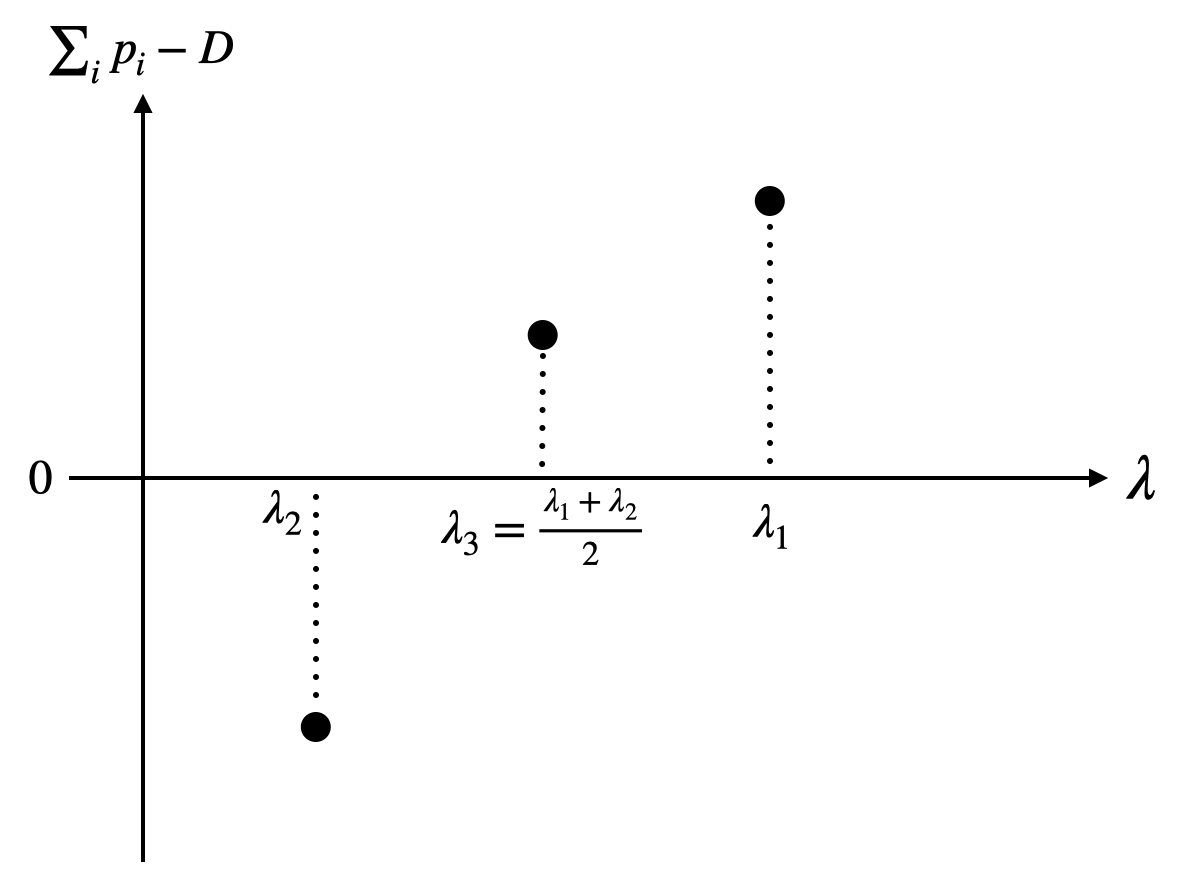}
    \caption{Linear interpolation of $\lambda$ in the lambda-iteration algorithm, adapted from \cite{wood2013power}. The next value $\lambda_4$ can be found by interpolating between $\lambda_2$ and $\lambda_3$. At each iteration, the difference between supply and demand $|\sum_i p_i - D|$ reduces. The algorithm terminates when the difference is below a tolerance $\epsilon$.}
    \label{methodology:fig:lambda_interpolation}
\end{figure}

The lambda-iteration method uses the Newton-Raphson method \cite{suli2003introduction} to search for $\lambda$, terminating when the difference between supply and demand $|\sum_i p_i - D| < \epsilon$, where $\epsilon$ is a small tolerance value (e.g. 1 MW). At each iteration, $\lambda$ is updated by interpolation between upper and lower bounds from previous estimates as shown in Figure \ref{methodology:fig:lambda_interpolation}. The power outputs are calculated using Equation \ref{methodology:eq:p_lambda} and the inequalities in Equations \ref{methodology:eq:lambda_geq} and \ref{methodology:eq:lambda_leq}. If $\sum_i p_i - D > 0$, then $\lambda$ is set to the midpoint of the two smallest values of $\lambda$ (in this case $\lambda_2$ and $\lambda_3$), otherwise the midpoint of the two largest values is used. 

The lambda-iteration is an efficient solution method for the ED problem and is an integral component of the power systems simulation environment described in Section \ref{ch1:env}, used to solve the ED problem for commitment decisions made by the decision-making agent. The resulting fuel costs are then use to evaluate the reward function in the UC MDP described in Section \ref{ch1:uc_mdp}. 

\section{Conclusion} 

This chapter forms the theoretical basis for the material presented in Chapters \ref{ch1}--\ref{ch3}, where we use tree search and RL to develop a scalable solution method for the UC problem. We use the actor-critic method PPO described in Section \ref{methodology:pg:ppo} to train an RL agent to solve the UC problem and use the trained policy to improve the efficiency of the tree search methods described in Section \ref{methodology:tree_search_algos}. Furthermore, throughout Chapters \ref{ch1}--\ref{ch3} we benchmark the RL-based solutions against traditional deterministic approaches using MILP, based on the methods described in Section \ref{methodology:mathop:milp}. The following chapter describes the power system environment used in this thesis and presents Guided UCS, an RL-aided tree search algorithm which is used to solve the UC problem. 

%% file: 04-chapter1/chapter1.tex
\section{Introduction} \label{ch1:introduction}


Real-world UC problems in large central-dispatching power markets may involve 1000s of generators \cite{chen2016improving}, requiring scalable solution methods. Such large problem sizes cannot not be easily solved by the tree search methods described in Section \ref{methodology:tree_search_algos} due to exponential growth of the search tree in the number of generators. As a result, existing tree search methods have been limited to small UC problems of up to 12 generators \cite{dalal2015reinforcement}. Practical challenges also limit the ability of model-free RL to scale to larger power systems, as shown by relatively small-scale studies reviewed in Section \ref{literature:rl}. Recent research has shown that RL can be exploited to significantly improve the efficiency of tree search algorithms, achieving tractability and superior performance in a number of challenging problem domains with large branching factors \cite{silver2016mastering, silver2018general, schrittwieser2020mastering, silver2017predictron, oh2017value}. In this chapter we present a scalable UC solution method using a novel RL-guided tree search algorithm. We show that whereas conventional tree search has exponential run time complexity in the number of generators, solution times are roughly constant in the number of generators when using guided tree search. We find that guided tree search results in negligible increases in operating costs as compared with conventional tree search without RL and outperforms industry-standard deterministic methods based on mixed-integer linear programming (MILP). 


In order to apply RL to the UC problem, this chapter describes an open-source power system simulation environment developed for this research. The UC problem is then formulated as an MDP, which can be simulated using the power system environment, allowing for RL agents to be trained to solve the UC problem by trial-and-error. In this chapter, we apply the tree search algorithm uniform-cost search (UCS), described in Section \ref{methodology:tree_search:ucs}, to the UC problem, evaluating performance in terms of operating costs and run time using the simulation environment. We show experimentally that UCS, while achieving low operating costs, exhibits exponential time complexity in the number of generators, limiting its application to real-world power systems. We then present \emph{Guided UCS}, which uses an `expansion policy' trained by model-free RL to reduce the breadth of the search tree. We show that operating costs do not significantly increase when applying an expansion policy trained by model-free RL despite removing branches, while the run time remains stable in the number of generators. We demonstrate the capability of RL to intelligently reduce the branching factor without degrading solution quality, making our approach scalable to larger power systems than studied in existing literature. Furthermore, Guided UCS is compared with industry-standard deterministic UC approaches, solved with MILP, and shown to achieve lower operating costs and better security of supply.

\subsection{Contributions}

This chapter makes the following contributions:

\begin{enumerate}
    \item The UC problem is formulated as an MDP and realised in an open-source simulation environment. The environment, which is used throughout this thesis, uses real demand and wind data from the GB power system and is described in detail in Section \ref{ch1:env}.
    \item Uniform-cost search (UCS) is used to solve the UC problem and shown to achieve operating costs that are competitive with deterministic MILP benchmarks. To the best of our knowledge, this is the first application of UCS to the UC problem. 
    \item  To improve the run time complexity of UCS in the number of generators, we present Guided UCS, an RL-aided tree search algorithm which uses a guiding `expansion policy' trained by model-free RL to reduce the branching factor.
    \item We train expansion policies using proximal policy optimisation (PPO) and a sequential feed-forward neural network architecture in the simulation environment for systems of between 5--30 generators. We use the trained policies to solve 20 unseen test UC problems model-free. To the best of our knowledge, this is the first application of policy gradient RL to solve the UC problem.
    \item Using Guided UCS to solve 20 unseen test problems, we conduct a parameter analysis, investigating the impact of breadth and depth parameters on run time, operating costs and schedule characteristics.
    \item Guided UCS is compared with UCS and shown to exhibit constant time complexity in the number of generators with no significant degradation of solution quality.
    \item Compared with the MILP benchmarks, Guided UCS is shown to achieve lower operating costs, loss of load probability and exhibit novel operational strategies.
\end{enumerate}

The rest of this chapter is organised as follows. Section \ref{ch1:env} describes the problem setup and power system environment used in this research. In Section \ref{ch1:uc_mdp}, the UC problem is formulated as a Markov Decision Process, suitable for applying RL methods. In Section \ref{ch1:ucs} UCS is described and applied to solve the UC problem. In Section \ref{ch1:guided_ucs}, we describe the RL-aided tree search algorithm, Guided UCS, and train expansion policies with model-free RL. The policies are applied in Section \ref{ch1:experiments}, where Guided UCS is used to the solve the UC problem. We conduct a parameter analysis of guided tree search, and compare performance with UCS and MILP benchmarks. In Section \ref{ch1:discussion} we discuss our findings and Section \ref{ch1:conclusion} concludes the chapter.

\section{Problem Setup \& Simulation Environment} \label{ch1:env}

This section describes the power system simulation environment (henceforth `environment') developed for this research. In contrast with existing power system environments such as PandaPower \cite{thurner2018pandapower}, ours is specifically designed for the UC problem and models a simplified, single-bus model of a power system, in line with most existing UC research. This comes at the cost of less accurate representation of transmission-related impacts of UC, such as transmission contingencies, reactive power support and line overloadings. The benefit of the single-bus model is that the computational cost of evaluating the environment is much lower. Our decision to neglect transmission constraints is in line with most existing research and industry practice, where post-solve AC load flow analyses are conducted to determine the feasibility and security of a UC solution \cite{knueven2020mixed}. 

The environment models stochastic demand and wind and can be used to generate scenarios reflecting the uncertainty and variability of real world power systems, making it suitable for stochastic UC research. The main function of the environment is to simulate the dispatch of generators and resulting operating costs given UC decisions inputted by the user (or RL agent), under scenarios of demand and wind generation. The simulation environment is available as an open-source Python package\footnote{\url{https://github.com/pwdemars/rl4uc}}. 

The cost-minimising UC problem represented in the environment is reflective of centrally-dispatching power markets such as those of North America. By contrast, GB's self-dispatching market structure means that no central operator solves a UC problem to determine the commitment of all generators. However, cost-minimising UC remains the most widely-studied setup, and the environment and solution methods presented in this thesis can be modified for profit-maximising problem setups \cite{abdi2021profit}.

This section begins by describing the main routine executed in the environment. We then describe the data used to realise the environment, defining generator specifications and forecasts for demand and wind. Lastly, the test problems used to evaluate the relative performance of UC methods are shown, along with a description of benchmark solutions to these problems using MILP. 

\subsection{Overview} \label{ch1:env:overview}

\begin{figure}
    \centering
    \includegraphics[width=\textwidth]{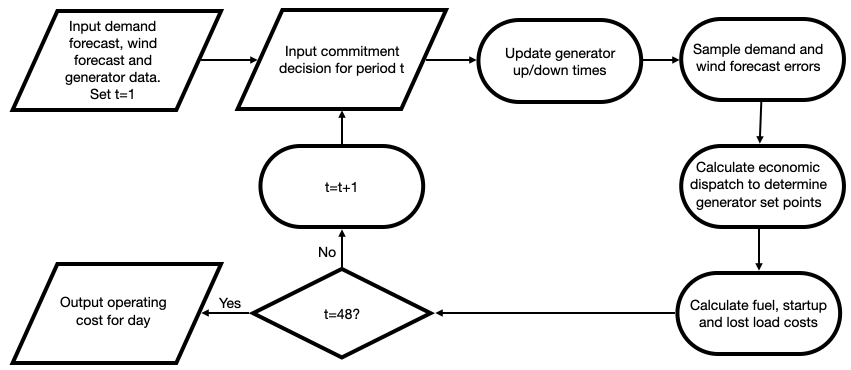}
    \caption{Flowchart of the simulation environment. The user inputs forecasts and generator data, and unit commitment decisions at each timestep of a 48-period day. The environment samples demand and wind scenarios and simulates dispatch by solving the economic dispatch problem. The environment outputs total operating costs at the end of the day.}
    \label{ch1:fig:env_flowchart}
\end{figure}

The environment models a power system of $N$ generators with 48 30-minute settlement periods per day, reflecting GB power market structure. Given data inputs of generator specifications and demand and wind generation forecasts, the environment repeats the following routine for each of the 48 settlement periods of the day (shown as a flowchart in Figure \ref{ch1:fig:env_flowchart}).

First, the user (or agent) inputs a commitment decision $\{0,1\}^N$, determining the on/off statuses of generators. This decision must satisfy generator constraints (see Generator Specifications below). The environment then updates the generator up/down times according to the commitment decision and samples forecast errors for demand and wind from stochastic processes. The `real' demand and wind generation are the sum of the forecast and forecast errors, and their difference is the net demand (wind is treated as negative demand). The environment solves the \emph{economic dispatch problem} \cite{wood2013power} to calculate the lowest-cost real-valued power outputs (set points) for the generators. The set points must meet the net demand if possible within the constraints of the online generators. If it is not possible to meet net demand, for instance if there is not enough capacity committed, then lost load is incurred. Finally, the operating costs are calculated as the sum of fuel costs, startup costs and lost load costs. This routine is repeated for each of the 48 decision periods, with forecasts rolling forward by 1 timestep at each iteration.

At the end of the episode, the environment returns the total operating cost for the entire day. For the same data inputs and commitment decisions, the environment may output different operating costs as different forecast errors are sampled. The environment can be used to evaluate operating costs of a unit commitment schedule under different scenarios of demand and wind.

We will now describe the key environment components in more detail. 

\subsubsection{Generator Specifications}

Generators in the simulation environment are specified by the following variables:

\begin{itemize}

\item $p_{\text{min}}, p_{\text{max}}$ (MW): minimum and maximum operating limits.
\item $a$ (\$/MWh$^2$),  $b$ (\$/MWh), $c$ (\$): quadratic coefficients of the fuel cost curve.
\item $u_0$: initial up/down time in settlement periods. 
\item $t_{\text{min}}^{\text{down}}$ $t_{\text{min}}^{\text{up}}$: minimum up/down times in settlement periods.
\item $c^s$ (\$): startup cost.

\end{itemize}

The generators in the environment do not have ramping constraints; in this respect, the simulated system studied is more flexible than real-world power systems. However, the solution methods presented in this thesis are equally applicable to generators with ramping constraints. 

The fuel cost for generator $i$ is a quadratic function of the variables $a, b, c$:

\begin{equation}
    C_i^f(p) = \frac{1}{2} z_i ( a_i(p_i)^2 + b_i(p_i) + c_i )
    \label{ch1:eq:cost_curves}
\end{equation}

where $z_i \in \{0,1\}$ is the generator commitment (on or off) and where $p_i$ is the real-valued power output in MW. The factor of $\frac{1}{2}$ is used to convert MW to MWh, given a settlement period length of 30 minutes.

At the beginning of each day, generators are initialised with up/down times (number of periods spent on/offline) defined by $u_0$. Positive values indicate up time, negative values indicate down time. The minimum up/down time constraints limit the extent to which generators can be committed/decommitted. An offline generator (that is with up/down time $u < 0$) cannot be committed until $u \leq -t_{\text{min}}^{\text{down}}$. Similarly, when $u > 0$, the generator must remain online until $u \geq t_{\text{min}}^{\text{up}}$. UC solutions that violate the minimum up/down time constraints are infeasible. 

Startup costs are incurred whenever a generator is committed (that is, in any settlement period when $u=1$). 

\subsubsection{Forecast Errors}

For each day, forecasts $\boldsymbol{d}$ and $\boldsymbol{w}$ for demand and wind are inputs into the environment and hence pre-determined. The environment is stochastic due to the inclusion of forecast errors, which are sampled from stochastic processes. A \emph{scenario} of demand and wind generation depends on demand forecast errors $\boldsymbol{x} \in \mathbb{R}^{48}$ and wind forecast errors $\boldsymbol{y} \in \mathbb{R}^{48}$, sampled from stochastic processes. The demand scenario is then $\bar{\boldsymbol{d}} = \boldsymbol{d} + \boldsymbol{x}$, and the wind scenario is $\bar{\boldsymbol{w}} = \boldsymbol{w} + \boldsymbol{y}$. Finally, given that wind is treated as negative demand, a \emph{net demand} is scenario $\bar{\boldsymbol{d}}_\text{net} = \bar{\boldsymbol{d}} - \bar{\boldsymbol{w}}$.

Forecast errors are modelled using autoregressive moving average (ARMA) processes, an approach that has previously been used in \cite{soder2004simulation, weber2009wilmar}. At settlement period $t$, the demand forecast error $x_t \sim X_t$ is sampled from:

\begin{align} \label{ch1:eq:arma}
    X_t = \sum_{i=1}^{p} \alpha_i X_{t-i}  + \sum_{i=1}^q \beta_i \epsilon_{d, t-i} + \epsilon_{d,t} 
\end{align}

where $p$ is the order of the autoregressive component and $q$ the order of the moving average component. $\alpha_i$ and $\beta_i$ are constant parameters and $\epsilon_{d,t}$ is a normally distributed random variable with mean 0 and standard deviation $\sigma$ (white noise). Similarly, wind forecast errors $y_t \sim Y_t$ are sampled from an ARMA process with different parameters and with white noise $\epsilon_{w,t}$.

\subsubsection{Economic Dispatch}

The UC problem deals only with binary decision variables, giving the on/off schedules of generators. In order to determine fuel costs, it is necessary to solve the economic dispatch (ED) problem, determining the real power outputs $\boldsymbol{p} \in \mathbb{R}^{N}$ that meet net demand at lowest cost. The ED problem is a convex optimisation problem and is solved with the lambda-iteration method \cite{wood2013power} in the simulation environment. We described the economic dispatch from and the lambda-iteration method in detail in Section \ref{methodology:mathop:ed}. The generator outputs $\boldsymbol{p}$ are used in Equation \ref{ch1:eq:cost_curves} to calculate the fuel costs. In some cases, an ED solution that meets net demand $\bar{d}_\text{net}$ is not possible due to the operating limits $p_{\text{min}}, p_{\text{max}}$ of the online generators. In these cases, we set $p = \{p_{\text{min}}, p_{\text{max}}\}$ for all online generators, depending on whether there is insufficient footroom (when net demand is low) or headroom (when net demand is high). The difference between net demand and generation from thermal plants is penalised at a value of lost load (see Equation \ref{ch1:eq:ll_cost} below). 


\subsubsection{Operating Costs}

For each settlement period, total operating costs  $C$ are calculated as the sum of fuel costs $C^f$; startup costs $C^s$; lost load costs $C^l$:

\begin{equation}
    C = C^f + C^s + C^l
    \label{ch1:eq:cost_function}
\end{equation}

Total operating costs for the day are calculated as the sum of period operating costs.

\paragraph{Fuel costs} Fuel costs for each generator are calculated according to the quadratic cost curves defined in Equation \ref{ch1:eq:cost_curves}: 

\begin{equation}
    C^f = \sum_{i=1}^N C^f_i
\end{equation}

\paragraph{Startup Costs} Startup costs are incurred whenever a generator is committed: 

\begin{gather}
    C^s = \sum_{i=1}^N \lambda_i c_i^s 
\end{gather}

where:

\begin{align}
    \lambda_i = 
    \begin{cases}
        1 & \text{if } u_{i}=1 \\
        0 & \text{otherwise}
    \end{cases}
\end{align}

\paragraph{Lost load costs} When available generators are unable to meet demand, a lost load cost is incurred: 

\begin{align}
    C^l = V l \\ 
    l = | \bar{d}_{\text{net}} - \sum_{i=1}^N p_i |
    \label{ch1:eq:ll_cost}
\end{align}

where $V$ (\$/MWh) is the value of lost load and $l$ (MWh) is the \emph{lost load}, the difference between supply $\sum_{i=1}^N p_i$ and net demand $\bar{d}_{\text{net}}$. Note that in this setup there is equal penalty for over-commitment and under-commitment of generation; in practice, the costs of over-commitment are likely to be substantially lower, and could be managed by wind shedding or other means apart from load shedding. As a result, the optimal commitments and reserve allocation strategies in this setup are likely to give greater priority to generator footroom than would be common in real power systems. 

\subsection{Data}

\begin{figure}
    \centering
    \includegraphics[width=0.5\textwidth]{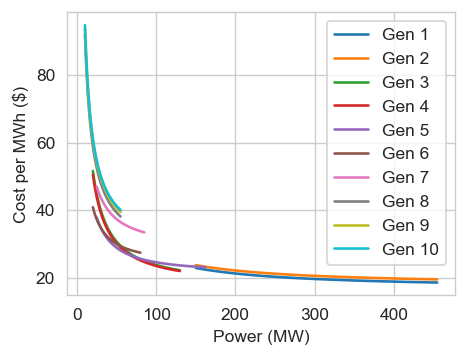}
    \caption{Quadratic cost curves for the 10 generators described in Table \ref{ch1:tab:gen_info} in \$ per MWh. Efficiency improves as load factor $\frac{p}{p_{\text{max}}}$ increases.}
    \label{ch1:fig:cost_curves}
\end{figure}

To define the generator variables described in Section \ref{ch1:env:overview}, we used data from Kazarlis et al. \cite{kazarlis1996genetic}. While this data is not recent (published in 1996), it is still widely-used as a benchmark power system \cite{carrion2006computationally, quan2015improved, panwar2018binary}; we reviewed deterministic UC research which has used the Kazarlis benchmark power system in Table \ref{literature:tab:duc_kazarlis}. In addition, the data provides complete descriptions of generator cost curves, whereas other data sources use piecewise linear approximations or assume constant efficiencies. The generator specifications from \cite{kazarlis1996genetic} are shown in Table \ref{ch1:tab:gen_info}. The quadratic cost curves for the 10 generators are shown in Figure \ref{ch1:fig:cost_curves}. Generators 1 and 2 most closely reflect baseload generation, having the lowest fuel costs, largest capacities and the most restrictive minimum up/down times (8 hours). Generators 8--10 are peaking plants, with small capacity, high fuel costs and short minimum up/down times (2 hours). Larger systems are created in the simulation environment by duplicating the 10 generators, an approach followed in existing research using this power system \cite{kazarlis1996genetic, carrion2006computationally, panwar2018binary, quan2015improved}. 

\begin{table}[t]
    \centering
    \resizebox{0.8\textwidth}{!}{
    \input{04-chapter1/tables/gen_info_10}}
    \caption{Generator specifications for the 10 generator problem, from \cite{kazarlis1996genetic}.}
    \label{ch1:tab:gen_info}
\end{table}

We set the value of lost load (VOLL) $V$ (Equation \ref{ch1:eq:ll_cost}) to be \$10,000 per MWh for both training and testing, set to represent the approximate VOLL for a range of customer types \cite{schroder2015value}. This large penalty reflects the potentially catastrophic outcomes of failing to meet demand in the real world. 

\begin{figure}[t]
    \centering
    \includegraphics[width=\textwidth]{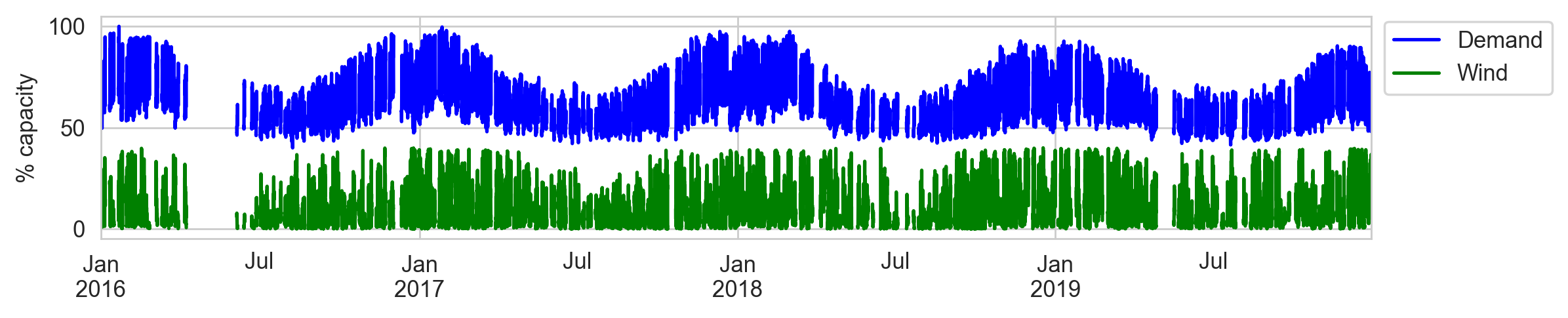}
    \caption{National Grid demand \cite{NG} and Whitelee wind generation data \cite{BMRS} used to define forecasts in the simulation environment. Demand and wind generation are scaled depending on the number of generators so are shown in terms of \% of total generator capacity. Incomplete days were removed, leaving 806 complete forecasts.}
    \label{ch1:fig:all_forecasts}
\end{figure}

For demand forecasts we used National Grid Demand Data \cite{NG} from 2016--2019. Demand is scaled linearly as a function of number of generators to be between 40--100\% of the total capacity $\sum_{i=1}^N p_{\text{max}, i}$. For wind forecasts, we used openly available data for Whitelee onshore wind farm \cite{BMRS}, chosen as a relatively large wind farm that operated continuously between 2016--2019. The output of a single wind farm is more volatile than the national wind generation, providing a diverse set of wind profiles for UC problems. In addition, using a single farm keeps the overall wind penetration roughly constant across the period 2016--2019, while GB-wide wind penetration increased significantly over the period. The wind generation data was scaled to be between 0--40\% of the total capacity of the generation mix. We found a significant number of incomplete days in the source data for either demand or wind. We omitted data for these days entirely, resulting in a total of 806 unique days of demand and wind forecasts. The full time series for both demand and wind, showing the omitted days, is displayed in Figure \ref{ch1:fig:all_forecasts}.

The $\alpha$ and $\beta$ parameters of the ARMA processes (defined in Equation \ref{ch1:eq:arma}) for demand and wind forecast errors were set manually and are identical for all power system sizes. Both were tuned to decay exponentially, such that there is a stronger correlation with more recent forecast errors. In both cases we set $p=q=5$, allowing $5$ steps of history to be accounted for in both autoregressive and moving average components of the ARMA process. The standard deviation parameters $\sigma$ were scaled proportionally to the number of generators, such that relative to the demand and wind generation, the level of uncertainty remained roughly constant for all problems.

\subsection{Test Problems and Benchmarks} \label{ch1:env:benchmarks}

In order to compare UC solution methods, we created a dataset of 20 test problems, sampled from the 806 complete forecasts. The remaining days form a training set of forecasts that were used to train RL agents. Whereas most UC research reports results for a single test problem such as the widely-studied demand profile in \cite{kazarlis1996genetic}, more statistically robust results can be achieved by evaluating solution methods on multiple test problems. 20 test problems ensures a wide range of days with different characteristics across multiple seasons while preserving a large set of training episodes. The 20 test problems (scaled for the 10 generator power system) are described in Table \ref{ch1:tab:test_problems} and visualised in Figure \ref{ch1:fig:all_test_problems}. The test problems exhibit a range of characteristics and daily wind penetration (wind generation as proportion of demand) ranges from 2.8\% to 36.8\%. 

\afterpage{

\clearpage

\begin{figure}
    \centering
    \includegraphics[width=1\textwidth]{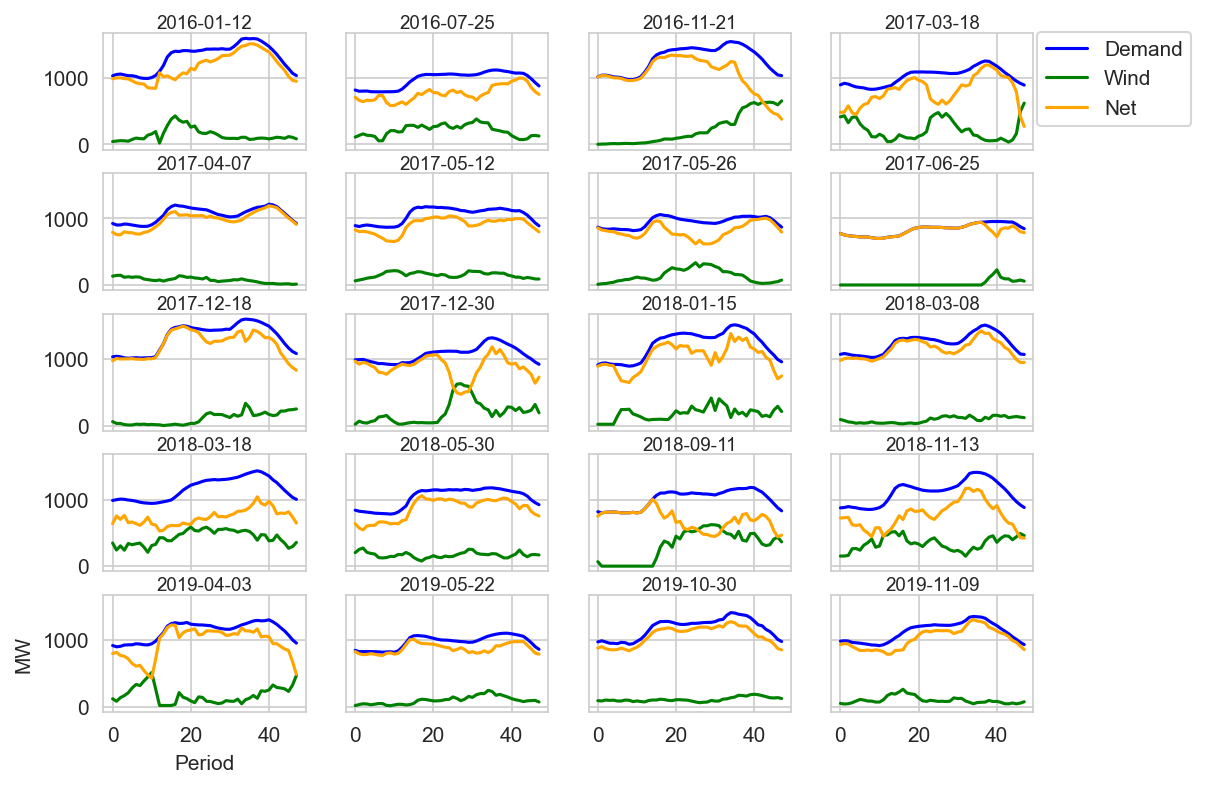}
    \caption{Unseen test problems, shown for the 10 generator problem.}
    \label{ch1:fig:all_test_problems}
\end{figure}

\begin{table}
    \centering
    \resizebox{0.5\textwidth}{!}{
    \input{04-chapter1/tables/test_problems_10}}
    
    \caption{Summary of test profiles for the 10 generator setting, visualised in Figure \ref{ch1:fig:all_test_problems}}
    \label{ch1:tab:test_problems}
\end{table}

\clearpage

}

In order to provide strong benchmark solutions to the 20 test problems, we used  mixed-integer linear programming (MILP) to solve a deterministic formulation \cite{morales2013tight} of the UC problem, as described in Section \ref{methodology:mathop:milp}. Deterministic UC solved with MILP is widely used in industry \cite{bertsimas2012adaptive}. The Power Grid Lib software package\footnote{\url{https://github.com/power-grid-lib/pglib-uc}} was used to formulate the deterministic UC problem and we used the open-source COIN-OR library using the branch-and-cut algorithm \cite{johnjforrest_2020_3700700} to solve the MILP. The specific formulation combines the MILP formulation described in \cite{morales2013tight} with the method for piecewise linear approximation of quadratic cost curves described in \cite{sridhar2013locally}.

We implemented two benchmarks, the first using a reserve constraint to manage uncertainties, the second assuming perfect foresight (i.e. all forecast errors being zero) and no reserve constraint. For the first benchmark, the reserve constraint was set to be 4 times the standard deviation of the net demand forecast errors, a common industry approach described in \cite{holttinen2008using}. We refer to this benchmark as MILP($4\sigma$). The standard deviation was determined empirically by sampling from the ARMA processes for demand and wind forecast errors. 

To probabilistically evaluate the quality of MILP($4\sigma$) solutions to the test problems, we applied the following Monte Carlo approach, which has also been employed in \cite{bertsimas2012adaptive, quan2014incorporating, dvorkin2014comparison, ruiz2009uncertainty}:

\begin{enumerate}
    \item Calculate the UC schedule using solution method (e.g. MILP, UCS, Guided UCS).
    \item Use the environment to calculate operating costs for 1000 scenarios of demand and wind.
\end{enumerate}

Step 2 involves repeatedly passing the UC schedule as an input to the environment (step 2 of the flowchart in Figure \ref{ch1:fig:env_flowchart}), and recording the operating costs on each iteration. This returns a distribution of operating costs over scenarios. 

For the second, perfect foresight benchmark, the reserve was set to 0. We call this benchmark MILP(perfect). This approach only considers the point forecast, with zero forecast errors. Hence, we do not evaluate this solution over the 1000 scenarios. 

\begin{figure}[t]
    \centering
    \includegraphics[width=\textwidth]{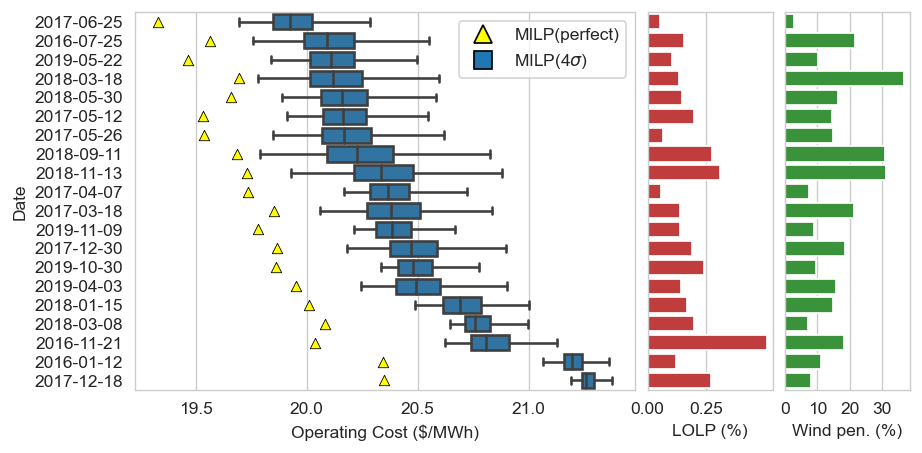}
    \caption{Operating costs for MILP benchmarks on the 20 test problems. The distribution of operating costs for MILP($4\sigma$), evaluated under 1000 scenarios of demand and wind generation are shown with outliers removed. The operating costs for the MILP(perfect) solution, which considers only the point forecast, is shown in yellow. Loss of load probability (LOLP) is shown for the MILP($4\sigma$) for each day, as well as the daily wind penetration.}
    \label{ch1:fig:benchmarks_boxplot}
\end{figure}

The results for MILP benchmarks for the 10 generator problem are plotted in Figure \ref{ch1:fig:benchmarks_boxplot}, shown in terms of cost per MWh net demand to control for variations in net demand under different scenarios. Also shown are loss of load probability (LOLP; average probability of a lost load event in any settlement period) for the MILP($4\sigma$) solution and the wind penetration of each test problem. There is significant variation in solution quality of MILP($4\sigma$) relative to MILP(perfect). For instance, MILP($4\sigma$) performs considerably worse on 2016-11-21, due to a high LOLP (roughly 0.5\%). This problem is characterised by a sharp increase in wind generation towards the end of the day, coinciding with decreasing demand, resulting in a rapid decrease in net demand (Figure \ref{ch1:fig:all_test_problems}). As a result, the UC solution is likely to encounter insufficient footroom constraints, with large lost load penalties according to Equation \ref{ch1:eq:ll_cost}. 

Note that the solver's optimality gap was set at 1\% for the experiments. Reducing the gap to 0.01\% caused a slight reduction in operating costs for the 10 generator problem (0.2\%), but increased operating costs by approximately 1\% for 20 and 30 generator cases. As both solutions employ the same reserve strategy and lost load contributes significantly to total operating costs, it is not surprising that reducing the gap does not strictly reduce expected costs. As a result, we adopt an optimality gap of 1\% throughout this thesis.

\bigskip

The power system simulation environment can also be used to represent a Markov Decision Process (MDP) formulation of the UC problem, with states, actions and rewards. In this context, an agent may be trained using RL to take actions in the environment that maximise the long-run expected reward, based on contextual information about the state. In the following section, we will describe the MDP formulation of the UC problem.

\section{Markov Decision Process Formulation} \label{ch1:uc_mdp}

\begin{figure}[t]
    \centering
    \includegraphics[width=0.8\textwidth]{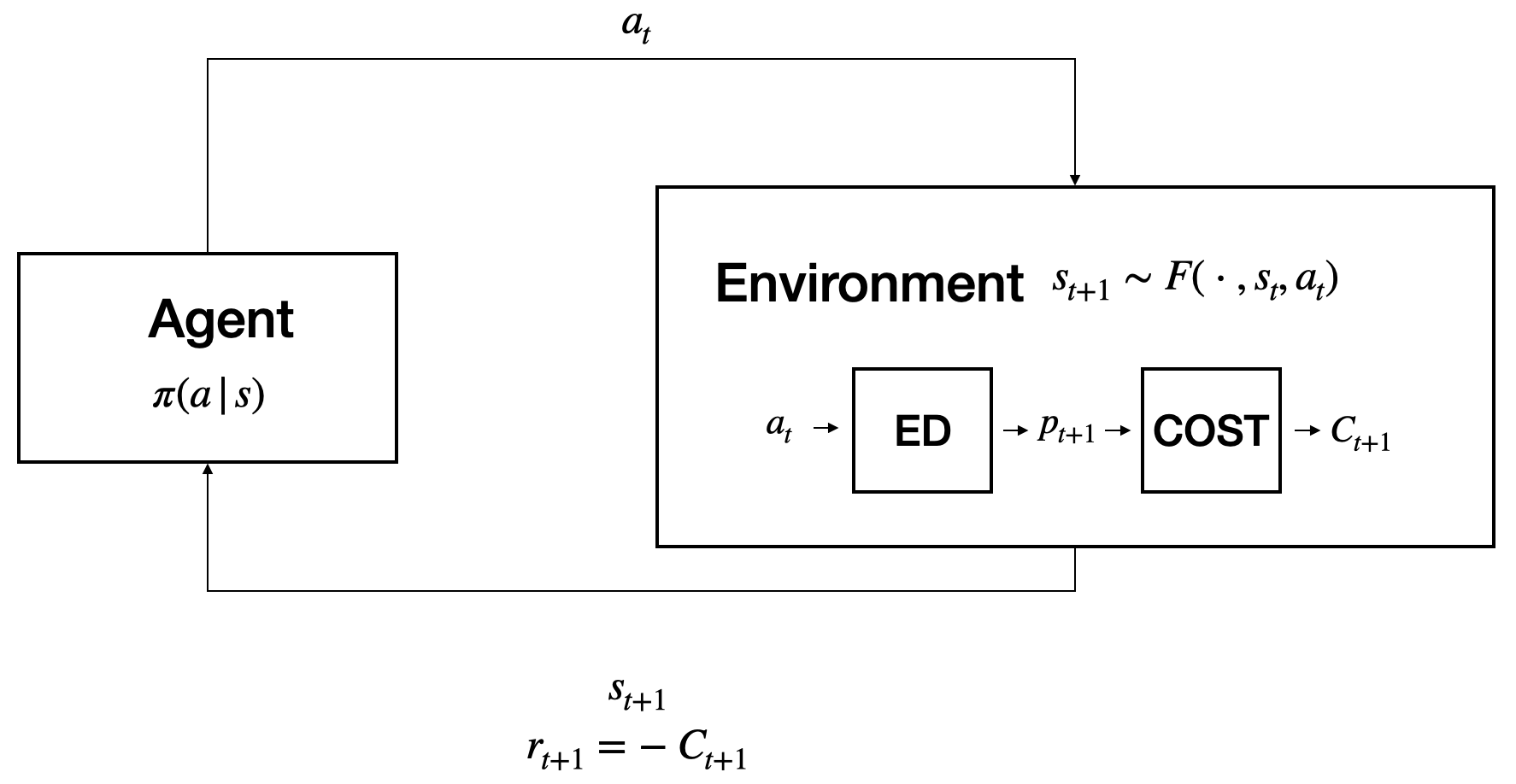}
    \caption{Agent-environment interaction in the UC MDP. The agent takes an action $a_t$, which is processed by the environment, returning a new state $s_{t+1}$ sampled from the transition function $F(s_{t+1}, s_t, a_t)$ and a reward $r_{t+1}$. The action is sampled from a policy $\pi(a_t|s_t)$, considering a partial observation of the state. To calculate the reward, the environment solves the economic dispatch (ED) and evaluates the cost function, as described in Section \ref{ch1:env:overview}. The reward is the negative operating cost: $r_{t+1} = -C_{t+1}$.}
    \label{ch1:fig:mdp_diagram}
\end{figure}

In Section \ref{ch1:env} we described the power system simulation environment used to train and evaluate RL agents. We also showed how the environment is used to evaluate MILP benchmark solutions to the 20 test problems. The environment can also be used to represent the UC problem as a partially-observable Markov Decision Process (MDP), which we will describe in this section. MDPs, which consist of states, observations, action, rewards and a transition function, were described in detail in Section \ref{methodology:background:mdp}. Formally describing the UC problem as an MDP allows for the application of RL methods, with an agent learning by trial and error in the simulation environment described in Section \ref{ch1:env}. The agent interacts with the environment by scheduling generators based on uncertain forecasts for demand and wind generation, aiming to minimise operating costs. A schematic of this agent-environment interaction in the UC MDP is shown in Figure \ref{ch1:fig:mdp_diagram}.


Using definitions of power system variables described in Section \ref{ch1:env}, we will now describe the components of the UC MDP.

\subsection{MDP Components}

The components of the UC MDP are shown in Table \ref{ch1:tab:mdp_components}. Here we will describe each component in more detail.

\begin{table}
    \centering
    \small
    \input{04-chapter1/tables/mdp}

    \caption{MDP components for the UC problem with $N$ generators and $T$ decision periods.}
    \label{ch1:tab:mdp_components}
\end{table}

\paragraph{States}

The state $s_t$ includes: generator up/down times $\boldsymbol{u}_t$; demand forecast $\boldsymbol{d}$; wind forecast $\boldsymbol{w}$; historical demand and wind forecast errors $\boldsymbol{x}_t, \boldsymbol{w}_t \in \mathbb{R}^p$; historical white noise samples for demand and wind forecast errors $\boldsymbol{\epsilon}_{d,t} \boldsymbol{\epsilon}_{w,t} \in \mathbb{R}^q$; the timestep $t$. $p$ and $q$ are the orders of autoregressive and moving average components of the ARMA($p,q$) processes. A state is terminal when $t=T$.  

\paragraph{Observations}

Observations include all elements of the state except forecast errors $\boldsymbol{x}_t, \boldsymbol{w}_t$ and white noise components $\boldsymbol{\epsilon}_{d,t} \boldsymbol{\epsilon}_{w,t}$. The MDP is therefore partially-observable. Omitting forecast errors from the observation preserves the day-ahead properties of the problem: the decision-maker cannot observe the forecast errors until after the UC problem has been solved. However, forecast errors are included in $s_t$ in order to preserve the Markov property in the MDP, and for completeness to ensure that the reward distribution $R(s_t)$ does not depend on a latent variable that is not included in the MDP.







\paragraph{Actions}

An action is a commitment decision that determines the on/off statuses of generators for the next timestep. An action is defined as an array $\boldsymbol{a}_t = [a_{1,t}, a_{2,t} ... , a_{N,t}]$, $a_{i,t} \in \{0,1\}$ for $N$ generators. The action space is therefore combinatorial, and has a total of $2^N$ unique actions. However, for a given state the set of of legal actions $A(s)$ is limited to those meeting the minimum up/down time constraints described in Section \ref{ch1:env:overview}. 

\paragraph{Rewards}

The reward is the negative total operating cost of the system: 

\begin{gather}
\label{ch1:eq:reward}
    r_t = -C_t
\end{gather}

where $C_t$ is the sum of fuel costs, startup costs and lost load costs (Equation \ref{ch1:eq:cost_function}). The fuel cost depends on the real-valued power outputs of the generators (Equation \ref{ch1:eq:cost_curves}), which are determined by solving the economic dispatch (ED) problem (see Section \ref{ch1:env:overview}) to satisfy net demand (which is stochastic). 


\paragraph{Transitions} \label{transitions}

Transitions consist of updating the generator up/down times and sampling demand and wind forecast errors using the ARMA processes. Given a commitment decision $a_{i,t}$ and generator status $u_{i,t}$ for generator $i$, the transition function for the generator status is:

\begin{align} 
    u_{i, t+1}= 
\begin{cases}
    u_{i,t} + 1,& \text{if } a_{i,t}=1 \text{ and } u_{i,t} > 0 \\
    1,& \text{if } a_{i,t}=1 \text{ and } u_{i,t} < 0 \\
    -1,& \text{if } a_{i,t}=0 \text{ and } u_{i,t} > 0 \\
    u_{i,t} - 1,& \text{if } a_{i,t}=0 \text{ and } u_{i,t} < 0 \\
\end{cases}
\end{align}

Forecast errors are sampled using the ARMA process described in Equation \ref{ch1:eq:arma}.

The transition function is stochastic due to the inclusion of forecast errors in the state. However, since these are not observed by the agent, the transition function can be considered deterministic with respect to the observations. This means that the MDP can be easily expressed as a search tree in the observation space, with nodes representing observations and edges representing actions. The search tree formulation is required in order to apply tree search methods such as uniform-cost search, which we describe in the next section. 




\section{Uniform-Cost Search} \label{ch1:ucs}

Having formulated the UC problem as an MDP, we will now apply tree search methods to solve the UC problem. In this section, a variation of the well-known uniform-cost search (UCS) algorithm \cite{russellnorvig}, described in Section \ref{methodology:tree_search:ucs} is applied to the UC problem. We begin by formulating the UC MDP as a search tree. The altered UCS algorithm is then described, before being applied to the UC test problems described in Section \ref{ch1:env:benchmarks}. 



\subsection{Search Tree Representation of the UC MDP} \label{ch1:methodology:search_tree}

\begin{figure}
    \centering
    \includegraphics[width=0.55\textwidth]{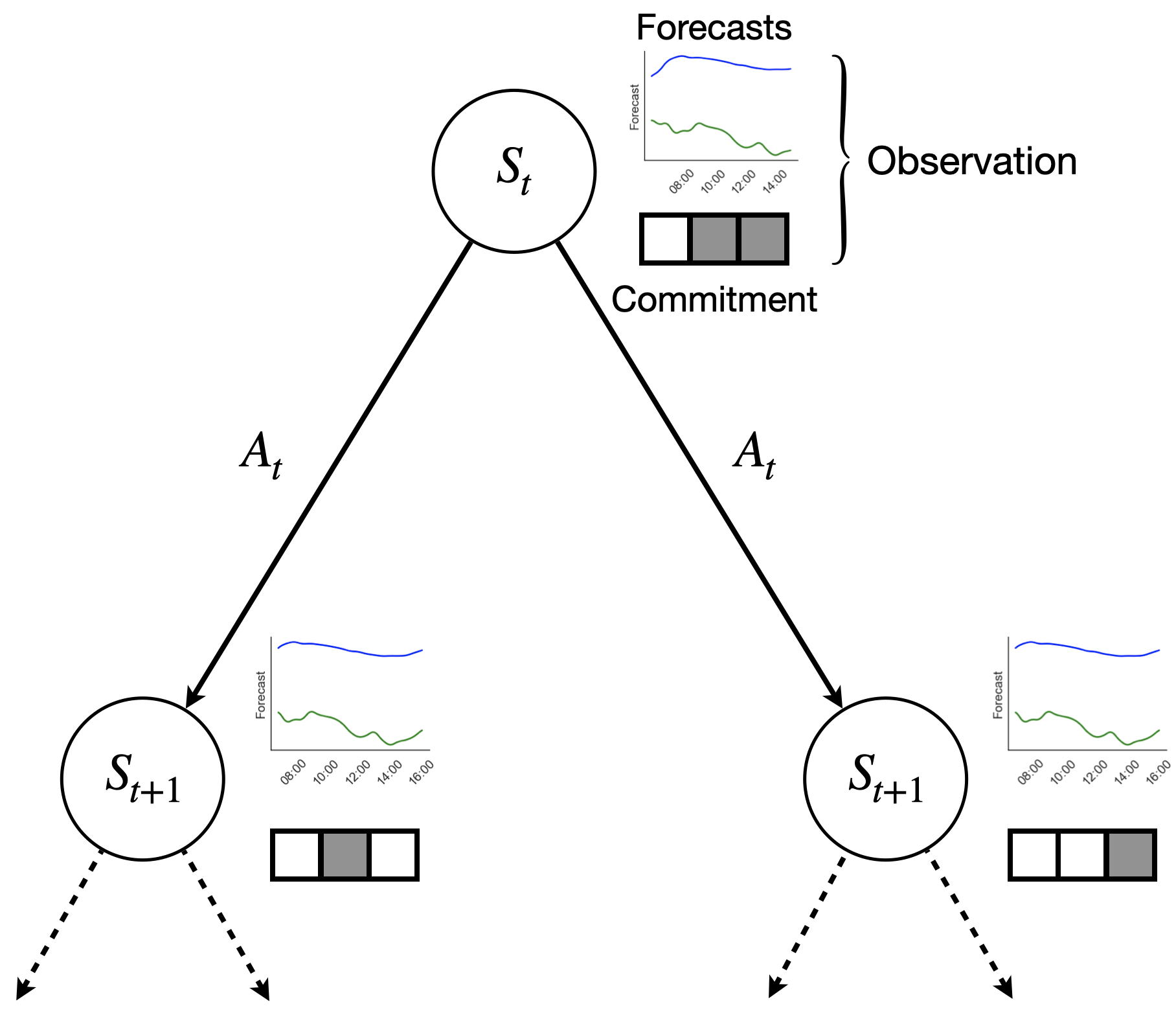}
    \caption{Search tree representing the UC MDP. Nodes represent observations, and edges represent actions. The cost of traversing an edge is the expected operating cost, estimated by a Monte Carlo method, simulating each transition $N_s$ times and calculating the mean. The time series at each node represent the demand and wind forecasts at that state, while the commitment is represented by blocks representing the commitment of three generators where grey/white refer to offline/online.}
    \label{ch1:fig:search_tree}
\end{figure}

As stated in the Section \ref{ch1:uc_mdp}, the MDP can be expressed as a simple search tree in the observation space, which we illustrate in Figure \ref{ch1:fig:search_tree}. A similar formulation is used in \cite{dalal2015reinforcement}, although only does not consider uncertain demand or wind generation in the MDP formulation. Each node in the search tree represents an observation (that is, forecasts $\boldsymbol{d}, \boldsymbol{w}$ and generator up/down times $\boldsymbol{u}_t$, and edges represent actions. Since the transition function is deterministic in this space, there is a one-to-one mapping from $(s_t,a_t) \rightarrow s_{t+1}$, which is required for the search tree representation. Traversing an edge on the tree incurs a cost, which is the negative reward in the UC MDP. A UC problem of $T$ decision periods and $N$ generators can be expressed as a search tree with depth $T$ and maximum branching factor $b = 2^N$. The branching factor is a theoretical maximum as in practice, some of the $2^N$ actions from any node will be illegal due to the minimum up/down time constraints. 

Maximising return in the UC MDP amounts to finding the cheapest cost path through the search tree, which can be achieved through tree search algorithms. The cost, however, is stochastic, as it depends on $x_t$ and $y_t$, the forecast errors. Hence, we are interested in determining the path of least \textit{expected} cost. To achieve this, we set the step costs $C(s)$ (the cost of traversing edge to state $s$) to an estimate for the mean cost of that transition. This is achieved by a Monte Carlo method: using a set of demand and wind forecast error scenarios $\mathcal{S} \in \mathbb{R}^{T \times N_s}$, sampled from the ARMA processes, we calculate the mean operating cost (negative reward) over $\mathcal{S}$:

\begin{equation}
    C(s_t) = -\frac{1}{N_s} \sum_{x \in \mathcal{S}_t} R(s_t,x)
    \label{ch1:eq:search_tree_cost}
\end{equation}

where $\mathcal{S}_t$ is the set of scenarios for timestep $t$, $x$ is a scenario in $\mathcal{S}_t$ and $R(s_t, x)$ is the reward function evaluated for state $s_{t+1}$ under scenario $x$. 

Each reward function evaluation requires solving the ED problem for a different scenario $x$, which may be computationally expensive if $N_s$ is large. However, the ED problem is a convex optimisation which can be solved very rapidly using the lambda-iteration method described in Section \ref{methodology:mathop:ed} and is parallelisable over scenarios. As a result, tree search algorithms using this Monte Carlo method have only $\mathcal{O}(N_s)$ complexity, which is preferable to the super-linear complexity in the number of scenarios of the stochastic optimisation methods reviewed in Section \ref{literature:suc}. 

\subsection{Algorithm: Real-Time UCS} \label{ch1:methodology:tree_search}

\begin{algorithm}[t]
\caption{Real-time method for solving the UC problem beginning in initial state $s_0$ using a tree search algorithm $f$, such as UCS. The algorithm $f$ solves the search tree up to a lookahead horizon $H$, returning a solution path (sequence of actions). The first action in the path is used to determine the root for the next sub-tree. The algorithm loops for $T$ settlement periods.} \label{ch1:algo:solve_day_ahead}
\begin{algorithmic}
\Function{RealTimeTreeSearch}{$s_0, f, H$}
  \State initialise solution schedule $\boldsymbol{a} \in \{0,1\}^{T \times N} $
  \State $s \leftarrow s_0$ 
  \For{$t$ in $0, 1, \dots , T-1$}
        \State $p \leftarrow $ solve tree rooted at $s$ with algorithm $f$ up to horizon $H$
        \State $a \leftarrow $ first action in solution path $p$
        \State $s \leftarrow $ state following $a$ from $s$
        \State $\boldsymbol{a}_{t} \leftarrow a$
  \EndFor
  \State \Return $\boldsymbol{a}$
\EndFunction
\end{algorithmic}
\end{algorithm}


Having formulated the UC MDP as a search tree, we use the uniform-cost search (UCS) algorithm \cite{russellnorvig} described in Section \ref{methodology:tree_search:ucs} to find the path of least expected cost through the tree. Numerous tree search algorithms exist which can be used to solve the UC problem. UCS is appropriate for being simple, heuristic-free and optimal \cite{russellnorvig}. Furthermore, unlike other simple tree search algorithms like breadth-first search and depth-first search, UCS is naturally applicable to trees with non-uniform step costs, such as in the UC problem. 

A notable candidate algorithm for solving the UC problem is Monte Carlo tree search (MCTS), which has been used in model-based RL methods including AlphaGo \cite{silver2016mastering, silver2018general}. MCTS is well-suited to zero-sum games domains where the objective is to maximise the probability of winning given an opponent's strategy, but is not a natural fit for conventional optimisation problems. MCTS has rarely been applied to stochastic optimisation problems, and is outperformed by mixed-integer approaches in \cite{bertsimas2017comparison}. We opt to use UCS as a more appropriate search algorithm designed for least cost path problems.

For a search tree with a constant branching factor $b$ nodes and goal node at depth $k$, UCS has time complexity $\mathcal{O}(b^k)$. For the UC problem with $T$ settlement periods, which has a maximum branching factor of $b = 2^N$, UCS algorithm has worst-case $\mathcal{O}(2^{NT})$ time complexity making it intractable even for very small power systems with 48 settlement periods. Hence, we use a \textit{real-time} \cite{korf1990real} approach to solve each UC problem. Real-time algorithms, which were discussed in Section \ref{methodology:tree_search:taxonomy}, do not exhaustively search to the end of the problem, but instead repeatedly solve reduced sub-problems, and take the best first action with respect to this sub-optimal solution. Real-time methods, while not optimal, are significantly less computationally expensive in many problems. In the real-time approach, UCS is used to solve $T$ sub-problems, one for each settlement period. Each sub-problem is limited by a fixed lookahead horizon $H \in \mathbb{N}, 1 \leq H \leq (T-t)$, where $H$ is a constant depth parameter. Once the sub-problem rooted at period $t$ is solved, the first action in the solution path is taken, determining the root node for the sub-problem at time $t+1$. This is repeated until the end of the day $T=t$. This routine is shown as pseudo-code in Algorithm \ref{ch1:algo:solve_day_ahead}. The \textsc{RealTimeTreeSearch} algorithm is modular and can be used with any tree search algorithm for finding the lowest cost path. Real-time UCS has $\mathcal{O}(2^{NH}T)$ complexity: it is linear in the number of settlement periods $T$ and exponential in both $H$ and $N$. Henceforth we refer to real-time UCS simply as UCS. 

Pseudocode for UCS was shown in Algorithm \ref{methodology:algo:ucs}. In order to implement real-time UCS, we assign all nodes at timestep $t+H$ to be goal nodes. This ensures that UCS will return the cheapest cost path to a node at the lookahead horizon. In addition, we use the Monte Carlo estimates of expected step costs described in Section \ref{ch1:methodology:search_tree} when calculating the path cost $c(n)$. In the limit of $H$ and $N_s$, UCS will minimise the expected operating cost over all scenarios by searching for the least expected cost path through the MDP. However, the exponential time complexity in the number of generators $N$ limits application of this approach to larger power systems.



\subsection{Application to Test Problems} \label{ch1:ucs:results}

We will now apply UCS to the test problems described in Section \ref{ch1:env:benchmarks} to empirically demonstrate exponential run time complexity and evaluate solution quality in comparison with MILP benchmarks described in Section \ref{ch1:env:benchmarks}. We set the depth parameter $H=2$ and applied UCS to power systems of 5--10 generators inclusive. Larger values of $H$ were not possible due to the exponential time complexity in this parameter. 

\begin{figure}
    \centering
    \includegraphics[width=0.5\textwidth]{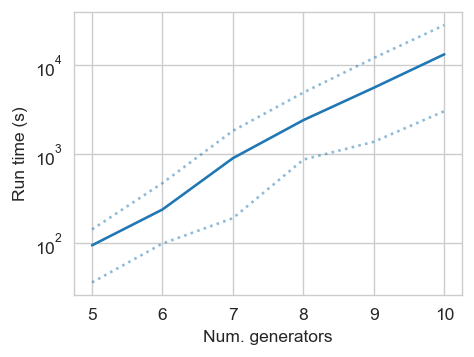}
    \caption{Run time of UCS with $H=2$ with increasing numbers of generators. Solid line shows mean daily run time; dotted lines show minimum and maximum. The straight line on the logged time axis indicates exponential run time complexity in the number of generators.}
    \label{ch1:fig:time_taken_ucs}
\end{figure}

A comparison of the costs and loss of load probability (LOLP) of the UCS and MILP approaches is shown in Table \ref{ch1:tab:ucs_milp_comparison}. We find that UCS achieves similar operating costs to MILP($4\sigma$), being 1.3\% more expensive in the 5 generator case, and 0.3\% in the 10 generator case. Compared with MILP(perfect), the solution for the point forecast with no reserve constraint, costs are 5.7\% and 6.9\% higher for 5 and 10 generator problems, respectively. Figure \ref{ch1:fig:time_taken_ucs} shows that UCS has an exponential run time complexity in the number of generators. Extrapolating the relationship to a system of 20 generators, the anticipated average run time is approximately $1\times10^6$ seconds per problem instance.

\begin{table}[]
    \centering
    \small
    \input{04-chapter1/tables/ucs_milp_comparison}
    \caption{Comparison of UCS with $H=2$ with MILP benchmarks from Section \ref{ch1:env:benchmarks}. UCS achieves similar operating costs and loss of load probability (LOLP) as compared with MILP($4\sigma$).}
    \label{ch1:tab:ucs_milp_comparison}
\end{table}

The relatively strong performance of UCS with a limited time horizon of $H=2$ (1 hour) indicates that tree search and direct sampling of the cost function under realisations of uncertainty to estimate expected costs (Equation \ref{ch1:eq:search_tree_cost}) is a strong method for managing uncertainty. However, the short-sightedness of this approach results in worse performance overall as compared with MILP($4\sigma$). With a less flexible generation mix, such as with longer minimum up/down times, the difference between short-sighted tree search and mathematical programming approaches is likely to be larger still. Above all, UCS is not a practical solution method due to the exponential complexity in the number of generators, preventing applications to larger power systems. In the next section, we will describe Guided UCS, which uses an RL-trained policy to reduce the branching factor, allowing for application to larger power systems and greater search depth.


\section{Guided Uniform-Cost Search} \label{ch1:guided_ucs}

The results in Section \ref{ch1:ucs:results} showed exponential time complexity of UCS in the number of generators. As a result, the application of this method to the UC problem is restricted to small power systems only. Even for the small systems of 5--10 generators, it was necessary to limit the search depth to $H=2$ as higher settings would not be complete in practical run times. While UCS was shown to be competitive with the MILP benchmarks, improvements to operating costs could be achieved by increasing $H$. In order to improve the run time complexity in number of generators, in this section we present Guided UCS, an RL-aided tree search algorithm that uses an RL-trained \emph{expansion policy} to reduce the breadth of the search tree. Learning by trial-and-error in the simulation environment, the expansion policy offers a rapid approximation of promising regions of the action space. The action space at any given node typically contains a large number of expensive or insecure commitment decisions, such as those which decommit baseload and those which do not commit enough capacity to meet the forecast net demand. The expansion policy is trained by RL to identify these actions as well as more complex, time-dependent properties of the action space as a function of the state variables. Guided by the expansion policy, the search tree of large power systems can be reduced to a much smaller size while preserving optimal or near-optimal solution paths, enabling tree search methods to be applied in larger problem instances.

In this section we will begin by describing \emph{guided expansion}, the method by which an expansion policy is incorporated into UCS to reduce the branching factor. We will then present our approach for training the expansion policy by model-free RL in the power system simulation environment. Finally, we will present the details of policies trained with model-free RL, which are used in experiments applying Guided UCS to test problems in Section \ref{ch1:experiments}. 

\subsection{Guided Expansion} \label{ch1:guided_ucs:guided_expansion}

\begin{figure}
    \centering
    \includegraphics[width=\textwidth]{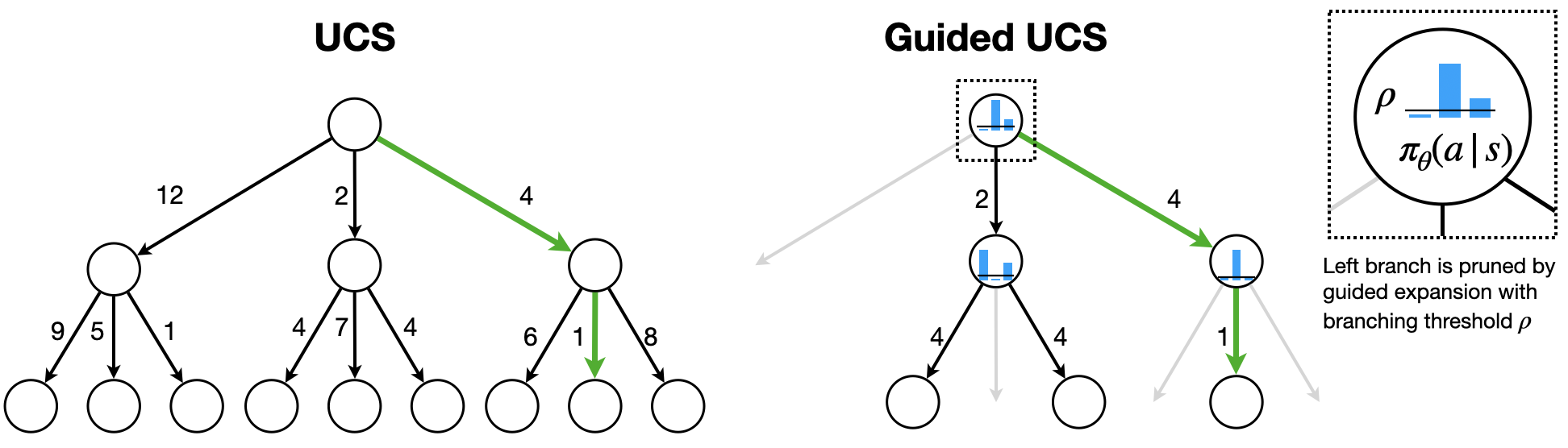}
    \caption{Comparison between UCS and Guided UCS algorithms, with a search depth $H=2$. While UCS considers the full search tree, Guided UCS uses a reduced search tree, with branches pruned by guided expansion (Equation \ref{ch1:eq:guided_expansion}). The histogram represents the distribution estimated by $\pi_{\theta}(a|s)$. From the root node, the left branch is pruned as its probability is less than the branching threshold $\rho$ (represents by a horizontal line on the histogram). Green line represents the lowest cost path.}
    \label{ch1:fig:guided_ucs_diagram}
\end{figure}

Guided UCS uses an \emph{expansion policy} $\pi(a|s)$ giving probabilities for the actions from state $s$ to prune low probability actions. We call this routine \emph{guided expansion}. Figure \ref{ch1:fig:guided_ucs_diagram} illustrates the difference between UCS and Guided UCS, showing how guided expansion is used to reduce the branching factor of the search tree. The expansion policy is used in Guided UCS to prune the left branch from the root. Note that in this illustrative example, both UCS and Guided UCS reach the same solution path, but Guided UCS is more efficient, requiring fewer nodes to be evaluated and expanded. However, Guided UCS may remove an optimal branch if the expansion policy $\pi(a|s)$ is inaccurate. By training the expansion policy with RL, we aim to ensure that $\pi(a|s)$ prunes only sub-optimal branches. 

Formally, Guided UCS follows Algorithm \ref{methodology:algo:ucs}, but only considers a subset of actions in the inner for loop over actions. Guided expansion reduces the complete set of actions $A(s)$ available from state $s$ to a subset of actions $A_{\pi}(s)$:

\begin{gather}
    A_{\pi}(s) = \{ a \in A(s) | \pi(a|s) \geq \rho \}
    \label{ch1:eq:guided_expansion}
\end{gather}

where $\rho$ is a branching threshold that controls the search breadth. The maximum number of nodes $M$ that can be added to the tree is therefore limited to $M\leq \frac{1}{\rho}$, since $\sum_{a \in A} \pi(a|s) = 1$. By preventing an exponential explosion in the branching factor with increasing number of generators, the worst case time complexity of Guided UCS is $\mathcal{O}(T(\frac{1}{\rho})^H)$, compared with $\mathcal{O}(2^{NH}T)$ for unguided UCS, presented in Section \ref{ch1:ucs}. As a result, Guided UCS does not exhibit exponential time complexity in the number of generators, making it feasible to consider real world application to larger power systems. For application to the UC problem, we additionally add the action that keeps all current generator commitments the same to $A_{\pi}(s)$. This `do nothing' action is always feasible as it does not change any generator commitments. 

The breadth and depth of the search tree are controlled by $\rho$ and $H$ respectively. There is a trade-off between these two parameters, as reducing $\rho$ and increasing $H$ both increase the run time of Guided UCS. Setting $\rho$ to be large results in a narrow search that can cause operating costs to increase. Similarly, a shallow search with low $H$ can also degrade performance due to short-sighted decision-making. In Section \ref{ch1:experiments:parameters} we demonstrate the impact of $H$ and $\rho$ parameter settings on run time, operating costs, and schedule characteristics for the UC problem.

The expansion policy can be defined with expert rules, trained by supervised learning on existing UC solutions, or trained by RL as in this research. Next, we will describe our approach for training the expansion policy with RL.

\subsection{Expansion Policy} \label{ch1:guided_ucs:policy}

\begin{figure}
    \centering
    \includegraphics[width=\textwidth]{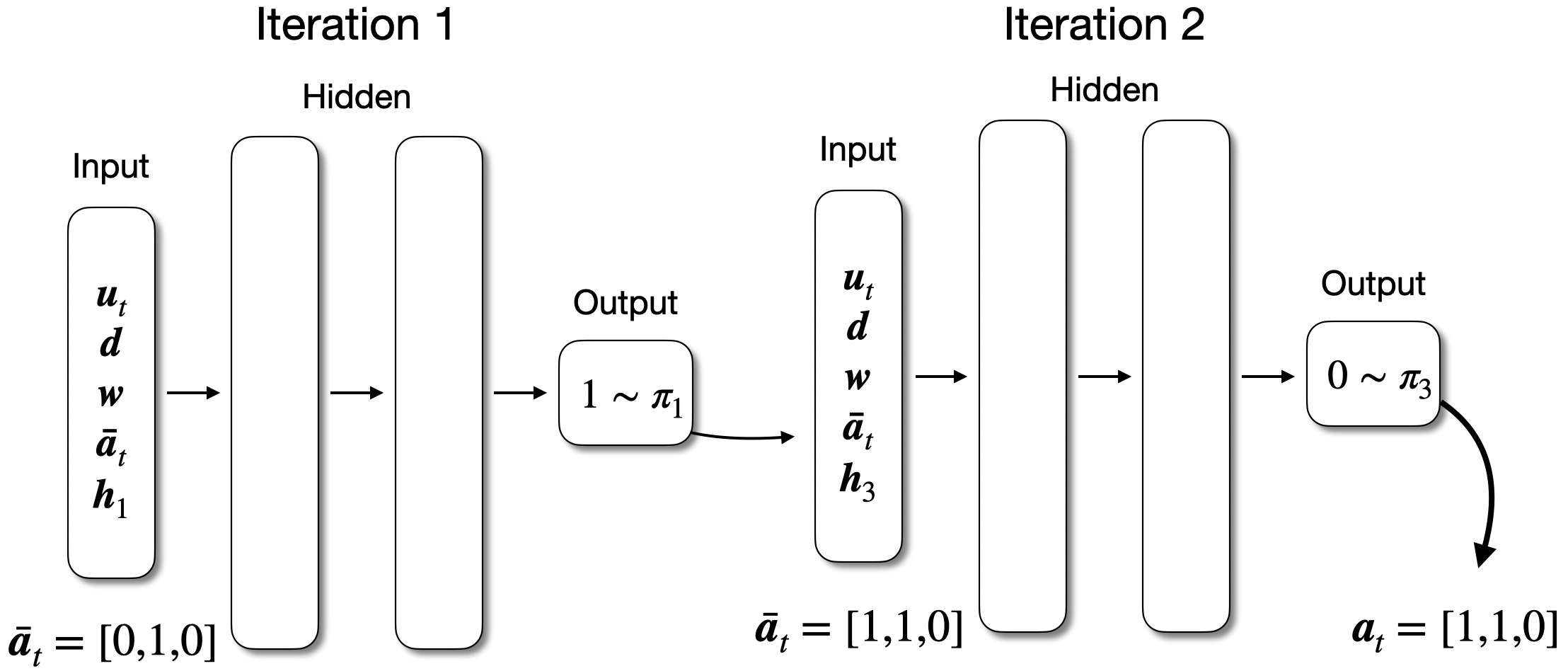} 
    \caption{Sequential feed-forward neural network architecture used to parameterise the expansion policy. Each generator commitment is classified sequentially with the current action sequence $\bar{a}_t$ used to estimate the following commitment. In the example, the second generator is constrained to remain on, so at the first iteration, $\bar{a}_t = [0,1,0]$. Commitment decisions for the unconstrained generators $i=\{1, 3\}$ are made in sequence, sampling from the distribution $\pi_i$ calculated using the neural network.}
    \label{ch1:fig:policy_diagram}
\end{figure}

The expansion policy is trained using the policy gradient RL method proximal policy optimisation (PPO) \cite{schulman2017proximal}, described in detail in Section \ref{methodology:pg:ppo}. The RL agent (i.e. the expansion policy) interacts with the environment described in Section \ref{ch1:env} to improve performance with respect to the reward function. As discussed in Section \ref{methodology:rl_taxonomy:value_policy} policy gradient methods have several advantages over value-based methods in certain problem domains. Policy gradient methods can naturally learn stochastic policies and are better suited to large action spaces \cite{sutton1998introduction}. The ability to learn a stochastic policy is essential in the context of training an expansion policy for Guided UCS, as the expansion policy should propose a diverse range of actions to add to the search tree in guided expansion. Entropy regularisation, discussed in Section \ref{methodology:pg:ppo}, can be used in policy gradient methods to further promote stochastic policies. PPO was chosen as a state-of-the-art policy gradient method that incorporates a clipped loss function which helps prevent catastrophic performance decreases. We optimise the entropy-regularised PPO objective given in Equation \ref{methodology:eq:ppo_entropy_objective}.  

In order to train an expansion policy in large action spaces, we use a sequential feed-forward neural network architecture, predicting a sequence of individual generator actions to create the joint commitment action. While recurrent neural network (RNN) architectures such as long short-term memory (LSTM) networks have excelled in sequence-based learning, most notably natural langauge processing (NLP) tasks such as speech-recognition \cite{graves2013speech}, simpler feed-forward neural networks are also competitive in language modelling \cite{dauphin2017language} and speech synthesis \cite{oord2016wavenet} and benefit from lower computational cost and generally better training stability \cite{miller2018stable}. 

Fully enumerating the actions at the output layer of the policy is not feasible due to the size of the action space. Parametrising the multi-dimensional action space with $N$ output nodes is also not appropriate due to the strong dependency of each generator's action probability on that of the other generators. Instead, the policy is parametrised as a binary classifier which sequentially predicts each value in the sequence $\boldsymbol{a} = [a_1, a_2,...,a_N]$ representing a commitment decision where $a_i \in \{0,1\}$ are sub-actions giving the commitment for generator $i$ (Figure \ref{ch1:fig:policy_diagram}). The output of the classifier at each iteration is passed as an input into the next forward-pass through the network, thus maintaining the history of generator commitments already decided. In addition, the input vector includes a one-hot encoding indicating the $a_i$ being classified on each forward pass as well as the observation. This parametrisation succeeds in preserving the interdependencies between generators while remaining tractable for larger power systems.

A disadvantage of this approach is that it is not possible to analytically compute the distribution $\pi(a|s)$. It is necessary to approximate this distribution in order to determine which actions meet the branching threshold $\rho$ in guided expansion (Equation \ref{ch1:guided_ucs:guided_expansion}). We use a Monte Carlo method to approximate the distribution: we estimate $\pi(a|s) = \frac{n_a}{N}$ where $n_a$ is the number of times $a$ was sampled and $N$ is the total number of samples.

\subsection{Training Details} \label{ch1:guided_ucs:policy_training}

Policies were trained for power systems of $N \in \{5, 6, 7, 8, 9, 10, 20, 30 \}$ generators using the method described in Section \ref{ch1:guided_ucs:policy}. Each power system is an instance of the environment described in Section \ref{ch1:env}, using the generator data from \cite{kazarlis1996genetic}. The trained policies were used as expansion policies in Guided UCS in the experiments described in the next section, Section \ref{ch1:experiments}. Here we provide technical details of the policy training.

\paragraph{Training Episodes} During training, the RL agent samples days at random from the training data, defining the demand forecast $\boldsymbol{d}$ and wind forecast $\boldsymbol{w}$ for the episode. The maximum daily wind penetration in training was 58\%; the minimum was 0.1\%. The RL agent therefore experiences more extreme levels of wind penetration than observed in the test problems where wind penetration is between 3\%--37\%. The initial generator up down times $\boldsymbol{u}_0$ are set randomly to encourage exploration of the state space. To speed up the early stages of training, an episode ends if the agent encounters lost load. 


\paragraph{Actor-Critic Parameters} A summary of the parameters used to train the 10, 20 and 30 generator policies is given in Table \ref{ch1:tab:policy_params}. Each policy was trained with 8 workers, with weights of the actor-critic neural network updated asynchronously after every 2000 forward passes through the policy network. Every update is an \textit{epoch}. This implementation differs from standard implementations of PPO due to the sequential parametristation of the policy: each forward pass through the network is recorded separately in the replay buffer, along with the complete encoding of the input vector (that is, including the one-hot encoding $\boldsymbol{h}$ and action draft $\bar{\boldsymbol{a}}$), such that for each timestep, multiple entries can be added to the replay buffer. As the input vector dimensions are different for the actor and critic (the actor uses the one-hot encoding $\boldsymbol{h}$ and draft action $\boldsymbol{\bar{a}}$), we used different architectures and replay buffers for the actor and critic networks. A grid search approach was used to determine the best performing combination of architectures, testing all combinations of the three architectures used in \cite{henderson2018deep}: (64, 64); (100, 50, 25); (400, 300), where ($x_i$) indicates the number of  nodes in hidden layer $i$. In addition, we set the discount factor $\gamma=0.95$. This assigns more credit to actions which are temporally close to the reward as compared with setting $\gamma=1$. We included an entropy bonus \cite{mnih2016asynchronous} for two reasons: first to encourage exploration in training; second, to prevent the policy from converging to a policy which strongly favours a single action, so as to maintain a diverse set of actions in Guided UCS.

\begin{table}[t]
    \centering
    \input{04-chapter1/tables/policy_params}
    \caption{Parameter settings for training the expansion policy using PPO.}
    \label{ch1:tab:policy_params}
\end{table}

\paragraph{Reward Transformation} 

We also transformed the reward function using $\hat{r} = \log(m/r)$, where $m$ is a constant used to scale the reward to around the range $[-1,1]$. The log transformation dampens extreme negative rewards resulting from lost load events, which can otherwise result in overly conservative behaviour.  

\paragraph{Observation Pre-Processing}

Observations were pre-processed before passing through the policy network to improve training stability. First, we capped the generator up/down times $u_i$ at the minimum up/down time, and scaled this to between $[-1,1]$ such that $-1$ indicates that the generator is offline and has satisfied its minimum down time constraint and likewise for the minimum up time constraint. This bounds the state space while exploiting symmetries in the state space such that there is no loss of information for the expansion policy. Second, the timestep was normalised by the episode length $\hat{t} = t/T$. Third, we scaled demand and wind variables by a constant factor, $1 / \sum_i p_{\text{max}}$. This aims to keep all state variables within roughly the same order of magnitude which is known to improve neural network training. Lastly, we truncated the forecasts $\boldsymbol{d}, \boldsymbol{w}$ to a forecast window of $k$ periods ahead of the decision period. This method benefits from reducing the size of the state vector and only presenting the most `relevant' forecast information. 

\subsubsection{Convergence Results} 

\begin{figure}[t]
    \centering
    \includegraphics[width=0.6\textwidth]{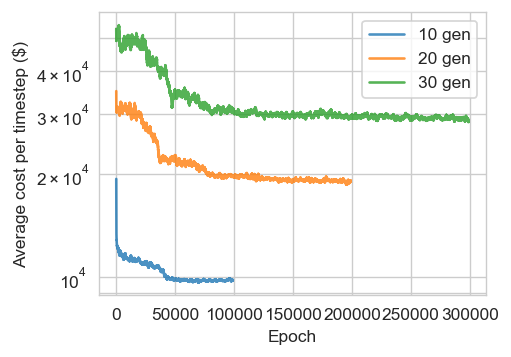}
    \caption{Average cost per timestep for 10, 20 and 30 generator policies during training. Plot shows a moving average over 1000 epochs. The 10 and 20 generator problems converged more quickly than the 30 generator problem. }
    \label{ch1:fig:training_plots} 
\end{figure}

Figure \ref{ch1:fig:training_plots} shows the convergence of 10, 20 and 30 generator policies in terms of operating cost per timestep. The agents improve rapidly at the beginning of training, learning to avoid lost load events. However, the average episode length is around 42 periods in all three problems by the end of training, indicating significant loss of load probability (episodes end when lost load is observed). 

We used the 10, 20 and 30 generator expansion policies to solve the 20 test problems `model-free', sampling directly from the distribution $\pi(a|s)$ for each state $s$, to evaluate the potential of the policies to solve the unseen problems without tree search. The results are compared with MILP benchmarks in Table \ref{ch1:tab:model_free_costs}. The model-free solutions are significantly more expensive, and have higher loss of load probability. This indicates that the policies are unable to provide good solutions to the UC test problems without tree search. In the next section, the expansion policies are applied in Guided UCS to solve the test problems.

\begin{table}[t]
    \centering
    \small
    \scalebox{0.9}{
    \input{04-chapter1/tables/model_free_costs}}
    \caption{Comparison of model-free solutions using expansion policies $\pi(a|s)$, with MILP benchmarks for the 20 test problems. The model-free solutions have much higher loss of load probability and hence higher operating costs. Average run times are at least one order of magnitude lower than MILP.}
    \label{ch1:tab:model_free_costs}
\end{table}

\section{Evaluating Guided UCS} \label{ch1:experiments}

Section \ref{ch1:guided_ucs} described Guided UCS, an RL-aided tree search algorithm that incorporates an expansion policy to reduce the branching factor of the search tree. In Section \ref{ch1:guided_ucs:policy_training} we trained expansion policies for systems of between 5--30 generators. Now, we will evaluate the performance of Guided UCS on the 20 test problems described in Section \ref{ch1:env:benchmarks}. As stated in Section \ref{ch1:guided_ucs:guided_expansion}, the depth and breadth parameters $H$ and $\rho$ in Guided UCS are important variables impacting the quality of the solution and the run time. First, we conduct a simulation study varying $H$ and $\rho$ to determine suitable parameters for the next experiments. We then compare Guided UCS with traditional (unguided) UCS for 5--10 generator problem instances, demonstrating that Guided UCS has better run time complexity in the number of generators while achieving similar solution costs. Last, we compare Guided UCS with the MILP benchmarks for 10, 20 and 30 generator problems in Section \ref{ch1:experiments:milp_comparison}, showing that Guided UCS achieves lower operating costs overall and exhibits novel schedule characteristics.

\subsection{Parameter Analysis} \label{ch1:experiments:parameters}

\begin{figure}
    \centering
    \includegraphics[width=0.6\textwidth]{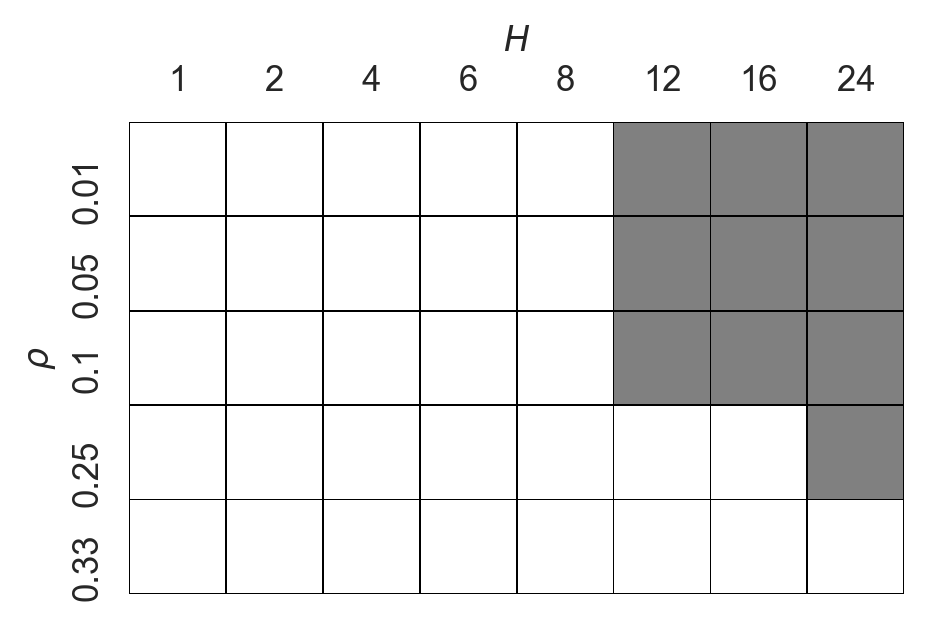}
    \caption{Parameter combinations for parameter analysis experiment. Those which did not complete in 24 hours are shaded in grey.}
    \label{ch1:fig:5gen_parameter_heatmap}
\end{figure}

The first experiment investigates the impact of search breadth and depth on performance. We considered the 5 generator problem only and set $H=\{1,2,4,6,8,12,16,24\}$ and $\rho=\{0.01,0.05,0.1,0.25,0.33\}$. Under some parameter combinations, a subset of the 20 test problems did not complete within a 24 hour time budget and these results are not reported. Figure \ref{ch1:fig:5gen_parameter_heatmap} shows the parameter combinations and those that did not complete within 24 hours (shaded in grey). 

In all experiments, we set $N_s = 100$, the number of scenarios used to calculated expected edge costs in Equation \ref{ch1:eq:search_tree_cost}. Here we will analyse performance as a function of $H$ and $\rho$ in terms of costs, run time and characteristics of the schedules. Using these results, we aim to determine parameter settings which achieve good solution quality within practical run times. 



\begin{figure}
\centering
\begin{subfigure}[b]{.8\linewidth}
\includegraphics[width=\linewidth]{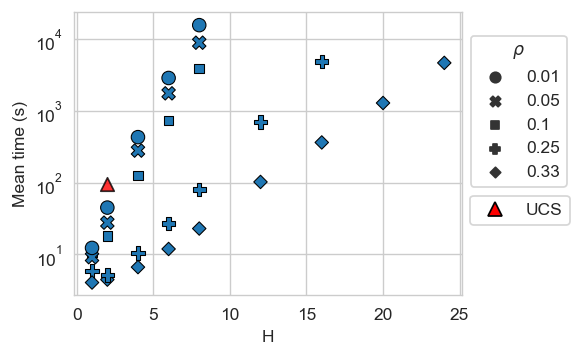}
\caption{}\label{ch1:fig:5gen_param_comparison:time_v_H}
\end{subfigure}
\begin{subfigure}[b]{.85\linewidth}
\includegraphics[width=\linewidth]{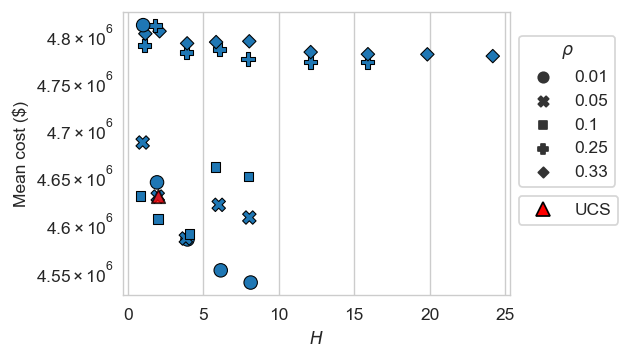}
\caption{}\label{ch1:fig:5gen_param_comparison:cost_v_H}
\end{subfigure}

\caption{Comparison of run time and cost for settings of $\rho$ and $H$ for the 5 generator problem. Figure \ref{ch1:fig:5gen_param_comparison:time_v_H} verifies that run time grows exponentially with $H$ for a fixed setting of $\rho$. The two largest settings of $\rho$ perform worst, even with large $H$. Figure \ref{ch1:fig:5gen_param_comparison:cost_v_H} shows that costs generally decrease with $H$ for fixed $\rho$. Performance of UCS (results from Section \ref{ch1:ucs:results}) is also shown to have similar nearly identical costs to Guided UCS with $H=2, \rho=0.05$. The lowest setting of $\rho=0.01$ achieves the lowest costs for fixed $H$ due to the wider search breadth, but Figure \ref{ch1:fig:5gen_param_comparison:time_v_H} shows this scales most quickly with run time.} 
\label{ch1:fig:5gen_param_comparison}
\end{figure}

Figure \ref{ch1:fig:5gen_param_comparison} summarises the performance of Guided UCS with different settings of $H$ and $\rho$ along side the UCS solutions with $H=2$ from Section \ref{ch1:ucs:results}. Figure \ref{ch1:fig:5gen_param_comparison:time_v_H}  verifies the exponential time complexity in $H$ for fixed $\rho$ (linear fits on log scale) and clearly shows the run time reduction achieved by Guided UCS at $H=2$ for all settings of $\rho$. The lowest settings of $\rho$ exhibit the steepest rise in run time with increasing search depth.

Figure \ref{ch1:fig:5gen_param_comparison:cost_v_H} compares operating costs as a function $H$ and $\rho$. The largest values of $\rho = \{0.25, 0.33\}$ (maximum branching factor of 4 and 3, respectively) consistently have among the highest operating costs consistently regardless of the depth $H$. Overall, the lowest costs were produced by setting $H=8$ and $\rho=0.01$ (maximum branching factor of 100), which also had the longest run time as shown in Figure \ref{ch1:fig:5gen_param_comparison:cost_v_H}. While the longer running parameter settings generally had lower costs, there is considerable variation between parameter combinations that result in similar run times: for instance, the $(H, \rho)$ parameter settings $(4, 0.01)$ and $(16,0.33)$ both have run times of around 400 seconds, the former's operating costs are around 4\% lower than the latter's. 

Guided UCS with $(H, \rho)$ settings of $(2, 0.05)$ and $(1, 0.1)$ both have similar operating costs to the UCS solution, but with lower run times (roughly 10 times faster in the latter case). Settings of $\rho=\{0.01, 0.05, 0.1\}$ all achieve consistently similar or lower costs to UCS for $H>1$, although $\rho=0.01$ is the only setting which improves monotonically with increasing depth $H$. However, the run time of $\rho=0.01$ is most sensitive to $H$, as shown in Figure \ref{ch1:fig:5gen_param_comparison:time_v_H}. 

\begin{figure}
    \centering
    \includegraphics[width=0.7\textwidth]{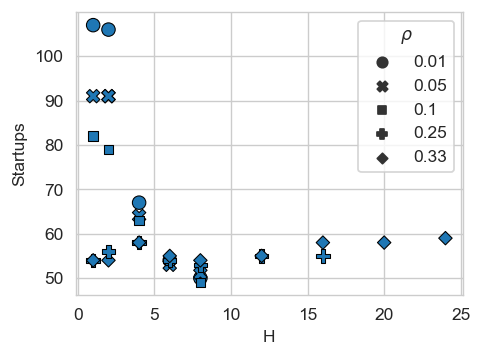}
    \caption{Total number of startups for the 20 unseen test problems with parameter settings of $H$ and $\rho$. Startups generally decrease with search depth up to $H=8$, after which we observe a small increase in startups.}
    \label{ch1:fig:5gen_startups_v_H}
\end{figure}

Varying $\rho$ and $H$ results in different schedule characteristics such as startup frequency and up times of generators. Startups were found to generally decrease with increasing $H$, shown in Figure \ref{ch1:fig:5gen_startups_v_H}. Small settings of $\rho$ usually have larger numbers of startups for fixed $H$, as a wider branching factor allows for greedier decision-making. Generator utilisation (proportion of periods spent online) also varies with $\rho$ and $H$, such as reduced utilisation of peaking plants when the depth $H$ is increased.

We determined that $\rho=0.05$ (maximum branching factor of 20) was the most suitable value to use for subsequent experiments, as it achieves consistently low operating costs and scales well with search depth relative to $\rho=0.01$. Using $\rho=0.05$, increasing $H$ as far as possible in practical computing times is likely to improve solution quality.

\subsection{UCS Comparison} \label{ch1:experiments:ucs_comparison}


Having conducted experiments to determine parameter settings for Guided UCS, in the second experiment we compare Guided UCS with unguided UCS (which does not use an expansion policy). In particular, we aim to determine whether Guided UCS succeeds in achieving sub-exponential time complexity in the number of generators, while achieving similar operating costs to UCS which exhaustively solves the search tree with no branches removed. Guided UCS with $\rho=0.05$ and $H=2$ (which was found to result in similar operating costs to UCS) was used to solve the test problems for power systems of 5--10 generators. The results are compared with UCS with $H=2$, as reported in Section \ref{ch1:ucs:results}. 

\begin{figure}[t]
    \centering
    \includegraphics[width=0.6\textwidth]{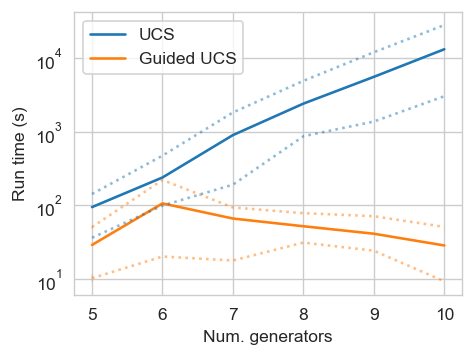}
    \caption{Mean computation time for guided and UCS from 5--10 generators. Dotted lines show the maximum and minimum time taken for a single problem. UCS run time increases exponentially with the number of generators for a fixed search depth, while Guided UCS shows no significant increase in run time.}
    \label{ch1:fig:time-taken}
\end{figure}

The run times of guided and UCS for systems of 5--10 generators are compared in Figure \ref{ch1:fig:time-taken}. Whereas the mean run time of UCS rises exponentially with the number of generators, run time for Guided UCS remains stable. For the 10 generator problem, the run time for Guided UCS is around 0.2\% that of UCS. By limiting the branching factor to $\frac{1}{\rho}$, the run time of Guided UCS slightly decreases after $N>6$, enabling application to larger power systems. 

Given the constant run time complexity of Guided UCS, the apparent downward trend in run time may be explained by challenges in tuning policy entropy for larger power systems. For larger action spaces, the number of actions meeting the branching threshold $\rho$ may be highly sensitive to the entropy of the expansion policy. If policy entropy is low, the policy approaches determinism and only a small number of actions are likely to be added to the search tree at each node. Similarly, if the policy entropy is too high, very few actions will be added, also resulting in low branching factors and hence shorter average run times. We address this issue explicitly in the next chapter using target entropy regularisation to solve the 100-generator problems in Section \ref{ch2:100gen}. 

The operating cost results for the 5 and 10 generator problems are summarised in Table \ref{ch1:tab:guided-vs-unguided}. Mean operating costs are very similar for both problem settings. Guided UCS achieves lower loss of load probability in the 10 generator problem. Guided UCS is successful in substantially reducing run time without notable increase in operating costs. Both methods exhibit substantial run time variability of roughly one order of magnitude between minimum and maximum run times.

\begin{table}[t]
    \centering
    \caption{Comparison of guided and unguided search for 5 and 10 generator problems.}
    \scalebox{0.7}{
    \input{04-chapter1/tables/guided_unguided_comparison}
    }
    \label{ch1:tab:guided-vs-unguided}
\end{table}

\begin{figure}
    \centering
    \includegraphics[width=0.8\textwidth]{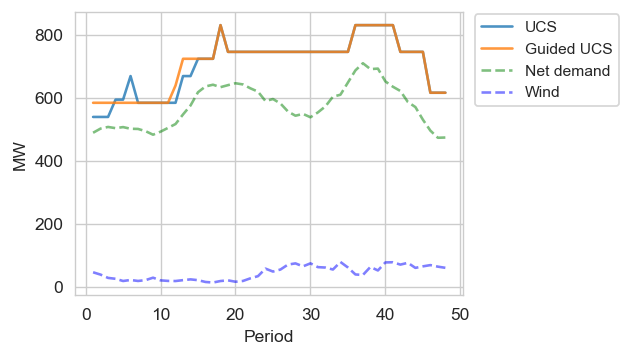}
    \caption{Comparison of committed capacity of Guided UCS and UCS schedules for the 2018-03-08 test problem with 5 generators. The generation floor is the sum of minimum operating outputs $p_{\text{min}}$ of committed generators. UCS makes more frequent commitment changes and operates tighter reserve margins.}
    \label{ch1:fig:guided_unguided_schedule}
\end{figure}

Schedules produced by guided and UCS were usually similar, but there were notable differences on some problems as demonstrated by Figure \ref{ch1:fig:guided_unguided_schedule}. In this example for the 5 generator problem, unguided search makes more frequent commitment changes and operates a tighter reserve margin. Guided UCS has longer periods of no commitment changes. Overall, Guided UCS used 11\% and 15\% fewer startups than UCS for 5 and 10 generator problems, respectively.

\subsection{MILP Comparison} \label{ch1:experiments:milp_comparison}

The results in Section \ref{ch1:experiments:ucs_comparison} show that Guided UCS with $H=2$ and $\rho=0.05$ has constant time complexity in the number of generators and achieves similar operating costs to exhaustive UCS with $H=2$. In our final experiment, Guided UCS is compared with the MILP($4\sigma$) and MILP(perfect) benchmarks for systems of 10, 20 and 30 generators. This experiment shows that Guided UCS is competitive with industry standard approaches, and performance in terms of operating costs does not deteriorate with increasing problem size. We set the breadth parameter $\rho=0.05$, allowing for a maximum branching factor of 20. While it was only possible to run UCS with $H=2$ due to the high computational cost of this approach, the lower run times of Guided UCS allowed the search depth to be increased to $H=4$ to achieve further reductions in operating costs. This decision was made on the basis of the parameter study in Section \ref{ch1:experiments:parameters}, where Figure \ref{ch1:fig:5gen_param_comparison:time_v_H} showed setting $\rho=0.05$ and $H=4$ gave a mean run time in the order of 100 seconds for the 5 generator problem, which is a practical time budget for real-world UC problems. Given constant run time complexity of Guided UCS in the number of generators, mean run times for the larger systems are likely to be similar with these parameter settings. Furthermore, given the run time variability of Guided UCS observed in Table \ref{ch1:tab:guided-vs-unguided}, maximum episode run times may be impractically large for larger values of $H$. 

The results are presented in Table \ref{ch1:tab:guided_vs_milp}. Guided UCS achieves lower operating costs than MILP($4\sigma$) for all three problem instances, with improvements of 0.33\%, 0.87\% and 0.45\% for 10, 20 and 30 generator problems respectively. Guided UCS has notably lower LOLP in all cases, which is reflected in lower standard deviation of operating costs over the 1000 realisations of demand and wind. The worst case (maximum) costs are also significantly lower, with the margin of improvement increasing from 6.3\% with 10 generators to 33\% for the 30 generator case. This is indicative of superior uncertainty management using Guided UCS as compared with the $4\sigma$ reserve strategy. Compared with MILP(perfect), which adopts no reserve constraint and is evaluated on the single point forecast scenario, Guided UCS is 6.2\%, 6.6\% and 8.0\% more expensive for 10, 20 and 30 generator problems, respectively. 

\begin{table}
    \centering
    \resizebox{\textwidth}{!}{
    \input{04-chapter1/tables/all_best_results_with_time}}
    \caption{Comparison of MILP and Guided UCS solutions for 10, 20 and 30 generator problems.}
    \label{ch1:tab:guided_vs_milp}
\end{table}

Mean run time of Guided UCS is of the order of 100 seconds in all cases, as predicted. Guided UCS is substantially slower than the MILP solutions, by around 1--2 orders of magnitude. As discussed in Section \ref{methodology:mathop:milp}, MILP solvers have benefited from a number of efficiency improving innovations in commercial and open-source solvers. For this research we used the open-source COIN-OR library's branch-and-cut algorithm \cite{johnjforrest_2020_3700700}, which implements methods such as cutting planes, parallelism and branching heuristics to very efficiently solve the MILP. Optimising the implementation of Guided UCS is beyond the scope of this thesis, but there are several avenues for improvement including parallel node evaluations, more efficient economic dispatch calculation \cite{chowdhury1990review} and implementation in a compiled language such as C++. As a result, it is not possible to address differences in absolute computational demands of MILP and Guided UCS by run time alone. 

\begin{figure}
    \centering
    \includegraphics[width=0.8\textwidth]{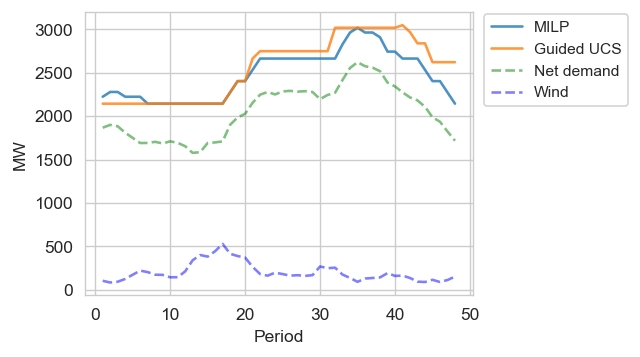}
    \caption{Committed capacity of Guided UCS and MILP($4\sigma$) solutions to the 2019-11-09 test problem (20 generators). Guided UCS makes more frequent use of actions making no commitment changes, thereby avoiding startup costs. The Guided UCS solution also employs larger reserve margins at the end of the day when forecast errors can be larger.}
    \label{ch1:fig:guided_milp_schedule}
\end{figure}

\begin{figure}
    \centering
    \includegraphics[width=\textwidth]{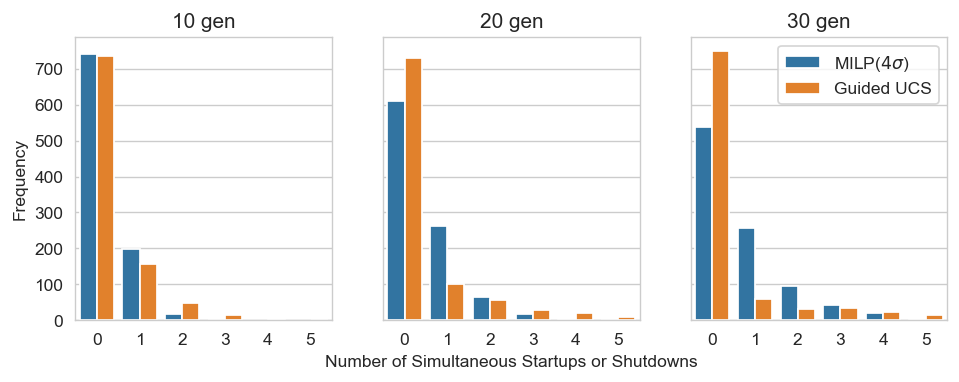}
    \caption{Frequency of actions by number of simultaneous startups or shutdowns, comparing Guided UCS and MILP$(4\sigma)$. The Guided UCS solutions have a longer tail of actions with multiple simultaneous commitment changes. For 20 and 30 generator problems, Guided UCS uses the `do nothing' action (0 commitment changes) more frequently than MILP.}
    \label{ch1:fig:switches}
\end{figure}

There were several differences in terms of schedule characteristics comparing MILP$(4\sigma$) with Guided UCS. Figure \ref{ch1:fig:guided_milp_schedule} shows example solutions for Guided UCS and MILP for the 20-generator test problem, 2019-11-09. Compared with the MILP solution, Guided UCS is characterised by longer periods of no commitment changes, and larger reserve margins at the end of the day when net demand uncertainty is greater. Actions taken by Guided UCS were more concentrated towards those which change the commitment of multiple generators at once, as illustrated in Figure \ref{ch1:fig:switches}. Guided UCS exhibits a `long tail' of actions changing multiple generator commitments at once, whereas MILP is concentrated towards actions with fewer simultaneous startups/shutdowns. For the 20 and 30 generator problems, we also found that the Guided UCS made more frequent use of the `do nothing' action, making no commitment changes.

\section{Discussion} \label{ch1:discussion}

Our results found that whereas UCS has exponential run time complexity in the number of generators, Guided UCS exhibited no significant increase in run time when increasing from systems of 5 to 10 generators. In addition, we found no significant deterioration in solution quality as measured by operating costs when using Guided UCS. Comparing with MILP, Guided UCS outperformed MILP($4\sigma$) for all problem sizes. Despite not employing a reserve constraint, Guided UCS achieved roughly half the LOLP as compared with MILP($4\sigma$), representing more secure operation.

While there are multiple possible sources of the improved solution quality of Guided UCS relative to mathematical programming, the most important in our experiments was improved management of uncertainty afforded using RL as compared with deterministic mathematical programming. Sampling directly from the environment enables the RL agent to develop a rich representation of uncertainty, whereas the deterministic MILP$(4\sigma)$ formulation is reliant on heuristics. The RL agent thus optimises directly for expected operating costs, providing more rigorous consideration of extreme scenarios that was evidenced by lower LOLP and better worst case operating costs. 

An additional advantage of RL is the ability to represent generator cost functions and constraints accurately, without linear approximations that are required in linear programming approaches. While this is unlikely to have contributed significantly to improved solution quality in this instance, in some cases linear functions may provide an inadequate representation of generator fuel cost curves, such as those exhibiting non-convex valve point effects \cite{victoire2005reserve}. 


The results in this chapter show that the policy learned by model-free RL was effective in intelligently selecting promising actions to add to the search tree. This was evident in the qualitative differences between schedules produced by Guided UCS and UCS. Guided UCS used fewer startups and took fewer greedy actions to minimise short-term costs; these actions were pruned by guided expansion. Guided UCS used novel strategies that differed qualitatively from those produced using MILP. Guided UCS tended to use more extreme actions, changing multiple generator commitments at once, with longer periods of no commitment changes. These actions may be difficult for human operators to identify and indicates scope for Guided UCS to be used as part of a decision support tool for system operators.

Guided UCS achieved lower LOLP as compared with MILP($4\sigma$), and lower worst case costs. The improvement in security of supply relative to MILP($4\sigma$) can be attributed in part to the Monte Carlo approach used to estimate the expected edge costs on the search tree described in Section \ref{ch1:methodology:search_tree}. This allows for costs to be evaluated under scenarios of demand and wind, thus anticipating possible lost load events. Whereas the MILP benchmark uses a heuristic reserve constraint, Guided UCS includes security of supply as part of the cost function, with a parameter (value of lost load $V$ in Equation \ref{ch1:eq:ll_cost}) that weights security relative to other costs. The ability to shape rewards to reflect societal value of security of supply, economic affordability and environmental sustainability is an important property of RL and tree search approaches to the UC problem, that is not easily afforded by mathematical programming methods. In Section \ref{ch3:curtailment}, we further investigate reward shaping in the UC problem by introducing carbon pricing.

A hybrid methodology combining strategies learned using RL with the advantages of deterministic mathematical programming such as a measurable optimality gap and fast solve times is a worthy topic of further research. Comparison of schedules produced by Guided UCS and MILP($4\sigma$) indicated that Guided UCS dynamically allocated reserves, reflecting increasing uncertainty throughout the day as the decision horizon increased. Incorporating the reserve margins of Guided UCS solutions into MILP formulations could provide a means of improving solution quality of deterministic mathematical programming under high uncertainty.

Guided UCS possesses attractive run time properties, with constant complexity in the number of generators $N$, linear complexity in the number of periods $T$ and linear complexity in the number of scenarios used to evaluate expected edge costs $N_s$. However, performance depends considerably on the quality of the expansion policy $\pi(a|s)$ and its ability to intelligently select promising actions to retain in the search tree. Nonetheless, our results did not find a deterioration in performance with increasing numbers of generators (Guided UCS outperformed MILP($4\sigma$) by the greatest margin in the 20 generator case), suggesting that the model-free RL approach employed to train the expansion policies was effective even in very large action spaces (up to roughly 1 billion actions in the 30 generator case). The sequential parametrisation of the policy was an effective approach in enabling scaling to larger power systems. The linear complexity of Guided UCS in $T$ may be valuable in future power markets as many transition towards higher frequency settlements. Finally, linear complexity in $N_s$ may also be a valuable characteristic in more complex stochastic environments which cannot be effectively reduced to a few scenarios for stochastic programming approaches reviewed in Section \ref{literature:suc}.

\subsection{Related Work}

Despite recent successes of RL in numerous challenging domains, until now only a small body of research has investigated applications of RL to the UC problem. Existing research \cite{jasmin2009reinforcement, dalal2015reinforcement, jasmin2016function, navin2019fuzzy, li2019distributed}, reviewed in detail in Section \ref{literature:rl}, has focused on small numbers of generators, in part due to the combinatorial action space that limits the application of existing RL methods `out-of-the-box'. Fuzzy Q-learning is used in \cite{navin2019fuzzy} to solve the widely-studied 10-generator Kazaris \cite{kazarlis1996genetic} UC problem. The results of this study are not directly comparable to those in this chapter, due to its use of a single demand profile and absence of uncertainty. In the most similar research \cite{dalal2015reinforcement}, tree search methods are applied to a system of 12 generators, which, to the best of our knowledge, is the largest prior study in this area. However, the problem considered is deterministic and does not consider generalisability to unseen problems. In subsequent related research, a larger power system is considered but the UC problem is simplified to a single commitment decision per day \cite{dalal2016hierarchical}. To the best of our knowledge, the work presented in this chapter is unique in considering generalisability to unseen profiles and training on multiple episodes, and is the largest simulation study of its kind.

\section{Conclusion} \label{ch1:conclusion}

In this chapter we formally described the UC problem as an MDP, and presented a power system simulation environment suitable for RL research in this area. We then showed how the UC problem can be formulated as a search tree and solved using the traditional planning algorithm uniform-cost search (UCS). This method is competitive in terms of cost with industry-standard MILP benchmarks and is suited to stochastic problems, but suffers from exponential time complexity in the number of generators. To improve the run time complexity in the number of generators we presented guided expansion, a method by which an RL-trained policy can be used to reduce the branching factor of a search tree. We applied this in Guided UCS, a guided tree search algorithm, with a policy trained with proximal policy optimisation (PPO). We conducted a parameter analysis to determine suitable values of the depth and breadth parameters for Guided UCS, considering run time, operating costs and schedule characteristics. 

Guided UCS was found to exhibit constant run time complexity in the number of generators, and achieved similar operating costs to UCS. Guided UCS was also shown to be competitive with the MILP benchmark employing a reserve constraint, resulting in lower operating costs and improvements to security of supply. Whereas existing research applying RL to the UC problem has been limited to small power systems \cite{jasmin2009reinforcement, dalal2015reinforcement, jasmin2016function, navin2019fuzzy, li2019distributed}, guided tree search was successful in outperforming deterministic MILP for problems of up to 30 generators. To the best of our knowledge, this is the largest application of RL to the UC problem in the literature. In addition, we found qualitative differences between schedules produced by Guided UCS and MILP, with Guided UCS using complex and unusual strategies that may be difficult for human operators to identify.

The principle of guided expansion is applicable in other tree search algorithms. In Chapter \ref{ch2}, we apply informed and anytime methods that leverage domain-specific knowledge of the UC problem to improve performance.


%% file: 04-chapter1/tables/gen_info_10.tex
\begin{tabular}{rrrrrrrrrr}
\toprule
 Gen ID &  $p_{\text{min}}$ &  $p_{\text{max}}$ &  $u_0$ &     $a$ &   $b$ &    $c$ &  $t_{\text{min}}^{\text{down}}$ &  $t_{\text{min}}^{\text{up}}$ &  $c^s$ \\
\midrule
      1 &               150 &               455 &     16 & 0.00048 & 16.19 & 1000.0 &                              16 &                            16 &   4500 \\
      2 &               150 &               455 &     16 & 0.00031 & 17.26 &  970.0 &                              16 &                            16 &   5000 \\
      3 &                20 &               130 &    -10 & 0.00200 & 16.60 &  700.0 &                              10 &                            10 &    550 \\
      4 &                20 &               130 &    -10 & 0.00211 & 16.50 &  680.0 &                              10 &                            10 &    560 \\
      5 &                25 &               162 &    -12 & 0.00398 & 19.70 &  450.0 &                              12 &                            12 &    900 \\
      6 &                20 &                80 &     -6 & 0.00712 & 22.26 &  370.0 &                               6 &                             6 &    170 \\
      7 &                25 &                85 &     -6 & 0.00079 & 27.74 &  480.0 &                               6 &                             6 &    260 \\
      8 &                10 &                55 &     -2 & 0.00413 & 25.92 &  660.0 &                               2 &                             2 &     30 \\
      9 &                10 &                55 &     -2 & 0.00222 & 27.27 &  665.0 &                               2 &                             2 &     30 \\
     10 &                10 &                55 &     -2 & 0.00173 & 27.79 &  670.0 &                               2 &                             2 &     30 \\
\bottomrule
\end{tabular}

%% file: 04-chapter1/tables/test_problems_10.tex
\begin{tabular}{llrrr}
\toprule
      Date & Day &  Wind (\%) &  $D_{\text{min}}$ (MW) &  $D_{\text{max}}$ (MW) \\
\midrule
2016-01-12 & Tue &       11.0 &                  990.9 &                 1593.1 \\
2016-07-25 & Mon &       21.7 &                  790.2 &                 1118.5 \\
2016-11-21 & Mon &       18.1 &                  972.5 &                 1545.3 \\
2017-03-18 & Sat &       21.3 &                  827.5 &                 1253.9 \\
2017-04-07 & Fri &        7.4 &                  880.2 &                 1214.2 \\
2017-05-12 & Fri &       14.4 &                  867.5 &                 1176.9 \\
2017-05-26 & Fri &       14.8 &                  817.4 &                 1058.7 \\
2017-06-25 & Sun &        2.8 &                  700.3 &                  954.1 \\
2017-12-18 & Mon &        8.1 &                 1011.6 &                 1599.9 \\
2017-12-30 & Sat &       18.4 &                  916.4 &                 1318.7 \\
2018-01-15 & Mon &       15.0 &                  893.3 &                 1513.2 \\
2018-03-08 & Thu &        7.0 &                 1024.1 &                 1509.8 \\
2018-03-18 & Sun &       36.8 &                  943.6 &                 1432.6 \\
2018-05-30 & Wed &       16.5 &                  782.9 &                 1177.1 \\
2018-09-11 & Tue &       30.8 &                  799.6 &                 1183.8 \\
2018-11-13 & Tue &       31.1 &                  852.6 &                 1409.8 \\
2019-04-03 & Wed &       15.6 &                  901.0 &                 1306.3 \\
2019-05-22 & Wed &       10.1 &                  819.0 &                 1104.8 \\
2019-10-30 & Wed &        9.5 &                  938.8 &                 1414.2 \\
2019-11-09 & Sat &        8.9 &                  918.5 &                 1356.7 \\
\bottomrule
\end{tabular}

%% file: 04-chapter1/tables/mdp.tex
\renewcommand{\arraystretch}{1.5}
\begin{tabular}{ p{2.5cm}p{9cm} } 
\toprule
\multirow{6}{*}{\textbf{States}} & $\boldsymbol{u}_t$: generator up/down times $\in \mathbb{Z}^N$  \\
& $\boldsymbol{d}$: demand forecast $\in \mathbb{R}^T$  \\ 
& $\boldsymbol{w}$: wind forecast $\in \mathbb{R}^T$ \\ 
& $\boldsymbol{x}_t$, $\boldsymbol{y}_t$: preceding demand and wind forecast  errors for ARMA($p,q$) processes $\in \mathbb{R}^p$\\ 
& $\boldsymbol{\epsilon}_{d,t}$, $\boldsymbol{\epsilon}_{w,t}$: preceding white noise samples for ARMA($p,q$) processes $\in \mathbb{R}^q$\\ 
& $t$: timestep $0 \leq t \leq T \in \mathbb{Z}$ \\ 
\midrule
\textbf{Observations} & $\{ \boldsymbol{u}_t, \boldsymbol{d}, \boldsymbol{w}, t \}$ \\
\midrule
\textbf{Actions} & $a_t$: commitment decisions $\{0,1\}^N$ \\
\midrule
\textbf{Rewards} & $r_t$: negative operating cost $\in \mathbb{R}$ \\
\midrule
\multirow{3}{*}{\textbf{Transitions}} & 

\begin{equation*}
u_{i,t+1} =
\begin{cases}
    u_{i,t} + 1,& \text{if } a_{i,t}=1 \text{ and } u_{i,t} > 0 \\
    1,& \text{if } a_{i,t}=1 \text{ and } u_{i,t} < 0 \\
    -1,& \text{if } a_{i,t}=0 \text{ and } u_{i,t} > 0 \\
    u_{i,t} - 1,& \text{if } a_{i,t}=0t \text{ and } u_{i,t} < 0 \\
\end{cases}
\end{equation*} \\
& $x_t \sim X_t$: sample demand forecast error (from ARMA) \\
& $y_t \sim Y_t$: sample wind forecast error (from ARMA) \\
\bottomrule
\end{tabular}

%% file: 04-chapter1/tables/ucs_milp_comparison.tex
\begin{tabular}{rlrrr}
\toprule
 Num. gens &         Version &  Mean cost (\$M) &  Std. cost &  LOLP (\%) \\
\midrule
         5 &   MILP(perfect) &             4.38 &       0.00 &      0.000 \\
         5 & MILP($4\sigma$) &             4.57 &       0.25 &      0.079 \\
         5 &             UCS &             4.63 &       0.31 &      0.074 \\
        10 &   MILP(perfect) &             8.82 &       0.00 &      0.000 \\
        10 & MILP($4\sigma$) &             9.40 &       1.02 &      0.180 \\
        10 &             UCS &             9.43 &       1.11 &      0.177 \\
\bottomrule
\end{tabular}

%% file: 04-chapter1/tables/policy_params.tex
\begin{tabular}{l|r|r|r}
     & \multicolumn{3}{c}{\textbf{Generators}} \\
    \textbf{Variable} & 10 & 20 & 30 \\
    \midrule
    Update & \multicolumn{3}{c}{PPO} \\
    \hline
    Reward transformation & \multicolumn{3}{c}{$\hat{r} = \log(m/r)$} \\
    \hline
    Clip ratio \cite{schulman2017proximal} & \multicolumn{3}{c}{0.1} \\
    \hline
    Entropy coefficient \cite{mnih2016asynchronous} & 0.05 & 0.001 & 0.0 \\
    \hline
    Actor architecture & 100, 50, 25 & 64, 64 & 64, 64 \\ 
    \hline
    Critic architecture & 64, 64 & 100, 50, 25 & 64, 64 \\
    \hline
    Epochs & 100,000 & 200,000 & 300,000 \\
    \hline
    Forecast window (periods) & \multicolumn{3}{c}{24} \\ 
    \hline
    Gamma & \multicolumn{3}{c}{0.95} \\
\end{tabular}

%% file: 04-chapter1/tables/model_free_costs.tex
\begin{tabular}{rlrrrr}
\toprule
 Num. gens &         Version &  Mean cost (\$M) &  Std. cost &  Mean time (s) &  LOLP (\%) \\
\midrule
        10 &      $\pi(a|s)$ &            22.00 &       5.54 &            0.2 &      4.045 \\
        10 & MILP($4\sigma$) &             9.40 &       1.02 &           19.1 &      0.180 \\
        10 &   MILP(perfect) &             8.82 &       0.00 &            2.3 &      0.000 \\
        20 &      $\pi(a|s)$ &            24.19 &       4.45 &            0.5 &      1.097 \\
        20 & MILP($4\sigma$) &            18.90 &       2.92 &            5.5 &      0.244 \\
        20 &   MILP(perfect) &            17.58 &       0.00 &           11.9 &      0.000 \\
        30 &      $\pi(a|s)$ &            34.87 &       7.81 &            0.8 &      1.020 \\
        30 & MILP($4\sigma$) &            28.53 &       4.94 &            8.0 &      0.291 \\
        30 &   MILP(perfect) &            26.31 &       0.00 &            6.8 &      0.000 \\
\bottomrule
\end{tabular}

%% file: 04-chapter1/tables/guided_unguided_comparison.tex
\begin{tabular}{rlrrrrrr}
\toprule
 Num. gens &    Version &  Mean cost (\$M) &  Std. cost &  Mean time (s) &  Max. time &  Min. time &  LOLP (\%) \\
\midrule
         5 & Guided UCS &             4.63 &       0.32 &           27.7 &       52.5 &        8.9 &      0.077 \\
         5 &        UCS &             4.63 &       0.31 &           94.1 &      141.8 &       35.9 &      0.074 \\
        10 & Guided UCS &             9.44 &       0.83 &           28.3 &       50.6 &        9.2 &      0.152 \\
        10 &        UCS &             9.43 &       1.11 &        13249.9 &    28272.4 &     3037.9 &      0.177 \\
\bottomrule
\end{tabular}

%% file: 04-chapter1/tables/all_best_results_with_time.tex
\begin{tabular}{rlrrrrrrr}
\toprule
 Num. gens &         Version &  Mean cost (\$M) &  Max. cost &  Std. cost &  Mean time (s) &  Max. time &  Min. time &  LOLP (\%) \\
\midrule
        10 &   MILP(perfect) &             8.82 &       8.82 &       0.00 &            2.3 &        5.6 &        1.5 &      0.000 \\
        10 &      Guided UCS &             9.37 &      18.15 &       0.84 &          807.3 &     1992.1 &       76.9 &      0.128 \\
        10 & MILP($4\sigma$) &             9.40 &      19.39 &       1.02 &           19.1 &      177.2 &        1.8 &      0.180 \\
        20 &   MILP(perfect) &            17.58 &      17.58 &       0.00 &           11.9 &      150.8 &        3.3 &      0.000 \\
        20 &      Guided UCS &            18.73 &      37.72 &       1.41 &          117.3 &      374.5 &       10.4 &      0.107 \\
        20 & MILP($4\sigma$) &            18.90 &      46.15 &       2.92 &            5.5 &        8.4 &        3.8 &      0.244 \\
        30 &   MILP(perfect) &            26.31 &      26.31 &       0.00 &            6.8 &       18.0 &        4.9 &      0.000 \\
        30 &      Guided UCS &            28.41 &      49.93 &       2.04 &          391.3 &      976.8 &      129.4 &      0.142 \\
        30 & MILP($4\sigma$) &            28.53 &      75.03 &       4.94 &            8.0 &       12.4 &        5.6 &      0.291 \\
\bottomrule
\end{tabular}

%% file: 05-chapter2/chapter2.tex
\section{Introduction}

Guided expansion, the key innovation of guided uniform-cost search (UCS) presented in the previous chapter, can be applied in a modular fashion to any tree search algorithm, creating a broader class of guided tree search methods. We focused on UCS as a simple, general-purpose algorithm that can be applied in any problem domain. As discussed in Section \ref{methodology:tree_search:taxonomy}, there exists a broad taxonomy of tree search algorithms with properties suiting different applications. Informed search methods can benefit from greater search efficiency by exploiting problem-specific knowledge but cannot be generally applied across different problem domains. In addition, informed search methods may lack the optimality guarantees of uninformed search methods. Similarly, anytime (interruptible) methods can mitigate run time variability and improve performance in time-constrained contexts at the expense of optimality guarantees. In this chapter we demonstrate that the application of informed and anytime search methods to the UC problem using the guided tree search approach improves performance relative to Guided UCS in terms of run time and operating costs.

Using the power system simulation environment introduced in Section \ref{ch1:env}, we demonstrate that guided tree search algorithms designed more specifically for the UC problem can achieve lower operating costs in similar run times, and exhibit practical benefits compared with the more general-purpose algorithm, Guided UCS. In particular, we show that heuristics based on simple priority list UC solution methods can be leveraged in informed search methods to substantially improve search efficiency, saving computational budget that can be used to increase the depth of search. Furthermore, an anytime algorithm exhibits practical benefits in mitigating the run time variability associated with Guided UCS, and allowing for computational resources to be fully exploited. 

These improvements, culminating in the informed, anytime algorithm Guided IDA* search, enable the application of guided tree search to larger problem instances of 100 generators. The larger action space of the 100 generator power system poses further challenges for policy training as policy entropy must converge at a suitable level for guided expansion. To manage this problem, we use a novel method of \emph{target entropy regularisation}, which penalises deviations of policy entropy from a specified value. Setting the target entropy as a function of the desired branching threshold $\rho$ and the number of generators unifies the elements of policy training and guided tree search, and is shown to improve policy convergence in practice.

\subsection{Contributions}

This chapter makes the following contributions. 

\begin{enumerate}
    \item We introduce two new guided tree search algorithms, Guided A* search and Guided IDA* search, applying the principle of guided expansion introduced in Section \ref{ch1:guided_ucs}. Both algorithms are informed, using a problem-specific heuristic to improve search efficiency. Guided IDA* is also anytime and we replace the depth parameter $H$ with a time budget, preventing high run time variability.
    
    \item Three heuristics based on a PL algorithm are introduced for application in Guided A* and IDA*: Naive Minimum Marginal Fuel Cost (MMFC); Naive Economic Dispatch (ED) and Constrained ED. The heuristics are analysed in terms of average run time, accuracy and admissibility and applied in Guided A* search to solve the 20 test problems from Section \ref{ch1:env:benchmarks} to evaluate improvements to search efficiency improvements as compared with Guided UCS. Constrained ED is found to be the most effective heuristic, reducing mean run time by between 64--94\% as compared with Guided UCS, without significant changes in operating costs. 

    \item Using the strongest heuristic (Constrained ED), the anytime algorithm Guided IDA* search is applied to the test problems and found to allow for deeper search on average by more consistently exploiting the computational budget. Costs are found to be between 0.4--1.0\% lower than Guided UCS while completing in similar run time.

    \item Guided IDA* search is applied to a 100-generator system, a significantly larger problem than previous research. A novel method of entropy regularisation based on \emph{target entropy} is used which is shown to improve training convergence and promote policies with a suitable level of entropy for guided expansion in high dimensional actions spaces. We find that operating costs are 0.14\% lower on average than the MILP benchmark using a reserve constraint. Guided IDA* achieves lower loss of load probability, and achieves lower expected operating costs than MILP in 15 out of 20 problems.
    
\end{enumerate}

In the next section, Guided A* search and Guided IDA* search algorithms are presented. In Section \ref{ch2:heuristics}, the three PL-based heuristics Naive MMFC, Naive ED and Constrained ED are described and analysed. In Section \ref{ch2:experiments} we compare Guided A* search using each of the three heuristics and apply the strongest heuristic (Constrained ED) in Guided IDA* search. In Section \ref{ch2:100gen}, Guided IDA* is applied to a larger power system of 100 generators. We discuss our findings in Section \ref{ch2:discussion} and Section \ref{ch2:conclusion} concludes the chapter.

\section{Informed and Anytime Algorithms} \label{ch2:methodology}

In Section \ref{ch1:guided_ucs} we presented Guided UCS, an RL-aided tree search algorithm. Guided UCS uses an expansion policy to determine a subset of branches to add to the search tree at each node. This process of guided expansion (Equation \ref{ch1:eq:guided_expansion}) can be applied in a modular fashion during the expansion phase of any tree search algorithm. We previously focused on UCS as a simple, heuristic-free algorithm that can be applied to search trees with non-uniform costs \cite{russellnorvig}. However, exploiting domain knowledge through \emph{informed} search algorithms can significantly improve the efficiency of tree search algorithms in practice while remaining optimal \cite{russellnorvig}. Furthermore, the significant run time variability of Guided UCS that was shown in Section \ref{ch1:experiments:ucs_comparison} and is investigated further for Guided A* in Section \ref{ch2:experiments:a_star} motivates the development of an \emph{anytime} algorithm - that is one which can be interrupted and return a solution, with the solution quality improving over time. We discussed the taxonomy of tree search algorithms, including informed and anytime algorithms, in Section \ref{methodology:tree_search:taxonomy}. 

In this section we apply guided expansion to A* search \cite{hart1968formal} (Guided A* search) and iterative-deepening A* search (Guided IDA* search) \cite{korf1985depth}. Guided A* is an informed search algorithm, while Guided IDA* is both informed and anytime. 

\subsection{Guided A* Search} \label{ch2:methodology:a_star}

\begin{figure}
    \centering
    \includegraphics[width=0.75\textwidth]{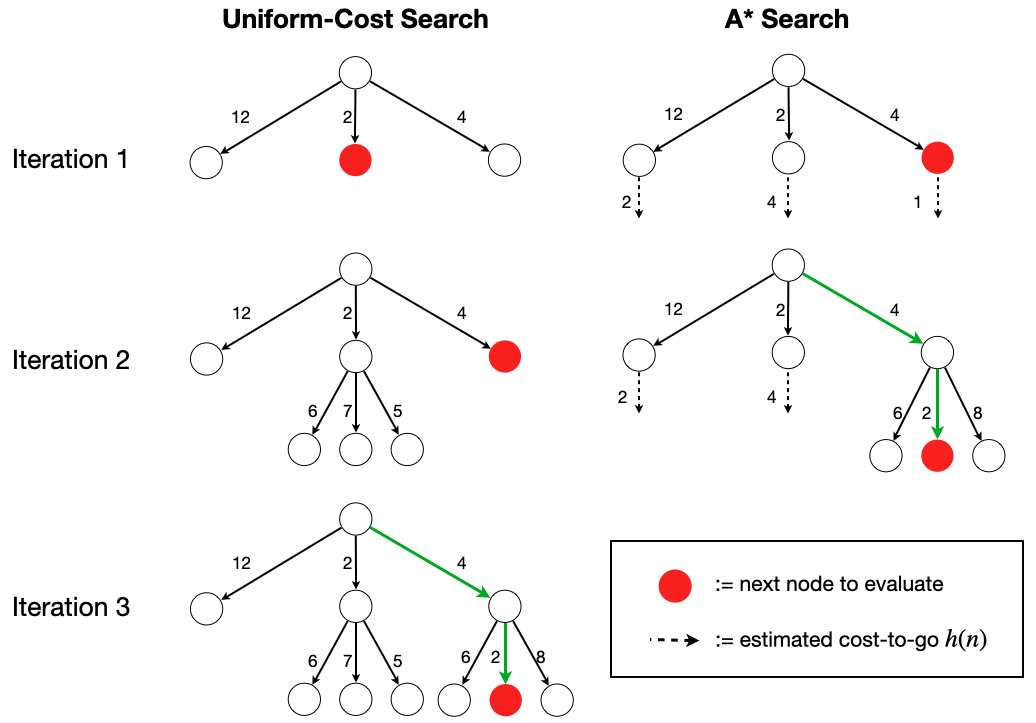}
    \caption{Comparison of uniform-cost search (UCS) and A* algorithms for a problem of depth $H=2$. Values on the search tree branches correspond to the step costs, while dotted show estimates of the cost-to-go $h(n)$. UCS takes three iterations to reach the solution path, while A* requires two. By expanding nodes in order of $g(n) + h(n)$, one fewer node evaluation is required for A* search.}
    \label{ch2:fig:a_star_diagram}
\end{figure}

First we present Guided A* search, in which guided expansion is applied to A* search \cite{hart1968formal}. A* search was described in detail in Section \ref{methodology:tree_search:a_star}, with pseudocode shown in Algorithm \ref{methodology:algo:a_star}. The A* search algorithm is similar to UCS, differing only in its ordering of the priority queue giving the next node to expand. Whereas UCS orders nodes by their path costs $g(n)$, A* orders nodes by:

\begin{equation}
    f(n) = g(n) + h(n)
\end{equation}

where $h(n)$ is a heuristic estimate of the optimal path cost from $n$ to a goal node (cost-to-go). In other words, while UCS chooses the next node to expand based on the cost to reach that node alone, A* search chooses based on the path cost plus an estimate for the remaining path cost to a goal node. A comparison between UCS and A* search, applied to a search tree with depth $H=2$, is shown in Figure \ref{ch2:fig:a_star_diagram}. By expanding nodes in order of $f(n)$, A* search requires one fewer node evaluation while still finding the optimal solution. 

In order to make A* search a guided tree search algorithm, guided expansion (Equation \ref{ch1:eq:guided_expansion}) is applied to select a subset of actions $A_{\pi}(s)$ for each state $s$, using an expansion policy $\pi(a|s)$. We discussed guided expansion in detail in Section \ref{ch1:eq:guided_expansion}. Using guided expansion, a reduced search tree is created with a smaller branching factor than the exhaustive (unguided) tree. A* search is then used to find the shortest path through the reduced tree. 

Unlike UCS, A* is not a general-purpose algorithm that is applicable `out-of-the-box' to any tree search problem as the heuristic $h(n)$ is problem-specific.\footnote{Excluding the case where $h(n)=0$ for all $n$, where A* is equivalent to UCS.} In order to apply Guided A* search to the UC problem, in Section \ref{ch2:heuristics} we develop three heuristics which estimate the cost-to-go (that is, the remaining operating costs from node $n$ to the search horizon). 

We apply the same real-time strategy described in Section \ref{ch1:methodology:tree_search}, Algorithm \ref{ch1:algo:solve_day_ahead}, whereby Guided A* is used repeatedly to solve $T$ sub-problems, with the first action in the solution being used to determine the root of the next sub-problem. As discussed in Section \ref{ch1:methodology:tree_search}, this prevents exponential run time complexity in the number of decision periods $T$. A* search  is least efficient when using no heuristic, i.e. $h(n)=0$ for all $n$, which is equivalent to UCS. As a result, the worst-case time complexities of A* and UCS are the same: $\mathcal{O}(2^{NH}T)$ without guided expansion, and $\mathcal{O}(T(\frac{1}{\rho})^H)$ with guided expansion using branching threshold $\rho$. However, we show in Section \ref{ch2:experiments:a_star} that with an appropriate heuristic, the absolute run times of Guided A* search are lower than Guided UCS (on average) due to improved search efficiency.


\subsection{Guided IDA* Search} \label{ch2:methodology:ida_star}

In this section we describe an \emph{anytime} algorithm, based on iterative-deepening A* (IDA*) search. As described in Section \ref{methodology:tree_search:taxonomy}, anytime (or interruptible) algorithms can be terminated at any point and return a solution. The development of an anytime algorithm for the UC problem is motivated by the results in Section \ref{ch1:experiments:ucs_comparison}, where we found the run time of Guided UCS is highly variable between episodes. In Section \ref{ch2:experiments:a_star}, we show that Guided A* exhibits similar characteristics, with over an order of magnitude separating the shortest and longest episode run times for fixed parameter settings. Furthermore, in Section \ref{ch1:experiments:parameters} we found that average run time of Guided UCS is highly sensitive to parameter choices. Run time depends on characteristics of the episode (such as demand variation) and cannot be easily predicted given a set of parameters. The depth parameter $H$ is difficult to tune as it has a significant impact on both solution quality and run time (to which it is exponentially related). Given the time-constrained nature of UC problems, there is motivation to develop an anytime algorithm for the UC problem which can be terminated when a time budget is spent (such as before market closure), rather than running to completion. 

Iterative deepening \cite{korf1985depth}, discussed in Section \ref{methodology:tree_search_algos:iterative_deepening}, is a general strategy that has been applied to a wide range of tree search algorithms and can be used to create anytime algorithms. The principle behind iterative deepening is to use a search algorithm to solve search trees of increasing search depth. A sub-optimal first action is found almost immediately by looking only one timestep ahead. Thereafter, the depth is increased at each iteration and the search is conducted again. The solution quality improves the longer the algorithm is run. The algorithm terminates once a stopping criterion (based on run time or solution quality) is met. 

We apply iterative-deepening to the A* search algorithm described in Section \ref{ch2:methodology:ida_star} (IDA* \cite{korf1985depth}). Pseudocode for IDA* is shown in Algorithm \ref{ch2:algo:ida_star}. Our implementation of IDA* replaces the depth parameter $H$ in A* and UCS with a time budget parameter $b$ (seconds). A* search is used to iteratively solve the sub-problem rooted at $r$ with a gradually increasing depth $H$. When the time budget $b$ has elapsed, the last solution is returned. A significant advantage of the Guided IDA* algorithm is that the depth parameter $H$ is replaced by the time budget $b$. In the context of UC, the time budget $b$ may easily be determined by market constraints, such as settlement period length. Using an anytime algorithm like Guided IDA* ensures that computational resources are fully exploited within time constraints. Note that Algorithm \ref{ch2:algo:ida_star} does not follow the same implementation as the popular IDA* algorithm described in \cite{korf1985depth}, where each iteration corresponds to a gradually increasing cost-related cutoff bound. 
 
\begin{algorithm}[t]
\caption{Anytime IDA* search algorithm for the UC problem from intial state $r$. A* search is run with progressively increasing search horizon $H$ until the time budget $b$ is spent.}\label{ch2:algo:ida_star}
\begin{algorithmic}
\Function{IDAStar}{$r, b$}
  \State $H \leftarrow $ 1
  \Repeat
    \State solution $\leftarrow$ \Call{AStar}{$r, H$}
    \State $H \leftarrow H+1$
  \Until{time budget $b$ is spent}
  \State \Return solution
\EndFunction
\end{algorithmic}
\end{algorithm}

Both Guided A* and Guided IDA* require a problem-specific heuristic $h(n)$ to be applied to the UC problem. In the next section we present three heuristics based on a priority list algorithm for estimating the cost-to-go $h(n)$ in Guided A* and Guided IDA*. 

\section{Heuristics for Unit Commitment} \label{ch2:heuristics}

As informed search algorithms, the A* and IDA* search algorithms described in Section \ref{ch2:methodology} require a problem-specific heuristic $h(n)$, which is used to estimate the lowest cost from node $n$ to a goal node. The choice of heuristic is an important decision that has significant impact on the effectiveness of informed search relative to uninformed methods \cite{russellnorvig}. 

In this section, we will begin by justifying our approach of using priority list (PL) algorithms as the basis for UC heuristics, with reference to the heuristic properties of run time, admissibility and accuracy outlined in Section \ref{methodology:tree_search:heuristic_properties}. We will then present three PL-based heuristics: Naive Minimum Marginal Fuel Cost (MMFC), Naive Economic Dispatch (ED) and Constrained ED. Finally, we will evaluate the run time, admissibility and accuracy of the three heuristic methods.

\subsection{Choice of Heuristic Approach}

In Section \ref{methodology:tree_search:heuristic_properties} we gave run time, admissibility and accuracy as important properties of heuristics impacting the efficiency improvements achieved by informed search (A* search) compared with uninformed search (UCS). Run time is significant since this impacts the extent to which efficiency improvements from fewer node evaluations are offset by the heuristic computation itself. Admissibility, which requires that the heuristic $h(n)$ should not over-estimate $h^*(n)$ is important as it is a criterion of optimality for A* search \cite{dechter1985generalized}. Finally, accuracy (that is, the average error between the estimated $h(n)$ and optimal $h^*(n)$) measures the heuristic's ability to effectively prune sub-optimal sub-trees. A perfectly accurate heuristic (where $h(n)=h^*(n)$) functions as an oracle, yielding maximal efficiency improvement by immediately identifying the optimal path; a naive heuristic (where $h(n)=0$) yields no efficiency improvements as A* is equivalent to UCS. Heuristic run time, admissibility and accuracy are often traded-off in practice. For instance, more complex heuristics may achieve higher levels of accuracy at the expense of run time or admissibility \cite{russellnorvig}.

There is no all-purpose approach to designing effective heuristics for a particular problem domain. Some widely-studied problems have well-established heuristics. For instance, in route planning problems, a common heuristic is the straight-line distance from $n$ to the destination \cite{hart1968formal, russellnorvig, sedgewick1986shortest, golden1978shortest}. Alternatively, expert pattern databases may be used in some problems, such as the Rubik's cube puzzle \cite{korf2001time}. Supervised learning has also been used to learn $h(n)$ for route planning problems \cite{wang2019empowering}. The choice of heuristic has a significant impact on the efficiency of informed search algorithms \cite{russellnorvig}. 

Due to the lack of research in applying tree search methods to the UC problem, no established UC-specific heuristics exist. In the UC problem, $h(n)$ aims to estimate the lowest expected operating costs from $n$ to a node at the search horizon. There is no `distance' measure that is analogous to the straight-line distance in route planning which can be applied in the UC problem. Furthermore, unlike problem like the Rubik's cube, there are innumerable states in the UC problem due to its continuous state space, so pattern database approaches cannot be easily applied. Supervised learning methods are possible in principle, but generating a large dataset would be computationally expensive. In addition, the supervised learning task of predicting future operating costs given a state would be very challenging, due to the highly non-linear and non-continuous operating cost function and the admissibility criterion would be difficult to satisfy.

Our approach is based on the priority list (PL) algorithms described in Section \ref{methodology:mathop:pl}. Improvements in MILP have made PL methods largely obsolete for practical UC problems, due to PL methods' lack of optimality guarantees and reliance on complex rules to fix constraints. However, PL methods have appropriate characteristics for designing heuristics. PL algorithms are very fast to compute, satisfying the run time property. In addition, the ability to relax constraints can be exploited to further reduce run time and increase admissibility by making cost estimates more optimistic. Lastly, PL algorithms are well-studied solution methods to the UC problem in their own right and are therefore capable of achieving a high degree of accuracy by approximating optimal schedules.

In the following section we develop three PL-based heuristics and compare their run time, admissibility and accuracy properties in Section \ref{ch2:heuristics:analysis}. In Section \ref{ch2:experiments:a_star} we evaluate efficiency improvements achieved in practice when each is applied in Guided A* search. 

\subsection{Priority List Heuristics} \label{ch2:heuristics:priority_list}

We will now present the three heuristic methods developed for the UC problem. The heuristics use different PL algorithms to generate an approximate UC schedule for the following periods up to the search horizon $H$. The operating costs of this schedule are then evaluated and used to approximate the optimal cost-to-go $h^*(n)$. All three heuristics are based on a PL ordering of generators by their economic efficiency at $p^{\text{max}}$, which is the minimum marginal fuel cost (MMFC) (\$ per MWh). The MMFC (in \$ per MWh) which we denote $q_i$ is the lowest point on the generator's fuel cost curve (e.g. Figure \ref{ch1:fig:cost_curves}) and found by evaluating the cost function for generator $i$ at $p_i^{\text{max}}$ and dividing by capacity: 

\begin{equation}  \label{ch2:eq:priority_list}
    q_{i} = \frac{a_i(p_i^{\text{max}})^2 + b_i(p_i^{\text{max}}) + c_i}{p_i^{\text{max}}}
\end{equation}

The PL orders generators in increasing order of MMFC $q_i$. The highest priority generator (first to be commmitted) is that which has the lowest $q_i$. In order to reduce computational costs and increase the heuristics' optimism (admissibility), we only consider fuel costs, ignoring start costs and lost load costs. With the PL ordered by $q_i$, we present three heuristics for estimating the cost-to-go $h(n)$. 

\subsubsection{Heuristic I: Naive MMFC}

The first heuristic, Naive MMFC, follows Algorithm \ref{methodology:algo:PL} described in Section \ref{methodology:mathop:pl}, committing generators without consideration for inter-temporal (only minimum up/down time in our problem setup) constraints in order of $q_i$ (Equation \ref{ch2:eq:priority_list}). For each period $t$ up to the search horizon (that is up to the depth of the search tree), generators are committed in increasing order of $q_i$ until: 

\begin{equation}
    \sum_{i \in K_t} p_i^{\text{max}} \geq D_t
    \label{ch2:eq:pl_cap}
\end{equation}

where $K_t$ is the set of generators committed at period $t$ and $D_t$ is the forecast demand. Note that no reserve constraint is enforced. By ignoring minimum up/down time constraints, the UC schedule produced by Naive MMFC is not guaranteed to be feasible but can be calculated very quickly.

To calculate the operating costs, we assume that all generators are fully-loaded ($p_i = p_i^{\text{max}}$) except the last committed (marginal) generator which is part-loaded and satisfies the remaining load. This means the operating level constraints $p_i^{\text{min}} \leq p_i \leq p_i^{\text{max}}$ of the marginal generator can be violated and the dispatch may be infeasible for this generator. As a result, we do not evaluate the quadratic fuel cost curves in Equation \ref{ch1:eq:cost_curves} to calculate operating costs, instead we take the dot product of generator dispatch vector $\boldsymbol{p}_t$ and the MMFC $\boldsymbol{q}$:

\begin{equation}
    C^{f}_t = \frac{1}{2} (\boldsymbol{q} \cdot \boldsymbol{p}_t)
\end{equation}

where the factor $\frac{1}{2}$ accounts for the settlement period interval of 30 minutes. By assuming all generators operate at maximum efficiency, this operating cost estimate is optimistic and hence likely to be admissible. The heuristic $h(n)$ is calculated as the sum of $C^{f}_t$ up to the search horizon $H$. 

\subsubsection{Heuristic II: Naive ED}

The second heuristic method, Naive ED, is slightly more advanced than Naive MMFC, since we use the lambda-iteration method described in Section \ref{methodology:mathop:ed} to solve the economic dispatch (ED) problem. This gives a more accurate estimate of operating costs, at the expense of increased run time. Like Naive MMFC, by ignoring the minimum up/down time constraints, the UC schedule is not guaranteed to be feasible. 

Naive ED generates a commitment schedule using the same method as for Naive MMFC, ignoring minimum up/down time constraints. Given this schedule, the ED problem is solved for each period $t$, and the resulting operating costs calculated using the quadratic fuel cost function (Equation \ref{ch1:eq:cost_curves}). 

\subsubsection{Heuristic III: Constrained ED}

The final heuristic, constrained ED, partially considers the minimum up/down time constraints of generators. Generators are committed in PL order, but must first serve their minimum up/down time constraints. That is, any generator that has been offline for less than its minimum down time in the state represented by the node $n$ must remain offline until it becomes available, with the same applying for minimum up time constraints. 

Once the schedule has been produced, the ED problem is solved for each period, in the same way as Naive ED. Compared with Naive MMFC and Naive ED, which do not consider up/down time constraints, Constrained ED is more expensive to compute due to the additional logic rules, but is able to recognise states where constraints are likely to be encountered. For instance, states where a base-load generator has recently been switched off will be recognised as having reduced capacity for the following periods, and less efficient generators may be required. Constrained ED is likely to give the most realistic UC schedule and the most accurate estimate of $h^*(n)$, the optimal cost-to-go. However, it is also the most complex and hence the slowest heuristic to calculate. As discussed in Section \ref{methodology:tree_search:heuristic_properties}, heuristic run time is an important property as it partially offsets efficiency improvements achieved by using an informed search method. 

\medskip

In summary, it should be emphasised that none of the three heuristics are guaranteed to offer feasible UC solutions as minimum up/down time constraints are not fully observed in any case. Constrained ED only considers the initial up/down time constraints, preventing generators which are constrained to remain on/off at node $n$ from being decommitted/committed before satisfying those constraints. Thereafter, generators can be cycled without obeying these inter-temporal constraints. As a result, none of the heuristics can be used on their own to solve the UC commitment problem without further modifications to satisfy constraints. 

Having presented the three heuristics, next we will analyse their properties in terms of run time, accuracy and admissibility properties described in Section \ref{methodology:tree_search:heuristic_properties}.

\subsection{Analysis of Heuristics} \label{ch2:heuristics:analysis}

We will now analyse the Naive MMFC, Naive ED and Constrained ED heuristics presented in the previous subsection in terms of run time, admissibility and accuracy. These properties were discussed in Section \ref{methodology:tree_search:heuristic_properties} as being important factors impact the efficiency improvements achieved by informed search relative to uninformed search.

In order to evaluate the admissibility and accuracy of the three priority list heuristic methods, we compare the predicted cost-to-go $h(n)$ with the optimal cost-to-go $h^*(n)$. Unguided UCS, which is guaranteed to produce optimal solutions, is used to calculate $h^*(n)$. The 20 test UC problem instances for the 5 generator problem described in Section \ref{ch1:env:benchmarks} were solved with UCS with $H=4$ (a larger value of $H$ was not practical due to exponential run time complexity of UCS), and the heuristics evaluated for each root node $n$ in the solution path. In total, this produced $48 \cdot 20 = 960$ values for $h(n)$ (for each heuristic) and $h^*(n)$. We then calculated admissibility as the proportion of nodes satisfying the admissibility criterion: 

\begin{equation}
    h(n) \leq h^*(n)
\end{equation}

We measured accuracy for the 960 estimates using the root-mean squared error. For $N$ estimates $h(n)$ and optimal $h^*(n)$, the RMSE is calculated using: 

\begin{equation}
    RMSE = \sqrt{\frac{1}{N} \sum_{i=1}^{N} (h^*(n_i) - h(n_i))^2 }
    \label{ch2:eq:rmse}
\end{equation}

In addition, we measured run time by sampling 1000 random nodes, and measuring the mean time taken to calculate $h(n)$ with a depth $H=4$. 

The run time measurements, along with admissibility and accuracy results are summarised in Table \ref{ch2:tab:heuristic_summary}. The Naive MMFC heuristic (which is the simplest PL heuristic) had the lowest run time, by roughly a factor of 3. Constrained ED was the slowest, with 11\% higher mean run time than Naive ED. Naive MMFC was 100\% admissible in our experiment, but was the least accurate, having the highest RMSE. All three heuristics succeed in making mostly admissible estimates: for Naive MMFC, 100\% of estimates are admissible; Naive ED achieves 95\% and Constrained ED 98\% admissibility.

\begin{table}[]
    \centering
    \input{05-chapter2/tables/heuristic_summary}
    \caption{Summary of run time, root mean squared error and admissibility (proportion of estimates where $h(n) \leq h^*(n)$) for the three PL heuristics.}
    \label{ch2:tab:heuristic_summary}
\end{table}

\begin{figure}
    \centering
    \includegraphics[width=\textwidth]{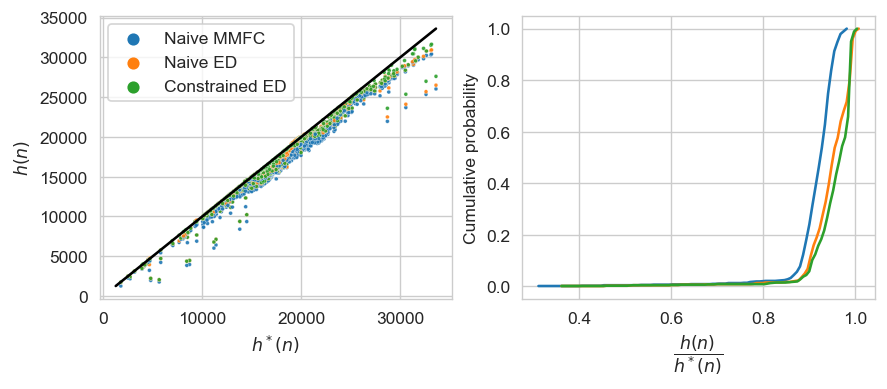}
    \caption{Admissibility of the three PL-based heuristics. Both plots use the same data: optimal cost-to-go $h^*(n)$ versus the heuristic $h(n)$ heuristic estimate (left) and cumulative distributions showing the proportion of admissible estimates (right).}
    \label{ch2:fig:heuristic_admissibility}
\end{figure}

The estimated and optimal cost-to-go $h(n)$ and $h^*(n)$ are plotted in Figure \ref{ch2:fig:heuristic_admissibility}. In the left-hand plot, points below the black line $h(n)=h^*(n)$ are admissible estimates $h(n) \leq h^*(n)$. Nodes with the highest $h^*(n)$ are generally under-estimated by the largest margin. It is likely that the optimal paths at these nodes include a relatively large loss of load probability (LOLP) whose costs are not considered by any heuristic and may be the cause of this gap. The same data is used to calculate cumulative distributions of $\frac{h(n)}{h^*(n)}$ in the right-hand plot. This plot clearly shows that Naive MMFC is the most optimistic and least accurate heuristic, and is only within 5\% of $h^*(n)$ (that is, $\frac{h(n)}{h^*(n)} > 0.95$) in 23\% of cases. By contrast, $\frac{h(n)}{h^*(n)} > 0.95$ in 60\% of cases for ED, and 70\% of cases for Constrained ED. 

In summary, the three heuristics exhibit different properties, and there is a clear trade-off between run time and accuracy with accuracy improving with increased run time. Naive MMFC is significantly faster but less accurate than the other two heuristics. Naive ED is the least promising heuristic as it has the lowest admissibility and lower accuracy than Constrained ED, with only a 0.19ms lower run time. Constrained ED is the most accurate and has high admissibility of 98\%, but is also more than 3 times slower than Naive MMFC. Having analysed the heuristic properties, in Section \ref{ch2:experiments:a_star} we evaluate the efficiency improvements achieved in practice by each heuristic when applied in Guided A* search. 

\section{Experiments} \label{ch2:experiments}

We now conduct two experiments applying heuristics developed in Section \ref{ch2:heuristics} to Guided A* search and Guided IDA* search using the power system defined in Section \ref{ch1:env}. We solve the test problems described in Section \ref{ch1:env:benchmarks} and compare performance with Guided UCS, which was shown in Section \ref{ch1:experiments:milp_comparison} to outperform MILP benchmarks. 

In the first experiment, each of the three heuristic methods is applied in A* search. We compare the run time of Guided A* using each heuristic with that of Guided UCS, finding Constrained ED to result in the largest run time reduction as compared with Guided UCS with no significant deterioration in solution quality. In the second experiment, we use IDA* with the Constrained ED heuristic and a varying time budget. We compare the performance of the time-limited anytime algorithm IDA* with the depth-limited A* and UCS search algorithms, finding IDA* to result in lower operating costs in comparable run times. Guided IDA* is shown to be the strongest guided tree search algorithm for the UC problem developed in this thesis, with practical benefits stemming from its anytime property. The algorithm achieves significant operating cost improvements as compared with MILP benchmarks and can be applied more effectively to larger power systems than fixed-depth search methods. We demonstrate this by solving 100-generator UC problems in Section \ref{ch2:100gen}.

\subsection{Guided A* Search} \label{ch2:experiments:a_star}

In the first experiment, Guided A* with $H=4$ was used to solve the 10, 20 and 30 generator test problems used in the previous chapter, described in Section \ref{ch1:env:benchmarks}. Each of the three PL heuristics (Naive MMFC, Naive ED and Constrained ED) described in Section \ref{ch2:heuristics} was used. We used the same expansion policies trained in Section \ref{ch1:guided_ucs:policy_training} for guided expansion. We compare run times and operating costs with the Guided UCS ($H=4$) results from Section \ref{ch1:experiments:milp_comparison}. 

The results are shown in Table \ref{ch2:tab:a_star_summary} and Figure \ref{ch2:fig:a_star_run_time}. All of the heuristics achieve significant run time reductions as compared with Guided UCS: greater than 86\%, 53\%, 75\% improvement for 10, 20 and 30 generator problems, respectively, with small deviations in operating costs ($\leq0.09\%$). The most effective heuristic for improving search efficiency in Guided A* was Constrained ED, where run times are reduced by 94\%, 64\% and 82\% for 10, 20 and 30 generator problems, respectively. Despite being the fastest heuristic to compute, Naive MMFC achieves the smallest efficiency improvements in all cases except the 30 generator problem, where Naive ED is the least effective. Due to some heuristic estimates being inadmissible, there are small cost differences between UCS and A* search with Constrained ED for 20 and 30 generator problems, indicating that unlike UCS, Guided A* search is not guaranteed to be optimal over the lookahead horizon. However, the deterioration in solution quality is small, with a maximum increase of 0.09\% in operating costs. The run times savings afforded by A* search as compared with UCS free computational budget that can be used to increase the search depth $H$ and compensate for the small deterioration in operating costs.

\begin{table}[]
    \centering
    \small
    \input{05-chapter2/tables/a_star_summary}
    \caption{Difference in mean run time and operating cost using Guided A* search with each of the three heuristic methods, compared with Guided UCS. Guided A* with all three heuristics achieves significant run time reductions, with only small changes in operating costs ($<0.1$\%.)}
    \label{ch2:tab:a_star_summary}
\end{table}

\subsubsection{Run Time Variability}

\begin{figure}
    \centering
    \includegraphics[width=\textwidth]{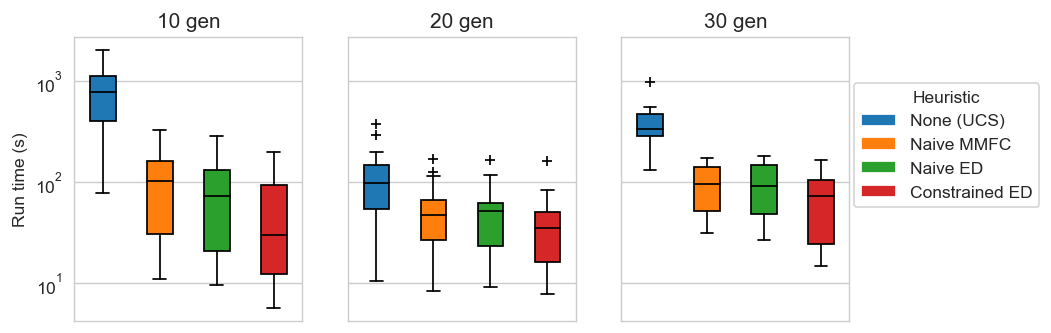}
    \caption{Run times (log-axis) of Guided A* search ($H=4$) with the three heuristic methods and using no heuristic (i.e. uniform-cost search, results in Section \ref{ch1:tab:guided_vs_milp}). All three heuristics achieve significant run time improvements relative to Guided UCS, with Constrained ED providing the largest speed-up.}
    \label{ch2:fig:a_star_run_time}
\end{figure}

In Section \ref{ch1:experiments:ucs_comparison}, Figure \ref{ch1:fig:time_taken_ucs} showed significant variation in episode run time using Guided UCS. Simple problem instances (those with a low branching factor using guided expansion) are solved quickly, while more challenging problems may take more than an order of magnitude longer. Figure \ref{ch2:fig:a_star_run_time} shows that Guided A* search exhibits similar characteristics, with episode run times typically varying by over an order of magnitude. 

\begin{figure}
    \centering
    \includegraphics[width=0.9\textwidth]{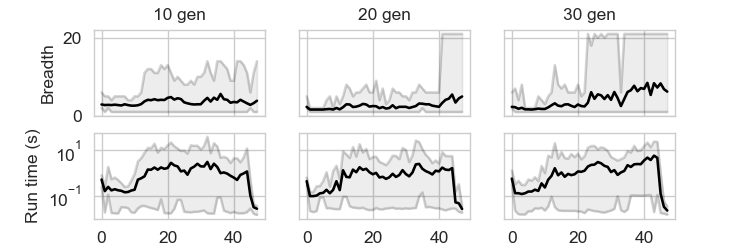}
    \caption{Mean, minimum and maximum search breadth at the root node (top row) and run time (bottom row) by period for A* search using the Constrained ED heuristic. Search breadth and run time are lower during early morning periods in all problem sizes, and larger later in the day. The sharp decline in run time at the end of the day in all instances is due to the truncated search horizon ($H<4$).}
    \label{ch2:fig:breadth_and_period_runtime}
\end{figure}

The variation in episode run time can be further explained by variation in period run times due to variable search breadth throughout the day. Figure \ref{ch2:fig:breadth_and_period_runtime} shows the variation in period run time and average search breadth (measured at the root of the search tree) with respect to decision period for different settings of $H$, using A* search with the Constrained ED heuristic. Search breadth is generally lower in the early morning periods, indicating the expansion policy used in guided expansion is relatively `certain' during these periods, with probability mass concentrated in a small number of actions. Later in the day the search breadth increases in all cases, although not uniformly between the 10, 20 and 30 generator problems. Mean search breadth in the 10 generator case roughly follows a typical demand profile, increasing in the morning (around period 14 or 7:00), decreasing in the middle of the day and increasing for the evening peak. By contrast, 20 and 30 generator problems had higher search breadth at the end of the day. This may be due to the larger uncertainties at the end of the day, due to the propagation of forecast errors. In addition, the 20 and 30 generator problems have larger action spaces as well as symmetries deriving from duplicate generators, which may lead to a more uniform distribution of probability mass over actions as compared with the 10 generator problem.

The search breadth trends are reflected in period run time trends, shown in the bottom row of Figure \ref{ch2:fig:breadth_and_period_runtime}. Periods with larger search breadth are solved more slowly due to the computational expense of using A* search to solve a broader search tree. As the number of nodes in the search tree scales with $b^H$, where $b$ is the branching factor, the period run time is highly sensitive to changes in $b$ caused by varying certainty of the expansion policy $\pi(a|s)$. Several orders of magnitude separate the mean run time in the early morning periods with the periods in the middle of the day. Solutions to the UC problems are usually characterised by more frequent commitment changes during peak periods to meet demand, contributing to greater uncertainty in decision-making and larger branching factors. The period run time variability is ultimately the cause of high variability in episode run time.

The run time variability of Guided A* motivates the anytime algorithm Guided IDA* developed in Section \ref{ch2:methodology:ida_star}. In the following section, we apply Guided IDA* search with the Constrained ED heuristic to the test problems.

\subsection{Guided IDA* Search} \label{ch2:experiments:ida_star}

Using the strongest heuristic, Constrained ED, we applied Guided IDA* to the test problems of 10, 20 and 30 generators with a time budget of $b=\{1,2,5,10,30,60\}$ seconds per period, constraining episode run time to a maximum of 48 minutes (2880 seconds). As with the previous experiment using Guided A* search, we used the expansion policies trained in Section \ref{ch1:guided_ucs:policy_training}. 

\begin{figure}
    \centering
    \includegraphics[width=0.7\textwidth]{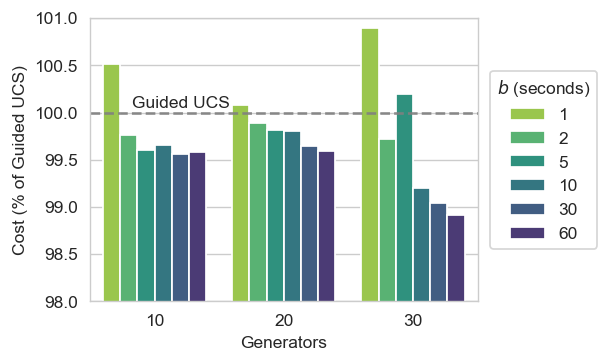}
    \caption{Cost saving of Guided IDA* with Constrained ED compared to Guided UCS. Operating costs generally decrease with increasing time budget. The largest improvements are found in the 30 generator case, where IDA* is 1.1\% cheaper than Guided UCS when $b=60$ seconds.}
    \label{ch2:fig:anytime_bars}
\end{figure}

Operating cost savings relative to Guided UCS solutions from Section \ref{ch1:experiments:milp_comparison} are shown in Figure \ref{ch2:fig:anytime_bars}. Costs were generally found to decrease with increasing time budget. For all three problems, budgets $b \geq 10$ seconds outperform Guided UCS. The largest savings were achieved in the 30 generator case, where costs were 1.1\% lower with $b=60$. However, in this case there was less consistent performance for lower time budgets, with operating costs increasing between $b=2$ and $b=5$. Compared with the deterministic MILP($4\sigma$) benchmarks from Section \ref{ch1:env:benchmarks}, IDA* achieved lower operating costs for time budgets $b \geq 2$ in all problem sizes.

IDA* with $b=30$ seconds is compared with Guided A* (using Constrained ED) and Guided UCS (from Section \ref{ch1:tab:guided_vs_milp}), both with $H=4$, in Table \ref{ch2:tab:ida_a_star_comparison}. We show $b=30$ (maximum run time of 24 minutes per problem) as this budget is most comparable in run time to Guided UCS. The results show that IDA* achieves lower operating costs than both Guided UCS and Guided A*, while mitigating the run time variability. 

\begin{table}[t]
    \centering
    \resizebox{\textwidth}{!}{
    \input{05-chapter2/tables/ida_a_star_comparison_formatted}}
    \caption{Comparison of IDA* ($b=30$ seconds), A* ($H=4$) and UCS ($H=4$) for 10, 20 and 30 generator problems.} 
    \label{ch2:tab:ida_a_star_comparison}
\end{table}

\begin{figure}
    \centering
    \includegraphics[width=0.9\textwidth]{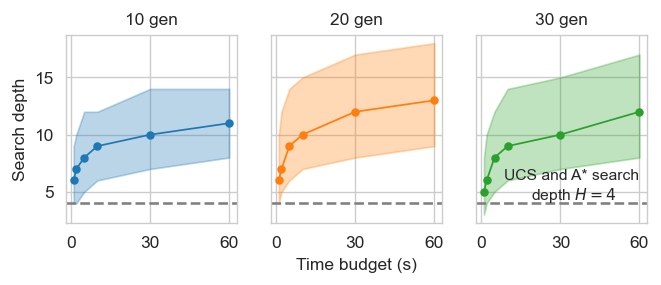}
    \caption{Variation of Guided IDA* search using Constrained ED heuristic. Solid line and points show the median search depth; shaded area indicates inter-quartile range. Dotted line shows $H=4$, the search depth used in Guided UCS and Guided A* search. The average search depth of Guided IDA* is significantly greater than these methods for all time budgets.}
    \label{ch2:fig:ida_star_depth_vs_g}
\end{figure}

Improvements in operating costs can be attributed to greater average search depths using Guided IDA* compared with Guided UCS and Guided A*, where search depth was fixed at $H=4$. Figure \ref{ch2:fig:ida_star_depth_vs_g} shows the median search depth $H$ using Guided IDA* with varying time budget $b$. Even with the lowest time budget of $b=1$ second per period, the median search depth is $H \geq 5$ and at larger time budgets is significantly higher than $H=4$ which was used for UCS and A*. Median search depth increases logarithmically with respect to the budget $b$, increasing by one for approximately each doubling of $b$. Retaining the search tree after each timestep means that search depth is inherited in subsequent periods, yielding significantly higher search depths on average for Guided IDA*. While search tree retention is not unique to Guided IDA*, the retained search tree in Guided A* and Guided UCS is used to reduce run time of solving subsequent periods rather than allow for deeper search. Whereas Guided A* adaptively spends computational resources (reflected in run time) on planning from more complex states, Guided IDA* adaptively reduces search depth. 

\begin{figure}
    \centering
    \includegraphics[width=0.55\textwidth]{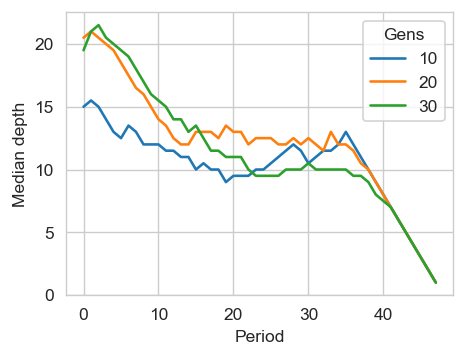}
    \caption{Median search depth for IDA* with $b=30$ seconds and Constrained ED heuristic. Deeper search is achieved in the early morning periods, where search breadth is comparatively narrow.}
    \label{ch2:fig:ida_star_depth_vs_period}
\end{figure}

Figure \ref{ch2:fig:ida_star_depth_vs_period} shows the variation in median search depth throughout the day for a budget $b=30$s. Similar to search \emph{breadth} of Guided A*, shown in Figure \ref{ch2:fig:breadth_and_period_runtime}, the figure shows that during the simpler early morning periods, search depth is relatively large, due to a lower branching factor and greater certainty in these periods. The large search tree created in early morning periods where median depth is greatest benefits periods later in the day due to search tree retention. 

\subsubsection{Redundant Computation in Iterative Deepening}

Iterative deepening algorithms necessarily involve a certain amount of additional computation compared with their non-iterative counterparts (such as IDA* and A*) due to the search tree being rebuilt at each iteration. As noted in \cite{russellnorvig}, in practice this is usually not an important consideration due to the exponential relationship between the depth of a generation and the number of nodes in that generation. 

\begin{figure}
    \centering
    \includegraphics[width=0.5\textwidth]{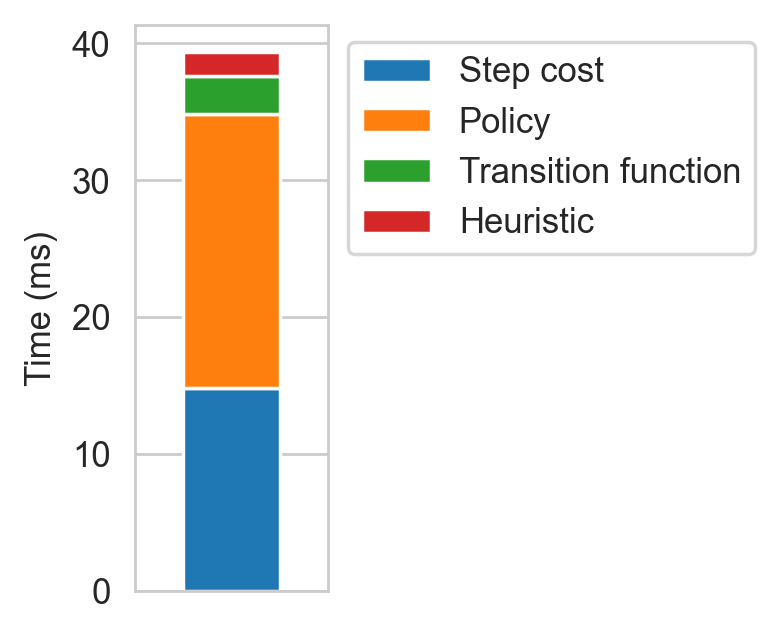}
    \caption{Composition of run time of the major routines of initial node evaluation in Guided IDA*. \emph{Step cost} is the economic dispatch calculations required to determine expected operating costs over net demand scenarios. \emph{Policy} is the neural network evaluation for guided expansion. \emph{Transition function} evaluates the system dynamics, advancing to a new state. \emph{Heuristic} (here using Constrained ED) is the only component which is evaluated when a node is revisited. When a node is revisited in IDA*, the computational cost is around 8\% of the first visit.}
    \label{ch2:fig:runtime_composition}
\end{figure}

Furthermore, applying Guided IDA* to the UC problem, repeated evaluation of nodes is not a significant computational concern, as most of the computational cost of a node evaluation can be retained after the first iteration. Figure \ref{ch2:fig:runtime_composition} shows the run time composition of a node evaluation consisting of 4 elements. The step cost (solving the ED problem for each of the net demand scenarios in $\mathcal{S}$ (see Section \ref{ch1:methodology:search_tree}) and evaluation of the expansion policy (required for guided expansion) constitute the majority of total run time. These components, as well as the transition function (i.e. taking a step in the simulation environment), only need to be computed on the first node evaluation in IDA*. The heuristic is the only component which needs to be re-evaluated after the first visit, which accounts for $<10$\% of the total run time.  

\section{100-Generator Problem} \label{ch2:100gen}

In Section \ref{ch2:experiments}, we showed that Guided A* search achieved better search efficiency by leveraging domain knowledge through a heuristic. We then showed that the anytime property of Guided IDA* means that further operating cost reductions can be achieved in practice by more effectively exploiting computational resources. These improvements invite the application of Guided IDA* to larger power systems in order to evaluate its potential for practical applications in regional or national-scale transmission networks. 

In this section we apply Guided IDA* to a power system of 100 generators, which to the best of our knowledge is the largest simulation study applying RL and/or tree search to the UC problem. Due to the high dimensionality of the action space, we use a novel form of entropy regularisation during training which aims to promote policies which match a pre-determined \emph{target entropy}. As opposed to maximum entropy RL techniques used previously in this research, the target entropy regularisation technique used here penalises policies whose entropy differs from a pre-determined target entropy. This ensures that policies achieves a level of stochasticity that is appropriate for guided tree search. In this section we show that Guided IDA* is capable of achieving operating costs that are competitive with industry-standard MILP approaches. 

\subsection{Target Entropy Regularisation} \label{ch2:100gen:target_entropy}

Guided expansion requires a stochastic policy $\pi(a|s)$ in order to build a search tree with several branches from each node. As described in Section \ref{methodology:pg:ppo}, policy entropy quantifies the randomness of a policy: high entropy policies are more stochastic than low entropy policies. In high dimensional action spaces, there are practical challenges in training a policy with a suitable level of policy entropy such that several actions satisfy the branching threshold $\rho$ in guided expansion, while ensuring the policy does not converge to a deterministic one. In Section \ref{ch1:guided_ucs:policy_training}, the entropy-regularised PPO objective function $J^{\text{PPO} + H}$ (Equation \ref{methodology:eq:ppo_entropy_objective}) was used to promote stochastic policies via an \emph{entropy bonus}. This method was found to be a successful approach for systems of up to 30 generators, with Guided UCS, A* and IDA* all outperforming MILP benchmarks when trained with this technique. As shown in Table \ref{ch1:tab:policy_params}, we decreased the level of entropy regularisation via the entropy coefficient $\beta$ in the entropy-regularised PPO objective function (Equation \ref{methodology:eq:ppo_entropy_objective}) as the number of generators increased. This reflected the increasing dimensionality of the action space, and thus the requirement for a lower entropy policy for guided expansion. However, as the dimensionality of the action space grows, tuning the level of entropy regularisation becomes increasingly difficult, and may even demand negative entropy bonus (i.e. $\beta < 0$) to produce satisfactory expansion policies. Furthermore, the final level of policy entropy for a given level of entropy regularisation may be difficult to predict, depending on the exploration of the policy and state spaces during training. Therefore, it is difficult to ensure a usable policy with a suitable entropy is produced during training.

To mitigate this problem of tuning $\beta$ in high-dimensional action spaces and control the final policy entropy level, we implement a novel entropy regularisation technique based on \emph{target entropy}. Rather than employing an entropy bonus, we introduce a term to the objective function based on the squared error of the policy entropy and the target entropy. Specifically, we use the following modified PPO objective function: 

\begin{equation}
    J^{\text{PPO} + H_T}(\theta) = \mathbb{E}[J^{\text{PPO}}(\theta) + \beta (H(\pi_{\theta}) - H_T)^2]
\end{equation}

where $J^{\text{PPO}}(\theta)$ is the PPO objective taken from Equation \ref{methodology:eq:ppo_objective} in Chapter \ref{methodology}, $\beta$ is a constant entropy coefficient controlling the level of regularisation and $H_T$ is the target entropy. Higher levels of $H_T$ promote more stochastic policies, and higher settings of $\beta$  penalise deviations from target entropy more strongly.

A heuristic method can be used to set the target entropy $H_T$ as a function of the number of generators $N$ and the branching threshold $\rho$, uniting the elements of policy training and guided tree search. For a given $\rho$, we aim to achieve an expected entropy such that the joint action probability $\pi(a = [a_1, a_2,...,a_N ] |s) = \rho$ on average, where $a_i \in \{0,1\}$ is the commitment decision for generator $i$. As we use a sequential policy architecture, described in Section \ref{ch1:guided_ucs:policy}, each sub-action $p = \pi(a_i|s)$ should be $p = \rho^{\frac{1}{N}}$ on average. Using the definition of the entropy of a random variable in Equation \ref{methodology:eq:entropy}, the target entropy as a function of $\rho$ and $N$ is:

\begin{gather}
    H_T = -p \log_2(p) - (1-p) \log_2 (1-p) \label{ch2:eq:entropy_eq} \\
    p = \rho^{\frac{1}{N}}
\end{gather}

In the following section we investigate the impact of $H_T$ and $\beta$ on policy training and convergence for the 100 generator problem.


\subsection{Training Details}

Expansion policies were trained for the 100-generator system using the method described in Section \ref{ch1:guided_ucs:policy_training}, using the target entropy regularisation described in Section \ref{ch2:100gen:target_entropy}. Using a grid search approach, we studied the impact of parameters $H_T$ and $\beta$ on policy training. 

The training parameters are shown in Table \ref{ch2:tab:100gen_policy_params}. The buffer size (representing the number of policy evaluations per epoch) was increased from 2000 in previous experiments to 5000. This reflects the fact that more forward passes are required to generate a single action using the sequential policy parametrisation described in Section \ref{ch1:guided_ucs:policy}. We also used a network architecture of two hidden layers with 128 and 64 nodes for both actor and critic architectures.  The total policy training time was approximately 30 hours over 8 CPU workers.

We demonstrate the impact of two target entropy regularisation parameters, target entropy $H_T$ and the entropy coefficient $\beta$, on policy training using a grid search approach. For target entropy, we set $H_T \in \{0.1567, 0.1919, 0.2648\}$, corresponding to $\rho \in \{0.01, 0.05, 0.1\}$ respectively according to Equation \ref{ch2:eq:entropy_eq}. For each setting of $H_T$, we trained a policy with entropy coefficient $\beta \in \{0.1, 1.0\}$, which was scaled linearly beginning at $\beta=0$ over 100,000 epochs to prevent premature convergence. The policies with target entropy regularisation were compared to a baseline trained without entropy regularisation (i.e. $\beta=0$).

\begin{table}[t]
    \centering
    \begin{tabular}{l|r}
        \toprule
        \bf{Variable} & \bf{Value} \\
        \midrule    
        Clip ratio & 0.1 \\
        Actor architecture & 128, 64 \\
        Critic architecture & 128, 64 \\ 
        Entropy coefficient $\beta$ & $\{0.1, 1.0\}$ \\
        Target entropy $H_T$ & $\{0.1567, 0.1919, 0.2648\}$ \\
        Buffer size  & 5000 \\ 
        Epochs & 300,000 \\ 
        Gamma & 0.95 \\
        \bottomrule
    \end{tabular}
    \caption{Parameters used to train 100-generator policies. Combinations of the target entropy regularisation variables $\beta$ and $H_T$ are used in a grid search studying policy convergence with respect to these parameters.}
    \label{ch2:tab:100gen_policy_params}
\end{table}

\begin{figure}
    \centering
    \includegraphics[width=\textwidth]{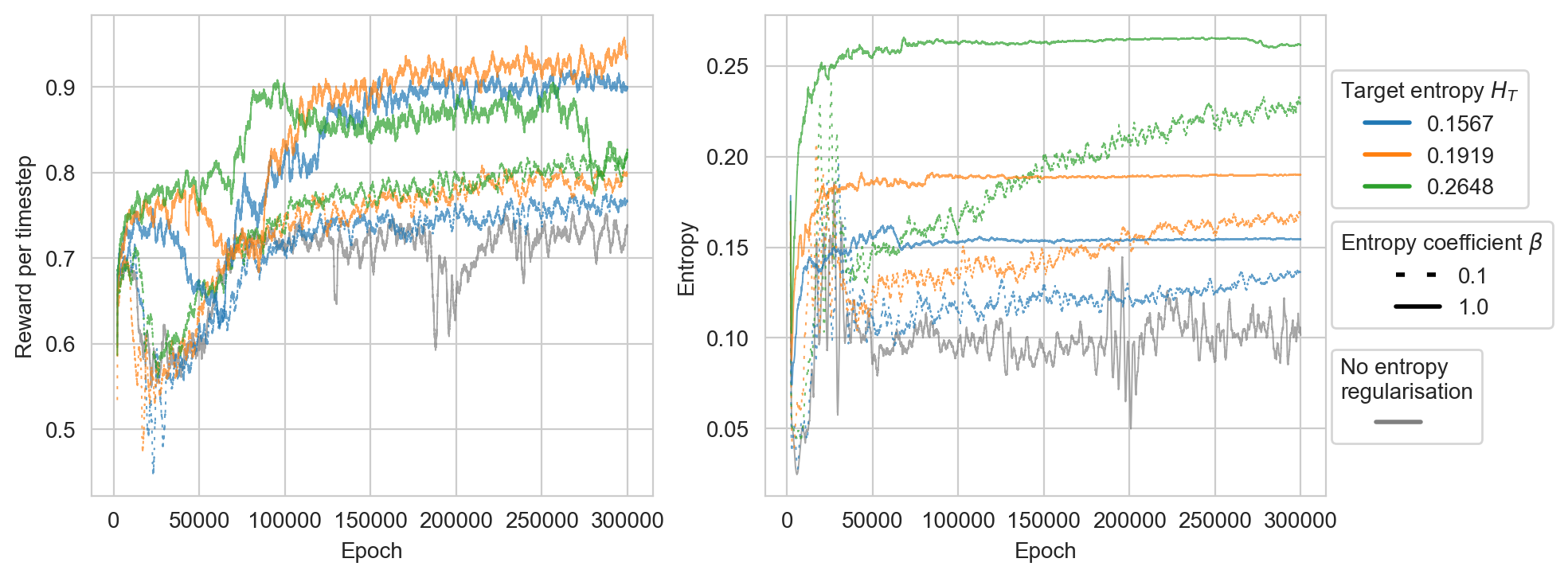}
    \caption{Convergence of 100 generator expansion policies using target entropy regularisation with varying $H_T$ and $\beta$. The grey line shows policy training with no entropy regularisation. The left plot shows reward per timestep while the right plot shows policy entropy. Both plots display a moving average over 2000 epochs.}
    \label{ch2:fig:100gen_training_curve}
\end{figure}

Figure \ref{ch2:fig:100gen_training_curve} shows the convergence of mean operating costs and entropy during training of the 7 expansion policies (6 using target entropy regularisation, 1 baseline with no regularisation). The policies with $\beta=1$ (solid lines) converge to a policy with stable entropy $H_T$, although there are significant variations in performance, such as the drop in performance towards the end of training for $H_T = 0.2648$. Those with a lower level of entropy regularisation $\beta=0.1$ (dotted lines) display gradually increasing entropy throughout training, and achieve lower average reward than those with $\beta=1$. The policy with $H_T = 0.1919$ and $\beta = 1.0$ was found to converge to the highest average reward, while the policy with no entropy regularisation was the worst performing. In addition, this baseline policy with no entropy regularisation converges to the lowest level of policy entropy, indicating a tendency in large power systems to converge to near-deterministic policies with entropy regularisation, which is not suitable for guided tree search.  

Overall, the policy trained with $H_T=0.1919$ and $\beta=1.0$ was found to achieve the highest average reward at the end of training. In the following section, we use this policy in Guided IDA* search to solve the 20 unseen test problems for the 100-generator problem.

\subsection{Results} \label{ch2:100gen:results}

Using the expansion policy trained with $H_T=0.1919$ and $\beta=1.0$ and using the Constrained ED heuristic, Guided IDA* search was used to solve the 20 unseen UC problem instances. The branching threshold was set to $\rho=0.05$, corresponding to the target entropy $H_T=0.1919$ set with Equation \ref{ch2:eq:entropy_eq}, and the time budget was set to $b=60$ seconds as in previous experiments. Comparison is made with the deterministic UC benchmarks MILP($4\sigma$) and MILP(perfect). 

\begin{figure}
    \centering
    \includegraphics[width=0.6\textwidth]{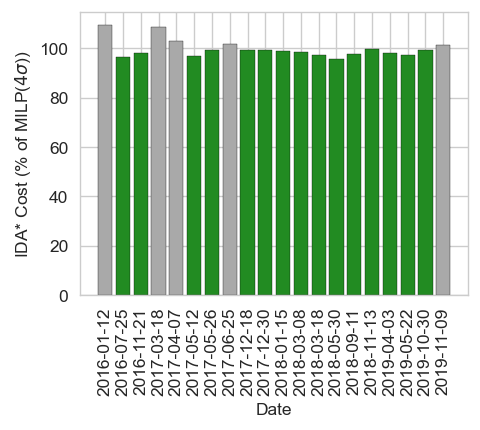}
    \caption{Day-by-day comparison of IDA* operating costs with MILP($4\sigma$). IDA* has lower mean costs than MILP on 15 out of 20 days (those in green).}
    \label{ch2:fig:100gen_comparison}
\end{figure}

The results are shown in Table \ref{ch2:tab:100gen}. Operating costs were 0.14\% lower using Guided IDA* search as compared with MILP($4\sigma$), and 11\% higher than MILP(perfect). Guided IDA* had greater security of supply than the reserve-constrained MILP($4\sigma$), achieving lower LOLP (0.19\% compared with 0.40\%) and reducing standard deviation in operating costs by a factor of 2. A comparison of mean operating costs between IDA* and MILP($4\sigma$) broken down into individual test problems is shown in Figure \ref{ch2:fig:100gen_comparison}. There is significant variation in expected Guided IDA* costs as a percentage of MILP($4\sigma$), ranging between 96\% to 109\%. Overall, IDA* has lower operating costs than MILP in 15 of the 20 days, marked in green in Figure \ref{ch2:fig:100gen_comparison}.
 
\begin{table}[t]
    \centering
    \resizebox{\textwidth}{!}{
    \input{05-chapter2/tables/100gen}}
    \caption{Comparison of Guided IDA*, MILP($4\sigma)$ and MILP(perfect) solutions to 100 generator test problems.}
    \label{ch2:tab:100gen}
\end{table}

\begin{figure}
    \centering
    \includegraphics[width=0.9\textwidth]{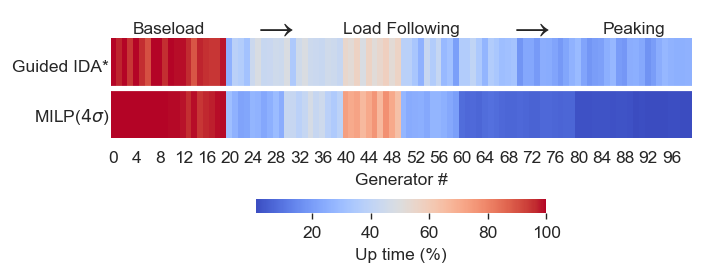}
    \caption{Proportion of periods spent online for generators in the 100-generator problem using Guided IDA* and MILP($4\sigma$). Small capacity peaking plants are shown at the right of the graph, with base-load at the left.}
    \label{ch2:fig:100gen_uptimes}
\end{figure}

Guided IDA* generally adopts more flexible operating patterns than MILP($4\sigma$). Figure \ref{ch2:fig:100gen_uptimes} shows the distribution of up times (proportion of periods spent online) for Guided IDA* and MILP($4\sigma$). Guided IDA* adopts lower utilisation rates for the 20 baseload generators and higher rates for peaking plants and also uses roughly twice as many startups as compared with MILP($4\sigma$). The generation patterns of MILP($4\sigma$) shown in Figure \ref{ch2:fig:100gen_uptimes} are more stratified as compared with Guided IDA*. This indicates similarities with a merit order or priority list commitment approach, whereas Guided IDA* adopts more nuanced strategies. Furthermore, Guided IDA* search uses more extreme actions than MILP($4\sigma$), changing up to 41 generator commitments simultaneously, compared with a maximum of 20 for MILP($4\sigma$). In addition, Guided IDA* made more use of the `do nothing' action, keeping all generator statuses the same. This follows similar results found in Section \ref{ch1:experiments:milp_comparison}, where we observed Guided UCS made more extreme commitment changes, followed by longer periods of no commitment changes.

\section{Discussion} \label{ch2:discussion}

Using guided expansion in a modular way, any tree search algorithm can be enhanced using an RL-trained agent as a guide. This chapter showed that the choice of tree search algorithm (UCS, A*, IDA*) is an important design decision in the broader class of guided tree search methods, and has a significant impact on performance. Using informed and anytime search methods yielded substantial performance improvements and practical benefits as compared with Guided UCS, and allowed for the successful application of Guided IDA* to larger UC problems of 100 generators.

\subsection{Advantages of Informed Search}

In the context of informed search, we showed that the heuristic is an important design decision that can substantially impact search efficiency. Since there is no single heuristic approach that is suitable for all problem domains, designing the heuristic for the UC problem requires domain expertise: our approach using priority list algorithms was based on UC literature that showed that these methods have properties that are suited to heuristic design such as short run time and the ability to relax constraints \cite{senjyu2003fast}. There is inevitably a trade-off between run time, accuracy and admissibility, but our results found that the most accurate and slowest heuristic (Constrained ED) was clearly the best performing in practice, reducing run time by up to 94\% as compared with Guided UCS with negligible impact on solution quality. There is scope for further research into more accurate heuristics that can further improve search efficiency. 

\subsection{Advantages of Anytime Search}

Anytime search with Guided IDA* was shown to outperform Guided A* for similar computational budgets, by maximising use of computational resources. Whereas some days were solved with very short run times by Guided A* due to narrow search breadth, Guided IDA* compensates by adaptively increasing the search depth in such situations. In practice, Guided IDA* achieves far greater search depths on average, as shown in Figure \ref{ch2:fig:ida_star_depth_vs_g}, resulting in lower operating costs for similar computational budgets. The anytime property of Guided IDA* is a significant practical advantage over UCS and A* for the UC problem, allowing for schedules to be reliably produced in time-constrained contexts. Both Guided UCS and Guided A* search methods have highly variable and unpredictable run times (Figure \ref{ch2:fig:a_star_run_time}) which makes the branching threshold and search depth parameters $\rho$ and $H$ difficult to tune in practice. In particular, due to the exponential complexity in $H$, the run times of Guided UCS and Guided A* are very sensitive to the search depth. In IDA*, $H$ is replaced with a time budget $b$, which can be set by knowledge of market constraints, such as the time to gate closure when bids and offers must be submitted. 

Increasing the time budget in IDA* generally resulted in operating cost reductions (Figure \ref{ch2:fig:anytime_bars}). However, in all three problems, even small time budgets $b \geq 2$ seconds per period were enough to outperform the MILP(4$\sigma$) benchmark and $b \geq 10$ outperformed Guided UCS. Further increasing the budget has a stabilising effect in addition to reducing costs. This is most clear for the 30 generator case, where costs fluctuate significantly between budgets of 1, 2 and 5 seconds per period, but increase steadily for larger budgets.

\subsection{Planning in Complex Decision Periods}

We observed analogous trends between the variations in period run time observed for Guided A* search (Figure \ref{ch2:fig:breadth_and_period_runtime}) and the depth variations when using IDA* (Figure \ref{ch2:fig:ida_star_depth_vs_period}), which reflect the relative uncertainty in decision-making during these periods. For Guided A*, run times were higher around the morning peak. By contrast, Guided IDA* conducts shallower search in periods with higher decision-making uncertainty where the branching factor is greater. There may be contexts where methods with adaptive run time, such as Guided A* search, are more appropriate than methods with adaptive depth (Guided IDA*); shallow search in complex decision periods using Guided IDA* may result in unreliable or insecure decision making. A hybrid approach, for instance enforcing a minimum search depth in IDA*, could also perform better in some contexts. 

\subsection{Scaling to Larger Power Systems}

Applying IDA* to the 100 generator problem, which has a very large action space of up to $2^{100}$ actions, we found Guided IDA* achieved operating costs that were 0.14\% lower than the MILP($4\sigma$) benchmark. While expected operating costs were similar to the MILP($4\sigma$) benchmark, Guided IDA* schedules were significantly more secure, with lower LOLP and variation in total operating costs. 

In previous experiments on smaller power systems, we used traditional entropy regularisation with an entropy bonus, using the objective function $J^{\text{PPO}+H}$ (Equation \ref{methodology:eq:ppo_entropy_objective}), to encourage exploration during training and an appropriate level of policy entropy for guided tree search. This was shown to be an effective strategy for systems of up to 30 generators but was not so effective for the 100-generator problem. Introducing the target entropy term to the PPO loss function, as described in Section \ref{ch2:100gen:target_entropy}, had a significant impact on policy training, with policies converging to higher average rewards when target entropy regularisation was used. Setting the target entropy $H_T$ based on the number of generators and the branching threshold $\rho$ was also shown to be a useful heuristic for uniting policy training and policy testing in large discrete action spaces.

In Figure \ref{ch2:fig:100gen_training_curve}, the smoothest convergence of mean reward was observed for policies with target entropy regularisation and $\beta=0.1$, although convergence was faster with $\beta=1.0$. All other policies exhibited sharp, temporary performance losses that indicate relatively unstable and unpredictable convergence properties which were more pronounced than for smaller systems. The practical challenges of training expansion policies in large action spaces invites further research, and improvements in expansion policy training could yield significant performance improvements. Overall, the application of Guided IDA* to the 100 generator problem is a significant milestone, and our results are promising for practical applications of RL-based methods to regional or national-scale electricity network operation. To the best of our knowledge, this is the largest application of RL and/or tree search to the UC problem in the existing literature.

\section{Conclusion} \label{ch2:conclusion}

In this chapter we presented two guided tree search algorithms, building on the method of guided expansion developed in Section \ref{ch1:guided_ucs:guided_expansion}: Guided A* search and Guided IDA* search. Both algorithms are informed search methods, using problem-specific knowledge to improve search efficiency at the expense of generality across problem domains. Guided IDA* is additionally anytime, which is practically advantageous in time-constrained contexts such as electricity markets.

We developed three heuristics based on priority list UC solution methods, which are used to rapidly approximate the optimal solution cost and are used by both algorithms to improve search efficiency relative to Guided UCS. Guided IDA* was shown to mitigate the run time variability of non-anytime algorithms, reaching greater search depths and reducing operating costs by up to 1.0\% relative to Guided UCS within similar time budgets. 

The improvements afforded by the informed and anytime search algorithms in this chapter allowed for Guided IDA* search to be applied to a larger power system of 100 generators. A novel technique of target entropy regularisation was used to improve policy convergence and unite policy training with tree search. Expected total operating costs were found to be lower than the MILP benchmark with a reserve constraint and Guided IDA* outperformed MILP on 15 out of 20 days, with a lower loss of load probability overall. These results demonstrate the potential for guided tree search methods to outperform existing industry methods in large-scale contexts.

%% file: 05-chapter2/tables/heuristic_summary.tex
\begin{tabular}{lrrr}
\toprule
     Heuristic &  Mean time (ms) &    RMSE &  Admissibility (\%) \\
\midrule
    Naive MMFC &            0.59 & 1528.27 &              100.00 \\
      Naive ED &            1.70 & 1134.88 &               95.00 \\
Constrained ED &            1.89 &  992.75 &               98.12 \\
\bottomrule
\end{tabular}

%% file: 05-chapter2/tables/a_star_summary.tex
\begin{tabular}{rlrr}
\toprule
 Generators &      Heuristic &  Time (\% of UCS) &  Cost (\% of UCS) \\
\midrule
         10 &     Naive MMFC &             13.55 &            100.00 \\
         10 &       Naive ED &             11.03 &            100.00 \\
         10 & Constrained ED &              6.41 &            100.00 \\
         20 &     Naive MMFC &             46.64 &             99.96 \\
         20 &       Naive ED &             45.20 &            100.08 \\
         20 & Constrained ED &             35.62 &            100.03 \\
         30 &     Naive MMFC &             24.19 &            100.01 \\
         30 &       Naive ED &             24.55 &            100.09 \\
         30 & Constrained ED &             17.74 &            100.08 \\
\bottomrule
\end{tabular}

%% file: 05-chapter2/tables/ida_a_star_comparison_formatted.tex
\begin{tabular}{rllrrrrrr}
\toprule
 Num. gens &          Method &      Heuristic &  Mean cost (\$M) &  Std. cost &  Mean time (s) &  Max. time &  Min. time &  LOLP (\%) \\
\midrule
        \multirow{4}{*}{10} &            IDA* & Constrained ED &             9.33 &       0.75 &         1086.3 &     1294.1 &      892.0 &      0.115 \\
         &              A* & Constrained ED &             9.37 &       0.84 &           51.8 &      195.8 &        5.7 &      0.128 \\
         &             UCS &           None &             9.37 &       0.84 &          807.3 &     1992.1 &       76.9 &      0.128 \\
         & MILP($4\sigma$) &           None &             9.40 &       1.02 &           19.1 &      177.2 &        1.8 &      0.180 \\ \midrule
        \multirow{4}{*}{20} &            IDA* & Constrained ED &            18.67 &       1.46 &         1099.6 &     1267.6 &      797.0 &      0.116 \\
         &              A* & Constrained ED &            18.74 &       1.44 &           41.8 &      159.3 &        7.7 &      0.112 \\
         &             UCS &           None &            18.73 &       1.41 &          117.3 &      374.5 &       10.4 &      0.107 \\
         & MILP($4\sigma$) &           None &            18.90 &       2.92 &            5.5 &        8.4 &        3.8 &      0.244 \\ \midrule
        \multirow{4}{*}{30} &            IDA* & Constrained ED &            28.14 &       1.96 &         1210.0 &     1359.7 &      986.0 &      0.110 \\
         &              A* & Constrained ED &            28.43 &       2.07 &           69.4 &      164.4 &       14.5 &      0.142 \\
         &             UCS &           None &            28.41 &       2.04 &          391.3 &      976.8 &      129.4 &      0.142 \\
         & MILP($4\sigma$) &           None &            28.53 &       4.94 &            8.0 &       12.4 &        5.6 &      0.291 \\
\bottomrule
\end{tabular}

%% file: 05-chapter2/tables/100gen.tex
\begin{tabular}{lrrrrrr}
\toprule
        Version &  Mean cost (\$M) &  Std. cost &  Mean time (s) &  Max. time &  Min. time &  LOLP (\%) \\
\midrule
  MILP(perfect) &            87.45 &       0.00 &           30.0 &       47.5 &       24.6 &      0.000 \\
    Guided IDA* &            96.65 &       9.73 &         2737.1 &     2782.0 &     2563.1 &      0.188 \\
MILP($4\sigma$) &            96.78 &      20.47 &           40.2 &       59.5 &       27.7 &      0.397 \\
\bottomrule
\end{tabular}

%% file: 06-chapter3/chapter3.tex
\section{Introduction}

Existing research has shown that RL algorithms can be used to achieve expert performance over a broad range of tasks, with little or no adaptation of the algorithm itself \cite{mnih2013playing, schrittwieser2020mastering, schulman2017proximal, haarnoja2018soft}. Compared with MILP approaches, which often require expert domain knowledge to develop a suitable mathematical formulation, flexibility across problem domains is a valuable property of RL that is beneficial for operating power systems with heterogeneous technologies for power generation, consumption, transmission and storage. Furthermore, RL has been shown to provide insights into solution techniques for complex problems \cite{mirhoseini2021graph, silver2016mastering, silver2018general, berner2019dota}, most famously shown in novel gameplay strategies used by AlphaGo \cite{silver2016mastering}. These studies have shown that RL is able to learn fundamentals of problem-solving across many problem domains without human intervention. Having demonstrated that Guided IDA* is an effective solution method for the UC problem in Chapter \ref{ch2}, in this chapter we extend the RL approach to solve two more advanced variants of the UC problem, considering carbon intensity and power system security.


The case studies reflect current challenges in the development of power systems and the integration of variable renewable energy \cite{heptonstall2021systematic}. First, we consider carbon pricing and an additional action to curtail wind generation. Increasing curtailment of variable renewable energy \cite{mararakanye2019renewable} poses a challenge to climate goals as the curtailed energy is usually replaced by fossil-fuel generators, but is now an essential part of power systems operation that must be optimised alongside other decisions. Additionally, carbon pricing is an important policy mechanism to achieve CO$_2$ emissions reductions, but entails changes to the objective function, thus requiring different operational strategies favouring lower carbon generation. In the second case study, we introduce generator outages to the environment. Outages are a crucial security consideration for system operators, currently handled by $N-x$ reserve criteria that protect against the largest loss of generation. However, current security practices must adapt to increasing penetration of variable renewables \cite{ela2011operating} and coincident outages \cite{murphy2018resource} which have been the cause of recent major blackouts such as the GB power network outage of 9 August 2019, which impacted over 1 million customers \cite{bialek2020does}. 

The experiments in this chapter show that Guided IDA* is a highly flexible methodology that can be applied to environments of arbitrary complexity without significant modifications beyond parameter tuning. It is also highly scalable in the number of scenarios $N_s$, which allows problems with large numbers of uncertain parameters to be approached probabilistically without heuristics.

\subsection{Contributions} 

This chapter makes the following contributions: 

\begin{enumerate}
    \item We present two variations on the power system simulation environment presented in Section \ref{ch1:env} and formulate corresponding MDPs for each. In the first environment we implement a \textbf{curtailment action} and a \textbf{carbon price}. In the second we introduce \textbf{generator outages} to the environment. These environments present novel and challenging tasks that can be used to benchmark RL approaches to the UC problem considering heterogeneous actions and stochastic generation. 
    
    \item Guided IDA* is used to solve unseen test problems with three levels of carbon price in the curtailment environment. We observe changes in schedule characteristics as the carbon price is increased, such as a reduction in coal generation, employment of gas as base-load and fewer startups. Our results show that Guided IDA* generalises across MDPs with different action spaces and is sensitive to changes in the reward function. 

    \item In the generator outages case study, Guided IDA* is shown to significantly outperform MILP benchmarks with $N-x$ reserve criteria. Guided IDA* adaptively allocates reserve margins that vary between $N-1$ and $N-4$ levels and achieves operating costs that are up to 1.9\% lower than the best performing MILP benchmark. Security of supply remains similar to that shown in previous experiments without outages. We show that Guided IDA* can be used to discover optimal reserve allocation strategies with uncertain generation availability without the use of heuristics. 
    
    \item We investigate the impact of the number of scenarios $N_s$ used to build the search tree on Guided IDA* solution quality in the outages case study. Our results show that with large numbers of uncertain parameters, increasing $N_s$ can be used to improve solution quality. Due to the linear time complexity in $N_s$ and anytime quality of Guided IDA*, we show that a large number of scenarios can be considered while constraining run time and maintaining deep search.
    
\end{enumerate}

This chapter is organised as follows. Section \ref{ch3:curtailment} covers the curtailment and carbon price case study. In this section we describe the problem environment, MDP formulation, experimental setup and the results of our experiments applying Guided IDA* to solve unseen test problems. Section \ref{ch3:robustness} follows the same structure for the generator outages case study. In Section \ref{ch3:discussion} we reflect on the results of both experiments, discussing implications for power systems operation. Section \ref{ch3:conclusion} concludes the chapter. 

\section{Case I: Wind Curtailment and Carbon Price} \label{ch3:curtailment}

In this case study, we modify the power system simulation environment described in Section \ref{ch1:env} to include an additional wind curtailment action, which can be used to temporarily reduce wind generation. In addition, we incorporate a carbon price into the environment's cost function and investigate the impact of carbon price levels on the operating strategies and CO$_2$ emissions of Guided IDA* schedules.

The effects of carbon pricing and curtailment are important topics of research for current and future power systems. Increasing levels of wind penetration have been accompanied by high rates of wind curtailment in several countries and transmission networks including Great Britain \cite{joos2018short}, China, Texas and Italy \cite{mararakanye2019renewable}. While in general the system operator aims to maximise wind penetration as it has no marginal cost, in some cases it is necessary to curtail wind generation in order to manage transmission network congestion or to ensure an adequate level of controllable reserve generation is available \cite{burke2011factors}. As a result, curtailment is an important power systems operational decision that can be used to benefit overall system security and can in some cases reduce operating costs and carbon emissions compared with other measures \cite{morales2021reducing, jacobsen2012curtailment}. Currently, curtailment decisions are often made by the system operator on an ad-hoc, manual basis \cite{krishnan2018resilient}. Studies have investigated automating short horizon curtailment decisions as remedial actions to maintain grid security \cite{krishnan2018resilient, liu2017decentralized}. In addition, some studies have investigated co-optimisation of curtailment alongside UC in a day-ahead context, with the aim of reducing total system operating costs \cite{psarros2019comparative, wang2016robust, alves2016wind, dvorkin2015wind, hozouri2014use}. However, curtailment decisions have not been introduced in RL studies for the UC problem. 

This case study shows that Guided IDA* is applicable to MDPs and environments with different characteristics to those studied in Chapters \ref{ch1} and \ref{ch2}. Compared with MILP approaches, introducing new actions to the MDP does not require modification of the solution method or manual reformulation of the problem. Furthermore, we analyse the impact of carbon pricing on operating strategies and curtailment rates of Guided IDA* solutions. We first describe the setup of the environment, modified from that described in Section \ref{ch1:env} to incorporate the wind curtailment action and carbon price. We then formalise by describing the modified MDP. Finally we present the results of experiments applying Guided IDA* to solve the 20 test problems. 

\subsection{Environment Setup} \label{ch3:curtailment:env}

To incorporate wind curtailment and carbon price, we modified the power system environment described in Section \ref{ch1:env}. We made two changes: first, we adjusted the fuel cost curves based on a carbon price. Second, we implemented an additional curtailment action, which is scheduled in a day-ahead context alongside generator commitment. The curtailment action and carbon price can be activated in a modular fashion in the open-source Python package developed for this research\footnote{\url{https://github.com/pwdemars/rl4uc}}. 

We adopt the approach used in \cite{dvorkin2015wind} of treating a wind curtailment action as a decision variable (i.e. action in the MDP). As we do not consider transmission network constraints, the purpose of the curtailment action is to maintain grid security by: (1) ensuring there is sufficient downward ramping capacity in cases of high wind generation and low demand; (2) reducing the net demand uncertainty to reduce reserve requirements. As wind is considered as negative demand in our problem setup, using the curtailment action has the effect of increasing net demand. The carbon price is reflected in adjusted fuel cost curves for the thermal power stations. The following sections describe the environment modifications in more detail.  

\subsubsection{Carbon Price-Adjusted Fuel Cost Curves}

In order to incorporate a carbon price, we adjusted the quadratic fuel cost curves (Equation \ref{methodology:eq:quadratic_fuel_cost}) for each generator to reflect the increased cost per unit of fuel combusted. Fuel types were not included in the original Kazarlis data \cite{kazarlis1996genetic} (Table \ref{ch1:tab:gen_info}) and were manually assigned to either coal, oil or gas. 

The generator fuel assignments are shown in Table \ref{ch3:tab:generator_fuel_assignment} and were decided based on properties of generators in the original data, which naturally formed three groups. The first group comprises the largest capacity (455 MW) generators 1 and 2, and was assigned the coal fuel type. These generators have properties that are most characteristic of base-load, with the largest startup costs, most restrictive up/down time constraints and lowest marginal costs. The second group includes the medium capacity (80--162 MW) load-following generators 3--7, which we assigned to gas. The final group includes generators 8--10 which are the smallest capacity (55 MW) and most flexible (shortest minimum up/down time constraints); this group was assigned to oil. As in previous experiments, to create the 30 generator power system used in this case study, we duplicated each generator three times.

\begin{table}[]
    \centering
    \begin{tabular}{cllcc}
        \toprule
        Generator & Fuel type & Characteristics & $EF$ [$\frac{\text{kgCO$_2$}}{\text{MMBTU}}$] & $FP$ [$\frac{\text{\$}}{\text{MMBTU}}$]  \\
        \midrule
        1--2 & Coal & Base-load & 95 & 1.30 \\ 
        3--7 & Gas  & Load-following & 54 & 2.69 \\
        8--10 & Oil  & Peaking & 73 & 3.18 \\
        \bottomrule
    \end{tabular}
    \caption{Generator fuel assignments, operational characteristics, carbon emissions factors $EF$ and fuel prices $FP$.}
    \label{ch3:tab:generator_fuel_assignment}
\end{table}

\begin{figure}
    \centering
    \includegraphics[width=0.7\textwidth]{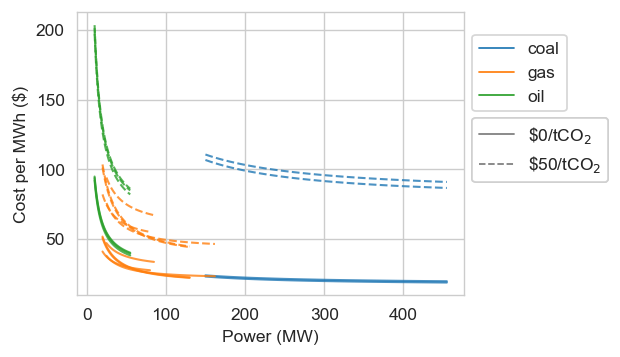}
    \caption{Fuel cost curves with a without a carbon price of \$50 per tCO$_2$ applied.}
    \label{ch3:fig:10gen_cost_curves_carbon}
\end{figure}

Having assigned fuel types, we adjusted the original fuel cost curves $C^f$ (Equation \ref{ch1:eq:cost_curves}) to account for the carbon price using the following method. The adjusted fuel cost curves for a carbon price of \$50/tCO$_2$ are shown in Figure \ref{ch3:fig:10gen_cost_curves_carbon}, which also shows the capacity and fuel cost curve characteristics of the three fuel types. Each curve is transposed upwards when the carbon price is applied, indicating higher total fuel costs. 

To calculate the adjusted fuel cost curves, original fuel cost curves $C^f(p)$ are decomposed into the product of heat rate $H(p)$ [BTU] (amount of fuel combusted as a function of generator output $p$) and the fuel price $FP$ ($\frac{\text{\$}}{\text{BTU}}$): 

\begin{equation}
    C^f(p) [\$] = H (p) [\text{BTU}] \times FP \Big[ \frac{\text{\$}}{\text{BTU}} \Big]
    \label{ch3:eq:fuel_cost_decomposition}
\end{equation}

Rearranging Equation \ref{ch3:eq:fuel_cost_decomposition} and subsituting the original quadratic fuel cost curves from Equation \ref{methodology:eq:quadratic_fuel_cost}, the heat rate curve is expressed in terms of the original cost curve coefficients $a,b,c$ and $FP$: 

\begin{align}
    H(p) &= \frac{C^f(p)}{FP} \\
         &= \frac{ap^2 + bp + c}{FP}
\end{align}

The heat rate curve can be used to calculate the carbon-adjusted price by multiplying by the sum of fuel price $FP$ and fuel-specific carbon price. The fuel-specific carbon price is the product of the fuel's CO$_2$ emissions factor $EF$ and the carbon price: 

\begin{equation}
    EF \Big[ \frac{\text{tCO$_2$}}{\text{BTU}} \Big] \times CP \Big[ \frac{\text{\$}}{\text{tCO$_2$}} \Big]
\end{equation}

In summary, for a generator $i$ with initial fuel cost curve coefficients $a_i, b_i, c_i$ (included in the original data \cite{kazarlis1996genetic}, Table \ref{ch1:tab:gen_info}), fuel price $FP_i$, emissions factor $EF_i$ and carbon price $CP$ (which are all manually-set constants), the carbon-adjusted fuel cost curve $C^f_{c,i}(p)$ can be calculated using the following equation: 

\begin{gather}
    C^f_{c,i}(p) = H_i(p) (FP_i + EF_i \times CP) \\
    C^f_{c,i}(p) = \frac{a_i m_i p^2 + b_i m_i p + c_i m_i}{FP} \\
    m_i =  FP_i + EF_i \times CP 
\end{gather}

For a given carbon price $CP$, the following variables (which are not defined in the original source data \cite{kazarlis1996genetic}) are required to adjust the fuel cost curves: 

\begin{itemize}
    \item Fuel price $FP$ for coal, gas and oil (\$ per BTU)
    \item Carbon emissions factors $EF$ for coal, gas and oil (tCO$_2$ per BTU)
\end{itemize}

We used 1996 (year of publication of \cite{kazarlis1996genetic}) average annual fuel prices for coal, gas and oil from \cite{EIA_data} to set the fuel prices $FP$ shown in Table \ref{ch3:tab:generator_fuel_assignment} (in \$/MMBTU). The emissions factors $EF$ are also shown in Table \ref{ch3:tab:generator_fuel_assignment} and are properties of the fuels. With these constants, $CP$ is included as a parameter in the power system environment and can be used to simulate operating costs under different levels of carbon pricing. When $CP=0$, the cost function is identical to that described in Section \ref{ch1:env:overview}. 

The adjusted fuel cost curves for carbon prices of \$0/tCO$_2$ and \$50/tCO$_2$ are shown in Figure \ref{ch3:fig:10gen_cost_curves_carbon}. The original fuel cost curves $CP$=\$0/tCO$_2$ show the three distinct generator groups that were identified. When the carbon price is applied, all curves are transposed vertically to reflect larger marginal fuel costs. Coal generators experience the largest increase in fuel costs due to having the largest CO$_2$ emissions factor $EF$. Furthermore, cost curves overlap significantly in the vertical axis with $CP$=\$50/tCO$_2$. For instance, oil units operating at maximum capacity $p = p_{\text{max}}$ have strictly lower marginal fuel costs than both coal power stations, but are more expensive when part-loaded. Therefore, optimal UC decisions are more sensitive to generator dispatch when the carbon price is applied.

\subsubsection{Curtailment Action}

To model wind curtailment scheduling, we added an additional curtailment action. The curtailment action is a binary decision which is scheduled in the same way as a generator commitment. As a result, UC schedules have increased dimensionality to $T \times (N+1)$, with the additional column indicating the curtailment decision. The curtailment action reduces wind generation by 50\%. A single discrete reduction was chosen to enable a straightforward search tree representation of the modified UC problem. Further discrete curtailment amounts could be used to enable greater flexibility. For continuous curtailment actions, tree search methods can be applied with techniques including progressive widening \cite{couetoux2011continuous}.  

\begin{figure}
    \centering
    \includegraphics[width=0.8\textwidth]{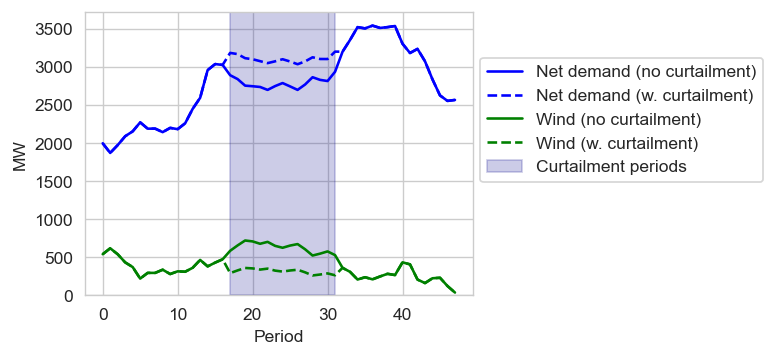}
    \caption{Example use of curtailment action. Curtailing wind during the afternoon increases net demand, preventing a reduction that might demand and shutdown and later startup of a generator before the evening peak.}
    \label{ch3:fig:example_curtailment_action}
\end{figure}

Figure \ref{ch3:fig:example_curtailment_action} shows the change in the point forecasts for net demand and wind when the curtailment action is applied. The curtailment action smoothes the net demand profile which can be used to prevent a shutdown or keep generators running at higher load factors.

\begin{figure}
    \centering
    \includegraphics[width=0.8\textwidth]{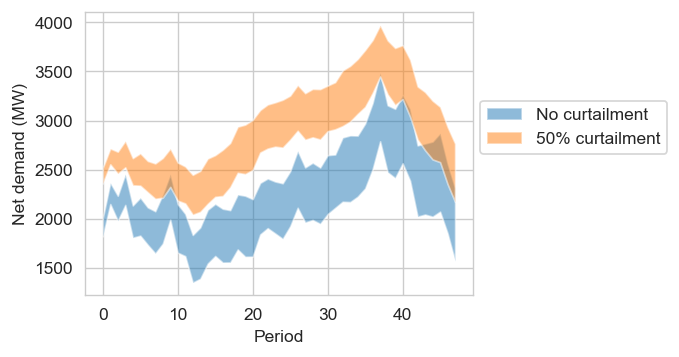}
    \caption{5--95\% quantile interval of net demand, with and without curtailment. The mean width of this interval is reduced by 18\% in this case when the curtailment action is applied.}
    \label{ch3:fig:example_curtailment_errors}
\end{figure}

As the curtailment action reduces wind generation by 50\%, the forecast errors are also assumed to reduce by the same factor, leading to a narrower distribution of wind generation and net demand realisations. The reduction in the 5--95\% quantile interval for net demand is shown in Figure \ref{ch3:fig:example_curtailment_errors}, where the curtailment action is applied uniformly throughout the day. The mean 5--95\% interval for net demand decreases by 18\% when the curtailment action is applied. In addition, the profile is noticeably smoother when the curtailment action is applied, due to the reduced wind generation variability. The curtailment action can therefore be used to reduce net demand variability and thus reduce spinning reserve requirements. 

\subsection{MDP Formulation} \label{ch3:curtailment:mdp}

Having described the changes made to the simulation environment, we will now formalise the changes by making necessary modifications to the UC the MDP described in Section \ref{ch1:uc_mdp}. The updated MDP contains two significant changes: the reward function is updated to reflect changes to the cost function, and the action definition is changed to incorporate the new curtailment action. Since there are no inter-temporal constraints on curtailment actions, the state and transition function definitions can be defined similarly to the original MDP in Section \ref{ch1:uc_mdp}. We will now describe each MDP modification in turn. 

\subsubsection{Reward Function}

The reward function reflects the updated cost function which includes the carbon-adjusted fuel costs $C^f_c$:

\begin{gather}
    r = -C \\
    C = C^f_{c} + C^s + C^l
\end{gather}

where $C^s$ and $C^l$ are startup costs and lost load costs respectively. 

\subsubsection{Action Definition}

The action definition is updated to include the curtailment action, and is a binary vector of length $N+1$ for $N$ generators:

\begin{equation}
    \boldsymbol{a} = [a_{1}, a_{2} ... , a_{N}, a_{c}] \quad a_{i} \in \{0,1\}
\end{equation}

where $a_c$ indicates the curtailment action. 

\subsubsection{State and Transition Function Definitions}

We will briefly describe the state and transition function definitions, although these are very similar to the original MDP. To reflect the impact of curtailment on demand, the wind generation forecast $w_{t}$ and forecast error $y_{t}$ are reduced by 50\% when the curtailment action is applied at the previous timestep.

\medskip 

The MDP can be represented as a search tree using the method described in Section \ref{ch1:methodology:search_tree}. The size of the action space $|\mathcal{A}|$ increases by a factor of two to $2^{N+1}$ with the additional curtailment dimension. In the following section, we will describe our experiments applying Guided IDA* to solve UC problems with carbon price and curtailment. 

\subsection{Experimental Setup and Policy Training}

The adjustments made to the environment described previously enable us to study the impact of carbon price on the solutions produced by Guided A* Search. Carbon pricing impacts the relative operating costs of different fuel types, leading to different optimal operating strategies. In addition, carbon price impacts the relative value of system security and renewables integeration, thus affecting the optimal curtailment rate.

Our results focus on the schedule characteristics at different levels of carbon price, rather than on operating costs in comparison with existing methods. Developing an MILP formulation of the UC problem including the curtailment action is beyond the scope of this thesis, and no MILP benchmark solutions were calculated. We investigate three carbon price levels $CP=\{0,25,50\}$ \$/tCO$_2$ and analyse solutions in terms of curtailment rate, CO$_2$ emissions, utilisation of fuel types and other system variables. Whereas previous chapters focused on scaling characteristics with the number of generators, in this chapter we exclusively examine the 30 generator problem. With the curtailment action, there are $2^{31} \approx$ 2 billion actions.


The changes to the environment and MDP mean that new expansion policies are required in order to apply Guided IDA*. We trained an expansion policy for each carbon price level using model-free RL using PPO, as described in Section \ref{ch1:guided_ucs:policy_training}. All parameters were the same as described in Table \ref{ch1:tab:policy_params} for the 30 generator problem except the number of epochs, which was increased to 500,000 in light of the increased problem complexity and larger action space. The wall clock training time over 500,000 epochs was approximately 26 hours, trained over 8 CPU workers as in previous experiments. We used the sequential policy parametrisation described in Section \ref{ch1:guided_ucs:policy_training} represented by a feed forward neural network. When predicting the action sequence as visualised in Figure \ref{ch1:fig:policy_diagram}, the curtailment action $a_c$ was predicted as the first value in the action sequence followed by the generator commitments $a_i, i \in \{1 \dots N\}$. 

\begin{figure}
    \centering
    \includegraphics[width=0.7\textwidth]{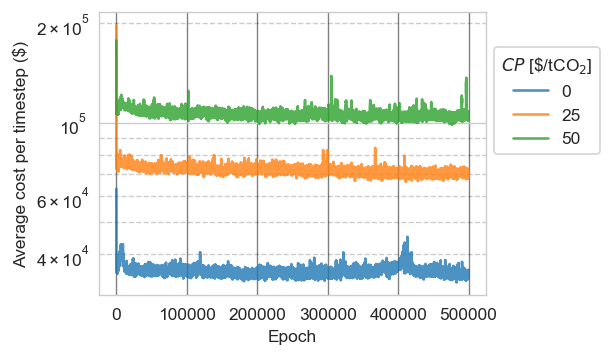}
    \caption{Convergence of expansion policies for the three carbon price levels. Each epoch represents 2000 policy evaluations. The policies trained with $CP>0$ converge to higher average operating costs due to the additional carbon costs.}
    \label{ch3:fig:curtailment_training}
\end{figure}

The convergence profiles of expansion policies for the three levels of carbon price are shown in Figure \ref{ch3:fig:curtailment_training}. The convergence to higher mean operating costs per timestep when $CP>0$ is explained by the additional carbon costs incurred. As with the policies trained in Section \ref{ch1:guided_ucs:policy_training}, large improvements are made in the early stages of training within the first 1000 epochs, as the policy learns to avoid lost load events. Thereafter, comparatively small improvements are made as the policy is fine-tuned to minimise fuel and startup costs. 

Using the trained expansion policies, we solved the 20 test episodes using Guided IDA* search, which was described in detail in Section \ref{ch2:methodology:ida_star}. The test episodes are identical to those described in Section \ref{ch1:env:benchmarks}, with the same demand and wind profiles. We set the time budget $b=60$s, the branching threshold $\rho=0.05$, number of scenarios $N_s=100$ and used the Constrained ED heuristic as in previous experiments (Section \ref{ch2:experiments:ida_star}). 


\subsection{Results} \label{ch3:curtailment:results}

Table \ref{ch3:tab:curtailment_results} summarises the results of our experiments using Guided IDA* to solve the 20 unseen test problems for each carbon price level. Increasing the carbon price from \$0 to \$25/tCO$_2$, there is a 22\% reduction in carbon emissions and a 115\% increase in total operating costs due to the additional carbon emissions costs. Higher total operating costs are inevitable under the carbon price rise in our problem setup, reflecting the cost burden of carbon pricing on consumers \cite{grainger2010pays}. Further increasing the carbon price to \$50/tCO$_2$ yields comparatively minor reductions in CO$_2$ emissions, whereas total operating costs increase by a further 49\%. Loss of load probability (LOLP) did not change substantially between \$0--25/tCO$_2$, but increased from 0.09\% to 0.22\% when increasing from \$25--50/tCO$_2$. The curtailment rate (defined as curtailed volume as a proportion of available volume) is lower at \$25 and \$50/tCO$_2$ due to the greater incentive to integrate wind generation and lower the average carbon emissions factor. The largest drop was observed between \$0--\$25/tCO$_2$, while there was a small relative increase between \$25--50/tCO$_2$. In the following sections we will analyse the changes to operational patterns under different levels of carbon price.

\begin{table}[]
    \centering
    \resizebox{\textwidth}{!}{
    \input{06-chapter3/tables/curtailment_results_formatted}}
    \caption{Comparison of Guided IDA* solutions to 30 generator test problems with curtailment action and varying carbon price.}
    \label{ch3:tab:curtailment_results}
\end{table}


\subsubsection{Changes to Generator Usage Patterns}

\begin{figure}
    \centering
    \includegraphics[width=0.7\textwidth]{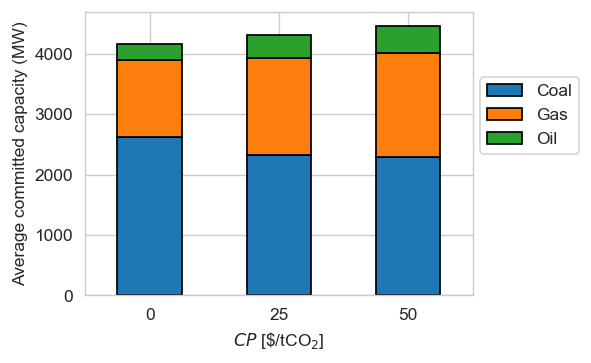}
    \caption{Committed capacity by fuel type for the three levels of carbon price. Gas and oil displace coal in terms of committed capacity as the carbon price increases. Total committed capacity also increases with carbon price.}
    \label{ch3:fig:fuel_committed_capacity}
\end{figure}

\begin{figure}
    \centering
    \includegraphics[width=\textwidth]{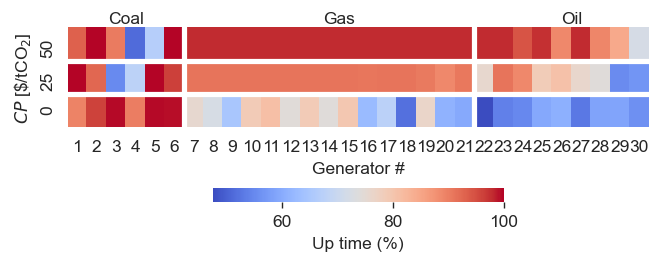}
    \caption{Up time of generators in the curtailment case study. Gas replaces coal as base-load generation, while oil shifts from peaking usage to predominantly load-following as the carbon price increases.}
    \label{ch3:fig:curtailment_heatmap}
\end{figure}

The utilisation rates and operational patterns of different fuel types were found to vary with carbon price $CP$. Average committed capacity disaggregated by fuel type is shown in Figure \ref{ch3:fig:fuel_committed_capacity}. Total committed capacity increased consistently with increasing $CP$, indicating larger reserve margins. This result is unexpected, as lost load costs are devalued relative to fuel-related costs when the carbon price is increased. The preference for different fuel types varied as expected with increasing $CP$: coal is displaced by gas and oil which have lower emissions factors. Figure \ref{ch3:fig:curtailment_heatmap} shows the generator-level utilisation, demonstrating the shift from coal to gas supplying base-load generation. In both the \$25 and \$50/tCO$_2$ settings, 2 of the 6 coal-fired power stations operate with significantly lower utilisation rates, whereas at \$0/tCO$_2$ all coal plants are online for $>$80\% of periods. At the highest carbon price level, all gas units have near 100\% utilisation rates. Similarly, oil is operated much more consistently when the carbon price is increased, representing a shift away from peaking use to load-following and even base-load in some cases. 

\begin{figure}
    \centering
    \includegraphics[width=0.6\textwidth]{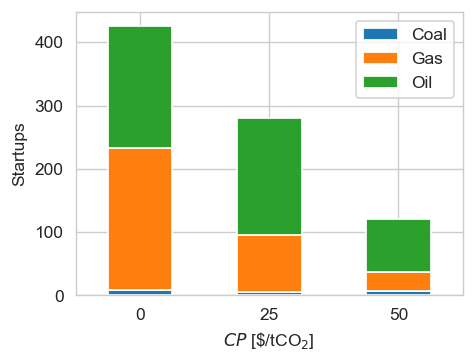}
    \caption{Total startups across the 20 test problems, disaggregated by fuel type. Coal startups remain roughly stable with $<$10 startups in all cases. Gas and oil startups decrease substantially as the carbon price is increases, as these fuel replaces coal as base-load generation.}
    \label{ch3:fig:curtailment_startups}
\end{figure}

Figure \ref{ch3:fig:curtailment_startups} shows the decline in total startups with carbon price, split by fuel type. Coal startups remain low for all three carbon price levels, with less than 10 startups. Gas startups decrease from 225 to 90 to 30 as the carbon price increases, adopting base-load operation patterns. Oil startups are not significantly reduced at a carbon price of \$25/tCO$_2$, but decrease substantially at \$50/tCO$_2$. Whereas Figure \ref{ch3:fig:curtailment_heatmap} shows that coal units are displaced in the merit order as the carbon price increases, the startups data show that coal still does not act flexibly at higher carbon prices. Fewer coal plants are committed on average, but these commitments tend to remain fixed for each episode. This is in part due to the large startup costs for these generators, as well long minimum up/down times of 8 hours which prevents them being committed for short periods such as the evening peak. 

\subsubsection{Impact of Carbon Price on Curtailment Rate}

Table \ref{ch3:tab:curtailment_results} shows that the curtailment rate varied with the carbon price, and was lowest at $CP=$ \$25/tCO$_2$. Combined with the relatively small decrease in CO$_2$ emissions between \$25--50/tCO$_2$, this indicates that there are diminishing returns from increasing $CP$ beyond \$25/tCO$_2$ in this problem setup. 

Furthermore, we observed differences in the circumstances under which the curtailment action was used at varying levels of $CP$. Figure \ref{ch3:fig:curtailment_boxplots} compares the distribution of forecast wind generation when the curtailment action is used versus when it is not for each of the carbon price levels. In the lowest carbon price case, there is a wider range of wind generation levels over which wind is curtailed, including periods of very high forecast wind generation. At carbon prices of \$25 and \$50/tCO$_2$, the curtailment action is used much more sparingly at higher wind penetration due to the greater value of zero-carbon generation relative to other sources. The median wind generation level at which the curtailment action is used thus decreases from 135 MW to 58 MW to 8 MW as $CP$ increases. Curtailment of wind at low penetrations is the least costly and is used to reduce net demand variability and probability of lost load due to suppressed net demand. However, the overall curtailment rate does not decrease after \$25/tCO$_2$. 

\subsubsection{Extreme Curtailment Events: Example of 2017-03-18}

\begin{figure}
    \centering
    \includegraphics[width=\textwidth]{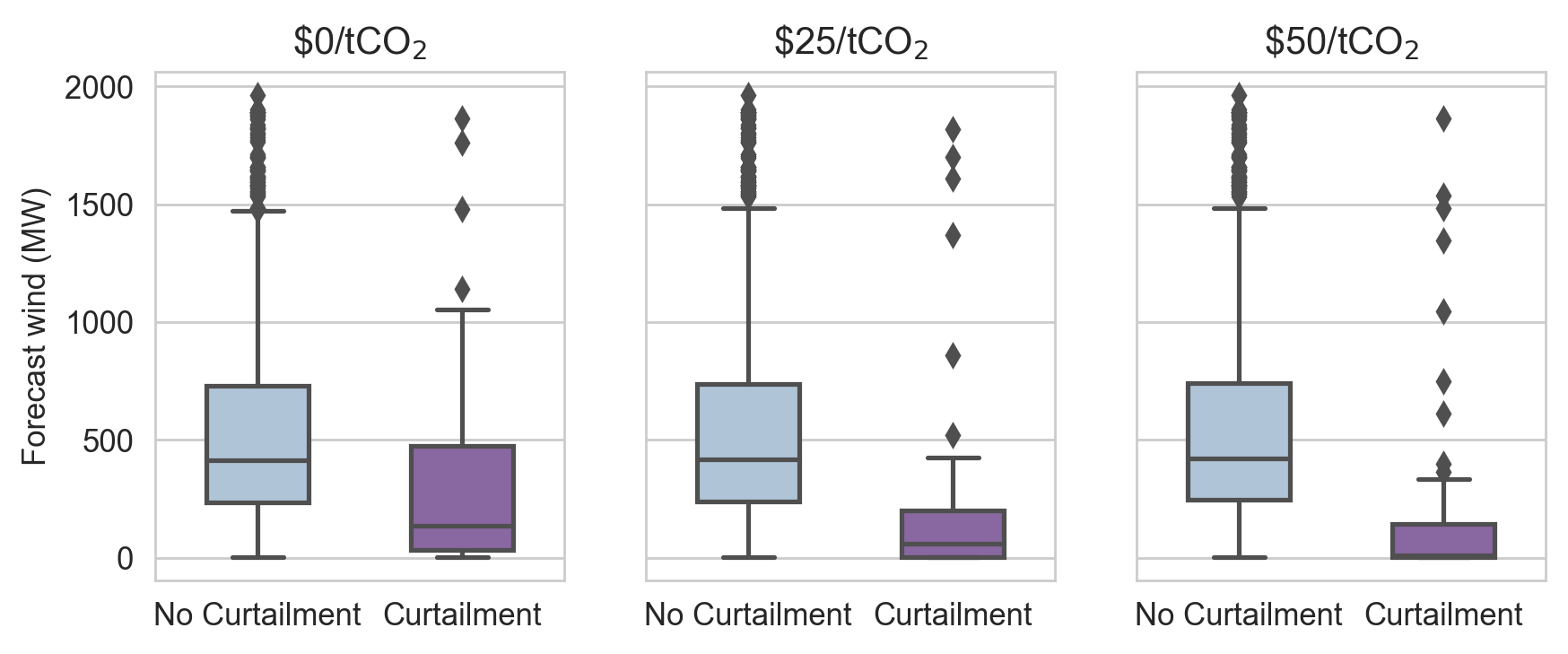}
    \caption{Distribution of forecast wind generation for periods when curtailment action is used versus when it is not. As the carbon price is increased, the curtailment action is used less frequently in periods of high wind generation.}
    \label{ch3:fig:curtailment_boxplots}
\end{figure}

\begin{figure}
    \centering
    \includegraphics[width=0.7\textwidth]{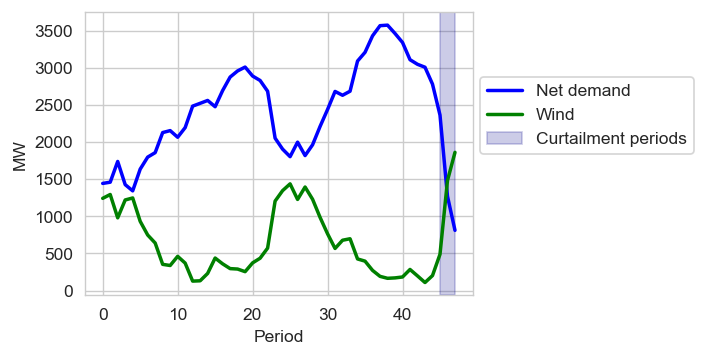}
    \caption{Curtailment action used by Guided IDA* in test problem 2017-03-18, with $CP=$\$50/tCO$_2$. Curtailment is used over 2 periods to mitigate the rapid decrease in net demand resulting from a spike in wind generation at the end of the day. Without the curtailment action during these periods, total operating costs for this schedule would be \$5108605, compared to \$4424294 when curtailment is used.}
    \label{ch3:fig:extreme_curtailment}
\end{figure}

Figure \ref{ch3:fig:curtailment_boxplots} indicates that at higher $CP$, curtailment volumes are dominated by a few extreme events, where forecast wind generation is $>1$ GW. Two of the three largest curtailment events with $CP=$\$50/tCO$_2$ occur at the end of the test problem 2017-03-18, visualised in Figure \ref{ch3:fig:extreme_curtailment}. Curtailment is used over two periods with forecast wind generation of 1.5 GW and 1.9 GW in order to dampen the large drop in net demand that occurs at the end of the day. A counter-factual analysis was conducted for this day, comparing the operating costs with and without the curtailment action. Without the curtailment action during these periods, mean operating costs for this schedule would be \$5,108,605, compared to \$4,424,294 when curtailment is used due to the lower LOLP. This shows economic benefits to using curtailment thanks to improved system security. 

\medskip

In summary, Guided IDA* exhibits significant changes in operating patterns as the carbon price $CP$ is increased. Fuel types play different roles in satisfying demand, with gas gradually transitioning to full base-load operation when $CP$=\$50/tCO$_2$. The curtailment action is used less frequently at higher $CP$, but is still used sparingly to manage uncertainties and improve system security.




\section{Case II: Generator Outages} \label{ch3:robustness}

In the second case study we introduced generator outages to the simulation environment described in Section \ref{ch1:env} in order to study the ability of Guided IDA* to learn robust strategies to handle generation losses. Protection against generator outages is typically achieved by including a criterion that is based on the \textit{single largest loss of infeed} \cite{NG_sqss}, commonly known as the $N-1$ criterion. The $N-1$ criterion generally offers acceptable levels of system security but may be insufficient to prevent lost load in cases of multiple coincident outages, which was the cause of several blackouts across North America and European power systems in 2003 \cite{andersson2005causes} and the 9 August 2019 blackout in the GB power system \cite{bialek2020does}. Similar $N-x$ approaches can be applied to protect against $x$ simultaneous outages, but may be overly conservative or encounter constraints around minimum operating levels of generators which requires the curtailment of variable renewable energy. Probabilistic approaches have been proposed to allocate reserve constraints in deterministic UC \cite{doherty2005new, bouffard2008stochastic}, but heuristic approaches such as $N-1$ are still widely-used \cite{NERC_reliability, bialek2020does}.  


This case study investigates the effectiveness of Guided IDA* for developing UC solutions that are robust to generator outages, a critical factor in power system stability. We begin by describing the simulation environment including modelling of generator outages and formulate the problem as an MDP. We then apply Guided IDA* and MILP to solve unseen test problems with outages. 

\subsection{Environment Setup}

In order to investigate the impact of generator outages, we modified the simulation environment described in Section \ref{ch1:env}. As with previous simulation environments, the environment with outages can be activated in the Python package developed for this research\footnote{\url{https://github.com/pwdemars/rl4uc}}. Each generator has a probability $\psi_i(u_{i,t-1})$ of failing when it transitioning from time $t-1$ to $t$, which is a function of the generator's up time $u_{i,t-1}$ at the previous timestep. Every generator begins the the episode available, and once it has failed it cannot be dispatched for the remainder of the episode. Furthermore, generators cannot fail during the first period of commitment (that is, when $u_{i,t-1} < 0$).

\begin{figure}
    \centering
    \includegraphics[width=0.7\textwidth]{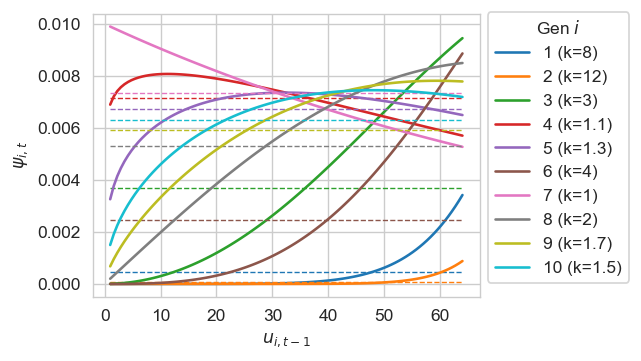}
    \caption{Outage rates $\psi_i$ as function of up time $u_{i, t-1}$. Outage rates are modelled with a Weibull with fixed scale parameter $\lambda=100$ and variable shape parameter $k$. The dotted lines show the mean outage rate over all periods. The two base-load generators 1 and 2 have the lowest forced outage rates. Most generators have generally increasing forced outage rates, while some have higher failure rates at the beginning of operation.}
    \label{ch3:fig:outage_rates}
\end{figure}

We use the Weibull distribution to model the failure rates of generators. The Weibull distribution is widely used for reliability analysis to represent failure rates \cite{mudholkar1993exponentiated, xie1996reliability, love1996utilizing} due to its generality, accommodating increasing, decreasing, and non-monotone failure functions. It is therefore well-suited to modelling failures of a heterogeneous generation mix. The two-parameter Weibull distribution is defined by a shape parameter $k$ and scale parameter $\lambda$. Failure rates are scaled to a realistic order of magnitude by setting $\lambda=100$ for all generators, leaving a single parameter $k$ which is unique for each generator. The $k$ parameters were assigned for generators to ensure a suitable level of heterogeneity and to achieve a weighted-equivalent forced outage rate (WEFOR) that approximately corresponds to real-world power systems. WEFOR measures the long-run average proportion of total generation capacity that is unavailable due to forced outages. The 5-year average WEFOR for North America, reported by the North American Electric Reliability Corporation, was 7.16\% between 2015--2019 \cite{NERC_WEFOR}. The outage rates as a $\psi_{i,t}$ function of generator up time $u_{i,t-1} > 0$ for the 10 generators described in Section \ref{ch1:env:overview} are shown in Figure \ref{ch3:fig:outage_rates}. 

\begin{figure}
    \centering
    \includegraphics[width=0.6\textwidth]{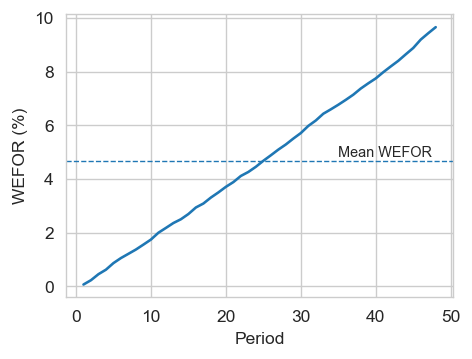}
    \caption{Weighted equivalent forced outage rate (WEFOR) as a function of decision period. Later periods have higher WEFOR, }
    \label{ch3:fig:wefor_by_period}
\end{figure}

We calculated the WEFOR by simulating outages over 1 year (365 episodes), where all generators are committed in every period. The WEFOR as a function of settlement period is shown in Figure \ref{ch3:fig:wefor_by_period}. Since generators cannot be repaired and must remain offline for the remainder of the episode after an outage, the WEFOR increases linearly throughout the day. In the final period of the day the WEFOR is 9.67\%. The average WEFOR over all periods was 4.68\% indicated by the dotted line in Figure \ref{ch3:fig:wefor_by_period}. While this is lower than 7.16\% reported by NERC \cite{NERC_WEFOR}, the comparatively large WEFOR in later periods indicates the potential for extreme outage scenarios under our modelling approach, posing a significant challenge for managing system security in the context of UC.

In the following section, we formalise the environment changes by modifying the UC MDP from Section \ref{ch1:uc_mdp}.

\subsection{MDP Formulation}

To formulate the UC MDP for the power system environment with outages, we modified the state and transition function definitions in the UC MDP from Section \ref{ch1:uc_mdp}. Actions and rewards remain unchanged.

\subsubsection{State Definition}

The state definition $s_t$ is modified to include an outages vector $\boldsymbol{z}_t \in \{0,1\}^{N}$, indicating which generators experience an outage at time $t$. We then include an additional availability vector $\boldsymbol{v}_t \in \{0,1\}^{N}$, indicating which generators have experienced an outage so far in the episode up to and including timestep $t$. Generators which are unavailable $v_{i,t} = 0$ cannot be dispatched. Like the forecast errors, neither $\boldsymbol{z}_t$ nor $\boldsymbol{v}_t$ are observed by the agent in the day-ahead setting. 

\subsubsection{Transition Definition}

The transition function $F(s_{t+1}, s_t, a_t)$ is updated in two steps. For each online generator, we sample an outage $z_{i, t}$ for each generator $i$, independently of previous periods or the current generator status:

\begin{equation}
    z_{i,t} \sim \text{Bern} (\psi_i (u_{i,t-1}))
    \label{ch3:eq:sample_outage}
\end{equation}

Second, the generator availability $v_{i,t}$ is updated: 

\begin{equation}
v_{t,i} =
\begin{cases}
    0 ,& \text{if } v_{t-1,i}=0 \\
    0 ,& \text{if } v_{t-1,i}=1 \text{ and } z_{i,t} = 1 \\
    1 ,& \text{if } v_{t-1,i}=1 \text{ and } z_{i,t} = 0 \\
\end{cases}
\label{ch3:eq:transition_availability}
\end{equation}

When $v_{t,i}=0$, the generator cannot be dispatched. In addition, it remains unavailable for the rest of the episode. The remaining components of the UC MDP are identical to those described in Section \ref{ch1:uc_mdp}. In the following subsection, we will describe how the UC MDP with outages can be represented as a search tree in order to apply tree search methods.

\subsection{Search Tree Formulation} \label{ch3:robustness:search_tree}

In Section \ref{ch1:methodology:search_tree} we described how the original UC MDP can be formulated as a search tree, with nodes representing states and edges representing actions. The step costs in the search tree are evaluated using a Monte Carlo approach described in Equation \ref{ch1:eq:search_tree_cost}, calculating mean operating costs over $N_s$ scenarios of net demand (demand minus wind generation). While the UC MDP with curtailment and carbon price could be represented as a search tree using precisely this method, the search tree representation for the outages case study must be modified to include outage scenarios. 




The general approach to calculating the expected step cost $C(s)$ of transitioning to state $s$ remains the same. We generate $N_s$ scenarios for the uncertain processes (net demand and outages), calculate the dispatch costs under each scenario and take the mean. However, we now must account for additional uncertain parameters corresponding to outages $\boldsymbol{z}$ and availability $\boldsymbol{v}$. 

 
Step costs are calculated using the following equation:

\begin{gather}
    C(s) = \frac{-1}{N_s} \sum_{k=1}^{N_s} R(s, x_k, \boldsymbol{z}_k (\boldsymbol{u}_{t-1}), \boldsymbol{v}_{t-1, k}) \\
    x_k = \mathcal{S}_{t,k} \quad \boldsymbol{z}_k(\boldsymbol{u}) = \mathcal{Z}_{*,\boldsymbol{u},k}
    \label{ch3:eq:outages_step_costs}
\end{gather}

The reward function $R(\cdot)$ represents the reward function evaluated for state $s$, net demand $x_k$ scenario, outage scenario $\boldsymbol{z}_k (\boldsymbol{u}_{t-1})$ and previous generator availabilities $\boldsymbol{v}_{t-1, k}$. $\mathcal{S} \in \mathbb{R}^{T \times N_s}$ represents the net demand scenarios, as described in Section \ref{ch1:methodology:search_tree}. $\mathcal{Z} \in \{0,1\}^{N \times u^\text{max} \times N_s}$ represents the outage scenarios. 

Generator outage scenarios are generated offline at the beginning of each episode and stored in a lookup table in the same way as net demand scenarios. This ensures that outages are fairly distributed across nodes. This is particularly important for the outages case study, as the extreme nature of outages and the large number of random parameters ($N+2$ including the wind and demand forecast errors) means that the distribution of outages over scenarios is likely to be highly variable. Whereas demand and wind forecast error distributions were a function of the timestep $t$, the outage probability distributions are functions of generator up time $u_{i, t-1}$. Hence, for each generator we create a lookup table with $N_s$ rows and $\max(T + u_{0,i}, T)$ columns, where $u_0,i$ is the initial up/down time of generator $i$ (this is the maximum up time of generator $i$ in a given episode). Each column is populated with $\{0,1\}$ sampled from the Bernoulli outage distribution in Equation \ref{ch3:eq:sample_outage}. The set of outage scenarios $\mathcal{Z}$ is the stack of all lookup tables for outage scenarios. 

Equation \ref{ch3:eq:outages_step_costs} calculates the empirical mean of operating costs over $N_s$ scenarios for a node at time $t$. For scenario $k \in \{1..N_s\}$, we use net demand scenario $k$ at timestep $t$. Then, for each generator, we use outage scenario $k$ at generator up/down time $u_{t-1, i}$, the up/down time of generator $i$ at the previous timestep. This scenario is retrieved from row $k$, column $u_{t-1, i}$ of the lookup table for generator $i$. The operating costs for this joint scenario of net demand and outages are then calculated. The expected costs are calculated by repeating this process for each of the $N_s$ scenarios and calculating the mean operating costs.

Having formalised the environment and MDP modifications and the search tree formulation, in the following section we will describe our experiments training and applying Guided IDA* to solve the UC problem instances.

\subsection{Experimental Setup and Policy Training} \label{ch3:robustness:experiments}

We will now describe the setup of experiments applying Guided IDA* and MILP with $N-x$ reserve criteria to solve unseen test problems. As with the curtailment experiments, in the generator outages problem we exclusively used the 30 generator system. We used the same split of training and test episodes as in previous experiments, holding out the 20 test episodes described in Table \ref{ch1:tab:test_problems} from the training data. 

\begin{figure}
    \centering
    \includegraphics[width=0.7\textwidth]{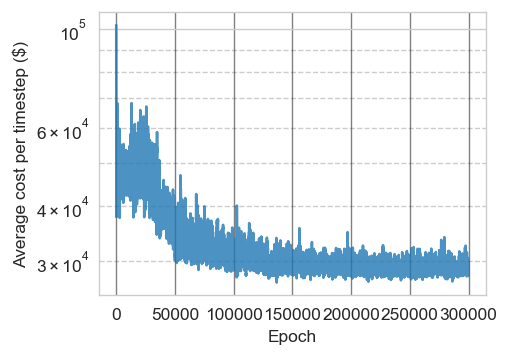}
    \caption{Convergence of expansion policy for the 30 generator environment with generator outages. The figure shows a moving average of operating costs per timestep over 100 epochs. The policy was trained by PPO using the method described in Section \ref{ch1:guided_ucs:policy_training}.}
    \label{ch3:fig:robustness_training}
\end{figure}

In order to apply Guided IDA*, we trained a new expansion policy by model-free RL with PPO, using the method described in Section \ref{ch1:guided_ucs:policy_training}. We used the same parameters used to train expansion policies in Chapter \ref{ch1}, described in Table \ref{ch1:tab:policy_params}. Convergence of the expansion policy for the outages environment is shown in Figure \ref{ch3:fig:robustness_training}. Compared with the expansion policies for the original MDP (Figure \ref{ch1:fig:training_plots}) and with curtailment and carbon prices (Figure \ref{ch3:fig:curtailment_training}), convergence is significantly slower, caused by the noisier reward function with respect to observations. Convergence speed may also be impacted by differences in the reward distribution, caused by greater frequency of lost load events due to generator outages. 

Guided IDA* was then used to solve the 20 unseen UC problem instances. To evaluate the schedules, we simulated the dispatch of 1000 scenarios as in previous experiments. Scenarios include random realisations of demand, wind generation and outages. We set the branching threshold $\rho=0.05$, the time budget $b=60$s and used the Constrained ED heuristic described in Section \ref{ch2:heuristics:priority_list}. In Section \ref{ch3:robustness:search_tree} we discussed the method of calculating expected step costs in the search tree considering scenarios of outages and net demand. In previous experiments which considered only demand and wind generation uncertainty, the number of scenarios $N_s$ was fixed at $N_s=100$. This sufficiently approximated the distribution of net demand. In the outages experiments, motivated by the greater number of random parameters, we ran Guided IDA* with $N_s = \{100,200,500,1000,2000,5000\}$. Higher $N_s$ allows for the expected step costs in the search tree to be more accurately estimated but increases the overall run time of each node evaluation, thus decreasing the search depth for fixed time budget $b$.

Guided IDA* was compared with MILP($N-x$) benchmarks which use $N-x$ reserve criteria. $N-1$ is a widely used reserve criterion that assures that enough capacity is available to protect against an outage of the largest generator. We implement $N-x$ criteria up to $x=4$. For the 30 generator problem, this corresponds to reserve margins of $\{455, 910, 1365, 1820\}$ MW for $x=\{1,2,3,4\}$, respectively.


\subsection{Results}

The operating costs and loss of load probability of Guided IDA* and MILP($N-x$) are compared in Table \ref{ch3:tab:robustness_results}. Operating costs are disaggregated into fuel costs, lost load costs and startup costs. 

\subsubsection{MILP Benchmarks}

Of the MILP benchmarks, MILP($N-2$) is found to achieve the lowest total operating costs. Fuel costs increase consistently with the reserve size as generators are forced to operate at lower efficiencies on average to satisfy the increased reserve requirements. Lost load costs decrease up to $N-3$, but increase thereafter due to increased probability of encountering minimum generation constraints (events of insufficient footroom to manage high wind generation outturn and/or low demand) with large reserve margins. The worst performing MILP benchmarks are those with the smallest and largest reserve margins: MILP($N-1$) has very high LOLP (2.62\%) and thus the highest lost load costs; MILP($N-4$) is the most conservative solution and thus has the highest fuel costs. 

\begin{table}[]
    \centering
    \resizebox{\textwidth}{!}{
    \input{06-chapter3/tables/robustness_results}}
    \caption{Comparison of Guided IDA* and MILP($N-x$) solutions in the generator outages case study.}
    \label{ch3:tab:robustness_results}
\end{table}

\subsubsection{Impact of $N_s$ on Operating Costs and Search Depth}

\begin{figure}
    \centering
    \includegraphics[width=0.7\textwidth]{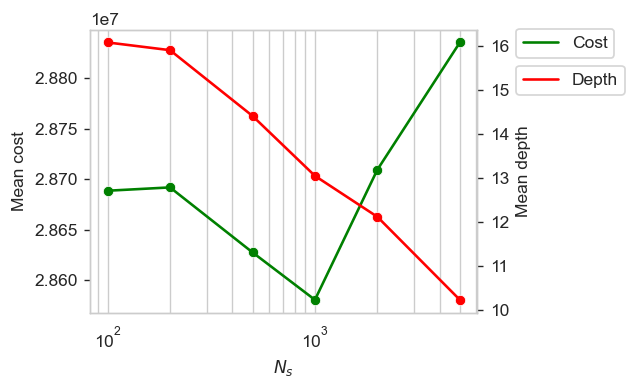}
    \caption{Mean total operating costs and search depth of Guided IDA* solutions with varying $N_s$. Search depth decreases with increasing $N_s$ due to the increased node evaluation time. The lowest operating costs are achieved with $N_s=1000$. At higher values of $N_s$, total operating costs increase due to shallower search depth.}
    \label{ch3:fig:robustness_depth_cost}
\end{figure}

All Guided IDA* versions achieve lower operating costs than MILP($N-2$), the best performing MILP benchmark. Figure \ref{ch3:fig:robustness_depth_cost} shows the change in operating cost and search depth with varying $N_s$. Search depth decreases monotonically with increasing $N_s$, due to the greater node evaluation time. The figure shows that for $N_s > 1000$, improvements stemming from more accurate estimates of expected step cost begin to be outweighed by the detrimental impact of shallower search depth. The highest settings of $N_s = \{2000, 5000\}$ had the greatest total operating costs. Guided IDA*($N_s=1000$) achieves the lowest total operating costs, which are 2.1\% lower than MILP($N-2$) and 0.4\% lower than Guided IDA*($N_s=100$). These cost savings are larger than those found in experiments without outages, reported in Section \ref{ch2:experiments:ida_star}, where Guided IDA* achieved operating costs that were 1.4\% lower than MILP($4\sigma$). Our results show comparable savings to those reported in the stochastic UC literature reviewed in Section \ref{literature:suc:duc_comparison} which found operating cost savings of approximately 1\% compared with deterministic methods \cite{tuohy2009unit, ruiz2009uncertainty, schulze2016value}. 

\subsubsection{Reserve Margins}

\begin{figure}
    \centering
    \includegraphics[width=0.7\textwidth]{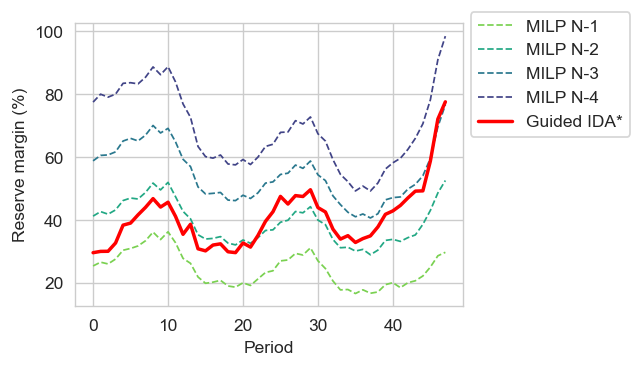}
    \caption{Average reserve margins by period for MILP with $N-x$ reserve criteria, and Guided IDA*($N_s=1000$). The figure shows average reserve as a proportion of net demand. Guided IDA* begins the day with reserve margins similar to MILP($N-1$), and finishes with reserve margins close to MILP($N-3$).}
    \label{ch3:fig:reserve_margins}
\end{figure}

Guided IDA* adaptively allocates reserve capacity to achieve low levels of LOLP without producing overly conservative schedules. The average reserve margins of Guided IDA*($N_s=1000$) for each settlement period are compared with the MILP benchmarks in Figure \ref{ch3:fig:reserve_margins}. The reserve margins of all methods follow a similar daily pattern, with the lowest reserve margins coinciding with morning and evening demand peaks. However, Guided IDA* increases its reserve commitment relative to the other methods throughout the day, beginning with similar margins to $N-1$, and finishing with roughly $N-3$ reserve margins. This reflects the greater probability of multiple coincident outages at the end of the scheduling period. 

Allocating constant reserve margins, the approach taken by MILP($N-x$) benchmarks, results in relatively poor performance for this problem setup. Our results show that Guided IDA* learns an adaptable reserve commitment that reflects properties of the environment. 

\medskip

Overall, Guided IDA* outperforms the MILP($N-x$) benchmarks for all settings of $N_s$, and allocates reserves in a more efficient way to protect against generator outages. Solution quality improved with increasing number of scenarios $N_s$ up to $N_s=1000$, but further increases had a detrimental impact on operating costs due to the shallower search depth.

\section{Discussion} \label{ch3:discussion}

Guided tree search is a highly adaptive framework for solving the UC problem. This chapter showed that complex problem settings incorporating co-optimisation of curtailment, carbon pricing and generator outages, can be reflected in the simulation environment and solved with Guided IDA* without significant changes to the solution method beyond limited parameter tuning. Guided tree search and RL approaches are not limited to linear objective functions or constraints, and can easily incorporate large numbers of stochastic variables in the environment. By contrast, MILP approaches would require complex, manual reformulations to incorporate co-optimisation of curtailment. As discussed in Section \ref{literature:duc:solution_methods}, designing effective MILP formulations for the UC problem is an active research topic, and can have significant impact on run time and/or solution quality. In addition, the traditional MILP deterministic approach of allocating reserve using $N-x$ criteria in the outages case was shown to be inadequate at trading off security of supply with expensive over-commitment of reserve. By contrast, Guided IDA* search is capable of handling stochastic decision making without explicit reserve constraints. 

\subsubsection{Curtailment and Carbon Price}

Including the curtailment action in the first case study demonstrated the potential for Guided IDA* to incorporate more heterogeneous decision-making alongside UC. This approach could be extended to other operational tasks such as demand-side response instructions and charge/discharge of storage assets. However, the results of this experiment indicate that including the curtailment action added considerable complexity that made the problem harder to solve. Comparing the results in Table \ref{ch3:tab:curtailment_results} for the \$0/tCO$_2$ case with those of experiments from the previous chapter (Table \ref{ch2:tab:ida_a_star_comparison}), overall operating costs were higher when the curtailment action was introduced, despite the reward function remaining the same. Improving the policy training component in this case study by implementing a different RL algorithm or policy network architecture could yield further cost reductions for the curtailment case. 

The Guided IDA* solutions employed unexpected operational strategies at higher carbon prices, which were used to develop insights regarding the problem setup. The increase in reserve margins with carbon price was surprising, and could be partially explained by the lack of flexible peaking capability when gas is employed as base-load, which could be the cause of greater reserve margins. We found that coal startups remained low, indicating that it was not used flexibly for peaking, such as during the evening peak. Furthermore, the curtailment action was not phased out completely even at the highest carbon price. Most of the curtailment volume at higher carbon prices was concentrated in a few extreme actions, such as the instance shown in Figure \ref{ch3:fig:extreme_curtailment}, where curtailment is used to manage a sharp drop in demand and increase in wind generation. Such extreme net demand swings are difficult to manage securely and the curtailment action is a useful and economic option in this instance. 


\subsubsection{Outages}

In the outages case study, MILP($N-x$) were shown to be outperformed by Guided IDA* which allocated reserve more dynamically. We experimented with varying the number of scenarios $N_s$ used to calculate expected step costs as a response to the increased number of random parameters in the outages problem. Our results showed that increasing $N_s$ to 1000 yielded operating cost reductions for the outages case study relative to lower values. Further increasing $N_s$ resulted in worsening of solution quality due to the shallower search depth. Linear time complexity in $N_s$ is a valuable property of Guided IDA*, allowing $N_s$ to be increased without large reductions in search depth for the same computational budget; average search depth was 10.2 at $N_s=5000$ compared to 16.1 with $N_s=100$. By contrast, large numbers of uncertain parameters pose problems for both scenario-based stochastic UC (reviewed in Section \ref{literature:suc}) and robust UC (reviewed in Section \ref{literature:ruc}) approaches. For scenario-based methods, solutions to stochastic programs with large numbers of scenarios are typically very computationally expensive and must usually be reduced substantially by scenario reduction methods \cite{wu2007stochastic}. For robust optimisation methods, outages (which are integer-valued) cannot be represented in a polyhedral uncertainty set. As a result, existing robust UC research has either used $N-x$ approaches, where a suitable value for $x$ must be assigned manually \cite{street2010contingency} or omitted outage-related security constraints completely \cite{jiang2011robust}. By contrast, the sampling approach adopted in Guided IDA* is highly scalable in the number of scenarios $N_s$, allowing rare contingencies to be considered probabilistically, without heuristic reserve constraints.

Figure \ref{ch3:fig:robustness_training} showed slower convergence of the expansion policy in the outages case study as compared with previous problems, reflecting the greater reward uncertainty. Further increasing the number of uncertain parameters (such as by considering disaggregated loads or wind farms) would be likely to further increase the training time required for Guided IDA*. As with the curtailment problem, there is scope to improve the policy training component for more complex problem domains. A useful property of Guided IDA* is that policy training algorithms can be substituted in a modular fashion, allowing for new state-of-the-art RL methods to be exploited in future.

\bigskip

In both case studies, Guided IDA* uncovered properties of the problem through its use of unexpected or novel operational strategies. There is scope to use such insights in hybrid methodologies or as a decision support tool. The reserve margins learned by Guided IDA* in the outages case study could easily be used to inform reserve constraints for deterministic UC approaches. Furthermore, Guided IDA* could be used to identify periods of low system security. Our results from the curtailment experiment found that Guided IDA* solutions exposed periods where curtailment significantly improved system security.


\section{Conclusion} \label{ch3:conclusion}

In this chapter we applied Guided IDA* search in two case studies, demonstrating the flexibility of this approach in heterogeneous power system contexts. In the first case study, we introduced a curtailment action and carbon price, finding that Guided IDA* responded dynamically to increasing carbon price in terms of curtailment rates and utilisation of different fuel types. We found that carbon emissions decreased by 22\% when increasing the carbon price from \$0--25/tCO$_2$, driven by lower curtailment rates and displacement of coal with gas and oil-fired power stations. In the second case study, we introduced generator outages and showed that Guided IDA* achieved up to 2.1\% lower total operating costs than MILP benchmarks with $N-x$ reserve criteria. Guided IDA* intelligently allocates reserves, reaching levels of loss of load probability that are similar to that achieved in the case with no generator outages. 

Our experiments show that guided tree search is a highly adaptive framework for solving UC problems in heterogeneous power systems. Other power systems decisions such as curtailment can be included in the simulation environment and co-optimised alongside UC and the reward function can be shaped to reflect societal relative values of security of supply and environmental considerations using a carbon price. Furthermore, the UC solutions produced by Guided IDA* exhibited unexpected properties that improved our understanding of the problem itself.

%% file: 06-chapter3/tables/curtailment_results_formatted.tex
\begin{tabular}{rrrrrrccc}
\toprule
 & & & & & & \multicolumn{3}{c}{Avg. committed (MW)} \\
\cmidrule[0.4pt](r{0.125em}){7-9}%
 \$/tCO$_2$ &  Cost (\$M) &  LOLP (\%) &  ktCO$_2$ &  Curtailment (\%) &  Startups &  Coal &  Gas &  Oil \\
\midrule
                 0 &       30.89 &       0.10 &   1638.27 &              2.01 &       426 &  2617 & 1273 &  276 \\
                25 &       66.39 &       0.09 &   1282.70 &              1.21 &       280 &  2329 & 1604 &  372 \\
                50 &       98.80 &       0.22 &   1253.66 &              1.30 &       121 &  2284 & 1726 &  451 \\
\bottomrule
\end{tabular}

%% file: 06-chapter3/tables/robustness_results.tex
\begin{tabular}{lrrrrrr}
\toprule
                 Method &  Total cost (\$M) &  Std. cost (\$M) &  Fuel (\$M) &  Lost load (\$M) &  Startups (\$M) &  LOLP (\%) \\
\midrule
 Guided IDA*($N_s=100$) &             28.69 &             0.86 &       28.22 &             0.32 &            0.15 &       0.08 \\
 Guided IDA*($N_s=200$) &             28.69 &             1.71 &       27.99 &             0.56 &            0.14 &       0.13 \\
 Guided IDA*($N_s=500$) &             28.63 &             0.95 &       28.13 &             0.35 &            0.15 &       0.09 \\
Guided IDA*($N_s=1000$) &             28.58 &             0.91 &       28.11 &             0.33 &            0.15 &       0.09 \\
Guided IDA*($N_s=2000$) &             28.71 &             0.92 &       28.17 &             0.40 &            0.15 &       0.09 \\
Guided IDA*($N_s=5000$) &             28.84 &             0.86 &       28.23 &             0.45 &            0.15 &       0.10 \\
            MILP($N-1$) &             43.83 &            15.67 &       26.70 &            17.00 &            0.13 &       2.62 \\
            MILP($N-2$) &             29.20 &             2.94 &       27.65 &             1.39 &            0.16 &       0.26 \\
            MILP($N-3$) &             29.74 &             1.15 &       28.82 &             0.74 &            0.18 &       0.11 \\
            MILP($N-4$) &             31.32 &             1.35 &       30.12 &             1.02 &            0.17 &       0.16 \\
\bottomrule
\end{tabular}

%% file: 07-conclusion/conclusion.tex
\section{Summary}

This thesis has shown that RL is a viable methodology for solving the UC problem, offering solution quality that is competitive with and often outperforms the state-of-the-art in deterministic mathematical programming methods. Novel guided tree search methods were used to solve UC problem instances with greater numbers of generators and more complex stochastic processes than have been addressed in the existing literature. 

In Chapter \ref{ch1} we developed guided tree search, a framework for applying RL to the UC problem using model-based planning and model-free RL methods, which is the methodological base for this thesis. In addition, we developed a novel benchmark simulation environment based on data from the GB power system, enabling the application of RL to solve the UC problem. Using guided tree search, the search space of generator commitments is intelligently reduced to a subset of promising actions, enabling the application of conventional tree search methods in practical computing times. We applied this methodology to uniform-cost search (UCS) for systems of up to 30 generators. Our results showed that Guided UCS achieved lower operating costs than deterministic UC approaches using MILP. Furthermore, we showed that the run time of Guided UCS remained roughly constant in the number of generators, with negligible impact of operating costs as compared with UCS without RL, despite the reduced branching factor of the search tree. In addition, the security of supply was significantly improved, with reserve commitments learned \emph{tabula rasa}. These experiments are the first to compare RL with MILP, the current state-of-the-art, and demonstrate the competitiveness of our approach in solving practical UC problems under uncertainty.
 
In Chapter \ref{ch2} we explored using more advanced search methods to address the variability in run time across UC problem instances and achieve greater search efficiency by employing domain-specific knowledge. A heuristic based on priority list solution methods \cite{senjyu2003fast} was employed in the informed search method Guided A* search, and found to reduce run times by up to 94\% as compared with Guided UCS with negligible impact on operating costs. This large efficiency improvement confirms the potential value of incorporating domain knowledge in RL algorithms to achieve tractability for challenging real-world problems \cite{dulac2019challenges, glavic2017reinforcement}. Guided IDA* employed the anytime strategy iterative deepening to tackle run time variability, enabling solution quality to be maximised within a fixed time budget. This enabled the application of Guided IDA* to a 100-generator problem, the largest in the existing literature. The only other study of comparable size used a simplified problem setup that eliminated intra-day commitment changes to achieve tractability for a problem of 99 generators, and cannot be directly compared with our experiments \cite{dalal2016hierarchical}. Our results show that guided tree search is competitive with MILP approaches for problems of this size, achieving similar (0.1\% lower) operating costs over 20 UC problem instances. In comparison with existing literature, this represents a significant step forward in the application of RL to practical UC problem instances at scale. In experiments of up to 30 generators, we found cost reductions of approximately 1\%, comparable to the improvements of the deterministic MILP methods currently used in practical contexts over Lagrangian relaxation \cite{carrion2006computationally}, which were found to yield large absolute cost savings following their adoption \cite{o2017computational}. Stochastic optimisation methods have also been shown to achieve similar improvements in operating costs of around 1\% \cite{ruiz2009uncertainty, bertsimas2012adaptive, tuohy2009unit} but are impractically expensive to run \cite{papavasiliou2014applying}. By off-loading most of the computational expense to training, RL offers the potential to substantially improve UC solution quality within practical computing times. 

In Chapter \ref{ch3} we adapted the simulation environment to incorporate advanced challenges in UC, introducing a carbon price, wind curtailment and generator outages. Our results showed that guided tree search methods can be applied to UC problem variants without reformulating the solution method. This allows for the application of RL to UC problems of arbitrary complexity, including stochastic environments with large numbers of uncertain parameters. In the generator outages case study, we showed that RL provides lower operating operating costs and higher levels of system security than MILP methods using $N-x$ reserve criteria. While there are concerns of trust surrounding safe RL \cite{garcia2015comprehensive} and the application of RL in critical infrastructure such as power systems, employing model-based methods guided tree search enables forward planning through consultation of a model, improving robustness and explainability \cite{dulac2019challenges}. In addition, our results showed that modifications to the reward function through a carbon price can incentivise operating patterns with lower carbon emissions. Using RL, decision-makers are able to adjust scheduling behaviour in line with current system priorities. In general, a qualitative analysis of RL solutions can be used to analyse properties of UC problems under consideration. The use of curtailment was indicative of periods of high uncertainty and possibility of load shedding; and the reserve margins employed by RL in the outages case study indicated robust commitments where heuristic methods were inadequate. The ability of RL to learn fundamentals of the problem domain can add value in decision support contexts where AI methods are used to inform human operators.

Machine learning methods have long been recognised as having significant potential to profoundly change and improve the operation of power systems \cite{wehenkel1998automatic, glavic2017reinforcement, kelly2020reinforcement}. The results of this thesis have shown that RL is capable of learning optimal control strategies in challenging power system contexts. As electricity networks become increasingly complex and uncertain, the optimality gap of traditional deterministic methods is likely to grow and the value of AI-augmented decision-making will further increase. This thesis has shown that RL is a competitive methodology for UC with the ability to bridge this gap and provide substantial cost reductions and practical benefits in the operation of future power systems.


\section{Limitations and Further Work}

This thesis has shown that RL can be applied in combination with tree search to solve the UC problem and outperform traditional mathematical optimisation approaches. However, there remain further opportunities to improve and extend this approach and address the limitations of our research. 

Comparisons with a broader array of mathematical programming methods would be beneficial for further understanding the limitations and advantages of RL-based approaches relative to other optimisation methods. The performance improvement of guided tree search as compared with the deterministic MILP formulations used in this thesis was due in large part to superior management of uncertainties, evidenced by lower levels of load shedding. Stochastic formulations, although relatively expensive to solve, may produce superior solution quality to guided tree search by capturing uncertainty across multiple scenarios. Furthermore, by sampling directly from the environment, RL and guided tree search do not require the objective function or constraints to be linear, as in linear programming methods. The advantages of this may be significant in contexts where the objective function cannot be well-approximated with piecewise linear functions. Our problem setup and open-source environment described in Section \ref{ch1:env} provides a valuable test-bed for further comparisons of UC solution methods.

We found no deterioration of solution quality in scaling from 5--30 generators relative to MILP. However, the margins of improvement in the 100-generator case were markedly smaller: 0.1\% lower operating costs as compared with c. 1\% in smaller problem instances. Target entropy regularisation was effective in unifying policy training and guided tree search and improving training convergence, but practical challenges remain in order to stabilise performance on large problems. Further improvements to Guided IDA* and applications to larger power systems of greater than 100 generators could be achieved by a more optimised implementation of this algorithm and greater computing resources. Using the anytime algorithm Guided IDA* ensures that run times do not become impractical for larger systems, but further experiments are required to assess the impact of increasing numbers of generators on solution quality. Since the guided tree search framework is modular in both the RL algorithm and the search algorithm, there is also potential for further variations which exploit new RL and planning methods which could also improve solution quality. A more efficient implementation of Guided IDA* would also help address the challenge of comparing computational requirements of guided tree search and MILP approaches, discussed in Section \ref{ch1:experiments:milp_comparison}.


This thesis focused exclusively on the day-ahead UC problem as the most widely-studied of UC problem variants. There is growing interest in UC for intraday markets, whose prominence as a share of total power traded is increasing \cite{neuhoff2016intraday}. The real-time algorithm presented in Algorithm \ref{ch1:algo:solve_day_ahead} that is used to implement all of the guided tree search approaches for the UC problem is well-suited to scheduling tasks with a rolling decision horizon, such as in intraday settings or system balancing. Intraday scheduling has also been shown to improve the overall power system efficiency and security through rolling horizon stochastic optimisation \cite{tuohy2009unit}. Compared with MILP approaches, guided tree search would exhibit more apparent run-time advantages in these contexts.

There are several avenues for improvement to the environment presented in this thesis. Adherence with the OpenAI Gym API \cite{brockman2016openai} could broaden access to the wider RL research community and enable straightforward benchmarking of state-of-the-art RL algorithms. As demonstrated in Chapter \ref{ch3}, RL is well-suited to solving UC problem variants through modification of the environment. Further extensions could accelerate research into pertinent topics in power systems operation, including the introduction of storage assets, transmission constraints, distributed renewables generation and demand-side response.
